\documentclass[longauth]{aa} 
\usepackage[switch]{lineno}
\usepackage{graphicx}
\usepackage{arydshln}
\usepackage{subcaption}
\usepackage{mwe}
\usepackage{appendix}
\usepackage{hyperref}
\hypersetup{colorlinks=true,linkcolor=blue,filecolor=blue,urlcolor=black,citecolor=blue}

\usepackage[normalem]{ulem}
\usepackage[colorinlistoftodos]{todonotes}

\usepackage{txfonts}
\begin{document}

   \title{Investigation of the correlation patterns and the Compton dominance variability of Mrk~421 in 2017}

    \author{\small
    MAGIC Collaboration:
    V.~A.~Acciari\inst{1} \and
    S.~Ansoldi\inst{2} \and
    L.~A.~Antonelli\inst{3} \and
    A.~Arbet Engels\inst{4}$^{,\ddagger}$\thanks{Corresponding authors are A.~Arbet Engels and D.~Paneque (\email{contact.magic@mpp.mpg.de}). \newline $^{\dagger}$ Also member of the MAGIC Collaboration \newline $^{\ddagger}$ Also member of the FACT Collaboration} \and
    M.~Artero\inst{5} \and
    K.~Asano\inst{6} \and
    A.~Babi\'c\inst{8} \and
    A.~Baquero\inst{9}
% ...
}
    
    \authorrunning{MAGIC Collaboration et al.}
    
    \institute {Instituto de Astrof\'isica de Canarias and Dpto. de  Astrof\'isica, Universidad de La Laguna, 38200, La Laguna, Tenerife, Spain
    \and Universit\`a di Udine and INFN Trieste, I-33100 Udine, Italy
    \and National Institute for Astrophysics (INAF), I-00136 Rome, Italy
    \and ETH Z\"urich, CH-8093 Z\"urich, Switzerland
    \and Institut de F\'isica d'Altes Energies (IFAE), The Barcelona Institute of Science and Technology (BIST), E-08193 Bellaterra (Barcelona), Spain
    \and Japanese MAGIC Group: Institute for Cosmic Ray Research (ICRR), The University of Tokyo, Kashiwa, 277-8582 Chiba, Japan
    \and Technische Universit\"at Dortmund, D-44221 Dortmund, Germany
    \and Croatian MAGIC Group: University of Zagreb, Faculty of Electrical Engineering and Computing (FER), 10000 Zagreb, Croatia
    \and IPARCOS Institute and EMFTEL Department, Universidad Complutense de Madrid, E-28040 Madrid, Spain
% ...
}

   \date{Received ...; accepted ...}

   \abstract
  {} 
  {A detailed characterisation and theoretical interpretation of the broadband emission of the paradigmatic TeV blazar Mrk~421, with special focus on the multi-band flux correlations.}
  {The dataset has been collected through an extensive multiwavelength campaign organised between 2016 December and 2017 June. The instruments involved are MAGIC, FACT, \textit{Fermi}-LAT, \textit{Swift}, GASP-WEBT, OVRO, Medicina and Mets\"ahovi. Additionally, four deep exposures (several hours long) with simultaneous MAGIC and \textit{NuSTAR} observations allowed a precise measurement of the falling segments of the two spectral components.}
  {The very-high-energy (VHE; $E>100$\,GeV) gamma rays and X-rays are positively correlated at zero time lag, but the strength and characteristics of the correlation change substantially across the various energy bands probed. The VHE versus X-ray fluxes follow different patterns, partly due to substantial changes in the Compton dominance during a few days without a simultaneous increase in the X-ray flux (i.e. orphan gamma-ray activity). Studying the broadband spectral energy distribution (SED) during the days including \textit{NuSTAR} observations, we show that these changes can be explained within a one-zone leptonic model with a blob that increases its size over time. The peak frequency of the synchrotron bump varies by 2 orders of magnitude throughout the campaign. Our multi-band correlation study also hints at an anti-correlation between UV/optical and X-ray at a significance higher than $3\sigma$. A VHE flare observed on MJD~57788 (2017 February 4) shows gamma-ray variability on multi-hour timescales, with a factor 10 increase in the TeV flux but only a moderate increase in the keV flux. The related broadband SED is better described by a two-zone leptonic scenario rather than by a one-zone scenario. We find that the flare can be produced by the appearance of a compact second blob populated by high energetic electrons spanning a narrow range of Lorentz factors, from $\gamma'_{min}=2\times10^{4}$ to $\gamma'_{max}=6\times10^{5}$.}
   {}

   \keywords{Galaxies: active -- BL Lacertae objects: individual -- Mrk 421
               }

   \maketitle
%
\section{Introduction}
Blazars belong to the group of jetted active galactic nuclei (AGNs) and constitute the most populated class of sources in the extragalactic very-high-energy (VHE; $E>100$\,GeV) sky\footnote{http://tevcat2.uchicago.edu/}. They host a super massive black hole ($10^6-10^9$ solar masses) surrounded by an accretion disc and display a pair of relativistic plasma jets flowing in opposite directions producing non-thermal radiation. The jet's axis is oriented at a small angle with the observer's line of sight, which gives rise to strong relativistic beaming effects of the radiation. Blazars are commonly divided in two families, the Flat Spectrum Radio Quasars (FSRQ) and BL Lac objects, depending on their optical spectra \citep{1995PASP..107..803U}. FSRQ exhibit strong emission lines in the optical band, while BL Lac type objects are defined by very weak lines or an absence of such features.\par 

The spectral energy distribution (SED) of BL Lac objects is dominated by non-thermal radiation from the jet and typically shows two continuous components \citep{2017MNRAS.469..255G}. Based on polarisation and spectral studies, it is commonly accepted that the low-energy component, peaking in infrared to X-rays, originates from synchrotron radiation by relativistic electrons and/or positrons in a magnetic field. The high-energy component peaks in the GeV-TeV regime and its origin remains under debate. The common scenario involves electron inverse-Compton (IC) scattering off the synchrotron photons emitted by the same population of electrons \citep[see for example][]{1992ApJ...397L...5M, 1998ApJ...509..608T, 2004ApJ...601..151K}. Such models are labelled as synchrotron self-Compton (SSC) models. In some cases, external Compton models, in which an additional target photon field for IC scattering is introduced, are better suited to described the SED of BL Lacs \citep[e.g.,][]{Madejski_1999, 2005A&A...432..401G, 2013ApJ...768...54B} also showed the viability of . More complex scenarios invoking hadronic processes can also present a viable explanation \citep[see for example][]{1993A&A...269...67M, 2013ApJ...768...54B, 2015MNRAS.448..910C}. The peak frequency of the low-energy component is frequently used to classify BL Lac type objects in further categories \citep{1995ApJ...444..567P, 2010ApJ...716...30A}. Low-frequency BL Lacs (LBL) are defined by a synchrotron peak frequency $\nu_s<10^{14}$\,Hz and BL Lacs  with $\nu_s>10^{15}$\,Hz are dubbed as high-frequency BL Lacs (HBL). Intermediate-frequency BL Lacs (IBL) have  $10^{14}\text{\,Hz}<\nu_s<10^{15}$\,Hz.\par 

Markarian 421 (Mrk~421; RA=11\textsuperscript{h}4'27.31'', Dec=$38^\circ$12'31.8'', J2000) is a HBL at redshift of 0.031 \citep{1991rc3..book.....D}. Owing to its brightness and proximity, the source can be well-detected on short time scales ($\lesssim 1$\,day) from radio to VHE with current instruments. After its first detection at VHE by the Whipple 10-m Telescope \citep{1992Natur.358..477P}, numerous multiwavelength (MWL) campaigns have been organised to characterise the broadband emission during individual flares as well as on longer time scales. Similarly to other HBLs, the source shows flux and spectral variability across its entire SED. The variability is most prominent in the X-rays and VHE and was observed down to sub-hour time scale \citep{1996Natur.383..319G, 2008ApJ...677..906F}. Using observations over a 14-year time period, \citet{2014APh....54....1A} derived a typical flux above 400\,GeV of about 50\% that of Crab Nebula flux unit\footnote{For a given energy threshold, C.U. is defined as the integral flux of the Crab Nebula above the threshold energy. We used here the Crab Nebula spectrum from \cite{2016APh....72...76A}} (C.U.). During low activity, the VHE flux can be as low as 10\% C.U., while during strong flaring events, it can reach more than 10 C.U., as observed in an outburst in 2010 \citep{2020ApJ...890...97A} and 2013 \citep{2020ApJS..248...29A}.\par 

Several studies reported correlated variability between VHE and X-ray emissions independently from the flux level, which is in agreement with standard SSC models \citep[e.g.,][]{2007A&A...462...29G,2008ApJ...677..906F,2016A&A...593A..91A,2016ApJ...819..156B}. The correlation is often parametrised by a simple power law, $F_{VHE} \propto F_{X-ray}^x$, where $F_{VHE}$ and $F_{X-ray}$ are the VHE and X-ray flux, respectively. While the existence of the correlation is well established, the index $x$ may show significant temporal variability: observations revealed sub-linear ($x<1$) as well as more-than-quadratic ($x>2$) behaviours \citep{2004ApJ...601..759T,2008ApJ...677..906F,2016ApJ...819..156B}. In fact, quadratic or more-than-quadratic trends are typically found during high states. $x$ also displays a strong dependency on the exact selection of the spectral bands \citep{2005A&A...433..479K,2008ApJ...677..906F,2016ApJ...819..156B,2020ApJS..248...29A}. These results underline the complex spectral properties of blazars, but potentially provide insights into the physical processes driving the broadband variability \citep[e.g.,][]{2005A&A...433..479K}. There is currently a lack of detailed investigation of the VHE and X-ray correlation in different energy bands during non-flaring activity, which would provide additional constraints on the emission mechanisms. In order to address this task, sensitive measurements with a dense temporal and wide energy coverage are crucial.\par 

The simplest emission models of blazars (called one-zone models) assume a cospatial particle population responsible for the SED above the infrared ($\gtrsim10^{13}$\,Hz). In this scenario, a correlation between UV/optical and X-ray photons is generally expected. The observed synchrotron flux is proportional to the product $\delta^4 \, n'_e \, B'^2$, where $\delta$ is the Doppler factor, $n'_e$ the number of electrons and $B'$ the magnetic field (here and in the following primed quantities refer to quantities in the plasma reference frame). Any change in the latter parameters would simultaneously affect the UV/optical and X-ray emissions. However, only sporadic indications of correlated variability were reported up to now. \citet{2015A&A...576A.126A} observed a first indication of an anti-correlation during the year 2009. On the other hand, \citet{2017MNRAS.472.3789C} did not report any correlated variability based on long-term observations from 2007 to 2015. It should be noted that these two bands have very different temporal behaviours, the X-ray emission being much more variable, which renders the detection of a correlation challenging. \par 

In this work, we present results from a MWL campaign organised between 2016 December and 2017 June. In order to provide an optimal energy and temporal coverage, we coordinated observations between a large number of instruments covering the emission from radio to VHE. The Florian Goebel Major Atmospheric Gamma Imaging Cherenkov telescopes (MAGIC) and First G-APD Cherenkov Telescope (FACT) carried out observations in the VHE regime. VHE data are complemented by  observations from the Large Area Telescope (LAT) on board the \textit{Fermi Gamma-ray Space Telescope} (\textit{Fermi}-LAT). Regarding the UV and soft X-ray emission, we organised many observations with the \textit{Neil Gehrels Swift Observatory} (\textit{Swift}) satellite in order to obtain a deep temporal coverage. Together with the \textit{Swift} schedulers team, we coordinated many of these observations to happen simultaneously (or close in time) to the VHE gamma-ray observations performed with MAGIC and FACT, in order to be able to properly characterise the temporal evolution of the low- and high-energy SED bumps of Mrk~421.  Moreover, we take advantage of four pointings of the \textit{Nuclear Spectroscopic Telescope Array} (\textit{NuSTAR}) to obtain a precise characterisation of the hard X-ray (${\gtrsim}10$\,keV) emission. The four pointings were coordinated to take place strictly simultaneously to the other instruments' observations. \textit{NuSTAR} is currently the most sensitive instrument measuring in the hard X-ray regime, corresponding to the high-energy end of the synchrotron spectrum of Mrk~421 \citep{2011ApJ...736..131A}. Hence, \textit{NuSTAR} probes the emission from the most energetic particles located in the jet. We investigate in detail the VHE versus X-ray correlation over different spectral bands along the entire MWL campaign. The \textit{NuSTAR} data significantly widen the energy coverage when combined with \textit{Swift} data and bring additional constraints to the theoretical model parameters. Furthermore, we study the X-ray versus UV/optical correlated variability.


\section{Instruments and analysis}
\subsection{MAGIC}
Mrk~421 belongs to the group of targets that the MAGIC telescopes monitor on a regular basis. The MAGIC telescopes form a system of two 17-m diameter imaging atmospheric Cherenkov telescopes (IACTs). They are situated at an altitude of 2231\,m above sea level, on the Canary Island of La Palma at the Roque de los Muchachos Observatory. The integral sensitivity for point sources observations above 220\,GeV is ($0.66 \pm 0.03$)\% C.U. in 50\,hours \citep{2016APh....72...76A}.\par

The MAGIC dataset presented in this work covers a ${\approx}6$-month period from MJD~57727 (2016 December 5) until MJD~57892 (2017 May 19). We analysed the data using the standard analysis tools from the MAGIC Analysis and Reconstruction Software (MARS) package \citep{zanin2013, 2016APh....72...76A}. A significant fraction of the observations were carried out with an increased night sky background light contamination (due to the presence of the moon), which directly affects the response of the telescopes \citep{2017APh....94...29A}. Due to the varying observing conditions, the data are split into several subsets depending on the level of the moon light contamination. Following \citet{2017APh....94...29A}, the analysis is then performed by adopting Monte Carlo simulations tuned to match the observing conditions of the different data subsets. The source was observed with zenith angles ranging from $9^{\circ}$ to $70^{\circ}$. After data quality selection, ${\approx}70$\,hours of observations were gathered over a total of 48 nights. The light curves are computed in two energy bands: 0.2-1\,TeV and $>1$\,TeV.\par

The MAGIC spectra are fitted above 100\,GeV with a log-parabola spectral model that is defined as follows:
\begin{equation}
\label{eq:logparabola_MAGIC}
    \frac{dN}{dE} = f_0 \left(\frac{E}{E_0}\right)^{-\alpha-\beta \log{\left(\frac{E}{E_0}\right)}}
\end{equation}
where $f_0$ is the normalisation constant, $\alpha$ the photon index and $\beta$ the curvature parameter. The normalisation energy $E_0$ is fixed at 300\,GeV. A simple power-law function defined as $\frac{dN}{dE} = f_0 \left(\frac{E}{E_0}\right)^{-\alpha}$ is applied in case a log-parabola is not preferred at a significance above $3\sigma$ (based on a likelihood ratio test). The best-fit parameters obtained can be found in Table~\ref{tab:MAGIC_spectral_param} from Appendix~\ref{appendix_magic} together with the fluxes integrated in the two energy bins defined for the light curve. All parameters are evaluated after correcting the spectra for extragalactic background light (EBL) absorption effects using the model of \citet{2011MNRAS.410.2556D}.\par 

Four nights have a longer observing time compared to the majority of the observations, which typically last between 40 and 60\,minutes. Each of these deep exposures is accompanied by simultaneous \textit{NuSTAR} and \textit{Swift} pointings. They took place on MJD~57757, MJD~57785, MJD~57813 and MJD~57840 (2017 January 4, 2017 February 1, 2017 March 1 and 2017 March 28). The MAGIC exposure during these nights varied from 2\,hours to 6\,hours depending on the date. For each epoch, light curves with 30-minute time bins are produced to study the correlation with the X-ray emission (see Sect.~\ref{sect:VHE_xray}).

\subsection{FACT}
The First G-APD Cherenkov Telescope (FACT) is an imaging atmospheric Cherenkov telescope with a mirror area of 9.5\,m${^2}$ \citep{2013JInst...8P6008A, 2014JInst...9P0012B}. It is located close to the MAGIC telescopes, on the Canary Island of La Palma at the Roque de los Muchachos Observatory. FACT measures gamma rays from several hundreds of GeV to about 10\,TeV. Operations are performed fully remotely and in an automatic manner. FACT is also pioneering the use of silicon-based photosensors (SiPM aka  Geiger-mode Avalanche Photo Diodes or G-APDs), allowing a robust and stable performance \citep{2014JInst...9P0012B}. The observing strategy is specifically tuned to achieve an unbiased long-term monitoring of TeV-emitting blazars.\par 

In this work, Mrk~421 observations between MJD~57720 (2017 November 28) and MJD~57899 (2017 May 26) are presented. The data quality selection is performed based on the cosmic-ray rate \citep{2017ICRC...35..779H}. The effect of the zenith distance on the cosmic-ray rate is taken into account following \cite{2017ICRC...35..612M} and \cite{2019APh...111...72B}. A total of 279\,hours of good-quality data spread over 86 nights is obtained.\par 

The analysis is performed following \cite{2015arXiv150202582D} using the standard FACT analysis software described in \citet{2010apsp.conf..681B}. Using the excess rate from the Crab Nebula, which is a constant source at TeV energies, the excess rate is corrected for its dependence on the zenith distance and the trigger threshold. The analysis threshold of the light curve computed using Monte Carlo simulation is about 0.95\,TeV.\par

\subsection{\textit{Fermi}-LAT}
The LAT instrument is a pair-conversion telescope on board the \textit{Fermi} satellite \citep{2009ApJ...697.1071A,2012ApJS..203....4A}. It monitors the gamma-ray sky in the 20\,MeV to $>300$\,GeV energy range with an all-sky coverage on a ${\sim}3$\,hr timescale. The analysis for this work was performed using the unbinned-likelihood tools from the \texttt{FERMITOOLS} software\footnote{https://fermi.gsfc.nasa.gov/ssc/data/analysis/} v1.0.10. We used the instrument response function \texttt{P8R3\_SOURCE\_V2} and the diffuse background models\footnote{http://fermi.gsfc.nasa.gov/ssc/data/access/lat/\\BackgroundModels.html} \texttt{gll\_iem\_v07} and \texttt{iso\_P8R3\_SOURCE\_V2\_v1}.\par 

We select all \texttt{Source} class events between 0.2\,GeV and 300\,GeV in a region of interest (ROI) with a radius of $15^\circ$ around Mrk~421. The events with a zenith angle $>100^\circ$ are discarded so that the contribution from limb gamma rays is reduced. A first unbinned analysis is performed over the entire considered time range, between MJD~57720 and MJD~57918. The source model includes all sources in the ROI from the fourth Fermi-LAT source catalog \citep[4FGL;][]{2020ApJS..247...33A}. During the fit, the normalisation and spectral parameters of sources within a radius of $10^{\circ}$ of Mrk~421 are left free to vary, while the remaining sources have their parameters fixed to the 4FGL values. Mrk~421 is modelled with a log-parabolic spectral function, $ \frac{dN}{dE} = f_0 \left(\frac{E}{E_0}\right)^{-\alpha-\beta \log{\left(\frac{E}{E_0}\right)}}$ with $E_0=1286.47$\,GeV as in the 4FGL catalog. The normalisations of the background components are left free to vary. After this first fit, sources which result in a detection test statistic \citep[TS;][]{1996ApJ...461..396M} of less than $15$ are removed from the model.\par

Using the simplified model, light curves with a 3-day binning are built in the 0.2-2\,GeV and 2-300\,GeV bands. In each time bin, the normalisation and the index $\alpha$ of Mrk~421 is left free to vary. The curvature parameter $\beta$ is fixed to 0.02, the value obtained from the fit over the entire period and is similar to the 4FGL catalog. For sources within $10^\circ$, only the normalisation is left free, while sources further than $10^\circ$ have all their parameters fixed to their 4FGL values. In case the target is detected with TS<5 an upper limit at 95\% confidence level is quoted after fixing the spectral index at $\alpha = 1.78$, the average value of the entire campaign.

\subsection{\textit{NuSTAR}}

The \textit{NuSTAR} observations in 2017 were performed in 4 epochs between MJD~57757 and MJD~57840 with a cadence of roughly one month, and in coordination with the observations performed with \textit{Swift} and the MAGIC telescopes. See Appendix~\ref{NuSTAR_tables} and Appendix~\ref{sect:MAGIC_nustar_lc} for the specific dates and times, and its relation with the MAGIC, \textit{Swift}, and optical observations. The raw \textit{NuSTAR} data were processed following the description in \citet{2016ApJ...819..156B}, but using the updated software and calibration packages NuSTARDAS (version 1.7.1), HEASoft (version 6.21), and CALDB (version 20170222). We used strict event filtering to eliminate high background fluxes during South Atlantic Anomaly passages (using the flags \texttt{saamode=strict} and \texttt{tenacle=yes} for the \texttt{nupipeline} processing script). The resulting exposures for the single-night observations are in the range 16--25\,ks.
The source counts were selected from a circular extraction region with a radius of 120$\arcsec$, while the background was sampled from the same focal plane detector excluding a circular region with a radius of 180$\arcsec$ centred on Mrk\,421, which is a bright X-ray soure and hence clearly detected by \textit{NuSTAR}. We binned the spectra ensuring that each energy bin exceeds the signal-to-noise ratio of 3. We also verified that a different selection of data processing parameters does not affect the downstream analysis in any significant way.

In addition to the analysis of data for each of the 4 epochs, we also separated the data per orbit as well as into 30-minutes bins for which we have strictly simultaneous coverage at VHE gamma-ray energies with the MAGIC telescopes. We find significant variability in the flux and spectral shape in each of the epochs, as shown in detail in Appendix~\ref{NuSTAR_tables} and Appendix~\ref{sect:MAGIC_nustar_lc}. The spectral analysis was performed in \texttt{Xspec} \citep{1996ASPC..101...17A}. The cross-normalisation between the Focal Plane Module A and B was assumed to be a free parameter in spectral fitting. The distribution of the cross-normalisation factor over the various spectral fits is tight with an average of 0.99 and a standard deviation of 0.02, which is firmly within the expectations of \textit{NuSTAR} \citep{2015ApJS..220....8M}. In all cases we employed the log-parabolic model with a pivot energy of 1\,keV to describe the spectra, finding significantly non-zero curvature parameter $\beta$ in longer integrations (e.g. see Table~\ref{tab:Nustar_spectral_param_nustar_sim}). For short observations, owing to the lower photon statistics, in some time intervals $\beta$ is statistically consistent with zero, implying that the spectra are consistent with a power law in those cases (see Appendix~\ref{NuSTAR_tables}). We provide fluxes calculated based on this model in the 3--7\,keV and 7--30\,keV bands, as the source is not significantly detected above the background level at energies higher than 30\,keV in short time integrations. All uncertainties are given at the 68\,\% significance level.

\subsection{\textit{Swift}-BAT}

The BAT light curves are generated using the BAT survey data, which are collected continuously by the spacecraft and are binned into ${\sim}300$\,s \citep{Markwardt07}. The survey data are processed using the standard BAT pipeline, \texttt{batsurvey}\footnote{https://heasarc.gsfc.nasa.gov/ftools/caldb/help/batsurvey.html}, which is part of the HEASoft tools and produces count rate data at the source location for each snapshot image in eight energy bands: 14-20, 20-24, 24-35, 35-50, 50-75, 75-100, 100-150, and 150-195\,keV.

The count rate for each image is combined into an eight-band light curve file, and the light curve is further binned into the desired time ranges (1 day, 3 days, and 6 days) and energy band (15-50\,keV). When performing the binning, errors are calculated with standard error propagation method using the “BKGVAR” column in the light-curve data, which measures the background variation at the source location.\par

A BAT spectrum is created using the eight-band information produced by \texttt{batsurvey} around MJD~57788.7, which corresponds to the peak activity in the \textit{Swift}-BAT (see Sect.~\ref{sect:MWL_lc} and Sect.~\ref{flare_modelling}). For this, we only use data when the source is on-axis relative to the BAT detector plane, in order to avoid complications due to different instrumental sensitivity at different incident angles. Once a spectrum is created, the corresponding instrumental response file is generated using the standard BAT tool, \texttt{batdrmgen}\footnote{https://heasarc.gsfc.nasa.gov/lheasoft/ftools/headas/batdrmgen.html}.

\subsection{\textit{Swift}-XRT}

The \textit{Swift}  X-ray Telescope \citep[XRT;][]{2005SSRv..120..165B} 
was used to characterise the emission from Mrk\,421 in the energy range from 0.3\,keV to 10\,keV. The \textit{Swift}-XRT observations were performed in the windowed timing (WT) readout mode, and the data were processed using the XRTDAS software package (v.3.5.0) developed by the ASI Space Science Data Center (SSDC), and released by the NASA High Energy Astrophysics Archive Research Center (HEASARC) in the HEASoft package (v.6.26.1). The calibration files from \textit{Swift}/XRT CALDB (version 20190910) were used within the  \texttt{xrtpipeline} to calibrate and clean the events.\par 

The X-ray spectrum from each observation was extracted from the summed cleaned event file. Events for the spectral analysis were selected within a circle of 20-pixel ($\sim$46 arcsec) radius, which encloses about 90 per cent of the point-spread function (PSF), centred at the source position. The background was extracted from a nearby circular region of 40-pixel radius. The ancillary response files (ARFs) were generated with the \texttt{xrtmkarf} task applying corrections for PSF losses and CCD defects using the cumulative exposure map.\par

The $0.3-10$\,keV source spectra were binned using the \texttt{grppha} task to ensure a minimum of 20 counts per bin, and then were were modeled in XSPEC using power-law and log-parabola models (with a pivot energy fixed at 1\,keV) that include a photoelectric absorption by a fixed column density estimated to be $N_{\rm H}=1.92\times10^{20}$\,cm$^{-2}$ \citep[][]{2005A&A...440..775K}. The log-parabola model typically fits the data better than the power-law model, and hence it was used to compute the X-ray fluxes in the energy bands $0.3-2$\,keV, and $2-10$\,keV, which are reported in Appendix~\ref{SwiftResults}. The fluxes were also computed in the $3-7$\,keV range in order to match the low-energy band of \textit{NuSTAR} and include them in the VHE versus X-ray correlation study in Sect.~\ref{sect:VHE_xray}. 

\subsection{\textit{Swift}-UVOT}
We selected the observations of the \textit{Swift} UV and Optical Telescope \citep[UVOT,][]{2005SSRv..120...95R} between 2016 December and 2017 June acquired in the UV filters W1, M2 and W2. We performed photometry on total exposures of each observations available in the official archive with the same apertures for source counts (the standard with 5$\arcsec$ radius) and background (mostly two circles of 16.5$\arcsec$ radii off the source) estimation. We executed the photometry task in the official software version included in the HEAsoft 6.23 package, by the HEASARC, and then applied the official calibrations \citep{2011AIPC.1358..373B} included in the more recent CALDB release (20201026). The source is on ``ghost wings'' \citep{2006PASP..118...37L} from the near star 51 UMa in most of the observations, so we checked the wing positions and the astrometry very carefully, excluding stray lights and support structure shadows. This quality control removed 4 UVOT images, leading to a final sample of 95 good-quality observations. The fluxes were dereddened considering a mean Galactic $E(B-V)$ value of 0.0123 mag \citep{2011ApJ...737..103S} and using the Galactic interstellar extinction curve from \citet{1999PASP..111...63F}.

\subsection{Optical}

In this paper, we only use photometry in the Cousins' R band. All the data were provided by the GLAST-AGILE Support Program \citep[GASP, e.g.][]{villata2008, villata2009, 2017MNRAS.472.3789C} of the Whole Earth Blazar Telescope\footnote{http://www.oato.inaf.it/blazars/webt/} \citep[WEBT, e.g.][]{villata2002, villata2006, rai2007,raiteri2017}. The observations cover the period from MJD~57716 to MJD~57917, and were provided by 24 telescopes in the following 22 observatories spread over the Northern Hemisphere: Abastumani (Georgia), ARIES (India), AstroCamp (Spain), Belogradchik (Bulgaria), Burke-Gaffney (Canada), Calar Alto\footnote{Calar Alto data was acquired as part of the MAPCAT project: \url{http://www.iaa.es/~iagudo/\_iagudo/MAPCAT.html}} (Spain), Crimean (Russia), Haleakala (US), Hans Haffner (Germany), KVA observatory (Spain), Lulin (Taiwan), McDonald (US), New Mexico Skies (US), Perkins (US), Rozhen (Bulgaria), Siena (Italy), Sirio (Italy), St.~Petersburg (Russia), Teide (Spain), Tijarafe (Spain), Vidojevica (Serbia), West Mountain (US).\par

The data reduction was performed according to standard prescriptions. To minimise offsets among the different data sets due to the presence of the host galaxy, a common aperture radius of 7.5 arcsec was used. Following \citet{nilsson2007}, we subtracted a host galaxy contribution of 8.2 mJy from the observed flux density and then corrected for a Galactic extinction of 0.033 mag according to the NASA/IPAC Extragalactic Database\footnote{https://ned.ipac.caltech.edu/} (NED).

\subsection{Radio}

The emission at radio frequencies was characterised with the single-dish telescopes at the Mets{\"a}hovi Radio Observatory, operating at 37 GHz, at the Owens Valley Radio Observatory (OVRO), that operates at 15 GHz, and the Medicina radio telescope at both 8~GHz and 24~GHz. The data from OVRO were retrieved from the instrument team web page\footnote{https://sites.astro.caltech.edu/ovroblazars/}, while the data from  Mets{\"a}hovi and Medicina were analysed following the prescription from \citet{1998A&AS..132..305T} and \cite{2020MNRAS.492.2807G}, and provided by the instrument teams specifically for this study. For these three single-dish radio instruments, Mrk\,421 is a point source and thus the measurements represent an integration of the full source extension. The size of the radio emitting region is expected to be larger than that of the region of the jet that dominates the X-ray and gamma-ray emission, known to vary on much shorter timescales than the radio emission.

%

\section{Multiwavelength light curves and spectral behaviours}
\label{sect:MWL_lc}
\begin{figure*}
   \centering
   \includegraphics[width=2.0\columnwidth]{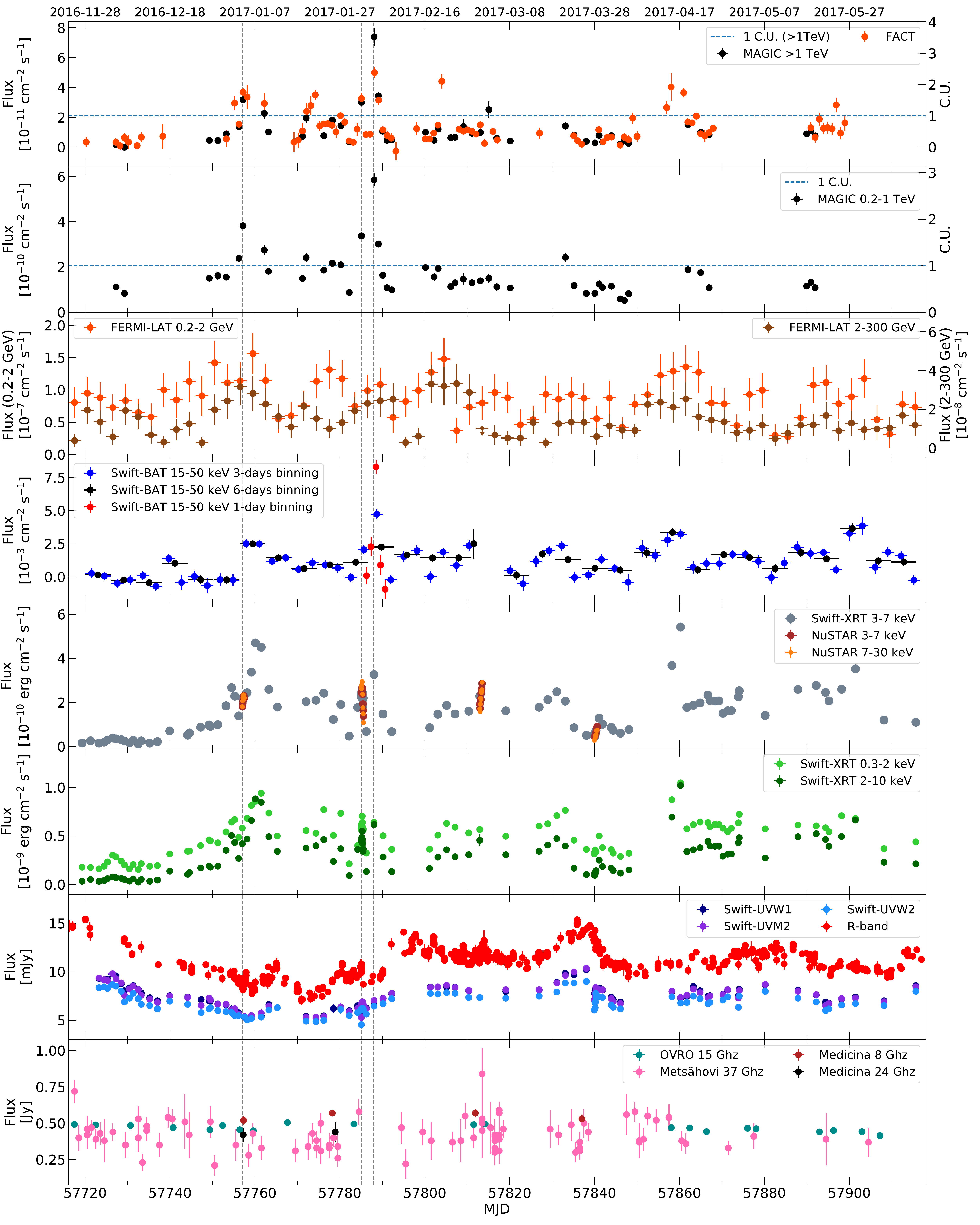}
   \caption{MWL light curves between MJD~57720 and MJD~57918. From top to bottom: MAGIC (0.2-1\,TeV \& $> 1$\,TeV) and FACT ($> 0.95$\,TeV), \textit{Fermi}-LAT (0.2-2\,GeV \& 2-300\,GeV), \textit{Swift}-BAT (15-50\,keV), \textit{NuSTAR} and \textit{Swift}-XRT (3-7\,keV \& 7-30\,keV), \textit{Swift}-XRT (0.3-2\,keV \& 2-10\,keV), UV/optical (\textit{Swift}-UVW1/UVM2/UVW2 \& R-band), OVRO (15\,GHz), Metsähovi (37\,GHz) and Medicina (8\,GHz and 24\,GHz). The dotted blue horizontal line in the first two panels from the top shows the 1\,C.U. flux in the $>1$\,TeV \& 0.2-1\,TeV bands, respectively. The \textit{Fermi}-LAT data have a 3-day binning. The \textit{Swift}-BAT light curve is computed in 3-day and 6-day binnings, as well as 1-day binning around the flare. The vertical black dashed lines highlight the dates that show an enhanced VHE activity and appear as outliers in the VHE versus X-ray correlation plots (see Sect.~\ref{sect:VHE_xray}).}
   \label{MWL}%
\end{figure*}

In Fig.~\ref{MWL}, we show the MWL light curves from radio to VHE between MJD~57716 and MJD~57918.\par

\subsection{Gamma rays}
\label{gamma_ray_lc_description}
\begin{figure}
   \centering
   \includegraphics[width=1.\columnwidth]{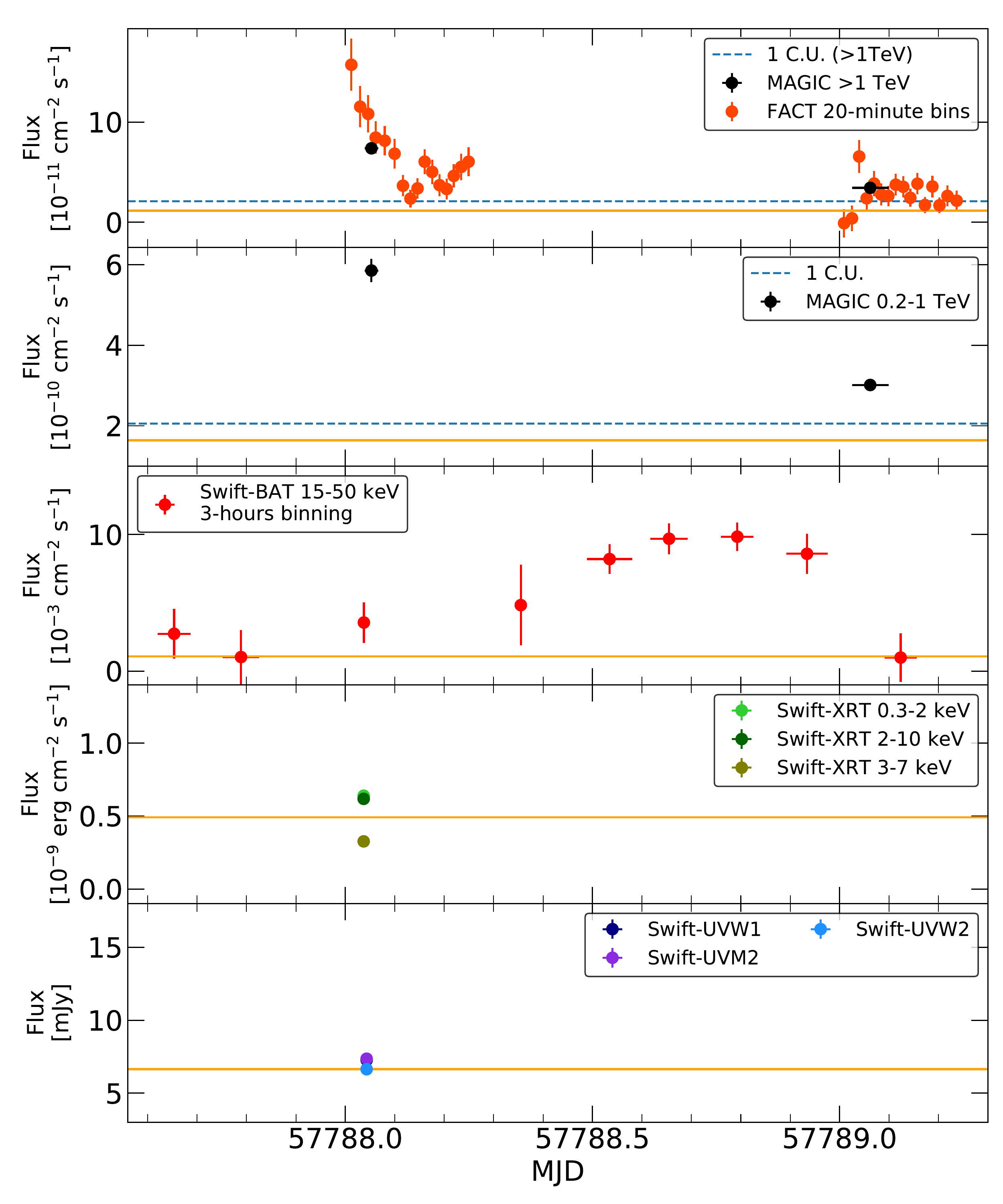}
   \caption{Zoom on the MWL light curves around the VHE flare on MJD~57788. The two first panels from the top show the FACT ($>0.95$\,TeV) and the MAGIC ($> 1$\,TeV \& 0.2-1\,TeV ) light curves. The FACT fluxes are computed on 20-minute binning while the MAGIC light curve is nightly averaged. On MJD~57788 and MJD~57789 the MAGIC exposures are ${\sim}$40\,min and ${\sim}$100\,min respectively, and are depicted with the width of the horizontal bars of the markers. The third panel from the top is the \textit{Swift}-BAT light curve with 3-hour binning. The two panels from the bottom show the \textit{Swift}-XRT (0.3-2\,keV \& 3-7\,keV \& 2-10\,keV) and \textit{Swift}-UVW1/UVM2/UVW2 fluxes. The horizontal orange lines show the average flux over the entire campaign in each energy regime. In the top panel, the orange line is the average from the MAGIC >1\,TeV light curve. In the two lowest panels, the orange line depicts the average in the \textit{Swift}-XRT 0.3-2\,keV and \textit{Swift}-UVW2 bands.}
    \label{flare_night_LC}%
\end{figure}

In the first two panels from the top, the MAGIC fluxes (0.2-1\,TeV \& >1\,TeV) are presented. For both energy bands the corresponding C.U. is depicted with a horizontal blue dotted line. The FACT light curve, whose analysis has an energy threshold of about 0.95\,TeV, is also shown in the top panel. From MJD~57755 to MJD~57790, the source is more active than usual. The flux is higher than 1\,C.U. for several nights, while \citet{2014APh....54....1A} derived a time-averaged flux over 14\,years of ${\approx}$0.5 C.U.. In particular, a bright flare with a duration of about a day is visible on MJD~57788. On that night, the averaged flux measured by MAGIC is ${\approx}3.5$\,C.U., while the FACT nightly flux is close to $2$\,C.U.. This difference is due to the longer exposure of the FACT observation combined with intra-night variability. The intra-night variability is illustrated in the FACT 20-minute binned light curve shown in Fig.~\ref{flare_night_LC}. The flux decays along the night with a halving time of about 1\,hour. A peak activity of ${\approx}7$\,C.U. is measured at the beginning of the observation. Interestingly, the VHE flare does not seem to be accompanied by a comparable burst in the \textit{Swift}-XRT pass band (0.3-10\,keV) as is typically observed for Mrk~421 \citep{2008ApJ...677..906F, 2015A&A...578A..22A,2020ApJ...890...97A,2020ApJS..248...29A}. During the VHE flare, the \mbox{0.3-2\,keV} flux remains close to the average of the campaign and the 2-10\,keV flux is only twice the average. As will be discussed later, this particular flare appears as an outlier in the VHE versus \textit{Swift}-XRT correlation. The \textit{Swift}-BAT daily light curve (red points in Fig.~\ref{MWL}) reveals a prominent flux increase close to the VHE flare. However, by using a finer binning (3-hours), Fig.~\ref{flare_night_LC} shows that the \textit{Swift}-BAT flux simultaneous to the VHE observations is near the typical state. The 15-50\,keV flux exhibits a clear increase by a factor $\sim 3$ only a few hours after the MAGIC and FACT observations. After MJD~57789, both the 15-50\,keV and VHE fluxes show a drop. Unfortunately, no VHE data are available during the period with the highest state measured by \textit{Swift}-BAT. \par

After MJD~57790, the VHE flux is mostly between ${\approx}$0.3 and ${\approx}$0.6 C.U.. The FACT light curve shows nevertheless that during the last 50 days of the campaign, the TeV flux was persistently higher than 0.6 C.U., including several nights with fluxes above 1 C.U..\par

The MAGIC observations unveil spectral variability on ${\sim}$day timescales. The index $\alpha$ lies between $2.80$ and $1.95$ depending on the night (see Table~\ref{tab:MAGIC_spectral_param}). The hardest spectrum coincides with the brightest VHE flare on MJD~57788 and the spectrum is best described by a log-parabola shape with $\alpha=1.95 \pm 0.04$ and $\beta=0.19 \pm 0.05$.\par

\begin{figure}
    \centering
    \begin{subfigure}[b]{0.497\textwidth}
    \includegraphics[width=1.\columnwidth]{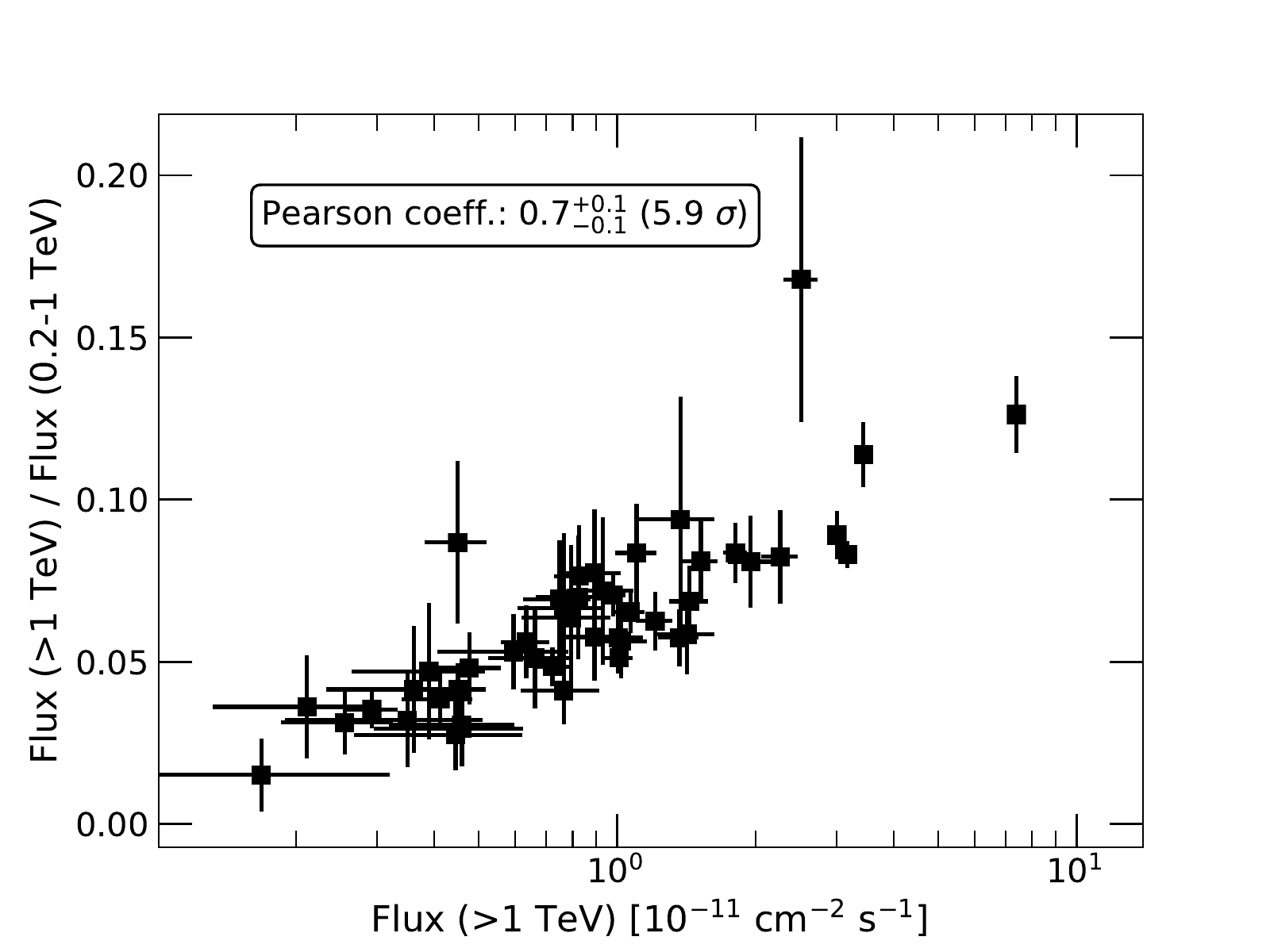}
    \end{subfigure}
    \centering
    \begin{subfigure}[b]{0.497\textwidth}
    \includegraphics[width=1.\columnwidth]{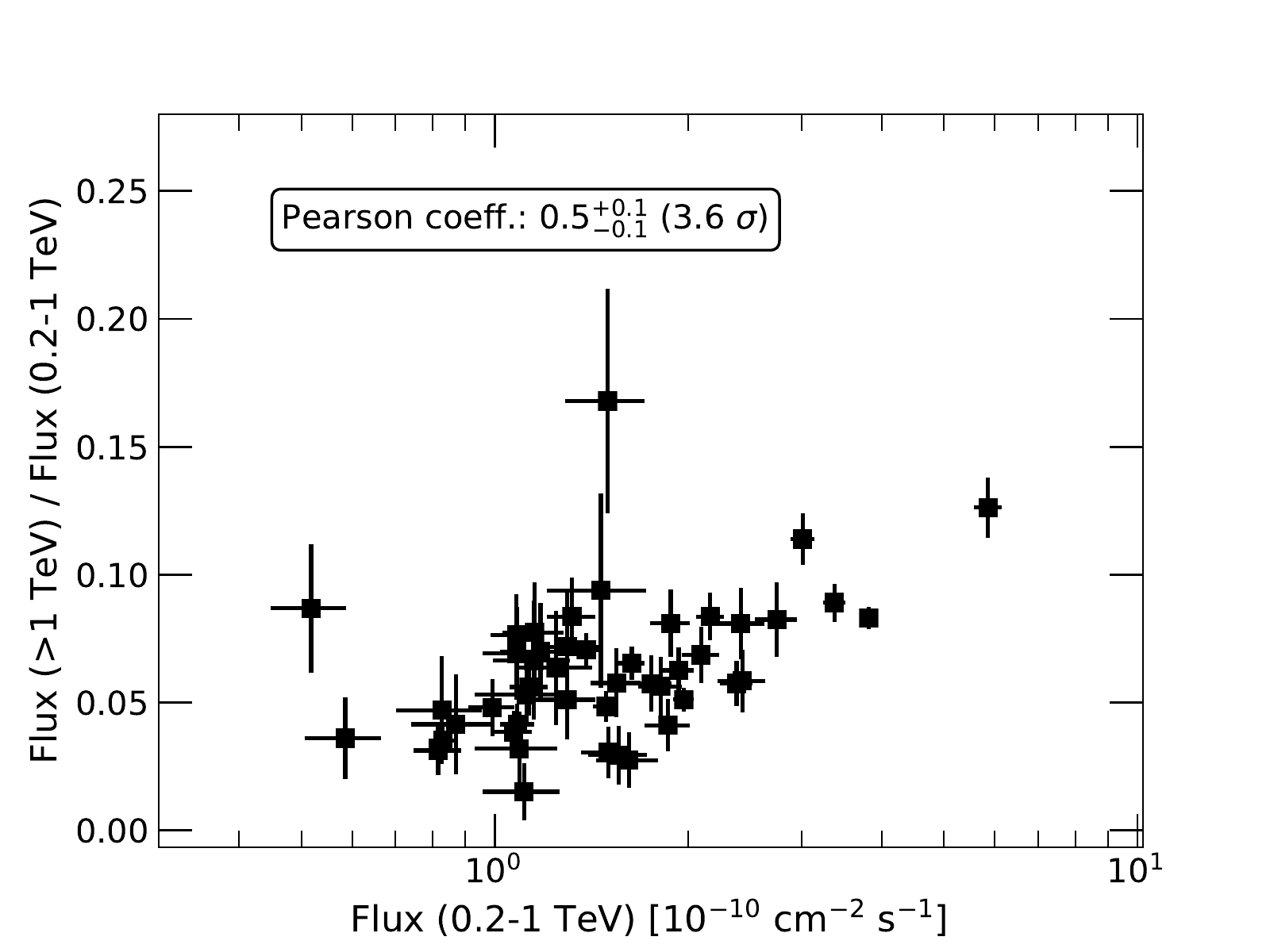}
    \end{subfigure}
    \caption{Hardness ratio ($F_{>1\text{\,TeV}}/F_{0.2-1\text{\,TeV}}$) versus $F_{>1\text{\,TeV}}$ (top figure) and versus $F_{0.2-1\text{\,TeV}}$ (lower figure) from the MAGIC measurements. The Pearson's coefficients and the corresponding significance \citep[following the prescription of][]{press2007numerical} are given in both plots.}
    \label{hardness_ratio_magic}%
\end{figure}

Fig.~\ref{hardness_ratio_magic} shows the hardness ratio (defined as the ratio between the $>1$\,TeV and the $0.2-1$\,TeV fluxes) obtained from the MAGIC observations. A clear harder-when-brighter behaviour is observed. When considering the 0.2-1\,TeV flux, the Pearson's coefficient is $0.5\pm0.1$ and the correlation significance, computed following the prescription of \citet{press2007numerical}, is $3.6\sigma$. When using the $>1$\,TeV flux, this behaviour is even more evident. The Pearson's coefficient is $0.7\pm0.1$ with a corresponding correlation significance of $5.6\sigma$. \citet{2020arXiv201201348M} also found a stronger harder-when-brighter behaviour in the $>1$\,TeV band during the years 2015 and 2016, when Mrk\,421 showed very low X-ray and VHE gamma-ray activity. However, unlike the 2015-2016 data, the data from 2017 do not show any saturation at the highest VHE fluxes.

The \textit{Fermi}-LAT light curves show no evidence for strong flaring episodes and have moderate flux variability in comparison to that observed at VHE. In the $0.2-2$\,GeV and the $2-300$\,GeV range, the fluxes are ${\sim}10^{-7}$\,cm$^{-2}$\,s$^{-1}$ and  ${\sim}10^{-8}$\,cm$^{-2}$\,s$^{-1}$ respectively, which are close to the quiescent state described in \citet{2011ApJ...736..131A}.\par

\subsection{X-ray}
\label{sec:X-ray}

In the fifth panel from the top of Fig.~\ref{MWL}, we show the light curves obtained during the four \textit{NuSTAR} observations (MJD~57757, MJD~57785, MJD~57813 and MJD~57840). The observations are simultaneous to MAGIC and \textit{Swift}. The data are binned orbit-wise, and the light curves are computed in two energy ranges, 3-7\,keV and 7-30\,keV. The first 3 observations show on average a relatively similar flux state, while during the last observation the flux is roughly 4 times lower. A flux variation by a factor $\approx$2-3 is present in each epoch, except on MJD~57757, where the flux varies only by about $30\%$ around $2 \times 10^{-10}$\,erg\,cm$^{-2}$\,s$^{-1}$. We estimate the flux doubling/halving time $t_{1/2}$ in the 7-30\,keV band (which shows the largest variability) using the prescription of \citet{Zhang_1999}. $t_{1/2}$ varies between 4\,hrs and 11\,hrs. It is on the night of MJD~57785 that the source shows the shortest variability. Such variability time scales are similar to those derived by \citet{2016ApJ...819..156B}. More details can be found in Appendix~\ref{NuSTAR_tables}, where the orbit-wise fluxes along with the exact values of $t_{1/2}$ are presented.\par 

The orbit-wise spectra are well described with a log-parabolic model and the results are also listed in Appendix~\ref{NuSTAR_tables}. Similarly to the \textit{Swift}-XRT spectral behaviour, the curvature parameter $\beta$ is not significantly dependent on the flux. In fact, one notes that the curvature on MJD~57740 is comparable to the rest of the observations despite a flux that is about 4 times lower.\par 

\begin{figure}[h!]
   \centering
   \includegraphics[width=1.\columnwidth]{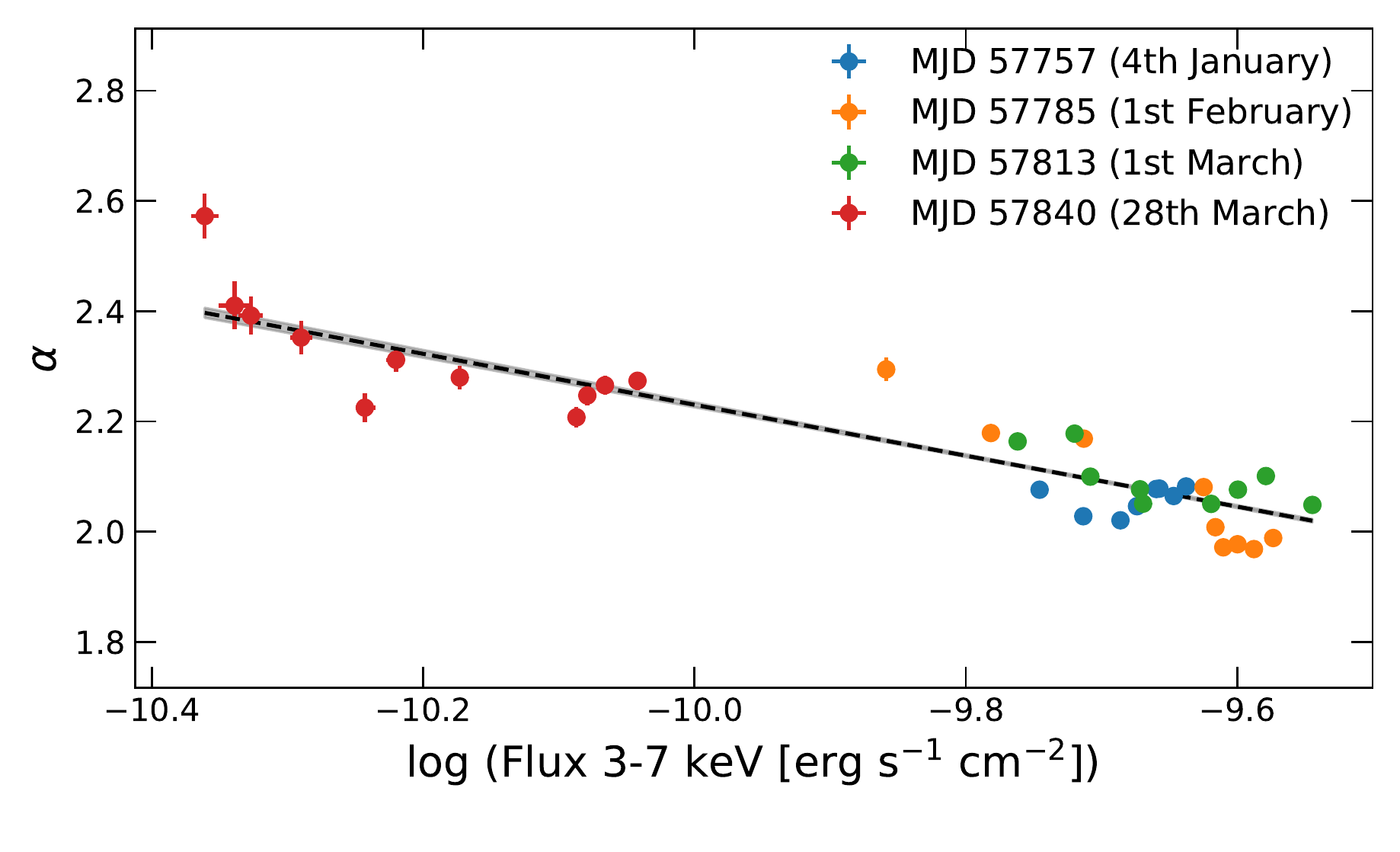}
   \caption{Log-parabola photon index $\alpha$ versus the 3-7\,keV flux of the orbit-wise \textit{NuSTAR} observations. $\alpha$ is fitted after fixing spectral curvature $\beta=0.22$ in the log-parabolic model. Each day is plotted with a different colour. The black line represents a linear fit, while the grey area is its uncertainty. The resulting slope is $-0.46 \pm 0.01$.}
    \label{NuSTAR_log_par}%
\end{figure}

Given the low variability in $\beta$, the spectral hardness versus the flux is studied by performing a second series of fits after fixing $\beta$ to $0.22$, which is the mean value of the orbit-wise spectra. The results are shown in Fig.~\ref{NuSTAR_log_par}. The low state on MJD~57840 is characterised by the softest spectra with $\alpha\approx2.3-2.6$. MJD~57757 and MJD~57813 show on average a very similar behaviour with $\alpha \approx 2.0-2.1$. It is on MJD~57785 that the hardness is the strongest and $\alpha \lesssim 2.0$ for most of the orbits. A linear fit to $\alpha$ versus the 3-7\,keV flux gives a slope of $-0.46 \pm 0.01$. This result is in good agreement with \cite{2016ApJ...819..156B}.

The X-ray observations from \textit{Swift}-XRT (sixth panel from the top of Fig.~\ref{MWL}) display a large variability amplitude. The fluxes lie between $F_{0.3-2\mathrm{keV}} \approx 2 \times 10^{-10}$\,erg\,cm$^{-2}$\,s$^{-1}$ and $F_{0.3-2\mathrm{keV}}\approx10^{-9}$\,erg\,cm$^{-2}$\,s$^{-1}$ in the 0.3-2\,keV energy range. In the 2-10\,keV band, they vary from $F_{2-10\mathrm{keV}} \approx 3 \times 10^{-11}$\,erg\,cm$^{-2}$\,s$^{-1}$ to $F_{2-10\mathrm{keV}} \approx 10^{-9}$\,erg\,cm$^{-2}$\,s$^{-1}$, which represents a change by more than a factor 30. The highest X-ray state is registered on MJD~57860 for all energy bands (0.3-2\,keV, 3-7\,keV and 2-10\,keV). A second peak flux is also visible in each band around MJD~57760, after a quasi monotonic increase over more than 40 days (from ${\approx}$MJD~57720 to ${\approx}$MJD~57760). An indication of an enhanced flux in the \textit{Swift}-BAT 3-day binned light curve is visible during these high-activity periods and the flux is about 3 times the campaign average. Unfortunately, no simultaneous VHE observations are available. We note that the flux level observed during those time periods remains moderately high compared to previous published works on Mrk~421 flares (see \citet{2019ApJ...877...26H} for a 12-year study of the Mrk~421 X-ray flux).\par

\begin{figure}
   \centering
   \includegraphics[width=1.\columnwidth]{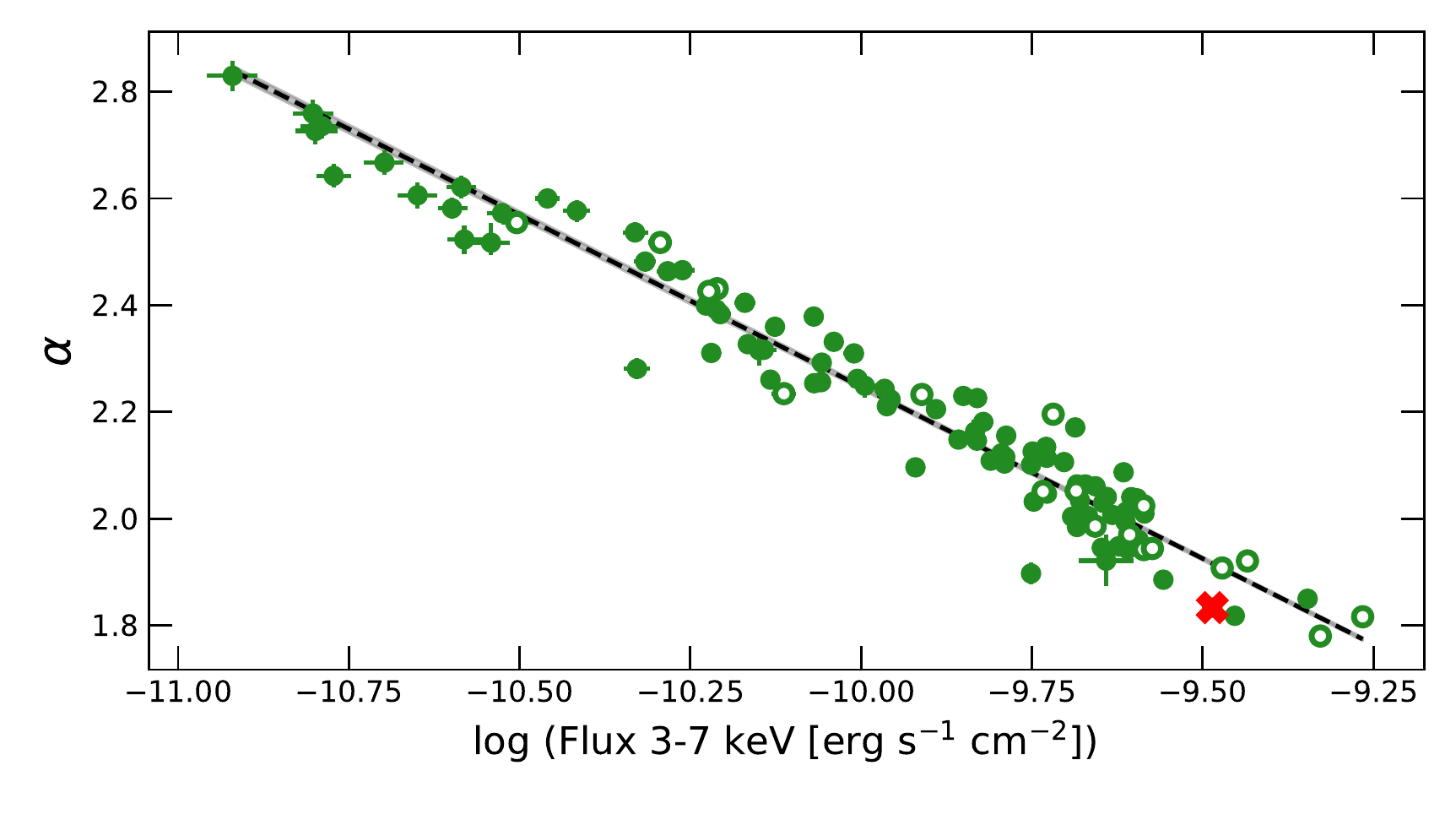}
   \caption{Log-parabola photon index $\alpha$ versus the 3-7\,keV flux from the \textit{Swift}-XRT observations. The log-parabolic fits are performed after fixing the spectral curvature $\beta=0.16$. The red cross represents the flare on MJD~57788 seen at VHE energies. The black line represent a linear fit, while the grey area (hardly visible in the plot) is its uncertainty. The slope of $\alpha$ versus the 3-7\,keV flux is $-0.64 \pm 0.01$. Hollow markers depict fits with a p-value below $5\times10^{-2}$, which are not considered in the linear fit.}
    \label{SwiftXRT_log_par}%
\end{figure}

Table~\ref{tab:swift_spectral_param} lists for each \textit{Swift}-XRT observation the spectral parameters of the best-fit log-parabolic and power-law models together with the fluxes in the 0.3-2\,keV, 3-7\,keV and 2-10\,keV bands. For the majority of them, the log-parabolic model provides a better description. The curvature parameter $\beta$ manifests a very weak dependence on the integral flux, differently from the index $\alpha$. Therefore, in order to study the spectral hardness as function of the flux, each observation is fitted again with a log-parabolic model with $\beta$ fixed to 0.16, which is the mean value of the dataset. This procedure removes the correlation between the two parameters of the model, which allows a more illustrative description of the evolution of the hardness versus flux to be obtained. The resulting $\alpha$ versus $F_{3-7\mathrm{keV}}$ is shown in Fig.~\ref{SwiftXRT_log_par}. A large variation of $\alpha$ is visible, ranging from ${\approx}1.8$ to ${\approx}2.8$, with a clear harder-when-brighter behaviour, as is typical in Mrk~421 \citep[][]{2015A&A...576A.126A, 2020arXiv201201348M} and HBLs in general \citep[][]{1998ApJ...492L..17P,2004ApJ...601..151K,2020MNRAS.496.3912M}. A linear fit results in a slope of $-0.64 \pm 0.01$, and corroborates the harder-when-brighter behaviour visible in the \textit{NuSTAR} data. For the linear fit we ignore the spectral fits resulting in a p-value\footnote{
The p-values are computed from the resulting $\chi^2$ and the corresponding degrees of freedom (dof).} below $5\times10^{-2}$ in order to remove the few nights (only 18 out of 107) that are not well described by the spectral model. These nights are depicted with hollow markers in Fig.~\ref{SwiftXRT_log_par}. One can see that they follow closely the general trend and do not show any particular trend or clustering. Thus, ignoring these measurements does not introduce any significant bias to our results. \par 

The observation corresponding to the VHE flare (MJD~57788) is characterised by a hard X-ray spectrum despite a moderate flux value ($F_{0.3-2\mathrm{keV}}$ and $F_{2-10\mathrm{keV}}$). The latter is shown as a red maker in Fig.~\ref{SwiftXRT_log_par} and the corresponding $\alpha$ is around 1.8, which is among the hardest spectra measured during the campaign.\par 

In the 15-50\,keV band, the \textit{Swift}-BAT observations do not reveal any particular flaring activity besides the one around MJD~57788 that is already discussed in Sect.~\ref{gamma_ray_lc_description}. Although being strongly variable (see Sect.~\ref{f_var_section}), the average flux is close to the value derived between 2008 and 2010 by \cite{2011ApJ...736..131A} when Mrk~421 showed quiescent activity. Using 3-day and 6-day binning, the average flux is at the level of ${\sim}10^{-3}$\,cm$^{-2}$\,s$^{-1}$.

\subsection{UV/optical}
The UV fluxes (from \textit{Swift}-UVOT) and the R-band fluxes (from GASP-WEBT) are shown in the seventh panel of Fig.~\ref{MWL}. They follow a very similar temporal evolution, which is expected given their proximity in energy. Interestingly, they show a continuous flux decay from ${\approx}$MJD~57720 to ${\approx}$MJD~57760, contrary to the almost monotonic increase in the X-rays during the same period. In addition, the low X-ray activity visible around MJD~57840 is accompanied with high fluxes in both the UV and R-band. This suggests an anti-correlation between X-ray and UV/optical over the campaign. This latter characteristic is investigated in Sect.~\ref{sect:synch_peak}.

\subsection{Radio}
The light curves from OVRO, Medicina and Mets{\"a}hovi in the bottom panel unveil no strong variability nor flaring episode. The average flux levels are ${\sim}0.5$\,Jy at 8\,GHz and 15\,GHz (Medicina and OVRO) and ${\sim}0.4$\,Jy at 24\,GHz and 37\,GHz (Medicina and Mets{\"a}hovi), respectively. Based on the ${\sim}5$-year OVRO and Metsähovi light curves presented in \citet{2015MNRAS.448.3121H}, this denotes a typical state during non-flaring activity.

\subsection{Multiwavelength variability}
\label{f_var_section} 

We study the broadband variability based on the \textit{fractional variability}, $F_{var}$, defined in \citet{2003MNRAS.345.1271V}. The uncertainty is computed following the strategy from \citet{2008MNRAS.389.1427P} and the implementation described in \citet{2015A&A...573A..50A}. $F_{var}$ quantifies the variance of the flux normalised to the mean value after subtracting the additional variance caused by the measurement uncertainties. $F_{var}$ is naturally affected by the instrument sensitivities, the binning and the flux sampling. Therefore, great care must be taken when comparing results from different telescopes. We refer the reader to \citet{2015A&A...576A.126A} and \citet{galaxies7020062} for a detailed study of the caveats inherent to the \textit{fractional variability} method.\par 

\begin{figure}[h!]
   \centering
   \includegraphics[width=1\columnwidth]{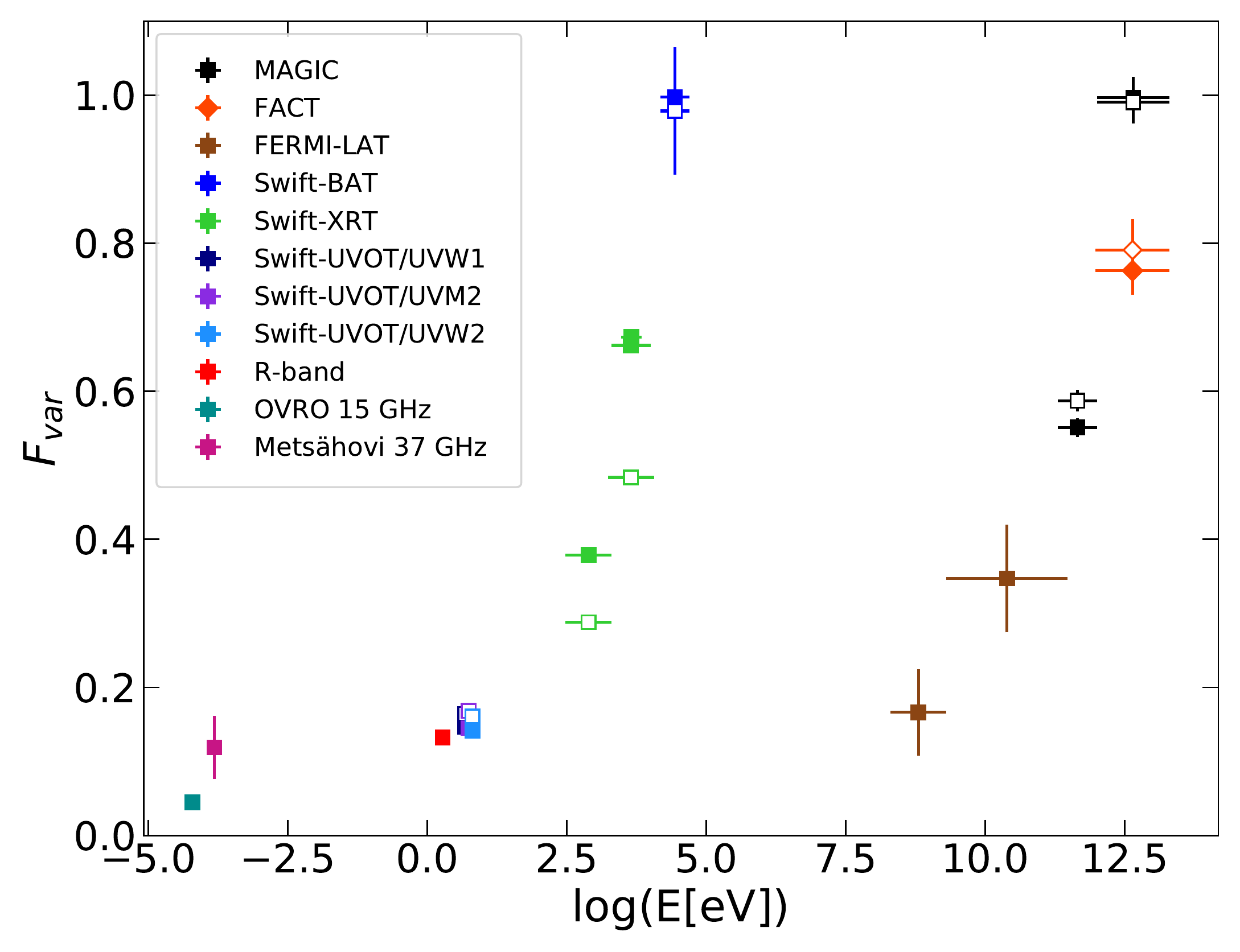}
   \caption{\textit{Fractional variability} $F_{var}$ obtained from the light curves shown in Fig.~\ref{MWL}. MAGIC, FACT, \textit{Swift}-XRT, \textit{Swift}-UVOT, R-band and radio fluxes are nightly binned. \textit{Fermi}-LAT and \textit{Swift}-BAT fluxes have a 3-day binning. Results from each instrument are plotted in different colours. The filled markers include all data. The hollow markers include VHE and \textit{Swift} data lying within a time window of 4\,hours from each other.} 
    \label{Frac_var}%
\end{figure}

The $F_{var}$ values are shown in Fig.~\ref{Frac_var} for each of the energy bands. They are computed using a nightly binning for the MAGIC, FACT, \textit{Swift}-XRT, \textit{Swift}-UVOT, R-band and radio light curves. Because of the limited sensitivity to detect Mrk\,421 on timescales of one day, for \textit{Fermi}-LAT and \textit{Swift}-BAT, we adopted a 3-day binning over which to integrate the data and compute the fluxes. Solid markers include all data from Fig.~\ref{MWL}. A discrepancy between the MAGIC (>1\,TeV) and FACT $F_{var}$ is visible and is explained by the different nightly averaged flux measured during the flare on MJD~57788 due to the different integration time, as previously mentioned. When the day of the flare is ignored, the $F_{var}$ values are fully compatible. All of the MAGIC/FACT/\textit{Swift}-XRT/BAT/UVOT measurements that are separated from one another by less than 4\,hours are considered and are shown with hollow markers. Here again, the \textit{Swift}-BAT fluxes are computed with a 3-day binning.\par

\begin{figure}[h!]
   \centering
   \includegraphics[width=1\columnwidth]{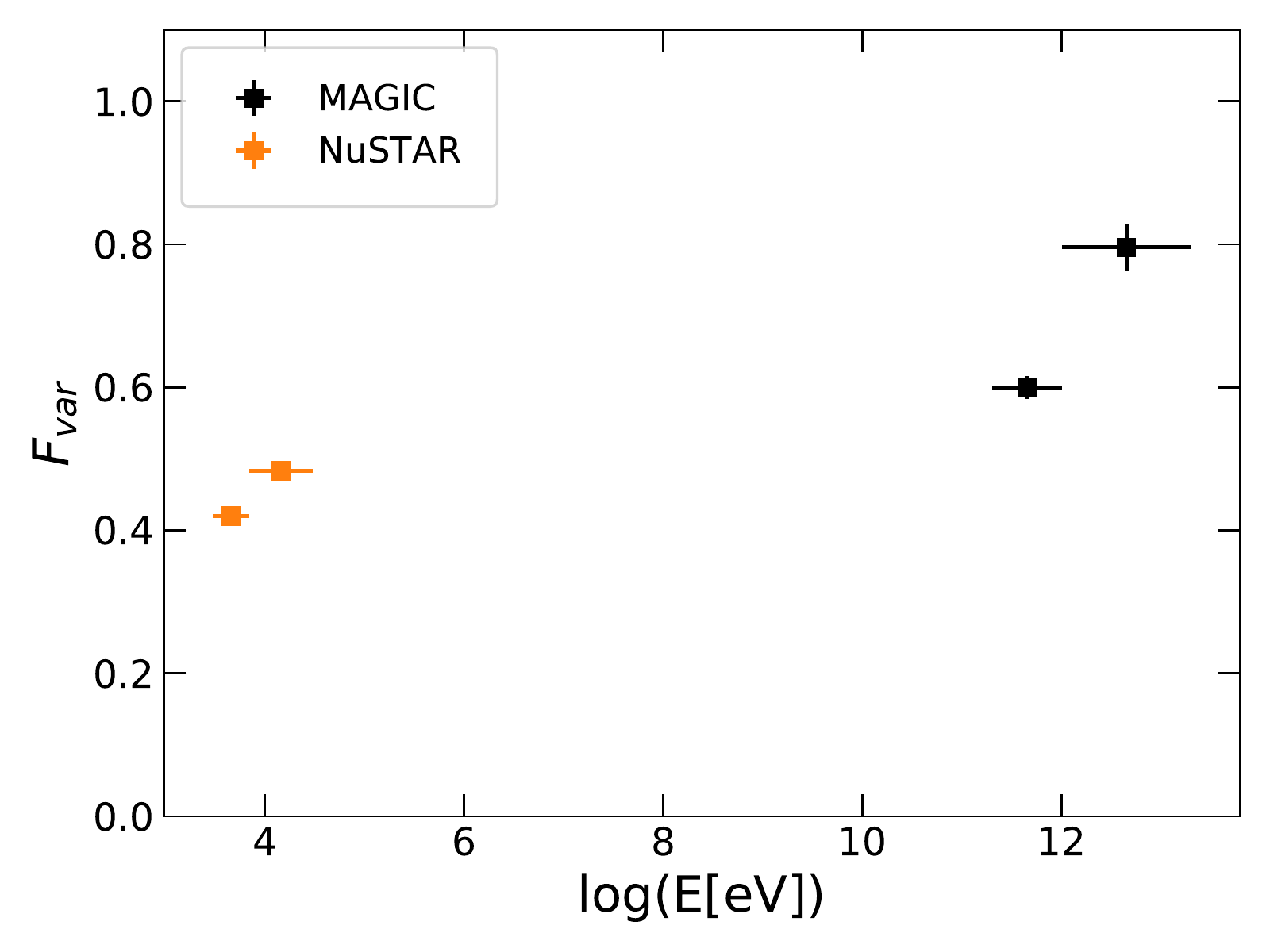}
   \caption{\textit{Fractional variability} $F_{var}$ from the simultaneous MAGIC-\textit{NuSTAR} observations performed on the four nights MJD~57757, MJD~57785, MJD~57813 and MJD~57840 (2017 January 4, 2017 February 1, 2017 March 1 and 2017 March 28). The $F_{var}$ values were computed with the fluxes determined on 30-minutes time bins that are reported in Appendix~\ref{sect:MAGIC_nustar_lc}.} 
    \label{Frac_var_nustar_magic}%
\end{figure}

Overall, a clear two-peak structure is visible. The first peak, with $F_{var} \approx 1$, occurs in the hard X-ray band (15-50\,keV). The second peak lies in the VHE band (around 1\,TeV), also with $F_{var} \approx 1$. The lowest variability is seen in the radio, UV/optical and 0.2-2\, GeV band and they all display $F_{var}<0.2$. \par 

In Fig.~\ref{Frac_var_nustar_magic}, the \textit{Fractional variability} is shown for the simultaneous MAGIC/\textit{NuSTAR} observations. Here, again, a significant increase of variability with energy is observed both in the X-rays and in the VHE band.\par

\begin{figure*}[h!]
   \centering
   \includegraphics[width=2.1\columnwidth]{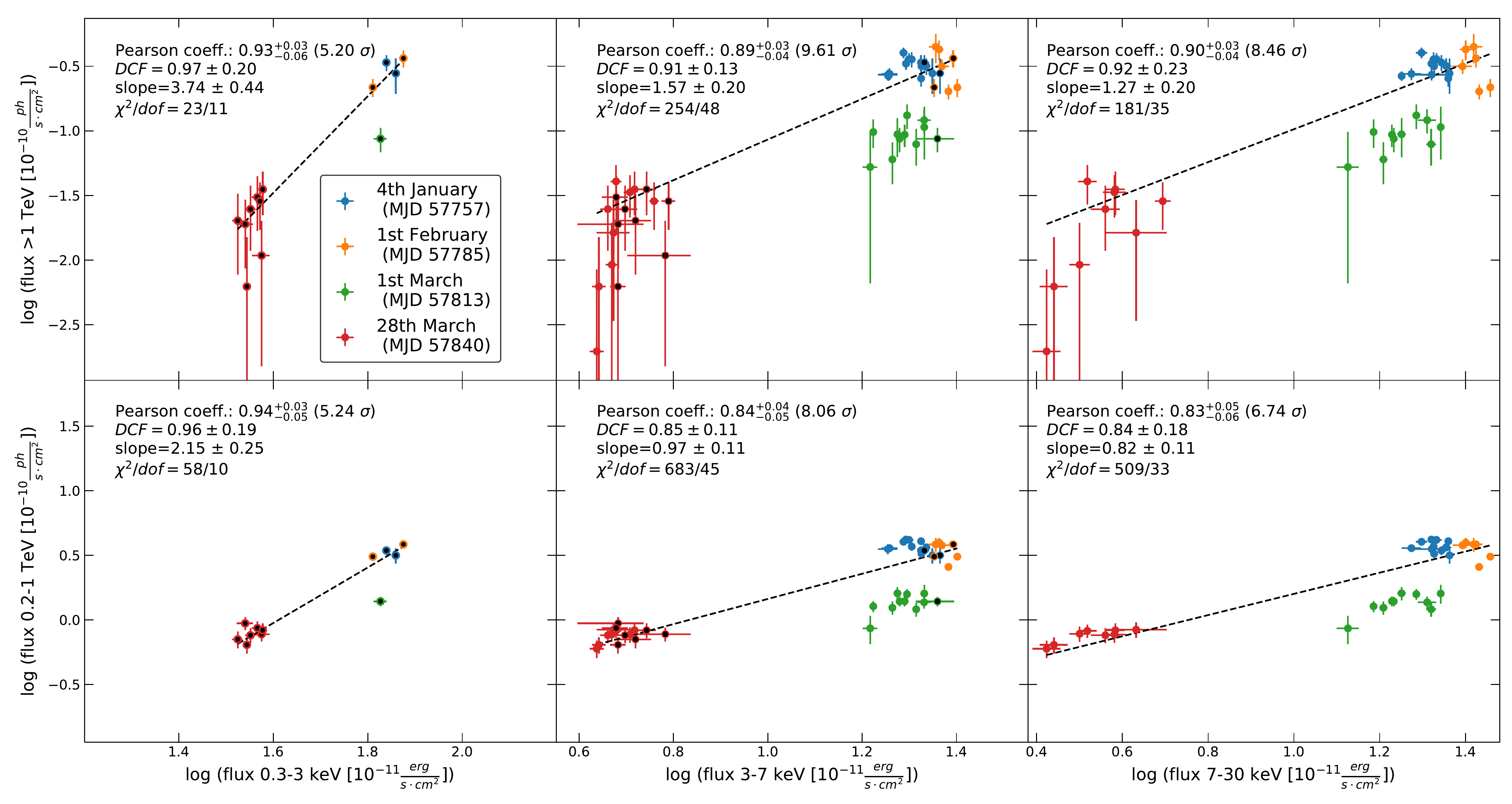}
   \caption{VHE flux versus X-ray flux during the simultaneous MAGIC/\textit{NuSTAR}/\textit{Swift} observations. Flux points are computed over time bins of 30 minutes. From left to right, the X-ray energy bands in the panels are 0.3-3\,keV, 3-7\,keV and 7-30\,keV. In the upper panels, the MAGIC fluxes are in the $>1$\,TeV band, while in the lower panels they are in the 0.2-1\,TeV band. Data points from \textit{Swift}-XRT are shown with black-filled markers. Fluxes corresponding to each day are plotted in a different colour. The results of the Pearson coefficients (and its significance in $\sigma$), DCF and the slope of the linear fits in the log-log plane (with the $\chi^2/dof$ values) are indicated in each of the corresponding subplots.}
    \label{NUSTAR_MAGIC}
\end{figure*}

The patterns reported above are common for Mrk\,421, and have been observed on multiple occasions \citep[][]{2015A&A...576A.126A, 2015A&A...578A..22A, 2016ApJ...819..156B}. The locations of the two peaks directly correspond to the falling edges of the two bumps noticeable in the SED of Mrk~421 \citep{2011ApJ...736..131A}. The low $F_{var}$ values (radio, UV/optical and MeV-GeV) on the other hand match the rising edges of the two SED bumps. In leptonic models, the rising segments of the SED bumps originate from less energetic electrons compared to the falling segments. Because the cooling rate of the particles due to synchrotron radiation is inversely proportional to their Lorentz factor ($t_{cool, synch} \propto 1/\gamma$), the variability is naturally expected to increase with energy around the two SED bumps. 

We investigate the flux correlations among the various energy bands that show significant variability, as reported in Fig.~\ref{Frac_var}, finding significant correlations only between the VHE gamma rays and the X-rays and between the X-rays and the UV/optical. The results from these studies are reported in Sect.~\ref{sect:VHE_xray} and \ref{sec:UV-X-ray}, respectively.

%

\section{Study of the VHE versus X-ray correlation}
\label{sect:VHE_xray}

\subsection{Correlation during the MAGIC/\textit{NuSTAR}/\textit{Swift} observations}

First, we investigate the VHE versus X-ray correlation making use of the strictly simultaneous MAGIC, \textit{NuSTAR} and \textit{Swift} observations performed during four days (MJD~57757, MJD~57785, MJD~57813 and MJD~57840). The combination of \textit{NuSTAR} and \textit{Swift}-XRT observations provides an X-ray coverage over 2 orders of magnitude, from 0.3\,keV to 30\,keV. Up to now, only a few published works have reported multi-band correlation studies extending into the hard X-rays (i.e., $\gtrsim10$\,keV). We note that in the previous works, the source was either probed during exceptionally low states \citep{2016ApJ...819..156B} or during flares \citep{2008ApJ...677..906F, 2020ApJS..248...29A}. In this paper, we aim at completing the picture by offering a view of the VHE versus X-ray correlation of Mrk~421 during a close-to-typical activity.\par

We build light curves for the MAGIC/\textit{NuSTAR}/\textit{Swift} strictly simultaneous observations with temporal bins of 30 min for all the instruments. This choice of binning allows sufficiently good statistics in all energy bands, in particular at VHE, where the measurement instruments typically have lower sensitivities compared to the X-ray telescopes. In the X-ray regime, the fluxes are calculated in 3 energy bands: 0.3-3\,keV, 3-7\,keV and 7-30\,keV. In the VHE regime, the fluxes are computed in the 0.2-1\,TeV and $>$1\,TeV bands. The resulting light curves are shown in Appendix~\ref{sect:MAGIC_nustar_lc}. No significant intra-night variability is found in the MAGIC light curves. On the contrary, the fluxes measured by \textit{NuSTAR} are significantly ($>5\sigma$) incompatible with a constant value for each epoch. For completeness, the light curves in Appendix~\ref{sect:MAGIC_nustar_lc} also include the densely sampled optical (R-band) data from GASP-WEBT, from which no clear variability pattern is visible.\par 

In Fig.~\ref{NUSTAR_MAGIC}, the flux-flux correlations are shown for all energy combinations. For each day, the data are plotted with a different colour. As mentioned above, the X-ray fluxes are quite similar between the first three pointings (MJD~57757, MJD~57785 and MJD~57813). The VHE fluxes, however, differ significantly between those nights, being ${\approx}1.6$\,C.U. for MJD~57757 and MJD~57785 and ${\approx}0.5$\,C.U. for MJD~57813. Such a feature points towards significant variability in the Compton dominance of the SED (that is defined as the relative luminosity of the high-energy SED component with respect to the low-energy SED component). The last observation, on MJD~57840, has the lowest VHE flux (${\approx}0.1-0.3$\,C.U.), which coincides with a lower X-ray activity level. \par

\begin{figure*}[h!]
   \centering
   \includegraphics[width=2.1\columnwidth]{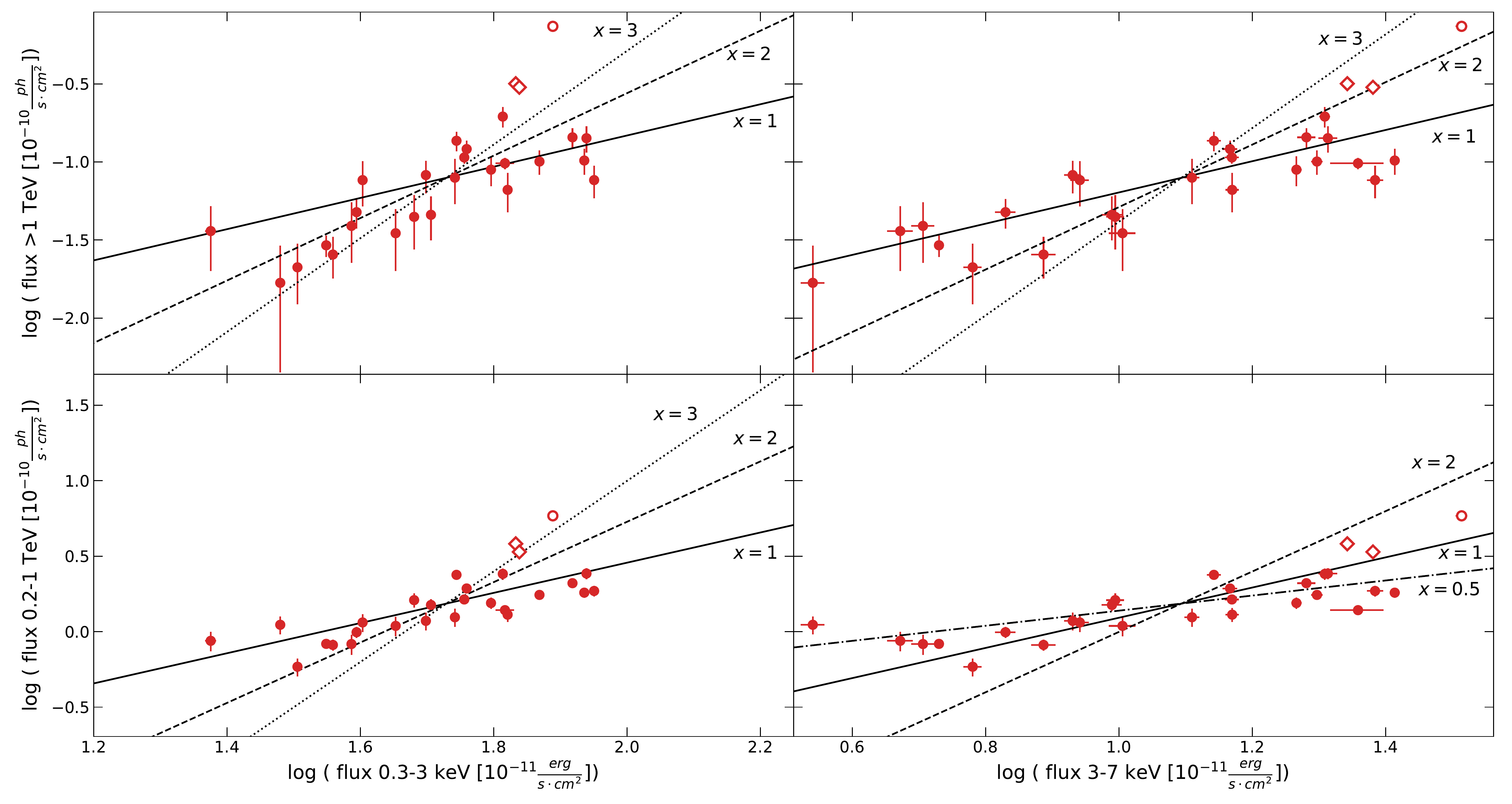}
   \caption{VHE flux versus X-ray flux over the MWL campaign using MAGIC and \textit{Swift}-XRT data. Data are nightly binned, and only pairs of measurement within 4\,hours are considered. MAGIC fluxes are in the $>1$\,TeV band (top panels) and in the 0.2-1\,TeV band (lower panels). \textit{Swift}-XRT fluxes are computed in the 0.3-3\,keV (left panels) and 3-7\,keV bands (right panels). The open red diamond markers highlight measurements on MJD~57757 and MJD~57785 (i.e., the dates of the first two simultaneous MAGIC/\textit{NuSTAR/}\textit{Swift} observations). The VHE flare (on MJD~57788) is plotted in open red circle marker. Black lines depict linear functions (in the log-log plane) with slopes $x=1,\,2,\,3$. In the 0.2-1\,TeV versus 3-7\,keV panel, because of the dynamical range of the data, the linear function with slope $x=3$ is replaced with one with slope of $x=0.5$. The Pearson coefficients, DCFs and the slopes from linear fits in the log-log plane are shown in Table~\ref{tab:correlation_results_lonterm}.}
    \label{swift_vs_MAGIC}
\end{figure*}

In order to quantify the correlation, we use the Pearson coefficient and the \textit{Discrete Correlation Function} \citep[DCF; ][]{1988ApJ...333..646E}. Compared to the Pearson coefficient, the DCF has the advantage of naturally including the measurement uncertainties. A linear fit in the log-log plane is used to further study the type of the correlation patterns. All the resulting numbers and the slopes of the linear fits (as well as the $\chi^2/$dof results) are included in the corresponding subplots of Fig.~\ref{NUSTAR_MAGIC}.\par 

Each Pearson coefficient shows a value above 0.8 with a significance higher than $5\sigma$. The strong correlation is further confirmed by the DCF, whose values are all inconsistent with 0 at a significance greater than $4\sigma$. At the lowest X-ray energies in the 0.3-3\,keV band, the \textit{Swift}-XRT coverage is much less compared to that of \textit{NuSTAR}. This prevents a direct comparison of the Pearson coefficients and the DCF obtained at 0.3-3\,keV with those obtained in the 3-7\,keV and 7-30\,keV bands. When restricting ourselves to the 3-7\,keV and 7-30\,keV panels, the DCF as well as the Pearson coefficients are compatible within statistical uncertainties between the different energy combinations. We stress that the correlations are strongly dominated by the day-by-day flux variations and no significant intra-night VHE versus X-ray correlation is found. \par

Regarding the correlation slopes, they vary significantly depending on the energies considered. The largest slope is derived for the >1\,TeV versus 0.3-3\,keV case and is consistent with a more-than-cubic trend. Conversely, the lowest slope is obtained for 0.2-1\,TeV versus 7-30\,keV, and indicates a sub-linear trend between these two energies.\par

It is interesting to note that a significant increase of the slope is visible at all X-ray energies when considering a higher VHE band (i.e., when moving from 0.2-1\,TeV to >1\,TeV in a given X-ray band). These trends indicate that the >1\,TeV flux has a stronger scaling with the X-ray flux compared to the 0.2-1\,TeV flux.\par

In a given VHE band, the slopes are compatible within uncertainties between the 3-7\,keV and 7-30\,keV bands, although the best-fit values seem to indicate lower slopes for the 7-30\,keV band. A significant decrease of the slope at higher X-ray energies was reported by \citet{2016ApJ...819..156B}, which we can not confirm in our work with the data at hand. We refrain from a comparison of the 0.3-3\,keV band with the other X-ray bands because of the significantly smaller amount of available measurements.\par

\subsection{Correlation over the entire MWL campaign}

In this section, we report the MAGIC versus \textit{Swift}-XRT correlation study over the full MWL campaign. In order to ease the comparison with the results shown in Fig.~\ref{NUSTAR_MAGIC}, \textit{Swift}-XRT fluxes are considered in the 0.3-3\,keV and 3-7\,keV bands. We correlate MAGIC and \textit{Swift}-XRT daily averaged fluxes by considering all pairs of observation falling within a time range of $\delta t=4$\,hours (28 pairs in total fulfill this criterion). The flux-flux plots are shown in Fig.~\ref{swift_vs_MAGIC}. \par 

The flare on MJD~57788 appears as a clear outlier (plotted with an open red circular marker). The $F_{>1\text{\,TeV}}$ and $F_{0.2-1\text{\,TeV}}$ are respectively ${\sim}10$ times and ${\sim}3$ times higher compared to the bulk of the measurements with comparable (and even slightly higher) $F_{0.3-3\text{\,keV}}$ values. When considering the 3-7\,keV band, the VHE flare indeed coincides with the highest X-ray flux among those that lie within $\delta t$ from a MAGIC observation. This corroborates the fact that in the X-rays the flare is characterised by a hardening of the spectrum without significant increase of the flux amplitude around 1\,keV. We stress however that the corresponding $F_{3-7\text{\,keV}}$ is only ${\approx}20\%$ greater than the second highest $F_{3-7\text{\,keV}}$ (which corresponds to MJD~57763), for which the $F_{>1\text{\,TeV}}$ is about 7 times lower. We note that for this particular night, the MAGIC and \textit{Swift}-XRT observations are simultaneous and overlap, as shown in Fig.~\ref{flare_night_LC}. This allows us to discard any bias due to the non-simultaneity of the observations. This is particularly important given the rather short duty cycle of the flare at VHE (flux halving time of a few hours, see Fig~\ref{flare_night_LC}).\par

Interestingly, the first two \textit{NuSTAR} nights on MJD~57757 and MJD~57785, which show an enhanced VHE activity in Fig.~\ref{NUSTAR_MAGIC}, also seem to stand out from the distribution. They are plotted with diamond open markers in Fig.~\ref{swift_vs_MAGIC}. The VHE flux is higher than the bulk of the data. Together with the MJD~57788 flare, these measurements suggest a change in the VHE versus X-ray correlation with respect to the typical behaviour.\par

\begin{table*}[h!]
\caption{\label{tab:correlation_results_lonterm}
 Results of the MAGIC versus \textit{Swift}-XRT correlations shown in Fig.~\ref{swift_vs_MAGIC}.} 

\setlength{\tabcolsep}{5pt} 
\renewcommand{\arraystretch}{2.} 
\centering
\begin{tabular}{l|ccc}   
\hline\hline
Energy bands & Pearson ($\sigma$) & DCF & slope ($\chi^2/dof$) \\
\hline

>1\,TeV vs 0.3-3\,keV& $0.75_{-0.10}^{+0.07}$ (4.9$\sigma$) & $0.79\pm0.18$ & $3.00\pm0.56$ (493/26)\\

0.2-1\,TeV vs 0.3-3\,keV & $0.75_{-0.10}^{+0.07}$ (4.9$\sigma$) & $0.77\pm0.17$ & $1.65\pm0.28$ (890/26)\\

\hdashline

>1\,TeV vs 3-7\,keV & $0.85_{-0.07}^{+0.05}$ (6.2$\sigma$) & $0.89\pm0.22$ & $1.80\pm0.19$ (219/26)\\

0.2-1\,TeV vs 3-7\,keV & $0.81_{-0.08}^{+0.06}$ (5.7$\sigma$) & $0.83\pm0.20$ & $1.00\pm0.11$ (472/26)\\

\hline
\end{tabular}

\end{table*}

The correlation is quantified following the same methodology described in the previous section. The Pearson coefficients, the DCF and the slopes from the linear fits are listed in Table~\ref{tab:correlation_results_lonterm}. The entire dataset (28 measurements) is considered. The Pearson coefficients are consistent within uncertainties between each energy combination and are all greater than $0.7$. Also, the significances are all above $4\,\sigma$. The DCF values are similar to the Pearson coefficients and do not significantly vary between the energy combinations. On the other hand, the slopes show distinct values depending on the energy band considered, similarly to the simultaneous MAGIC/\textit{NuSTAR}/\textit{Swift} observations. A roughly cubic trend is obtained for >1\,TeV versus 0.3-3\,keV, while an almost linear trend is visible for 0.2-1\,TeV versus 3-7\,keV. Finally, a close-to-quadratic relation is found for 0.2-1\,TeV versus 0.3-3\,keV and for >1\,TeV versus 3-7\,keV. Within uncertainties, the slopes are consistent with those obtained from the MAGIC/\textit{NuSTAR}/\textit{Swift} nights (in the 0.3-3\,keV and 3-7\,keV bands).\par 

Similarly to the correlation during the MAGIC/\textit{NuSTAR}/\textit{Swift} nights, Fig.~\ref{swift_vs_MAGIC} reveals that the correlation slope increases at higher VHE energies in a given X-ray band. This pattern is also verified in Table~\ref{tab:correlation_results_lonterm}, where the the best-fit slopes in the >1\,TeV band are systematically greater than in the 0.2-1\,TeV band.\par    

The $\chi^2/dof$ values indicate a poor fit and implies that a $F_{VHE} \propto F_{X-ray}^{x}$ parametrisation is too simple to describe the relation between these two bands. Nonetheless, the best-fit slopes still provide information about the main trend in the relation between the fluxes from these two energy bands. The above mentioned results suggest steep correlations maintained over monthly time scales.\par

\begin{figure*}[h!]
   \centering
   \includegraphics[width=2.1\columnwidth]{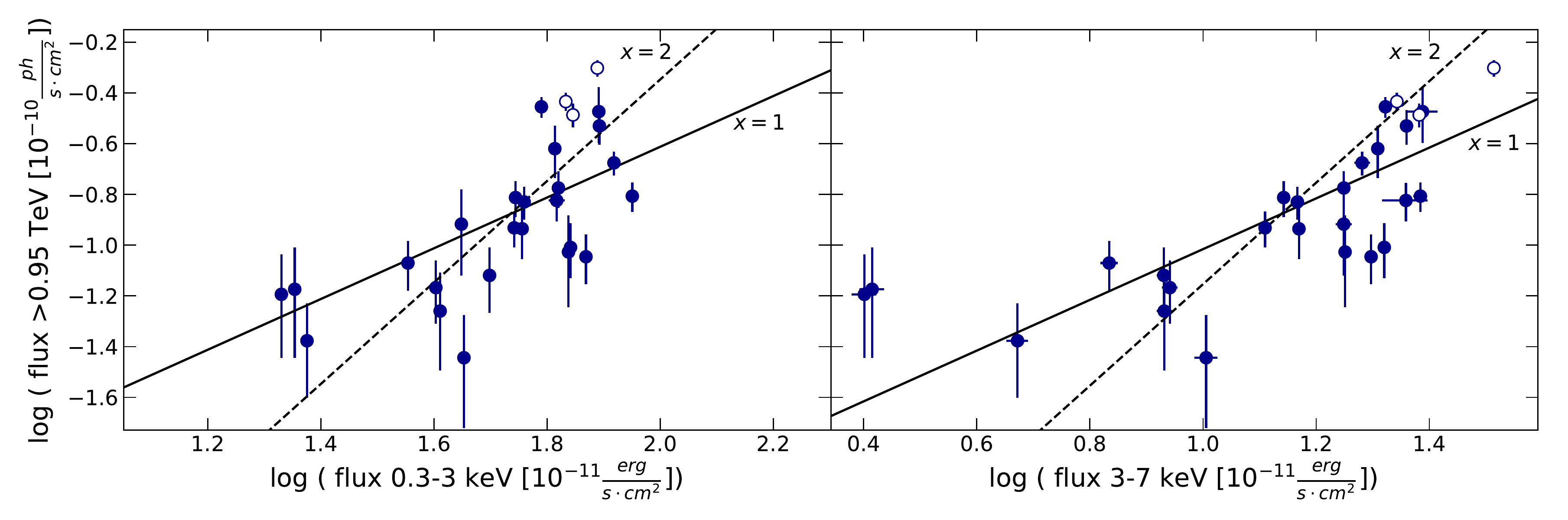}
   \caption{VHE flux versus X-ray flux over the MWL campaign using FACT and \textit{Swift}-XRT data. Data are nightly binned, and only pairs of measurement within 4\,hours are considered. FACT fluxes are computed above $0.95$\,TeV and only measurements with a signal-to-noise ratio $>2$ are taken into account. \textit{Swift}-XRT fluxes are computed in the 0.3-3\,keV (left panel) and 3-7\,keV bands (right panel). The open blue markers highlight measurements on MJD~57757 and MJD~57785 (i.e., the dates of the first two simultaneous MAGIC/\textit{NuSTAR/}\textit{Swift}) and the VHE flare (on MJD~57788). Black lines depict linear functions (in the log-log plane) with slopes $x=1$ and $x=2$. The Pearson coefficients, DCF and slope are shown in Table~\ref{tab:correlation_results_lonterm}. The Pearson coefficients, DCF and slopes from linear fits in the log-log plane are shown in Table~\ref{tab:correlation_results_lonterm_fact}.}
    \label{swift_vs_FACT}
\end{figure*}

\citet{2005A&A...433..479K} and \citet{2010A&A...510A..63K} argued that in one-zone SSC models one would expect in general linear or sub-linear trends between the X-ray and VHE emission in HBLs. One should note that more-than-linear correlations may still occur in some conditions, but demand rather specific evolution of the model parameters \citep[see][]{2005A&A...433..479K}. This is particularly true when one considers the emission above the peaks of the SED components, which is typically the case for the >1\,TeV and 3-7\,keV bands in Mrk~421 (see Sect.~\ref{sect:synch_peak}). From this point of view, under the assumption of a one-zone SSC scenario, it is surprising to derive over the full campaign a slope of $1.80 \pm 0.19$ for >1\,TeV versus 3-7\,keV fluxes. However, from a simple quality check, we found that the previously mentioned three outliers (marked in open red markers in Fig.~\ref{swift_vs_MAGIC}), which show enhanced gamma-ray activity, significantly influence the best-fit slope towards higher values even though they only represent about 10\% of the measurements. When ignoring them, the correlation agrees with a linear trend. The mentioned outliers seem to follow a different pattern in the log-log plane, implying that a physical regime different from a more typical behaviour was probed. Therefore, these results suggest a temporal dependence in the correlated variability between the X-ray and VHE bands.

In Fig.~\ref{swift_vs_FACT}, we show the results of the correlation between FACT and \textit{Swift}. As was done previously, only pairs of observations that occurred within a time window of $\delta t=4$\,hours of one another are considered. The fluxes are nightly binned and only FACT measurements with a signal-to-noise ratio $>2$ are taken into account. The results are listed in Table~\ref{tab:correlation_results_lonterm_fact}. The three outliers MJD~57788 (the VHE flare), MJD~57757 and MJD~57785 are depicted with open blue symbols in Fig.~\ref{swift_vs_FACT}. The derived correlation significance is $\gtrsim 4\sigma$ in all energy bands. The DCF and Pearson's coefficients are in good agreement with those obtained using the MAGIC data.

\begin{table*}[h!]
\caption{\label{tab:correlation_results_lonterm_fact} Results of the FACT versus \textit{Swift}-XRT correlations shown in Fig.~\ref{swift_vs_FACT}.} 
\setlength{\tabcolsep}{5pt} 
\renewcommand{\arraystretch}{2.} 
\centering
\begin{tabular}{l|ccc}     
\hline\hline
 Energy bands & Pearson ($\sigma$) & DCF & slope ($\chi^2/dof$) \\
\hline
>0.95\,TeV vs 0.3-3\,keV& $0.70_{-0.12}^{+0.09}$ (4.2$\sigma$) & $0.75\pm0.20$ & $1.60\pm0.45$ (248/25) \\

\hdashline

>0.95\,TeV vs 3-7\,keV & $0.74_{-0.11}^{+0.08}$ (4.7$\sigma$) & $0.80\pm0.19$ & $1.25\pm0.18$ (124/25) \\

\hline
\end{tabular}

\end{table*}
%

\section{Investigation of the behaviour of the low-energy component in the SED}
\label{sect:synch_peak}

\subsection{Synchrotron peak frequency}
By definition, the low-energy component of the SED of HBLs peaks at frequencies above $10^{15}$\,Hz. In the case of Mrk~421, the peak frequency generally lies between ${\sim}10^{17}$\,Hz and ${\sim}10^{18}$\,Hz \citep{2007A&A...466..521T}, making it an extreme synchrotron blazar \citep{2020NatAs...4..124B}. However, strong variability has been reported in recent works. For instance, \citet{2016ApJ...819..156B} derived a peak located at frequencies $\lesssim10^{16}$\,Hz during particularly low states, while during flares, values close to $10^{19}$\,Hz were reported by \citet{2009A&A...501..879T}. Motivated by the large X-ray flux variability amplitude shown in Fig.~\ref{MWL}, which is more than a factor 30 in the 2-10\,keV band, we carry out an in-depth study of the low-energy component.\par 

\begin{figure*}[h!]
   \centering
   \includegraphics[width=2\columnwidth]{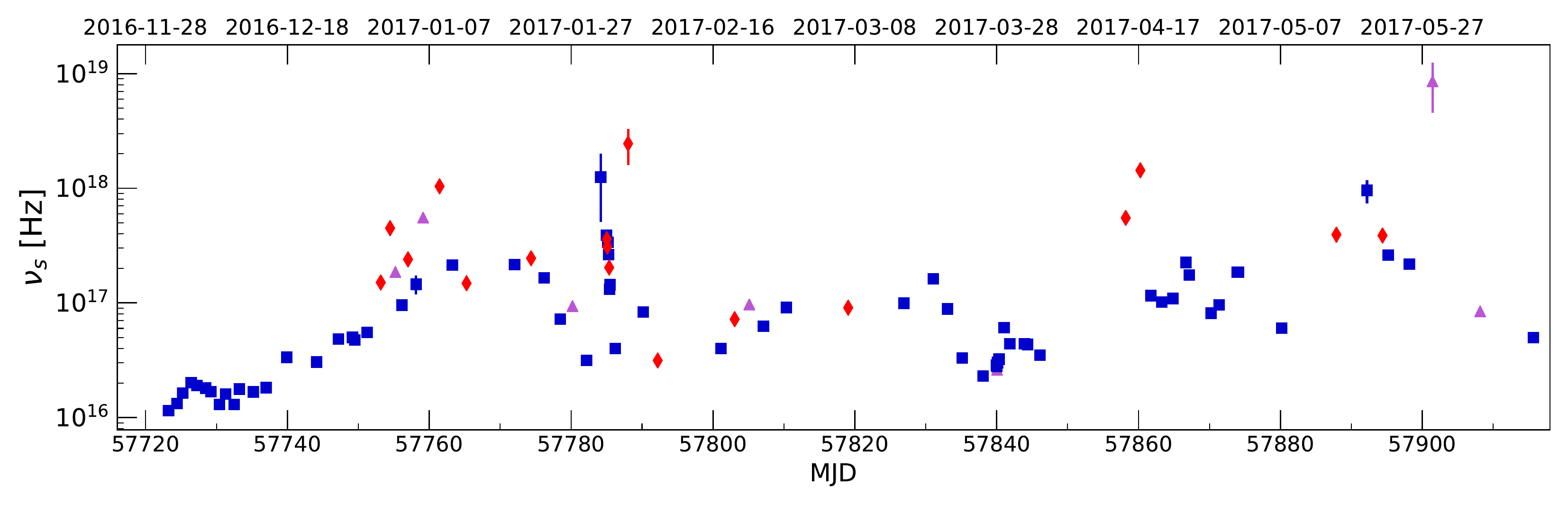}
   \caption{Evolution of the synchroton peak frequency $\nu_s$ versus time throughout the MWL campaign. The peak frequencies are obtained by fitting a log-parabola to simultaneous \textit{Swift}-XRT, \textit{Swift}-UVOT and \textit{Swift}-BAT observations. Blue square markers indicate fits that result in a p-value above $10^{-3}$. Data points in violet triangles correspond to a p-value between $10^{-3}$ and $10^{-4}$, and red diamond markers have a p-value lower than $10^{-4}$.}
    \label{nu_peak_LC}
\end{figure*}

\begin{figure}[h!]
   \centering
   \includegraphics[width=1\columnwidth]{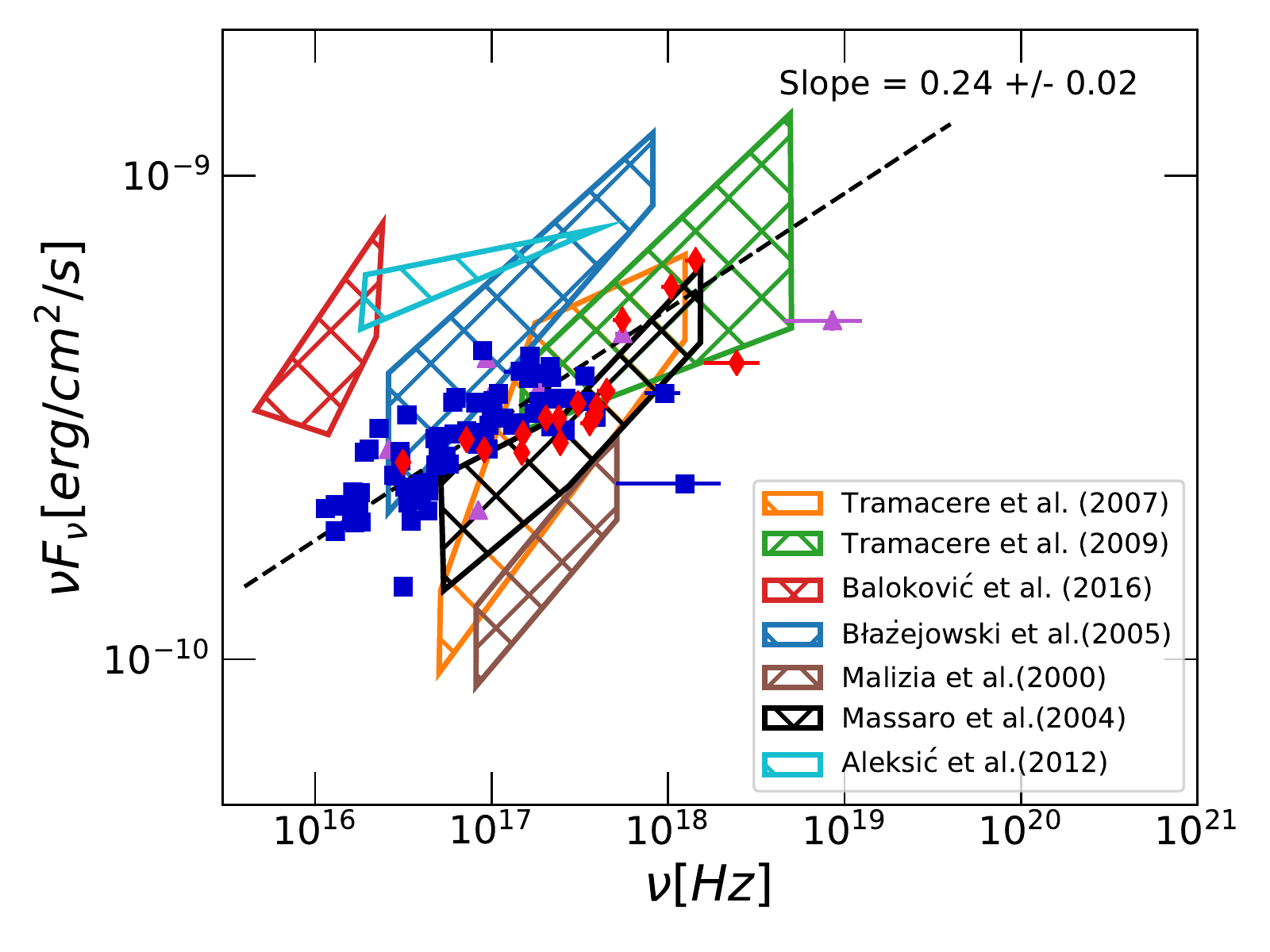}
   \caption{Synchrotron peak frequencies throughout the MWL campaign obtained from log-parabola fits to simultaneous \textit{Swift}-XRT, \textit{Swift}-UVOT and \textit{Swift}-BAT observations. Blue square markers indicate fits that result in a p-value above $10^{-3}$. Data points in violet triangles correspond to a p-value between $10^{-3}$ and $10^{-4}$, and red diamond markers have a p-value lower than $10^{-4}$. The black dotted line is a linear fit to the data and the obtained slope is indicated on the plot. The dashed areas represent the ranges of archival synchrotron peak frequencies, and are taken from Fig.~12 of \citet{2016ApJ...819..156B}. The respective references are given in the legend.}
    \label{synch_peaks}%
\end{figure}

\begin{figure}[h!]
   \centering
   \includegraphics[width=1\columnwidth]{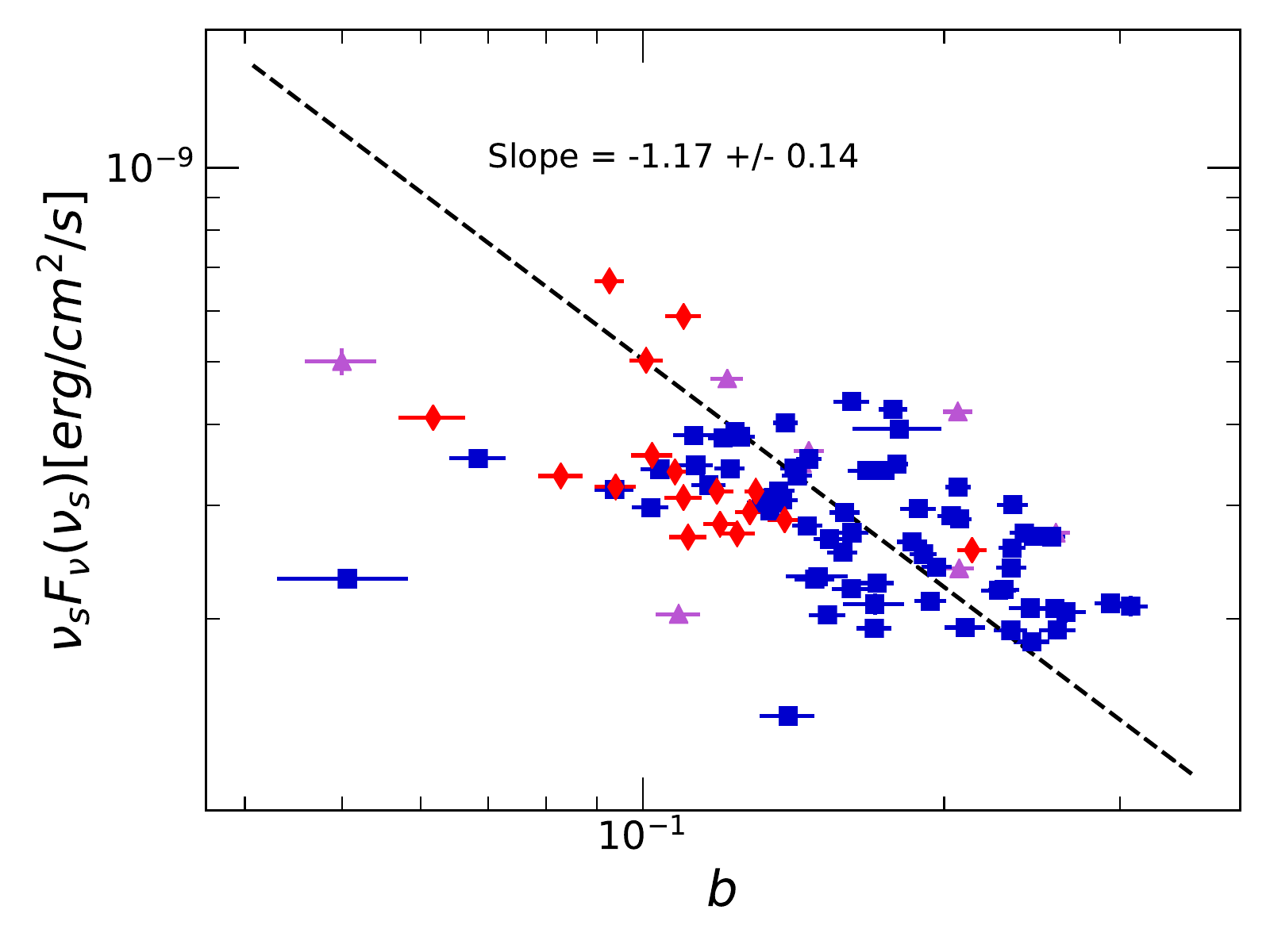}
   \caption{$\nu_{\rm s} F_\nu (\nu_{\rm s})$ versus the curvature parameter $b$ throughout the MWL campaign obtained from log-parabola fits to simultaneous \textit{Swift}-XRT, \textit{Swift}-UVOT and \textit{Swift}-BAT observations. As in Fig.~\ref{synch_peaks}, blue square markers indicate fits that result in a p-value above $10^{-3}$. Data points in violet triangles correspond to a p-value between $10^{-3}$ and $10^{-4}$, and red diamond markers have a p-value lower than $10^{-4}$. The black dotted line is a linear fit to the data, and the obtained slope is indicated on the plot.}
    \label{amplitude_vs_beta}%
\end{figure}

\begin{figure*}[h!]

        \centering
        \begin{subfigure}[b]{0.497\textwidth}
            \centering
            \includegraphics[width=\textwidth]{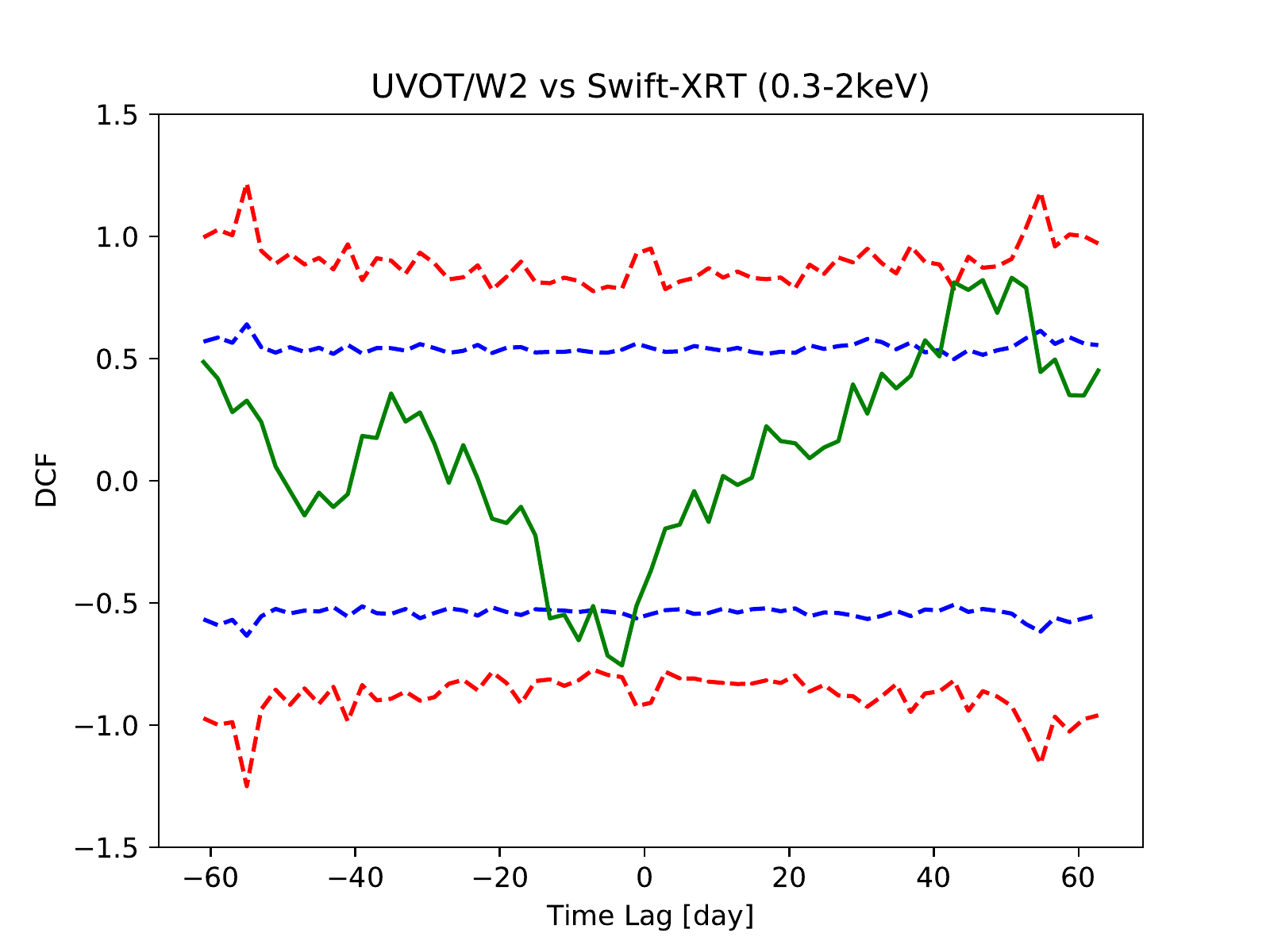}
        \end{subfigure}
        \begin{subfigure}[b]{0.497\textwidth}  
            \centering 
            \includegraphics[width=\textwidth]{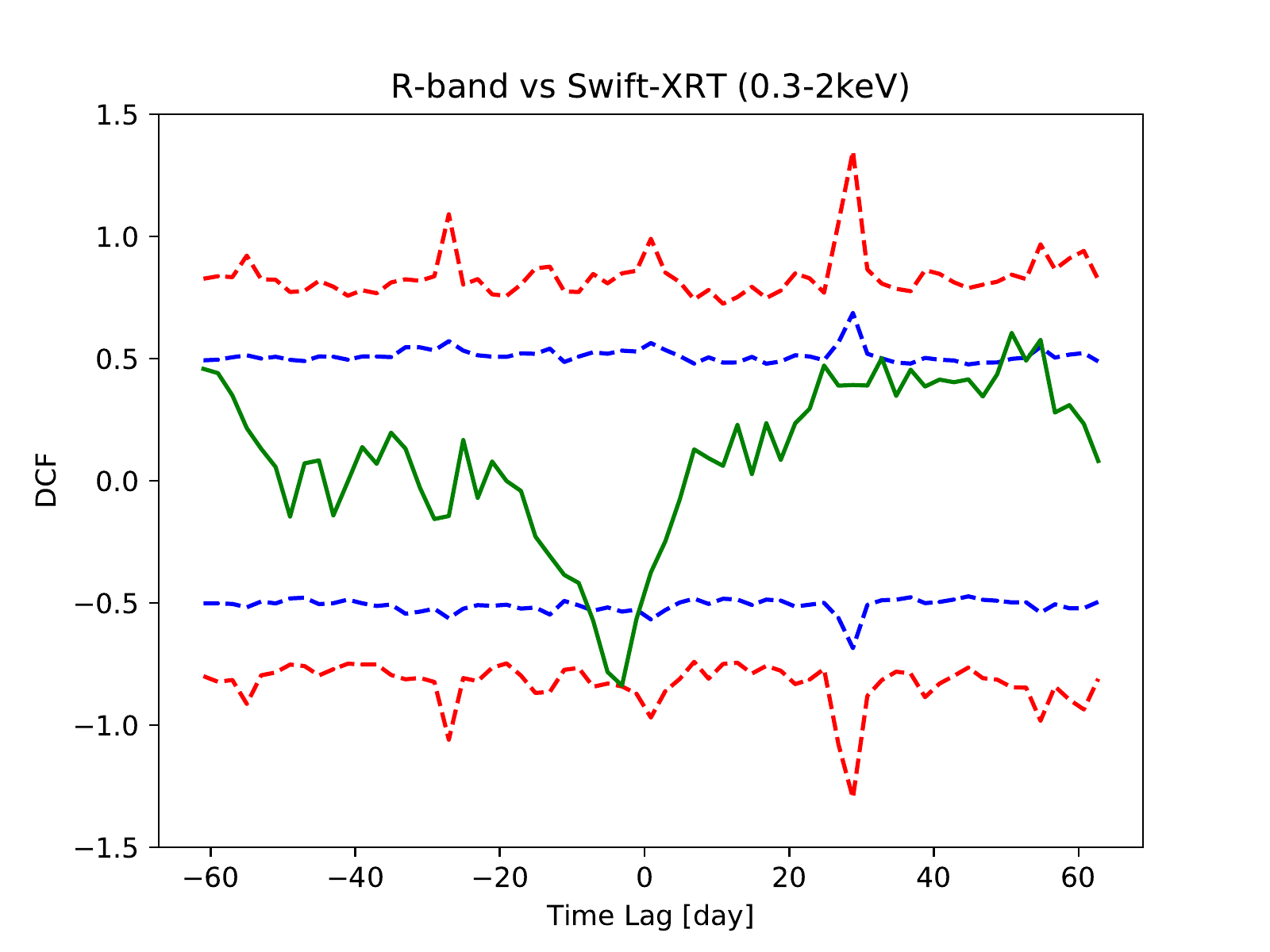}
        \end{subfigure}
       
        \centering
        \begin{subfigure}[b]{0.497\textwidth}
            \centering
            \includegraphics[width=\textwidth]{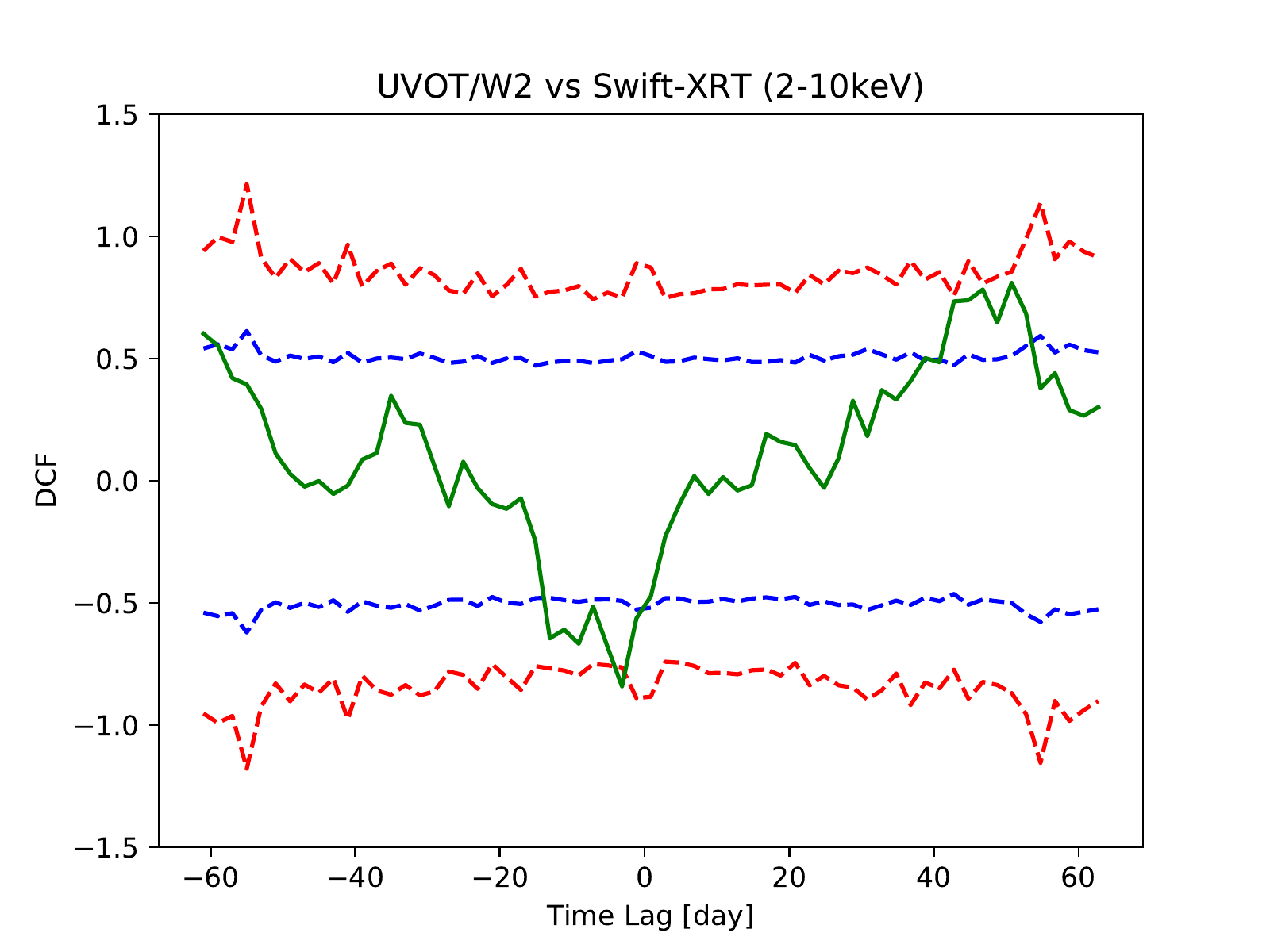}
        \end{subfigure}
        \begin{subfigure}[b]{0.497\textwidth}  
            \centering 
            \includegraphics[width=\textwidth]{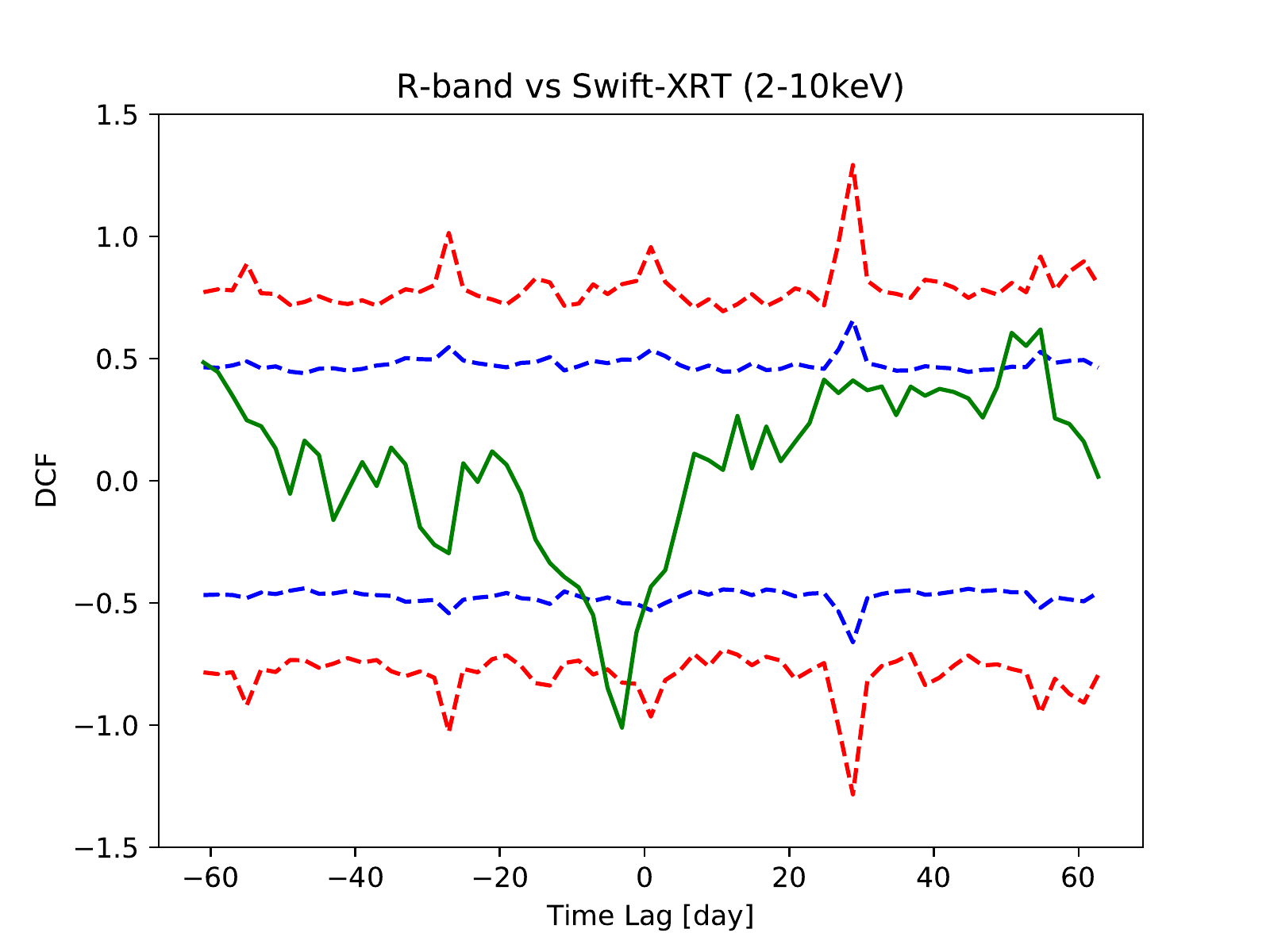}
        \end{subfigure}
        \caption{DCF between X-ray (0.3-2\,keV and 2-10\,keV) and UV/optical (\textit{Swift}-UVW2 and R-band) during the MWL campaign. The blue- and red-dashed-lines indicate the $2\sigma$ and $3\sigma$ confidence intervals estimated from the Monte Carlo simulations, as described in the text.} 
        
\label{DCF_xray_optical}
\end{figure*}

From optical to X-rays, the SED is usually well described with a log-parabola shape \citep{2007A&A...467..501T}:
\begin{equation}
\label{eq:logparabola}
    \nu F_\nu (\nu) = f_{0} \cdot 10^{-b \cdot (\log_{10}(\nu / \nu_{\rm s}))^{2}} \,\rm{erg\,cm}^{-2}\,s^{-1},
\end{equation}
where $\nu_{\rm s}$ is the peak frequency, $b$ the curvature and $f_{0}$ the normalisation constant. Eq.~\ref{eq:logparabola} is equivalent to Eq.~\ref{eq:logparabola_MAGIC}, with the former now having the peak frequency as a free parameter. In order to locate the peak frequency throughout the campaign, we fit Eq.~\ref{eq:logparabola} to all available pairs of simultaneous \textit{Swift}-XRT and \textit{Swift}-UVOT observations. Most of the time, the \textit{Swift}-XRT and \textit{Swift}-UVOT instruments are operating together, resulting in a total of 93 pairs of observations. Furthermore, we complement the SED with \textit{Swift}-BAT data. For this, we exploit the orbit-wise light curve in the 15-50\,keV range and compute a 6-hour averaged photon count rate (in cm$^{-2}$ s$^{-1}$) centred at each of the simultaneous \textit{Swift}-XRT and \textit{Swift}-UVOT observations. The \textit{Swift}-BAT requires a longer integration time because of its limited sensitivity to detect Mrk\,421 at hard X-rays in short timescales. The 6-hour averaged count rates are converted to a flux point (in erg cm$^{-2}$ s$^{-1}$) using the prescription of \cite{Krimm_2013}. In Fig.~\ref{nu_peak_LC}, the resulting locations of the peak frequencies, $\nu_{\rm s}$, are plotted versus time. We show in blue square markers the results from fits that have a p-value above $10^{-3}$, in violet triangles the results from fits with p-values between $10^{-3}$ and $10^{-4}$ and in red diamond markers the results from fits with p-values below $10^{-4}$. Around 70\% of the fits have a p-value above $10^{-3}$, and only 17 fits (less than 20\%) yielded a p-value below $10^{-4}$, indicating that Eq.~\ref{eq:logparabola} provides an acceptable parametrisation of the SED in most of the cases. In Fig.~\ref{synch_peaks}, the resulting locations of the peak frequencies, $\nu_{\rm s}$, are reported in the $\nu F_\nu (\nu) - \nu$ space. Archival synchrotron peak frequencies are depicted with dashed areas in Fig.~\ref{synch_peaks}.\par

Fig.~\ref{synch_peaks} reveals a strong variability of $\nu_{\rm s}$. The peak can be as low as ${\sim}10^{16}$\,Hz, similarly to the particularly low activity reported by \citet{2016ApJ...819..156B}. The highest $\nu_{\rm s}$ are ${\sim}10^{18}$-$10^{19}$\,Hz. Such variations roughly cover the entire range of $\nu_{\rm s}$, which has been observed up to now in the case of Mrk~421. The interesting aspect of our result stems from the fact that this large variability amplitude only happened within a ${\approx}6$-month period. The strong flux variability amplitude in the 0.3-2\,keV and 2-10\,keV bands over the MWL campaign (see Sect.~\ref{sect:MWL_lc}) is therefore accompanied by particularly large shifts of the synchrotron bump.\par

A linear fit is performed in the log-log space of $\nu_{\rm s} F_\nu (\nu_{\rm s})$ versus $\nu_{\rm s}$. The slope obtained is $0.24 \pm 0.02$, indicating a significant increase of the X-ray flux with higher peak frequencies. A very similar slope ($0.25 \pm 0.02$) is derived when only the log-parabola fits with a p-value above $10^{-3}$ (blue data points) are considered. This result is consistent with the results reported in Section~\ref{sec:X-ray}, and agrees with the well-known harder-when-brighter behaviour previously reported for Mrk\,421 and other HBLs \citep{1998ApJ...492L..17P,2004ApJ...601..151K,2015A&A...576A.126A}. However, the scatter of the data over the main trend shows that there are many days when the synchrotron emission of Mrk~421 did behave substantially differently. This may be caused by different particle or environment conditions in the main emitting region, or because of the presence of additional components that, during short periods of time, contribute significantly to the overall synchrotron emission of Mrk~421. Either way, the $\nu_{\rm s} F_\nu (\nu_{\rm s})$ versus $\nu_{\rm s}$ data for these days appear as outliers in the the main trend with slope of about 0.2, shown in Fig.~\ref{synch_peaks}.\par

Fig.~\ref{amplitude_vs_beta} shows $\nu_{\rm s} F_\nu (\nu_{\rm s})$ versus the curvature parameter $b$ as defined in Eq.~\ref{eq:logparabola}. The data tend to indicate a lower curvature with increasing $\nu_{\rm s} F_\nu (\nu_{\rm s})$ values, although this anti-correlation remains weak and some outliers exist. A linear fit yields a slope of $-1.17 \pm 0.14$. A very similar slope is obtained when one considers only the log-parabola fits having a p-value above $10^{-3}$ (blue data points). A decrease of the curvature with the flux may be indicative of an evolution of the flux driven by stochastic electron acceleration, as demonstrated by \citet{2011ApJ...739...66T}.

\subsection{UV/optical versus X-ray anti-correlation}
\label{sec:UV-X-ray}

Within standard one-zone SSC models, which assume a single emission zone responsible for the SED (from optical to VHE), a positive correlation between the UV/optical and X-ray is generally expected. Changes in parameters describing the emission zone environment (for instance Doppler factor, electron density or the size of the region) would modify the synchrotron emission at each energy, and thus, lead to correlated variability between the UV/optical and X-ray emissions. Nevertheless, the very different temporal behaviour between these two energy bands renders the detection of a correlation difficult. Fig.~\ref{Frac_var} shows that $F_{var}$ is more than two times higher in the X-rays than in the UV/optical. Hence, for any given flux change in the X-rays, the correlated UV/optical flux variation would be much suppressed. On the other hand, the large variability of the synchrotron peak frequency discussed may facilitate the detection of correlation patterns.\par

We cross-correlate UV/optical with X-ray fluxes by selecting all possible combinations between the following light curves: optical (R-band), \textit{Swift}-UVOT, \textit{Swift}-XRT(0.3-2\,keV) and \textit{Swift}-XRT(2-10\,keV). Regarding \textit{Swift}-UVOT, we consider data from the W2 filter. Given their proximity in frequency, fluxes in the three \textit{Swift}-UVOT filters are strongly correlated and all of them yield very similar results. The choice of the W2 filter is further motivated by the slightly larger amount of measurements compared to W1 and M2, providing a better coverage of the source. As in Sect.~\ref{sect:VHE_xray}, we use the DCF \citep{1988ApJ...333..646E} to quantify the correlation. It is known that the significance of the correlation assessed directly from the DCF uncertainty given in \citet{1988ApJ...333..646E} can be overestimated \citep{2003ApJ...584L..53U}. This is especially the case for red-noise light curves that are regularly found in blazar observations. In this work, a more careful statistical treatment employing Monte-Carlo light curve simulations is undertaken. The details of the procedure can be found in Appendix~\ref{anti_corr_significance}.\par 

The DCF coefficients between \textit{Swift}-XRT (0.3-2\,keV) and \textit{Swift}-XRT (2-10\,keV) on one hand, and \textit{Swift}-UVOT/W2 and R-band on the other hand, are shown in Fig.~\ref{DCF_xray_optical}. The green full lines depict the DCF results obtained with the data, while the blue and red dotted lines are the $2\sigma$ and $3\sigma$ significance bands (derived by simulations), respectively. Fig.~\ref{DCF_xray_optical} reveals a negative peak with DCF$\approx$-0.75 and DCF$\approx$-1 located between $-3$ and $-5$\,days. This clear feature is visible in the cross-correlations derived for all the energy bands. The significance is at the level of $3\sigma$ for \textit{Swift}-XRT (0.3-2\,keV) versus \textit{Swift}-UVOT/W2, \textit{Swift}-XRT (2-10\,keV) versus \textit{Swift}-UVOT/W2 and \textit{Swift}-XRT (0.3-2\,keV) versus R-band. For \textit{Swift}-XRT (2-10\,keV) versus R-band the significance is above $3\sigma$. These results represent evidence for an anti-correlation between UV/optical and the X-rays.\par 

The multi-instrument LCs from Fig.\ref{MWL} show that, during the time interval from about MJD~57720 and about MJD~57760, there is an overall decrease in the UV/optical fluxes with time, while the opposite occurs with the X-ray fluxes, that show an overall increase with time. Such a clear trend does not seem to be apparent for time intervals after MJD~57760, and hence it may happen that the above-mentioned time range of about 40 days dominates the marginally significant anti-correlation reported in Fig.~\ref{DCF_xray_optical}. To investigate this, we repeat the study and compute the DCF as well as its significance after excluding all data taken prior to MJD~57760. The results are displayed in Fig.~\ref{DCF_xray_optical_after_57760} of Appendix~\ref{sect:DCF_before_after_57760}, and show that the anti-correlation disappears (significance drops below $3\sigma$) for all the bands probed.\par 

Additionally, we also evaluate the correlation using only data within the time interval MJD~57720--57760. The results are shown in Fig.~\ref{DCF_xray_optical_before_57760} of Appendix \ref{sect:DCF_before_after_57760}. This time, the significance is above $3\sigma$ for the cross-correlations from all energy bands (the DCF values are also higher), confirming the overall dominance of the time interval MJD~57720--57760 in the cross-correlation results. This observation suggests that these two bands are not persistently related to each other (as occurs between the VHE gamma rays and the X-rays in Mrk~421), but instead show a degree of correlation that changes with time, and is significant only during month timescales.

%

\section{Broadband SED and modelling of the strictly simultaneous MAGIC/\textit{NuSTAR}/\textit{Swift} observations}
\label{sect:SED_modelling}

\begin{figure*}
    \centering
    \begin{subfigure}[b]{0.497\textwidth}
        \centering
        \includegraphics[width=\textwidth]{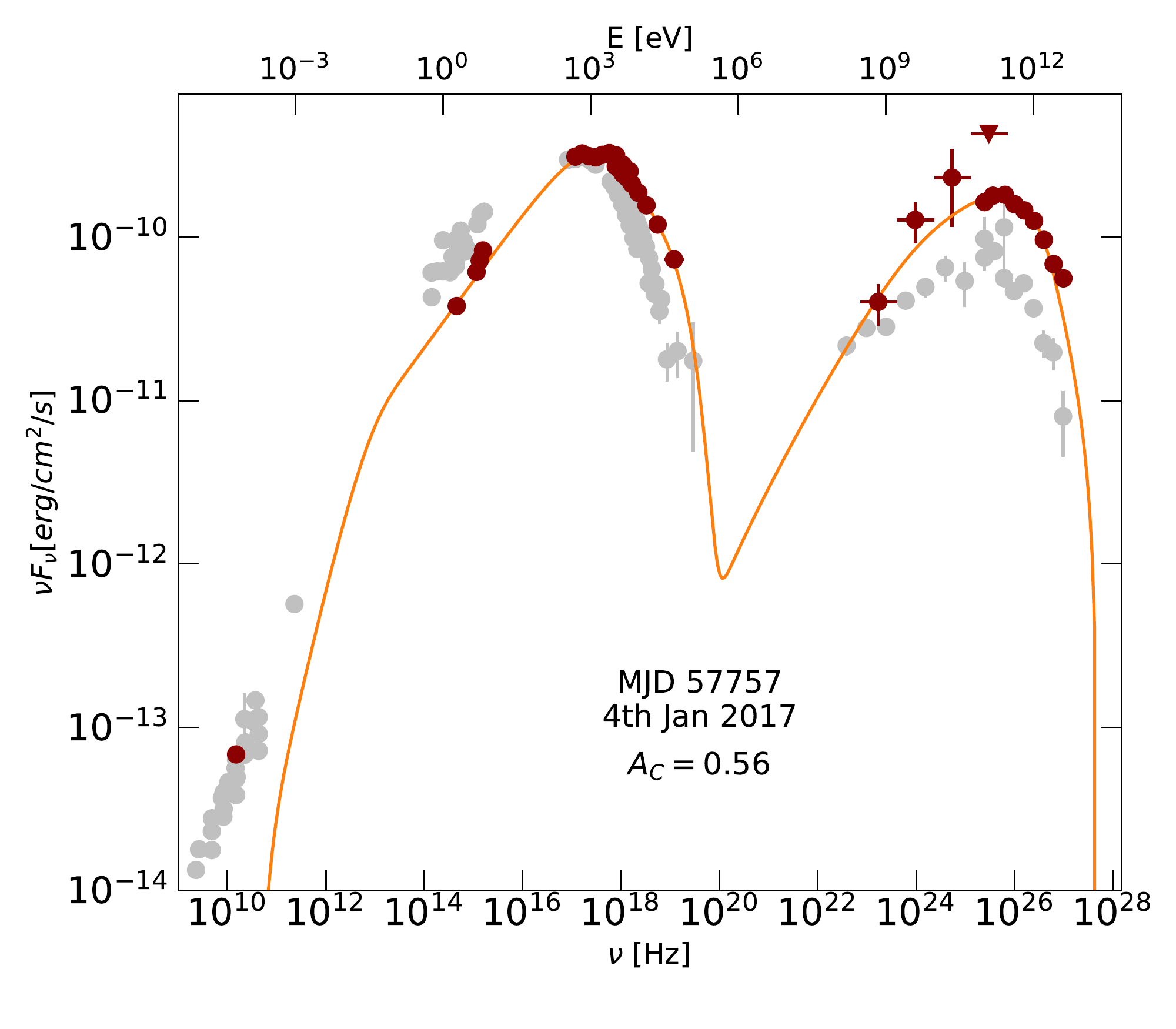}
    \end{subfigure}
    \begin{subfigure}[b]{0.497\textwidth}  
        \centering 
        \includegraphics[width=\textwidth]{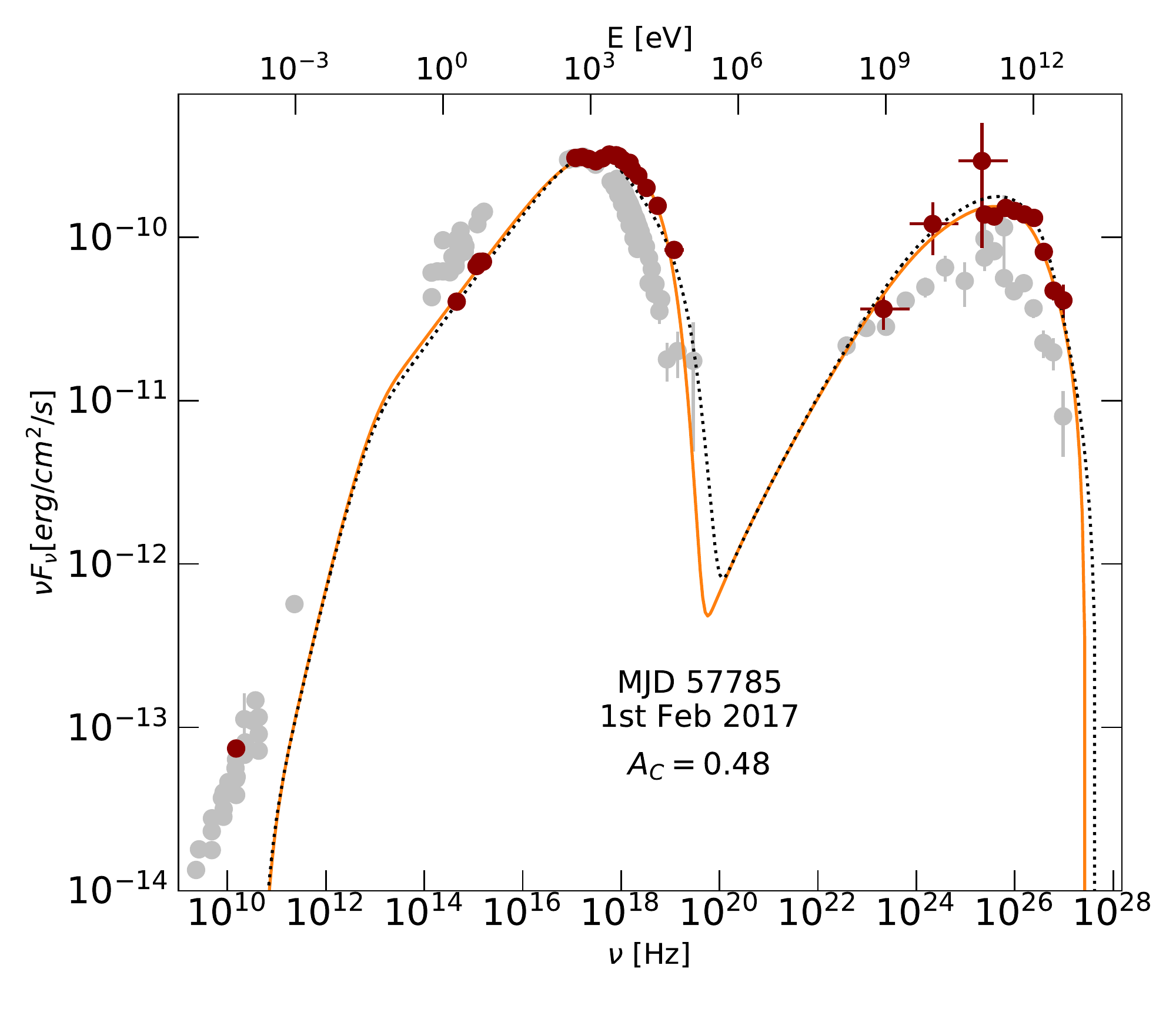}
    \end{subfigure}
    
    \begin{subfigure}[b]{0.497\textwidth}   
        \centering 
        \includegraphics[width=\textwidth]{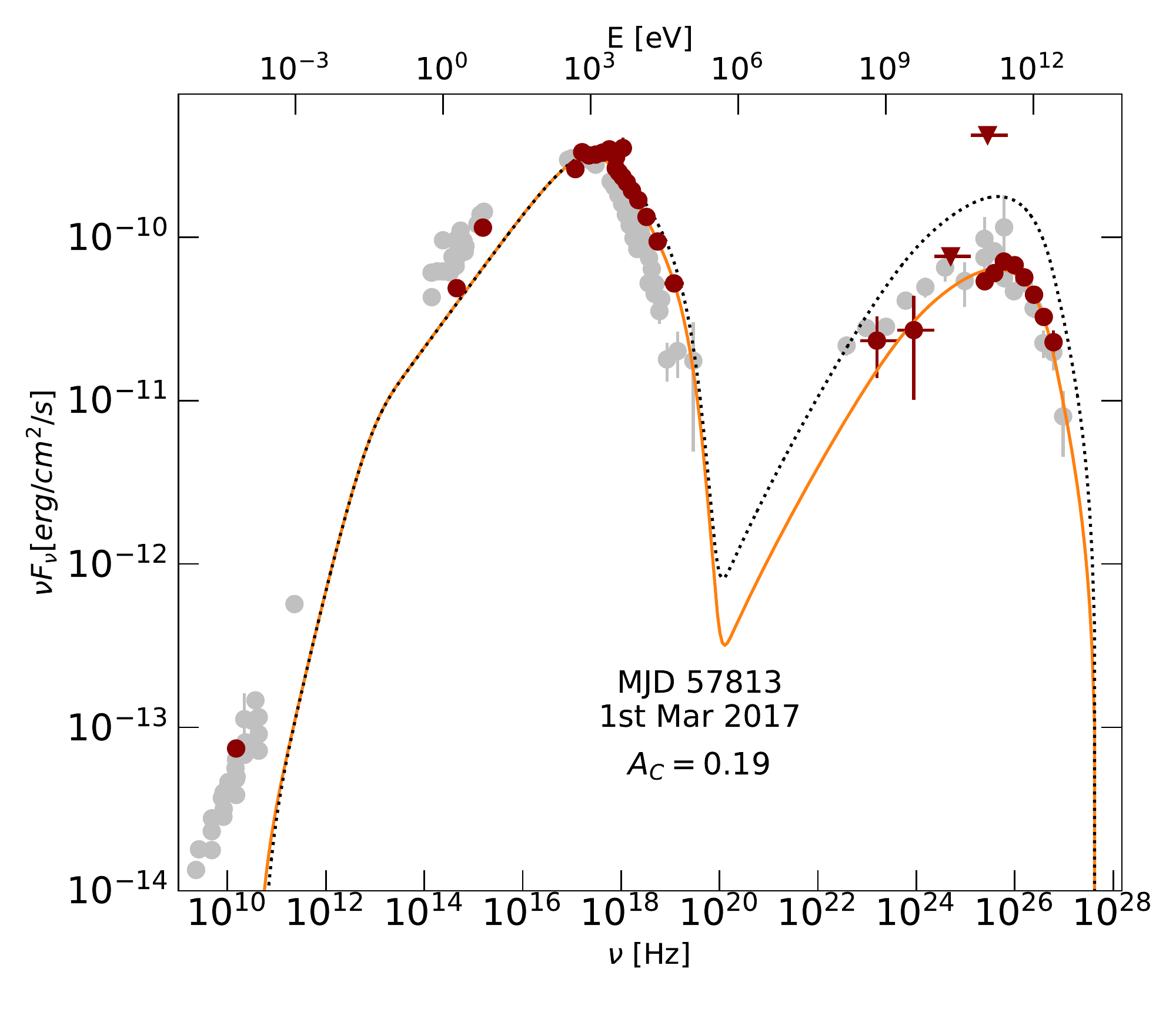}
    \end{subfigure}
    \begin{subfigure}[b]{0.497\textwidth}   
        \centering 
        \includegraphics[width=\textwidth]{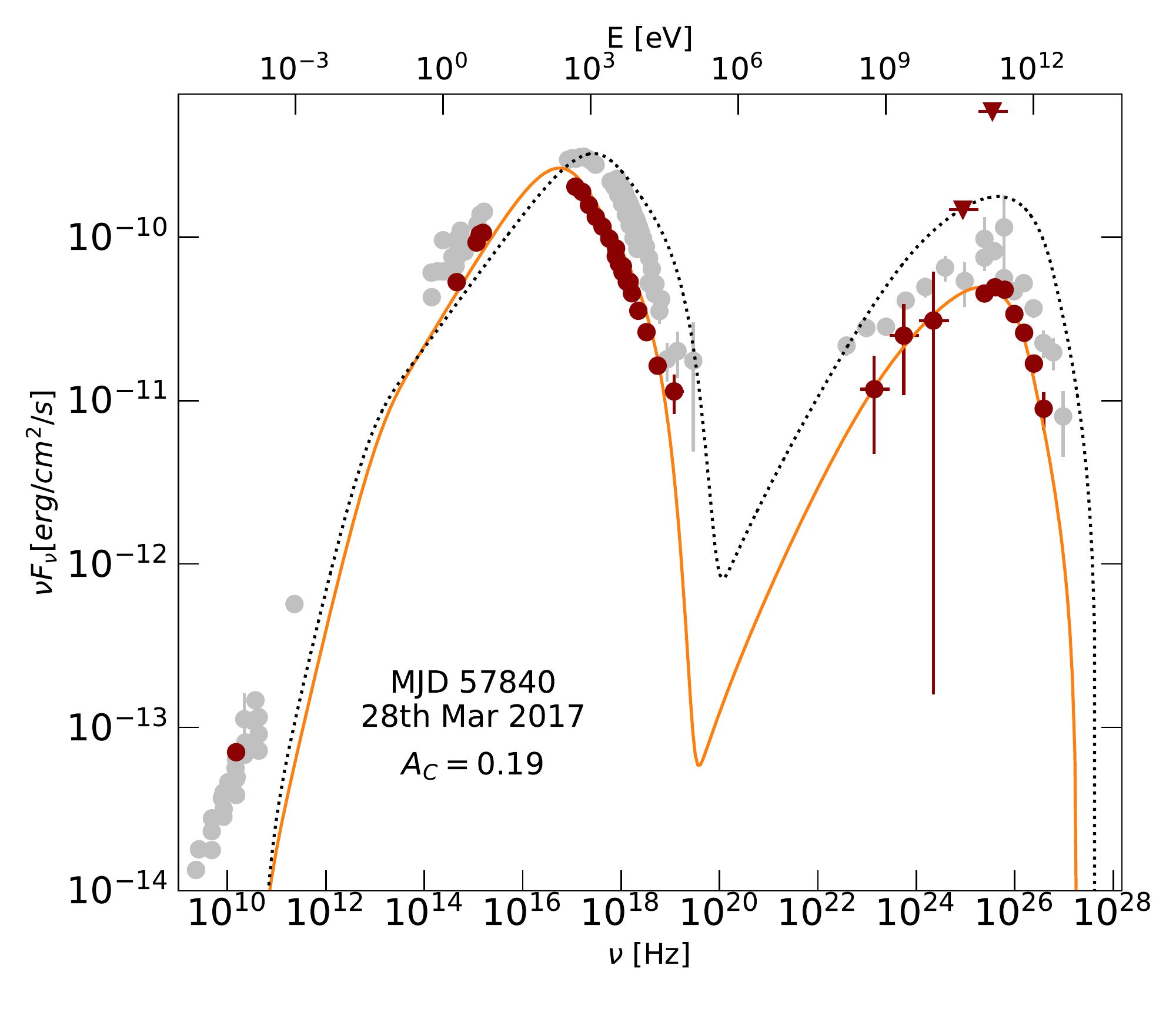}
    \end{subfigure}
    \caption{SED of the four simultaneous MAGIC/\textit{NuSTAR}/\textit{Swift} observations (MJD~57757, MJD~57785, MJD~57813, MJD~57840). The data points and obtained one-zone SSC models are plotted with brown markers and orange lines, respectively. The parameters of the one-zone SSC models are shown in Table~\ref{tab:ssc_parameters}. For comparison purposes, the one-zone SSC model of the first pointing (MJD~57757) is plotted with black dotted lines in all the other subplots. The Compton dominance parameter $A_{C}$ determined from the model is indicated on each panel. Archival data representing the typical Mrk~421 state from \cite{2011ApJ...736..131A} are shown in grey.} 
    \label{NuSTAR_nights_sed}
\end{figure*}

We build four broadband SED snapshots from radio to VHE around the strictly simultaneous MAGIC/\textit{NuSTAR}/\textit{Swift} observations. The excellent coverage in the X-rays, combined with the HE-VHE data, offer unprecedented time-resolved SEDs during rather typical states for Mrk~421.\par 

At VHE, the MAGIC spectra are averaged over each observing night. The effects of EBL absorption are corrected for by applying the EBL model of \citet{2011MNRAS.410.2556D}. The resulting best-fit parameters of a log-parabola spectral shape are shown in Table~\ref{tab:MAGIC_spectral_param_nustar_sim}. A log-parabola model is preferred with respect to a simple power-law model at a significance higher than 5$\sigma$ (based on a likelihood ratio test) for the four spectra. The \textit{NuSTAR} and \textit{Swift} spectra are computed over the exact same time range as the MAGIC observations, limiting possible biases due to non-simultaneity. We note though that  the \textit{NuSTAR} observations consist of continuous exposures, while for \textit{Swift} the observations are shorter and scattered over the MAGIC window (see Fig.~\ref{NUSTAR_MAGIC_sim_LC_first} to Fig.~\ref{NUSTAR_MAGIC_sim_LC_fourth} in Appendix~\ref{sect:MAGIC_nustar_lc}). The \textit{NuSTAR} best-fit spectral parameters assuming log-parabolic models are shown in Table~\ref{tab:Nustar_spectral_param_nustar_sim}. In the high energy (HE) band, owing to the limited ability to significantly detect Mrk\,421 in this energy range, the \textit{Fermi}-LAT spectra are averaged over 3 days, which provides a good compromise between simultaneity and spectral resolution. Optical (R-band) data are not strictly simultaneous, but the closest observation in time is chosen, which results in a time difference of at most one day. Because of the low variability in the optical band reported in Fig.~\ref{MWL} and Fig.~\ref{Frac_var}, they can be considered as a good proxy for the simultaneous emission in the optical band.\par

\begin{table*}[h!]
\caption{\label{tab:MAGIC_spectral_param_nustar_sim} MAGIC spectral parameters of the simultaneous MAGIC/\textit{NuSTAR}/\textit{Swift} observations obtained from a log-parabola fit according to Eq.~\ref{eq:logparabola_MAGIC}.} 
\centering
\begin{tabular}{ l c c c c c c}     
\hline\hline 
 Night &  & $f_{0}\,[10^{-10} \mathrm{cm}^{-2} \mathrm{s}^{-1} \mathrm{TeV}^{-1}]$ & $\alpha$ & $\beta$ & $\chi^{2}$/dof \\  
\hline\hline   
MJD~57757 (4\textsuperscript{th} January 2017) &  & $12.1\pm0.2$ & $2.15 \pm 0.02$ & $0.28 \pm 0.03$ & 10.9/13 \\
MJD~57785 (1\textsuperscript{st} February 2017) &  & $10.6\pm0.3$ & $2.09 \pm 0.03$ & $0.36 \pm 0.06$ & 34.5/12 \\ 
MJD~57813 (1\textsuperscript{st} March 2017) &  & $4.75\pm0.16$ & $2.09 \pm 0.04$ & $0.47 \pm 0.09$ & 14.0/10 \\
MJD~57840 (28\textsuperscript{th} March 2017) &  & $2.87\pm0.13$ & $2.49 \pm 0.05$ & $0.50 \pm 0.13$ & 10.3/8 \\
\hline 
\end{tabular}
\tablefoot{The normalisation energy is 300\,GeV.} 
\end{table*}

\begin{table*}[h!]
\caption{\label{tab:Nustar_spectral_param_nustar_sim} \textit{NuSTAR} spectral parameters of the simultaneous MAGIC/\textit{NuSTAR}/\textit{Swift} observations obtained from a log-parabola fit.} 
\centering
\begin{tabular}{ l c c c c c}     
\hline\hline 
 Night &  & $\alpha$ & $\beta$ & $\chi^{2}$/dof \\  
\hline\hline   
MJD~57757 (4\textsuperscript{th} January 2017) &  & $2.08 \pm 0.03$ & $0.19 \pm 0.02$ & 34.8/31 \\
MJD~57785 (1\textsuperscript{st} February 2017) &  & $1.87 \pm 0.05$ & $0.28 \pm 0.03$ & 30.0/30 \\ 
MJD~57813 (1\textsuperscript{st} March 2017) &  & $2.06 \pm 0.04$ & $0.25 \pm 0.02$ & 28.7/31 \\
MJD~57840 (28\textsuperscript{th} March 2017) &  & $2.51 \pm 0.10$ & $0.15 \pm 0.06$ & 38.9/36 \\
\hline 
\end{tabular}
\tablefoot{The normalisation energy is 1\,keV.} 
\end{table*}

\begin{table*}[h]
\caption{\label{tab:ssc_parameters}Parameters of the SSC models obtained for each MAGIC/\textit{NuSTAR}/\textit{Swift} simultaneous observing epoch.}
\centering
\begin{tabular}{l c c c c c}     
\hline\hline
 Parameters  & MJD~57757 & MJD~57785 & MJD~57813 & MJD~57840 \\
   & 4\textsuperscript{th} January 2017 & 1\textsuperscript{st} February 2017 & 1\textsuperscript{st} March 2017 & 28\textsuperscript{th} March 2017 \\
\hline
\hline
$\Gamma_b$ & 25 & 25 & 25 & 25 \\
$B'$ [10$^{-2}$~G]  & 6.1 & 7.0 & 6.1 & 10.0 \\
$R'$ [10$^{16}$~cm] & 1 & 1 & 1.65 & 1.33 \\
$U'_e$ [10$^{-2}$~erg cm$^{-3}$] & 1.1 & 1.0 & 0.24 & 0.22\\
$\alpha_1$ & 2.2 & 2.2 & 2.2 & 2.0\\
$\alpha_2$ & 3.8 & 3.1 & 3.9 & 4.0\\
$\gamma'_{min}$ [10$^3$] & 1.0 & 1.0 & 1.0 & 1.0\\
$\gamma'_{br}$ [10$^5$] & 2.1 & 1.4 & 2.1 & 0.8\\
$\gamma'_{max}$ [10$^6$] & 1.5 & 0.9 & 1.5 & 0.6\\
$U'_B/U'_e$ & $1.4 \times 10^{-2}$ & $2 \times 10^{-2}$ & $6.1 \times 10^{-2}$ & $1.8 \times 10^{-1}$\\
\hline
\end{tabular}
\tablefoot{See text in Sect.~\ref{sect:SED_modelling} for the description of each parameter.} 
\end{table*}

The broadband SEDs are shown with brown data points in Fig.~\ref{NuSTAR_nights_sed}. The SED data describing a quiescent state (averaged over 5 months) reported in \citet{2011ApJ...736..131A} are also plotted, for comparison purposes, with grey full dots. A slight mismatch is visible between \textit{Swift}-XRT and \textit{NuSTAR} in their overlapping region (around $10^{18}$\,Hz). For the X-ray spectra from MJD~57757, MJD~57785 and MJD~57840, the mismatch is always less than 15\%, which is within the systematic uncertainties of these two instruments \citep{2017AJ....153....2M}. On MJD~57813 (lower left panel), the mismatch is somewhat larger. The last \textit{Swift}-XRT point, the one with the largest discrepancy, is $34\% \pm 17\%$ off from the lowest-energy point of the \textit{NuSTAR} spectrum. We note that this remains at the level of $2\sigma$, and thus may well be due to a statistical fluctuation. Moreover, the X-ray flux variability on hour timescales and the longer exposures of \textit{NuSTAR} (see light curves in Appendix~\ref{NuSTAR_tables}) may also contribute to this small mismatch between the X-ray fluxes measured by these two instruments.\par

The SEDs show distinct spectral and flux characteristics across the whole spectrum. While the first three observations (MJD~57757, MJD~57785 and MJD~57813) display a comparable X-ray emission with a synchrotron peak frequency located around $10^{17}$\,Hz, the last observation shows a much softer and fainter spectrum as well as a synchrotron peak shifted to lower frequencies by an order of magnitude, at ${\sim}10^{16}$\,Hz.\par

As mentioned in Sect.~\ref{sect:VHE_xray}, the first two observations on MJD~57757 and MJD~57785 have an enhanced VHE activity compared to that of MJD~57813, despite the three nights having a comparable synchrotron emission up to the hard X-rays. Fig.~\ref{NuSTAR_nights_sed} further shows that the enhanced VHE activity is also reflected in the \textit{Fermi}-LAT data, revealing an enhancement of the whole IC intensity. The relative luminosity between the two components in the SED can be quantified using the Compton dominance $A_C = L_{IC, peak}/L_{synch, peak}$ \citep{2013ApJ...763..134F}, where $L_{IC, peak}$ is the luminosity at the IC peak and $L_{synch, peak}$ the luminosity at the synchrotron peak. The excellent coverage in the X-rays and gamma rays allow these two quantities to be precisely determined. For MJD~57757 and MJD~57785, $A_C\approx0.6$ and $A_C\approx0.5$, respectively. On the other hand, for MJD~57813, $A_C\approx0.2$, which is roughly three times lower. We stress that these three nights are characterised by similar spectral properties both in X-ray and VHE (see Table~\ref{tab:MAGIC_spectral_param_nustar_sim} and Table~\ref{tab:Nustar_spectral_param_nustar_sim}). This points towards a simple difference in the relative strengths of both spectral components without spectral changes. For MJD~57840, we find $A_C\approx0.2$. \par

We adopt a stationary one-zone SSC model \citep{1996ASPC..110..436G, 1998ApJ...509..608T, 2004ApJ...601..151K} to interpret the four simultaneous SEDs. Within this scenario, the SED from the infrared is dominated by a single emitting zone. The low-energy component is due to synchrotron radiation by relativistic electrons, while the high-energy component originates from IC scattering on the synchrotron photons by the exact same electron population. This simple leptonic scenario was already applied to Mrk~421 and successfully described the SED on a wide range of flux states \citep{2008ApJ...686..181F, 2012A&A...542A.100A, 2016ApJ...819..156B}. Compared to most of the published studies, we benefit from an excellent coverage in the hard X-ray band. The combination of optical, \textit{Swift}-UVOT, \textit{Swift}-XRT and \textit{NuSTAR} observations fully encompasses the synchrotron component over ${\sim}4$ orders of magnitude in frequency, which brings stronger constraints on the electron energy distribution (EED).\par 

In this work, we assume an emitting zone consisting of a spherical blob with radius $R'$ (primed quantities refer to quantities in the plasma reference frame) that is moving downstream of the jet with a bulk Lorentz factor $\Gamma_b$. The blob is homogeneously filled with electrons and is embedded in a homogeneous magnetic field $B'$. We assume a jet axis at an angle $\Theta = 1/\Gamma_b$. The advantage of this assumption is to reduce the number of degrees of freedom because, in this configuration, the Doppler factor $\delta$ becomes equal to $\Gamma_b$. For simplicity, the EED is characterised by a broken power-law (BPL), which is a very common parametrisation in blazar modelling:
\begin{equation}
    N'(\gamma')= \begin{cases}
    N'_0\, \gamma'{}^{-\alpha_1}, \quad \gamma'_{min}<\gamma'<\gamma'_{br}\\
    N'_0\, \gamma_{br}'{}^{\alpha_2-\alpha_1} \gamma'^{-\alpha_2}, \quad \gamma'_{br}<\gamma'<\gamma'_{max}\\
    \end{cases}
\end{equation}
where $N'_0$ is a normalisation constant. The corresponding electron energy density is given by $U'_e$ (in [erg cm$^{-3}$]). The dimensionless parameters $\gamma'_{min}$, $\gamma'_{br}$ and $\gamma'_{max}$ are defined as the minimum, break and maximum Lorentz factor, respectively. Because of the rather poor constraint on $\gamma'_{min}$ provided by the data, we fix it to $10^3$ to decrease the numbers of degrees of freedom and to be consistent with previous works on Mrk~421 \citep{2011ApJ...736..131A}. $\alpha_1$ and $\alpha_2$ are the index below and above the break Lorentz factor, respectively. We note that a pure power-law EED, which has the advantage of having fewer degrees of freedom, clearly fails at describing the SED, especially in the X-ray regime. Consequently, such a simpler functional form is not a viable solution, and the X-ray data require a significant break in the EED. Internal photoabsorption of the VHE photons caused by the interaction with synchrotron radiation is included following \citet{2008ApJ...686..181F}. Synchrotron self-absorption occurring in the radio domain is implemented following the prescription in \citet{2011hea..book.....L}. We employ routines from the open-source software \texttt{naima} to compute the synchrotron and IC interactions\footnote{A jupyter Python notebook describing how the routines from \texttt{naima} are used to compute the synchrotron and inverse Compton emission is available at \url{https://github.com/Axelarbetengels/Mrk421-2017-campaign-paper}} \citep{naima}. Here, as in all of the models presented in this study, we did not use a minimisation strategy, such as that provided in the \texttt{naima} package, but rather, we performed a simple “eye-ball fit” where we tried to find a parameterisation with sensible model parameters that agreed with the broadband SED.\par 

Prior to fitting the model, we fix the Doppler factor, $\delta$, to $\delta=25$. This is representative of the value typically found for HBLs \citep{2010MNRAS.401.1570T}. The size of the emitting region is constrained by the light crossing time $R' \leq \delta t_{var} c /(1+z)$, where $t_{var}$ is the variability time scale in the observer frame. Depending on the night, $t_{var} \approx 4-11$\,hours (see Appendix~\ref{NuSTAR_tables}), which is the flux halving time seen in the orbit-wise \textit{NuSTAR} light curves.\par 

In general, the model is able to describe well the data from optical to VHE for all the observing epochs. We remark that, on MJD~57813, a mismatch in the UV is apparent despite a good match from the X-rays to VHE. The UVOT measurement is under-predicted by a factor ${\sim}2$. Such a discrepancy may be attributed to the simplicity of the considered model, and could be better described by an EED with an additional break (two breaks instead of one). Alternatively, a higher $\gamma_{min}$ (close to $10^4$) together with a harder $\alpha_1$ also represents a viable solution to describe the UV/optical emission (not only for MJD~57813, but also for the other three nights), although this would be at the cost of an underestimation of the MeV-GeV SED measured by \textit{Fermi}-LAT. Finally, one may also consider an additional region that contributes to the emission between the UV and the soft X-ray bands. 

Moreover, as is typically found for blazars when considering one-zone SSC models, the radio flux is largely underestimated due to synchrotron self-absorption. Radio emission is most likely coming from regions outside of the inner jet, that have broader and complex morphology not included in our simple model \citep{2006ApJ...646..801G,2014A&A...571A..54L}. In our work, we mostly concentrate on emissions from optical to VHE, whose flux is believed to be dominated by a smaller region inside the jet.\par 

The \textit{NuSTAR} spectra extending to ${\approx}40$\,keV allow the parameters of the higher-energy part of the EED (i.e. $\alpha_2$ and $\gamma'_{max}$) to be well-constrained. In HBLs, these two parameters are usually difficult to constrain and show large degeneracy with typical SEDs covering the synchrotron emission only up to the soft and medium X-ray band \citep[see for example][]{2017A&A...603A..31A}. We find a value of $\sim10^{6}$ for $\gamma'_{max}$ for each of the four models, with their precise values all lying within a factor of two of each other.\par 

Regarding the break Lorentz factor $\gamma'_{br}$, we find that the values are in rough agreement with the expected cooling break $\gamma'_{br, exp}$ that is obtained by balancing the synchrotron cooling time scale with the fiducial adiabatic cooling time scale of $R'/c$ related to the expansion of the emitting zone, $\gamma'_{br, exp}=\frac{6 \pi m_e c^2}{\sigma_T B'^2 R'}$ \citep{1998ApJ...509..608T, 2001APh....15..121M}, where $\sigma_T$ is the Thomson cross section and $m_e$ the electron mass. Indeed, the modelling yields values that are at most a factor ${\approx}3.5$ away from $\gamma'_{br, exp}$. We note also that for most observing epochs the change of the index at the break, $\Delta \alpha = \alpha_2-\alpha_1$, is higher than the canonical synchrotron cooling break ($\Delta \alpha =1$) expected in models with a homogeneous emitting region \citep{2011hea..book.....L}. For instance, we find $\Delta \alpha \approx 1.6$ and $\Delta \alpha \approx 1.7$ for MJD~57757 and MJD~57813, respectively. On MJD~57840, $\Delta \alpha \approx 2$. As mentioned above, the spectral shape of the EED is well-constrained by the data and fixing $\Delta \alpha =1$ during the fit of these particular epochs would result in a worse description of the SED. A large spectral break is in fact a recurrent result found in the modelling of SED and has been already reported for Mrk~421 \citep{1998ApJ...509..608T,2011ApJ...736..131A,2016ApJ...819..156B}. This points to a more complex and less homogeneous emitting region. The canonical break condition may be loosened and larger values of $\Delta \alpha$ can be explained by considering velocity gradients in the jet \citep{2005A&A...432..401G}, for example.\par 

As an alternative to the BPL model for the EED, we also investigated a log-parabolic model with a low-energy power-law branch \citep[LPPL;][]{2006A&A...448..861M, 2009A&A...501..879T}. Several works have shown that such curved distributions may be produced through stochastic acceleration \citep{1962SvA.....6..317K, 2009A&A...501..879T, 2011ApJ...739...66T} or via an energy-dependent acceleration probability process \citep[EDAP;][]{2004A&A...422..103M}. We found that a LPPL satisfactorily describes the four simultaneous MAGIC/\textit{NuSTAR}/\textit{Swift} observations, with very similar results as those obtained with a BPL EED. The curvature of the LPPL model was also in good agreement with the observed curvature in the synchrotron SED (as derived in Sect.~\ref{sect:synch_peak}). The data do not show a clear preference between the BPL and the LPPL model. 

\section{Broadband SED and modelling of the intriguing VHE flare on MJD~57788}
\label{flare_modelling}

The flare detected on MJD~57788 deserves special treatment. It is characterised by a strong VHE flux increase: the FACT observation on the day before results in a flux of ${\approx}0.4$\,C.U., while a peak activity of ${\approx}7$\,C.U. is measured during the outburst. The VHE flux decays during the night on ${\sim}$hour time scale (see Fig.~\ref{flare_night_LC}). During the MAGIC observations the >1\,TeV flux is ${\approx}3.5$\,C.U.. The X-ray counterpart in the 0.3-10\,keV band is much less evident. In fact, Fig.~\ref{swift_vs_MAGIC} shows that the $>1$\,TeV flux level is about 10 times higher than other nights with a comparable 0.3-10\,keV flux. The finely-binned \textit{Swift}-BAT 15-50\,keV light curve displays interesting features in Fig.~\ref{flare_night_LC}. Simultaneous to the MAGIC highest state, the 15-50\,keV flux is around $3.5 \times 10^{-3}$\,cm$^{-2}$\,s$^{-1}$, while in the following hours (where we lack simultaneous VHE data) the flux is higher by a factor ${\sim}3$. This suggests a renewed flaring activity on ${\sim}$MJD~57788.7 after the decaying phase around MJD~57788. In the UV/optical and HE, no significant flux enhancement is visible around the outburst (see Fig.~\ref{MWL}).\par

\begin{figure}[h!]
   \hspace{-0.4cm}
   \includegraphics[width=1.1\columnwidth]{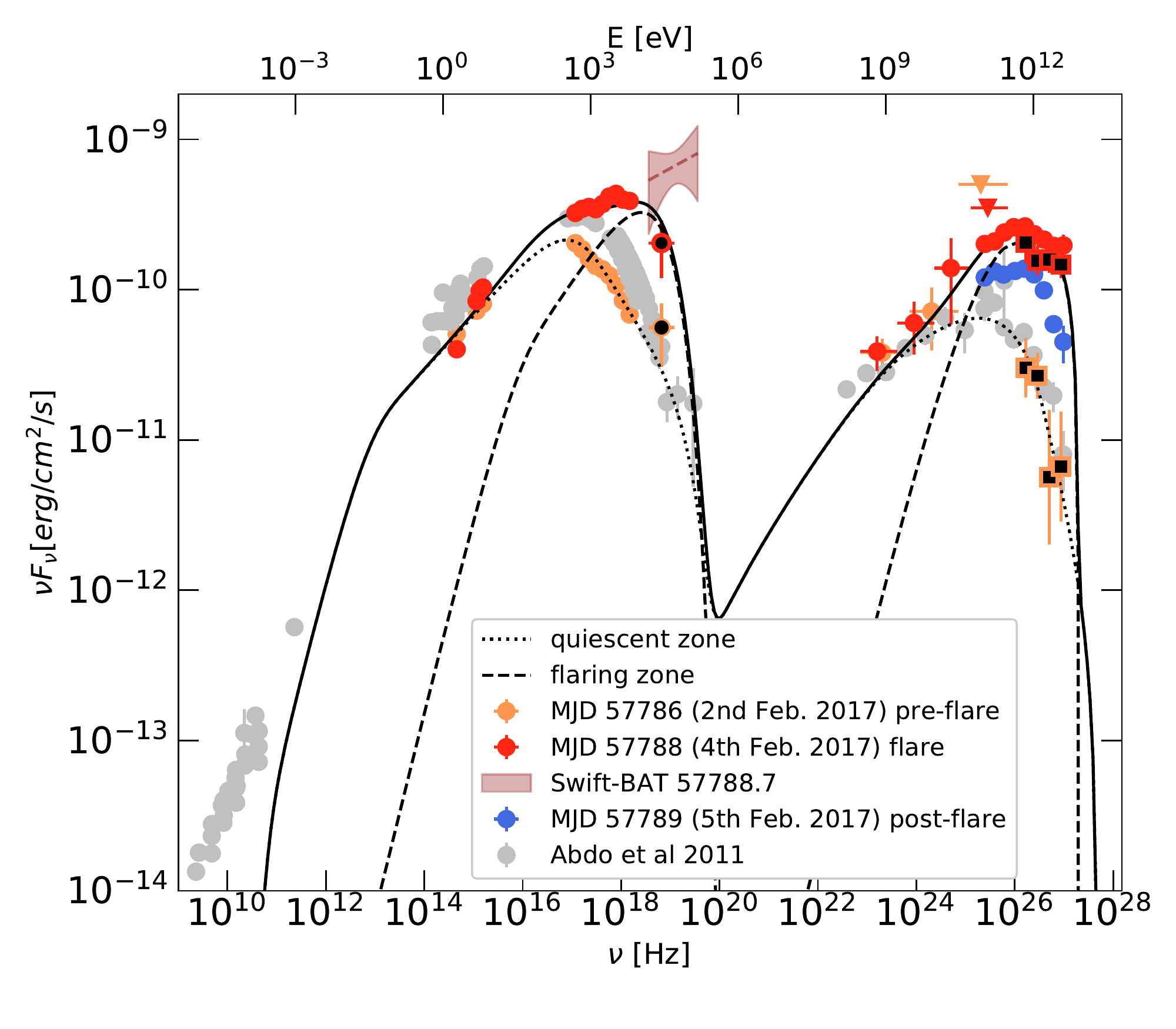}
   \caption{Simultaneous broadband SEDs of MJD~57786 (pre-flare state), MJD~57788 (flare), and MJD~57789 (post-flare). \textit{Fermi}-LAT spectral points are integrated over 3~days around the VHE measurements. VHE data with square black-filled markers are obtained from FACT observations, while X-ray data in black-filled markers are from \textit{Swift}-BAT. The FACT SED for the pre-flare state is averaged from MJD~57786 to MJD~57787. The full black line is the two-zone model for the MJD~57788 flare state. The black dotted line represents the emission from the quiescent zone while the dashed line is the one from the flaring zone. The model parameters are listed in Table~\ref{tab:ssc_parameters_2_zone}. Archival data representing the typical Mrk~421 state from \cite{2011ApJ...736..131A} are shown in grey.}
    \label{flare_sed}%
\end{figure}

Fig.~\ref{flare_sed} presents SEDs that illustrate the broadband evolution from MJD~57786 (2017 February 2) to MJD~57789 (2017 February 5). The SED on MJD~57786 is built around the closest \textit{Swift} observation in time before the flare and is dubbed the "pre-flare" state. In the VHE band, we use the FACT data averaged between MJD~57786 and MJD~57787 to improve the statistics since no significant flux variability is visible between these two nights (see Fig.~\ref{MWL}). The SED of the flare is plotted in red markers and uses the strictly simultaneous \textit{Swift}-XRT/\textit{Swift}-BAT/MAGIC observations. The butterfly from the \textit{Swift}-BAT best-fit power-law model around MJD~57788.7 is shown, which corresponds to the renewed activity in hard X-rays. As shown in Fig.~\ref{flare_night_LC}, no strictly simultaneous measurements are available at other wavebands. Finally, in blue markers we show the MAGIC SED measured on MJD~57789 (labeled as "post flare"), for which no simultaneous MWL data are available within less than a day. In each SED, the \textit{Fermi}-LAT data are averaged over three days centred on the VHE observation. Optical data are not strictly simultaneous but are located less than a day away from the other wavebands. For comparison the typical state of Mrk~421 \citep{2011ApJ...736..131A} is shown in grey. \par

\begin{table}[h!]
\caption{\label{tab:ssc_parameters_2_zone}
Parameters of the 2-zone SSC model shown in Fig.~\ref{flare_sed} during the flare of MJD~57788.} 
\centering
\begin{tabular}{l c c c }     
\hline\hline
 Parameters & quiescent zone & flaring zone\\
\hline
\hline
$\Gamma_b$ & 25 & 25 \\
$B'$ [10$^{-2}$~G]  & 6.1 & 16.5 \\
$R'$ [10$^{16}$~cm] & 1.6 & 0.1 & \\
$U'_e$ [erg cm$^{-3}$] & $3.4 \times 10^{-3}$ & $3.4 \times 10^{-1}$ \\
$\alpha_1$ & 2.3 & 2.0 \\
$\alpha_2$ & 4.0 & -- \\
$\gamma'_{min}$ [10$^3$] & 1.0 & 20 \\
$\gamma'_{br}$ [10$^5$] & 1.3 & -- \\
$\gamma'_{max}$ [10$^6$] & 1.5 & 0.6 \\
$U'_B/U'_e$ & $4 \times 10^{-2}$ & $3 \times 10^{-3}$\\
\hline
\end{tabular}
\tablefoot{See text for the description of each parameter. The EED of the quiescent zone follows a broken power-law (BPL) with indices $\alpha_1$ and $\alpha_2$ before and after the break Lorentz factor $\gamma'_{br}$. In the case of the flaring zone, a simple power-law function (with index $\alpha_1$) is adopted.} 
\label{tab:ssc_parameters_2_zone}
\end{table}

The softness of the pre-flare X-ray spectrum (\textit{Swift}-XRT power-law index of ${\approx}2.4$) suggests a synchrotron peak frequency around $10^{17}$\,Hz (${\approx}0.4$\,keV). An evident hardening occurs on MJD~57788, and the power-law index measured by \textit{Swift}-XRT is ${\approx}1.9$. The combined \textit{Swift}-XRT and \textit{Swift}-BAT measurements indicate that the synchrotron peak frequency is located at ${\sim} 10^{18}$\,Hz ($\sim4$\,keV). At VHE, the spectrum is hard. The flux at the highest energies ($>1$\,TeV) is one order of magnitude higher than the pre-flare level. The IC peak frequency $\nu_{IC}$ is around 0.4--0.5\,TeV, which is among the highest values for Mrk~421 \citep{2015A&A...578A..22A}. The SEDs reveal no substantial increase in the UV/optical, where the flux varies by only 15-20\%. The ${\approx} 2\times 10^{17}$\,Hz (${\approx}1$\,keV) flux remains at the level of the typical Mrk~421 activity (grey points in Fig.~\ref{flare_sed}). Also, the flux at the low-energy peak frequency increases only by a factor ${\sim}2$, while that at the high-energy peak frequency seems to exhibit a factor ${\sim}4$ enhancement.\par

The question is what scenario could generate such a sudden change in the Compton dominance with respect to the pre-flare state assuming a one-zone SSC mechanism. In the synchrotron regime, the observed hardening without strong changes in the peak amplitude suggest modifications of spectral parameters from the high-energy part of the EED, i.e., parameters related to electrons emitting synchrotron photons significantly above $\nu_s$. In particular, it suggests a hardening of $\alpha_2$ and/or an increase of $\gamma_{max}'$ or $\gamma'_{br}$ to push $\nu_s$ to higher energies. The shift by ${\sim}$one order of magnitude in $\nu_s$ constrains $\gamma'_{br}$ to change by a factor ${\sim}3$ at most ($\nu_s \propto \gamma_{br}'{}^2$), while the spectral hardening of ${\sim}0.5$ seen by \textit{Swift}-XRT constrains a hardening of $\alpha_2$ by ${\sim}1$ at most. However, mainly because of the onset of the Klein-Nishina suppression, these modifications are not sufficient to explain the large increase of the VHE flux. This is especially true at $\gtrsim1$\,TeV (i.e., deep in the Klein-Nishina regime) where the spectrum is particularly hard and the flux is about a factor of 10 higher than in the pre-flare state. Furthermore, one should note that the high-energy electrons emitting synchrotron photons above $\nu_s$ radiate IC emission above ${\sim}$1-10\,TeV, while the flare is already clearly visible at the low-energy end of the VHE spectra, around a few 100\,GeV.\par 

Modifications of the electron density would be needed to enhance the IC luminosity. This would require a rather fine tuning of other parameters (e.g., the magnetic field or the blob radius) in order to keep the UV/optical and ${\approx}1$\,keV flux at an almost constant level. Since changes in the spectral shape of the EED are not sufficient to explain the hardness at VHE as explained above, a variation of $\delta$ would also be required to push the IC component to higher energies. We conclude that a one-zone SSC model demands too much fine-tuning to explain the transition in state from MJD~57786 to MJD~57788 and hence is not a good scenario.\par

Another scenario would be the appearance of a second emitting zone, in addition to the quiescent zone responsible for the pre-flare state. The variability time scale (${\sim}$hours) constrains the second emitting zone to be more compact, i.e., $R'\lesssim10^{15}$\,cm. The absence of a UV/optical flare also suggests a more energetic and narrower EED. In this way, the second zone would suddenly dominate in the hard X-ray domain and in the VHE regime, and would remain subdominant in the rising segments of the two SED components, which would naturally explain the observations. This two-zone SSC scenario was employed to describe a flaring activity of Mrk\,421 in 2010, as reported in \citet{2015A&A...578A..22A}.

We tried to perform a similar modelling by assuming two spherical emission zones spatially separated such that their respective radiation fields and particle populations do not interact with each other. In this scenario, the quiescent zone is responsible for the broadband emission during the pre-flare state on MJD~57786 (as well as the previous days), while the flaring zone dominates the hard X-ray and >GeV emission during the flare. The EED for the quiescent zone follows a BPL distribution, while in the case of the flaring zone a simple PL function is adopted. Just as in Sect.~\ref{sect:SED_modelling}, the jet axis angle is $\Theta$=1/$\Gamma_{b}$ such that $\Gamma_{b}=\delta$. To limit the number of degrees of freedom, the bulk Lorentz factor of the flaring zone is equal to the one of the quiescent zone ($\Gamma_{b}=25$). The resulting models are shown in Fig.~\ref{flare_sed}. The corresponding parameters are listed in Table~\ref{tab:ssc_parameters_2_zone}. Both the pre-flare and the flare are well described by the models, although the UV/optical slope during the flare seems slightly harder than predicted by the model. The data indeed provide an indication of a hardening in the UV/optical during the transition from the pre-flare to the flare state, yet the flux difference is only at the level of 15-20\%, which remains a rather mild effect. As mentioned in the previous paragraph, the flaring zone is characterised by a more energetic and narrower EED with respect to the quiescent zone to ensure that it dominates only in the falling edge of two SED components. The lowest electron energies of this second region are constrained by the measured emission at $\sim$keV and $\sim$100~GeV, while the highest electron energies are constrained by the multi-TeV emission measured by MAGIC and the lack of strong X-ray emission above 20~keV, as indicated by the {\it Swift}-BAT measurement simultaneous to the XRT and MAGIC spectra (see Fig.~\ref{flare_sed}). The model that we use to successfully describe the spectral measurements uses $\gamma'_{min}=2 \times 10^4$ and $\gamma'_{max}=6 \times 10^5$, and  the radius of the flaring zone is $R'=10^{15}$\,cm, consistent with the rapid variability on hour time scales at VHE.\par 

It can be seen from Fig.~\ref{flare_night_LC} that the SED on MJD~57788 lies in the decay phase of the flare. Fig.~\ref{flare_sed} includes the \textit{Swift}-BAT butterfly a few hours later on MJD~57788.7, during which a rapid renewed activity in the hard X-rays (15-50\,keV) is observed. The best-fit power-law index is $\Gamma=1.82 \pm 0.42$, implying a synchrotron peak frequency roughly in the \textit{Swift}-BAT pass-band. On MJD~57789, the measured \textit{Swift}-BAT fluxes are very low (compatible with no signal), and the VHE flux is around 1.5 C.U.. Given the fast variability and the lack of strictly simultaneous MWL data during these two epochs, we do not model the emission.\par

%

\section{Discussion}
\subsection{VHE versus X-ray correlation}
\label{sect:X-ray_vhe_correlation_discussion}
We confirm the existence of a significant correlation between the X-ray and the VHE emission, hence indicating a cospatial origin. Additionally, the \textit{fractional variability} displays a two-peak shape with the highest variability both in the X-rays and VHE. These results suggest a single population of particles responsible for the emission in those bands. We remark that when using the simultaneous MAGIC/\textit{NuSTAR}/\textit{Swift} observations, $F_{var}$ is significantly higher at VHE gamma rays than at X-rays, as clearly shown in Fig.~\ref{Frac_var} and Fig.~\ref{Frac_var_nustar_magic}, as well as in Fig.~\ref{NUSTAR_MAGIC} and Fig.~\ref{NuSTAR_nights_sed}. These results differ from what was reported during the very low activity from the first months in the year 2013 \citep{2016ApJ...819..156B}, but they are in agreement with the variability patterns measured during the extremely low activity in the observing campaign from the year 2016 \citep[see Fig.~6 in][]{Mrk421_2015-16}. On the other hand, during large X-ray and VHE gamma-ray activity, the fractional variability at VHE gamma rays is equal to or lower than that measured at X-rays \citep{2020ApJ...890...97A,2020ApJS..248...29A}.\par

The correlations show a complex behaviour when different energy bands are compared. In fact, the data point to a correlation slope that depends on the spectral band (see Fig~\ref{NUSTAR_MAGIC} and Fig~\ref{swift_vs_MAGIC}). Thanks to the sensitive MAGIC measurements, we see a trend of higher slope with increasing VHE energies (for any given X-ray band). Such a behaviour was already noted by \citet{2020ApJS..248...29A}, when the source was strongly flaring. Here, the source was probed at a close-to-typical state. This pattern suggests a more direct relation of the >1\,TeV flux with the X-rays compared to the 0.2-1\,TeV flux. Considering a SSC model and assuming a generic magnetic field strength of $0.1$\,G as well as a Doppler factor of $25$, the ${\gtrsim}1$\,keV flux is dominated by electrons with Lorentz factors of $\gamma' \approx 2 \times 10^{5}$ \citep{1998ApJ...509..608T}. Through IC processes, these same electrons dominate the $\gtrsim$\,TeV flux. Given the X-ray energies considered in this work, the observed trend is somewhat expected and is consistent with pure leptonic models.\par 

For most of the energy combinations, the correlations are close to linear or even less during the MAGIC/\textit{NuSTAR}/\textit{Swift} observations. Nevertheless, we also find steep slopes. The correlation is quadratic or cubic in the VHE versus $0.3-3$\,keV case. \cite{2005A&A...433..479K} evaluated the VHE/X-ray correlation in BL Lac type objects within a one-zone SSC model using different scenarios for the parameters evolution. In most of the cases, the VHE versus X-ray correlation is expected to follow a linear trend rather than a quadratic trend (or cubic). It should be noted that a more-than-linear correlation may still be possible but under rather specific conditions and/or evolution of the parameters. One of the main reasons is that the IC processes responsible for the VHE emission are expected to occur in the Klein-Nishina regime, where the interaction cross section of electrons with seed photons above the synchrotron peak frequency $\nu_s$ is strongly suppressed. As a result, the VHE-emitting electrons do not efficiently up-scatter their self-produced synchrotron photons, but photons with lower energies, in the UV/optical. The criteria for having an IC luminosity and peak frequency affected by Klein-Nishina suppression is $\gamma'_{br} \nu_{s} \geq \delta\frac{3 m_e c^2}{4 h}$ \citep{1998ApJ...509..608T}. The SEDs in Fig.~\ref{NuSTAR_nights_sed} show synchrotron peak frequencies in the range of $\nu_s \sim 10^{16} - 10^{17}$\,Hz and the break Lorentz factors obtained from the modelling are $\gamma'_{br} \approx 0.8 \times 10^{4}-2 \times 10^{5}$. Thus, the condition for having significant Klein-Nishina suppression is fulfilled at least for the three first observations. However, \cite{2005A&A...433..479K} argue that steeper correlation may also be obtained under a specific choice of the spectral bands, even taking into account Klein-Nishina effects. A quadratic and more-than-quadratic trend can occur when selecting X-ray energies close to or below the synchrotron peak, which is the case for the 0.3-3\,keV band in Mrk~421. In conclusion, in the specific case of VHE versus 0.3-3\,keV, the very steep correlation is possibly caused by the selection of the spectral bands. Moreover, Mrk~421 showed a displacement of the synchrotron peak frequency of two orders of magnitude throughout the campaign (see Fig.~\ref{sect:synch_peak}), and this may also contribute to the modifications in the correlation slopes.\par

Over the full MWL campaign, the best-fit slopes indicate a roughly cubic relation for >1\,TeV versus 0.3-3\,keV, while an almost quadratic trend is seen for 0.2-1\,TeV versus 0.3-3\,keV and >1\,TeV versus 3-7\,keV (see Fig.~\ref{swift_vs_MAGIC} and Table~\ref{tab:correlation_results_lonterm}). In the latter situation (>1\,TeV versus 3-7\,keV), for the reasons mentioned in the previous paragraph, a long-term quadratic correlation is rather surprising since we are considering synchrotron flux emitted above $\nu_{s}$ (at least for the vast majority of the measurements, see Fig.~\ref{synch_peaks}). The steep slope cannot be attributed to the choice of an X-ray band below the peak, as proposed above. Nonetheless, the quadratic slope is not representative of the average behaviour during the campaign. The best-fit slope is biased towards higher values by a small fraction of measurements ($10\%$), including two of the simultaneous MAGIC/\textit{NuSTAR}/\textit{Swift} observations and the short flare on MJD~57788. These three observations show a significantly enhanced VHE activity compared to nights with similar X-ray flux, and they appear as clear outliers in Fig.~\ref{swift_vs_MAGIC}. With the exception of these three days with large Compton dominance, the overall trend is well consistent with a ${\sim}$linear relation, as can be seen in Fig.~\ref{swift_vs_MAGIC}.\par 

The mentioned outliers suggest sudden emission phases (on time scales of about a few days) that show very different correlation patterns compared to more typical states. This translates into a particularly large scatter of the measurements in the correlation plots. Overall, the simple description of the correlation in the form $F_{VHE} \propto F_{X-ray}^x$ is overly simplistic, as indicated by the poor $\chi^2/dof$ values. The reason for the scatter, that can also be seen at lower X-ray and VHE fluxes in Fig.~\ref{swift_vs_MAGIC}, may be the consequence of several additional processes contributing to a dominant process driving the overall variability. A large scatter could also be explained by different emitting regions (from different parts of the jet) that dominate the X-rays and VHE bands for a given time period. It is important to stress that, in this study, we are not considering a specific flaring episode (that is most likely caused by the same emitting zone in the jet), but rather the nightly averaged flux spread over monthly time scales that possibly includes several small flaring episodes from different components of the jets. These regions may undergo changes in their physical environments, namely related to $B'$, $R'$, $\delta$ or $U'_e$, which could result in a disparate Compton dominance in the SED \citep{2010A&A...510A..63K}. For instance, a modification of the blob radius and/or electron density would modify the synchrotron photon target density for the SSC processes and increase or decrease the Compton dominance. A different Compton dominance naturally induces a large scatter in the VHE versus X-ray plot, giving rise to an apparent break in the correlation slopes. The modelling of the first three simultaneous MAGIC/\textit{NuSTAR}/\textit{Swift} observations (on MJD~57757, MJD~57785 and MJD~57813), which can indeed be explained by the variation of a few parameters related to the environment (see later in Sect.~\ref{sect:modeling_discussion}), is consistent with this hypothesis. On the other hand, while a one-zone SSC model is able to explain the MAGIC/\textit{NuSTAR}/\textit{Swift} observations, a deeper investigation of the outlier related to the MJD~57788 flare challenges the simple leptonic scenario (see Sect.~\ref{flare_modelling}).\par

\subsection{UV/optical versus X-ray anti-correlation}
\label{sect:optical_X-ray_anti_correlation_discussion}

We find an anti-correlation between UV/optical and X-ray at a significance level of above $3 \sigma$ over the entire MWL campaign (see Sect.~\ref{sec:UV-X-ray} for details). We also find that the strength and the significance of the anti-correlation is dominated by the observations during the first 40 days of the multi-instrument campaign (i.e., data taken before MJD~57760). This indicates that the anti-correlation between these two bands is not persistent; it varies over time and may become significant only over month timescales. A first indication of an anti-correlation was reported by \cite{2015A&A...576A.126A}, also for Mrk~421. The anti-correlation was marginally significant over a few months timescales, and over time lags ranging from 0 to -20\,days, in agreement with what is found here. It was, however, not observed in other multi-instrument campaigns when Mrk~421 did not show strong flaring activity in X-rays \citep{2016ApJ...819..156B,Mrk421_2015-16}. During a bright outburst of Mrk~421 in 2006 June, which lasted about 2 weeks, \citet{2008A&A...486..721L} reported a hint of positive correlation between the optical and X-ray fluxes. However, the particularly low significance of the correlation did not allow a conclusive claim. In 2010 February, during the brightest VHE gamma-ray activity of Mrk~421 detected to date, a marginally significant positive correlation of about 3$\sigma$ was observed between the VHE gamma rays and the optical flux in the data taken on 2010 February 17, the day with the highest VHE gamma-ray flux \citep{2020ApJ...890...97A}. Unfortunately, there were no X-ray observations simultaneous to the VHE observations, and hence the X-ray versus optical correlation could not be studied for that night. But owing to the very tight correlation that Mrk~421 always shows between the X-ray and the VHE gamma-ray bands, it is reasonable to assume that, on that night with outstandingly large VHE gamma-ray activity, the X-rays and optical emission may have been positively correlated. On the other hand, when considering the few-week-long dataset from 2010 February, no correlation is observed between the X-ray and optical fluxes, in contrast to what was reported in \citet{2008A&A...486..721L}. Moreover, in 2013 April, during the second brightest VHE gamma-ray activity from Mrk~421 detected to date, which included several tens of hours of strictly simultaneous optical, X-ray and VHE gamma-ray observations, Mrk~421 did not show any correlation between X-ray and optical, despite showing large variability and a high degree of correlation between VHE and X-rays \citep{2020ApJS..248...29A}.\par 

To the best of our knowledge, Mrk~421 is the only BL Lac type object where an anti-correlation between these two segments of the SED has been observed to date. This second instance of this characteristic, which we presented and described here with better sampled observations than those presented in \citet{2015A&A...576A.126A}, suggests that this anti-correlation pattern, visible over a few month timescales, is a recurrent feature with possibly a real physical origin in the synchrotron emission of BL Lacs. \par

The first and direct implication is that the synchrotron emission is dominated by a cospatial population of electrons, at least from the UV/optical to the X-ray. One may interpret the anti-correlation as being due to a change in the efficiency of the electron cooling processes. Indeed, assuming a constant acceleration timescale, a stronger cooling shifts the overall EED towards lower energies, resulting in a reduction of the emission in the X-rays, while the UV/optical flux increases. The electron cooling (expected to be dominated by synchrotron radiation in case of HBL where the emitted synchrotron power is larger or at least comparable to the gamma-ray component, see \cite{2010A&A...519A...9S}) may be increased by a stronger magnetic field $B'$. Alternatively, as suggested by \cite{2015A&A...576A.126A}, the EED may also be shifted towards lower or higher energies in case of changes in the acceleration efficiency.\par 
  
The large variability of the synchrotron peak frequency during the campaign further supports changes in the cooling or acceleration mechanisms. For a given electron population, the synchrotron peak frequency evolves as $\nu_s \propto \gamma_{br}'{}^2 \cdot B' \cdot \delta$ \citep{1998ApJ...509..608T}. The variability of $\nu_s$ by about two orders of magnitude, that is reported in Section~\ref{sect:synch_peak}, disfavours $\delta$ as being the main driving parameter, as it would imply a variation by about 2 orders of magnitude, leading to unphysical and much larger values usually found in TeV BL Lac type objects \citep[e.g.,][]{2010MNRAS.401.1570T}. Moreover, a change in $\delta$ would lead to a positive correlation, in contradiction to the observed anti-correlation pattern. Therefore, the large changes in the peak synchrotron frequency are likely produced by changes in $B'$ and $\gamma'_{br}$, that are in turn parameters linked to acceleration and cooling mechanisms. However, while the shift in the synchrotron peak frequency is typical in Mrk~421, and particularly extreme in the 2017 observing campaign, the correlations between optical and X-rays are extremely rare. It indicates that the $\nu_s$ shifts are in most of the cases produced by the appearance and disappearance of emission at hard X-rays without affecting substantially the low-energy emission (e.g., optical and below). This could be produced, for instance, by the acceleration and cooling of the highest-energy electrons without a substantial change in the lowest-energy electrons, or perhaps by the time-variable contribution of an additional component that dominates the emission at the hard X-rays \citep[e.g.,][]{2015A&A...578A..22A}. Only in a few cases (e.g., during the first 40 days of this campaign, as shown in Fig.~\ref{DCF_xray_optical_before_57760}) there seems to be a tight relation between the optical-emitting electrons and the X-ray-emitting electrons, which could be caused by a full shift of the entire EED to higher or lower energies. As shown in Fig.~\ref{nu_peak_LC}, the synchrotron peak frequency $\nu_s$ did increase continuously by about two orders of magnitude (from $10^{16}$~Hz to $10^{18}$~Hz) during the first $\sim$40 days of the campaign, when the optical vs X-ray anti-correlation is most evident. Such a large increase in $\nu_s$ cannot be produced by a simple change of the magnetic field, since this would imply an extreme variation by two orders of magnitude of $B'$. Hence, under the assumption that a single zone is responsible for both the optical and X-ray emission, one must have the contribution from acceleration processes (possibly in shocks and turbulence) and cooling processes (e.g. synchrotron, inverse-Compton, adiabatic cooling) to shape an EED whose effective radiation yields the continuous increase in X-ray emission (by a factor of $\sim$10) and in $\nu_s$ (by a factor of $\sim$100), while decreasing the overall emission at optical/UV frequencies (by a factor of $\sim$2).\par 

\subsection{Interpretation of the emission during the simultaneous MAGIC/\textit{NuSTAR}/\textit{Swift} observations}
\label{sect:modeling_discussion}
The broadband emission during the simultaneous MAGIC/\textit{NuSTAR}/\textit{Swift} observations reveals intriguing behaviours. The most striking feature is visible when comparing the first three observations: while the synchrotron flux level is very similar, the corresponding Compton dominances $A_C$ are significantly different. Namely, on MJD~57757 and MJD~57785 $A_C$ is about three times higher with respect to MJD~57813. In fact, MJD~57757 and MJD~57785 appear as clear outliers in the correlation plots of Fig.~\ref{swift_vs_MAGIC}. We note that the X-ray and VHE spectral parameters between those nights are very similar as shown in Table~\ref{tab:MAGIC_spectral_param_nustar_sim} and Table~\ref{tab:Nustar_spectral_param_nustar_sim}. Within one-zone SSC models, this points towards an electron population with similar spectral characteristics. Hence, the difference in the Compton dominance $A_C$ does not seem to be caused by acceleration and/or cooling mechanisms, as this would inevitably impact the particle distribution and result in a variation of the spectral properties.\par  

We interpret the simultaneous MAGIC/\textit{NuSTAR}/\textit{Swift} observations within a simple SSC model, which successfully describes the data. Benefiting from a wide energy coverage in the X-rays thanks to the combined \textit{NuSTAR}/\textit{Swift}-XRT data, the EED is rather well constrained. According to the expectations mentioned above, the EED spectral parameters are comparable between the first three nights. Furthermore, the magnetic field is almost constant between those nights, and $B'\approx0.06-0.07$\,G. In the Klein-Nishina regime, as is most likely the case here (see Sect.~\ref{sect:X-ray_vhe_correlation_discussion}), the ratio of the peak frequencies of the two emission components relates as $\frac{\nu_s}{\nu_{IC}^2} \propto \frac{B'}{\delta}$. Under the assumption of a roughly constant Doppler factor, the absence of significant variability in $\nu_s$ and $\nu_{IC}$ thus supports the constant behaviour of the magnetic field. The divergence in $A_C$ is reconciled by a change in the emission zone radius $R'$ and the electron energy density $U'_e$: on MJD~57813, $R'$ is increased by a factor $1.65$ and $U'_e$ is reduced by a factor ${\approx}4$ compared to MJD~57757 and MJD~57785. The lower electron density reduces the target synchrotron photon field, resulting in a reduction of the IC flux. It is important to note that the total number of electrons in the emitting zone, which is proportional to $U'_e R'^3$, is almost constant between the three nights. In this sense, the difference in the modelling parameters is dominated by a simple increase of the radius, which could happen due to a natural expansion of the emitting region, in the absence of sufficient pressure to constrain it to a given physical size. Under the latter hypothesis of an adiabatic expansion (without significant particle loss) and that the same emitting blob is responsible for the emission during the MAGIC/\textit{NuSTAR}/\textit{Swift} observations, it is then difficult to attribute the break in the EED to the cooling of the electrons (estimated by equating the adiabatic expansion time scale with the synchrotron losses, see Sect.~\ref{sect:SED_modelling}). Indeed, given that the MAGIC/\textit{NuSTAR}/\textit{Swift} observations are separated by $\approx30$\,days, the adiabatic expansion must also happen on similar time scales, which would result in a cooling break located at much lower Lorentz factors than the values of $\gamma'_{br}$ found in the modelling. Adiabatic expansion on ${\sim}$weekly-monthly time scales may still be possible if one assumes that the emitting blob has a size similar to the cross-sectional radius of the jet and that the blob expansion is driven by its movement downstream in a conical jet structure. In this scenario, the adiabatic expansion time scale (co-moving frame) is given by $t'_{ad}=\frac{3}{2}\frac{R'}{\theta \Gamma_b c}$, where $\theta$ is the half-opening angle of the jet \citep{2003A&A...406..855M}. Assuming $\Gamma_b \sim 10$ and $\theta\sim0.1^\circ$, which is in agreement with studies of parsec scale studies of blazar jets \citep{2005AJ....130.1418J}, one can obtain a value for $t'_{ad}$ which matches the weekly-monthly time scales measured in the observer’s frame. The break in the EED may thus be attributed to the intrinsic properties of the acceleration processes. On MJD~57840, the synchrotron component is clearly shifted towards lower energies, which points towards a change in the EED. Accordingly, the break Lorentz factor obtained is lower with respect to the other three MAGIC/\textit{NuSTAR}/\textit{Swift} nights, possibly indicating a decrease in the efficiency of the acceleration processes.\par 

In all models we use a minimum Lorentz factor fixed to the fiducial value $\gamma'_{min}=10^3$. At such low energies, the dominant cooling process is most likely the adiabatic one, which may lead to a decrease of the electron energy below \mbox{$\gamma'_{min}=10^3$.} The fact that the same high value of $\gamma'_{min}$ is used for the four models, that are separated in time by ${\approx}30$\,days, suggests a continuous re-acceleration of the electrons in the emitting zone. \citet{2011ApJ...736..131A} and \citet{2016ApJ...819..156B} also modelled the broadband emission of Mrk\,421 (for timescales of months and days) with $\gamma'_{min} \sim 10^3$, and stated the need for in-situ electron re-acceleration to fully explain these data sets.   

The one-zone SSC modelling suggests a significant change of the Compton dominance caused by a modification of only a few parameters related to the jet environment. This provides a natural cause of the rather large scatter in the flux-flux correlation plots in Fig.~\ref{NUSTAR_MAGIC} and Fig.~\ref{swift_vs_MAGIC}.\par

The change in the relative strength of the two SED components could also be interpreted within the framework of external Compton models, in which an additional target photon field for the IC radiation is invoked. A possible approach would be the spine-layer model developed in \citet{2005A&A...432..401G}. In the latter, the jet is assumed to be composed of a central part, the spine, surrounded by a layer. The bulk Lorentz factor of the spine ($\Gamma_S$) exceeds that of the layer ($\Gamma_L$). The synchrotron and IC emissions of the spine itself dominate over that of the layer. Due to their relative motions, the radiation of one region as observed in the reference frame of the other is seen boosted by a factor ${\sim}\Gamma'^2$, where $\Gamma'=\Gamma_S \Gamma_L (1-\beta_S \beta_L)$. This boosted radiation provides additional seed photons to each region and their respective IC luminosities are significantly enhanced, in particular the one from the spine. In this scenario, changes of $\Gamma_L$ or the radiation density of the layer would therefore qualitatively explain the variability in the Compton dominance without modifying the synchrotron luminosity. One further advantage of the spine-layer model over the SSC model is being able to reach a system that is close to equipartition, i.e. $U'_e \sim U'_B$, as shown by \citet{2016MNRAS.456.2374T}. We note on the other hand that adding a second region (the layer) doubles the number of free parameters and the data at hand are not sufficient to constrain the model well (differently from the SSC one-zone model).\par

\subsection{Theoretical interpretation of the intriguing VHE gamma-ray flaring activity on MJD~57788}

The bright VHE flare on MJD~57788 reaches a peak flux of $\approx7$\,C.U., but it only shows a moderate activity in the X-rays. At lower energies, the UV/optical flux exhibits low variability ($\approx 20\%$) and no substantial increase with respect to the day prior to the flare. These MWL characteristics are difficult to explain with a one-zone SSC scenario, and lead us to propose a two-zone SSC scenario to describe the broadband behaviour. As described in Sect.~\ref{flare_modelling}, the two zones are spatially separated so that they do not interact with each other, with one region dominating the regular (quiescent or slowly varying) broadband emission, and another zone that dominates the flux enhancement observed at hard X-rays and VHE gamma rays.

Within this theoretical scenario, the non-variable UV/optical and MeV/GeV emission could be produced in a shock-in-jet component of relatively typical size dimensions ($R'\sim10^{16}$cm), while the X-rays and VHE gamma rays, that show large variability during these days, could be dominated by the emission from a region that is smaller by about one order of magnitude ($R'\sim10^{15}$cm). This small region could originate in the base of the jet and produce an EED characterised by a very high minimum Lorentz factor $\gamma'_{min}$ and with a narrow range of energies  ($\gamma'_{max}/\gamma'_{min}=30$). The value $\gamma'_{min}$ needs to be above $10^4$ to avoid the overproduction of the keV flux and the 0.1~TeV flux. A similar two-zone SSC scenario was also used in \cite{2015A&A...578A..22A} to describe the temporal evolution of the broadband emission  of Mrk~421 during a 2-week flaring activity in 2010 March. This flaring activity contained several days with narrow SED peaks, and the two-zone scenario with a narrow EED (from $\gamma'_{min}=3\times10^{4}$ to $\gamma'_{max}=6\times10^{5}$) could describe the shape of the (narrow) X-ray and VHE bumps better than the one-zone scenario.\par 

These narrow EEDs may arise through stochastic acceleration by energy exchanges with resonant Alfven waves in a turbulent medium yielding to the production of quasi-Maxwellian distributions of particle energies \citep{1985A&A...143..431S,2008ApJ...681.1725S,2014ApJ...780...64A}. An alternative way to produce narrow EEDs is through the emission resulting from an electromagnetic cascade initiated by electrons accelerated to a narrow range of energies in a magnetospheric vacuum gap, as proposed by various authors \citep{2011ApJ...730..123L,2016A&A...593A...8P,Wendel2021}, and used successfully to explain an extremely narrow spectral component detected (at marginally significant level of 3--4$\sigma$) in the VHE spectrum of Mrk~501 during a large flaring activity in 2014 July \citep{2020A&A...637A..86M}. Another scenario that could lead to narrow EEDs is magnetic reconnection, which has been invoked as an efficient particle acceleration process in AGN jets \citep{1992A&A...262...26R,2010MNRAS.402.1649G,2013MNRAS.431..355G}. Blobs of magnetised plasma containing high-energy particles could be formed in the reconnection regions of jets and lead to high-energy emission, as proposed in the semi-analytic model from \cite{2016MNRAS.462.3325P} and demonstrated in by dedicated particle-in-cell simulations \citep{2019MNRAS.482...65C,2020MNRAS.492..549C}. This theoretical framework was recently used to describe the temporal and spectral properties of the multiband flares that Mrk~421 showed in 2013 April \citep{2020ApJS..248...29A}. Through magnetic reconnection, the dissipated magnetic energy would be converted into non-thermal particle energy, hence leading to a decrease in the magnetic field strength $B'$ for increasing gamma-ray activity, and hence leading to a ratio $U'_B/U'_e$ as low as $10^{-3}$, which is needed to explain the measured broadband SED from MJD~57788.\par 

The requirement for a narrow EED is linked to our assumption that the distribution follows a power law with index $\alpha_1=2$ (see Table~\ref{tab:ssc_parameters_2_zone}). It is a generic assumption common in blazar modelling that is supported by the prediction of magnetic reconnection \citep[e.g.,][]{2014ApJ...783L..21S,2015MNRAS.450..183S} and also by standard shock wave acceleration mechanisms \citep{2019MNRAS.485.5105C}. As an alternative to the narrow EED in the flaring zone, a distribution of electrons spanning a wide range of Lorentz factors but identified by a hard power-law index less than 2 could reproduce the SED. In such a configuration, the flaring zone would remain subdominant in the UV/optical and the MeV/GeV regime even if the EED extends below $\gamma'_{min}\sim10^3$. \citet{2014ApJ...783L..21S} reported clear evidence that magnetic reconnection mechanisms can easily generate electron distributions following a power-law index harder than 2 in the regime where the magnetisation of the blob is larger than 10.\par


\section{Summary}

We have reported on a dense MWL observing campaign of Mrk~421 carried out between 2016 December and 2017 June. The MWL dataset comprises more than 10 instruments, providing information from radio (with OVRO, Medicina and Mets{\"a}hovi) to VHE gamma rays (with FACT and MAGIC), and including various instruments covering the optical and UV bands (e.g. GASP-WEBT and \textit{Swift}-UVOT), X-ray bands (\textit{Swift}-XRT and \textit{Swift}-BAT and \textit{NuSTAR}) and GeV gamma rays (with \textit{Fermi}-LAT). Owing to the inclusion of \textit{NuSTAR} data, we obtained a precise characterisation of the hard X-ray emission thought to originate from the high-energy tail of the same population of electrons dominating the VHE gamma-ray emission. This helped us to interpret the measurements within a standard SSC scenario.\par

The fractional variability versus energy showed the typical double-bump structure, observed in other campaigns. However, in contrast to many other MWL campaigns, when using strictly simultaneous data, the variability in the VHE gamma-ray domain was measured to be larger than that in the X-ray energy range. We found that the VHE gamma rays and X-rays are positively correlated with no time lag, but the strength and characteristics of the correlation change substantially across the various energy bands probed. The multi-instrument light curves and the broadband SEDs showed a large increase in the gamma-ray activity without a clear counterpart in the X-ray range. These {\em orphan} VHE gamma-ray activities, present in only a few of the observations (less than 10\%), yielded quadratic and cubic dependencies in the VHE versus X-ray flux relations. Removing these few measurements, the relations become linear or sub-linear, in agreement with previous observations \citep{2015A&A...578A..22A,2020ApJS..248...29A,Mrk421_2015-16}, and easier to explain with standard SSC models. We showed that a one-zone SSC scenario with an expanding blob (change of the size of the emission blob without changing the number of electrons) can explain the decrease in the Compton dominance of the broadband SEDs during the MAGIC/\textit{NuSTAR}/\textit{Swift} observations, which show orphan gamma-ray activity.\par

The manuscript reports a substantial {\em harder-when-brighter} behaviour in both the gamma rays and X-rays, including displacements in the peak of the synchrotron bump by more than 2 orders of magnitude in energy. We also show an anti-correlation between UV/optical and X-ray at a significance above $3\sigma$. This is the second time that such a trend is observed in Mrk~421, hence indicating that it is a repeating feature with a real underlying physical mechanism. This might be due to a change in the efficiency of the electron cooling or acceleration processes.\par

The manuscript also discusses an intriguing VHE flare observed on MJD~57788. In the synchrotron regime, the latter coincides with a clear hardening of the 0.3-10\,keV spectrum with respect to the pre-flare state, but the flux around ${\sim}$1\,keV remains close to the typical Mrk~421 quiescent activity. At VHE, the flare is strong and the flux above 1\,TeV is roughly 10 times higher than the typical quiescent activity. Within simple one-zone SSC models, this broadband behaviour is difficult to reproduce without a fine tuning of the parameters. We therefore interpret the flare as being caused by the appearance of a more compact second blob of highly-energetic electrons that span a relatively narrow range of energies (from $\gamma'_{min}=3\times10^{4}$ to $\gamma'_{max}=6\times10^{5}$) and could have been produced by stochastic acceleration, by magnetic reconnection or by electron acceleration in the magnetospheric vacuum gap, close to the supermassive black hole.\par

\begin{acknowledgements}
The journal referee is gratefully acknowledged for a constructive list of remarks that helped us improve the contents and clarity of the manuscript. \\
The MAGIC Collaboration  would like to thank the Instituto de Astrof\'{\i}sica de Canarias for the excellent working conditions at the Observatorio del Roque de los Muchachos in La Palma. The financial support of the German BMBF, MPG and HGF; the Italian INFN and INAF; the Swiss National Fund SNF; the ERDF under the Spanish Ministerio de Ciencia e Innovaci\'{o}n (MICINN) (FPA2017-87859-P, FPA2017-85668-P, FPA2017-82729-C6-5-R, FPA2017-90566-REDC, PID2019-104114RB-C31, PID2019-104114RB-C32, PID2019-105510GB-C31,PID2019-107847RB-C41, PID2019-107847RB-C42, PID2019-107847RB-C44, PID2019-107988GB-C22); the Indian Department of Atomic Energy; the Japanese ICRR, the University of Tokyo, JSPS, and MEXT; the Bulgarian Ministry of Education and Science, National RI Roadmap Project DO1-268/16.12.2019 and the Academy of Finland grant nr. 320045 is gratefully acknowledged. This work was also supported by the Spanish Centro de Excelencia ``Severo Ochoa'' SEV-2016-0588, SEV-2017-0709 and CEX2019-000920-S, and "Mar\'{\i}a de Maeztu” CEX2019-000918-M, the Unidad de Excelencia ``Mar\'{\i}a de Maeztu'' MDM-2015-0509-18-2 and the "la Caixa" Foundation (fellowship LCF/BQ/PI18/11630012) and by the CERCA program of the Generalitat de Catalunya; by the Croatian Science Foundation (HrZZ) Project IP-2016-06-9782 and the University of Rijeka Project 13.12.1.3.02; by the DFG Collaborative Research Centers SFB823/C4 and SFB876/C3; the Polish National Research Centre grant UMO-2016/22/M/ST9/00382; and by the Brazilian MCTIC, CNPq and FAPERJ.\\
The important contributions from ETH Zurich grants ETH-10.08-2 and ETH-27.12-1 as well as the funding by the Swiss SNF and the German BMBF (Verbundforschung Astro- und Astroteilchenphysik) and HAP (Helmoltz Alliance for Astroparticle Physics) are gratefully acknowledged. Part of this work is supported by Deutsche Forschungsgemeinschaft (DFG) within the Collaborative Research Center SFB 876 "Providing Information by Resource-Constrained Analysis", project C3. We are thankful for the very valuable contributions from E. Lorenz, D. Renker and G. Viertel during the early phase of the project. We thank the Instituto de Astrofísica de Canarias for allowing us to operate the telescope at the Observatorio del Roque de los Muchachos in La Palma, the Max-Planck-Institut f\"ur Physik for providing us with the mount of the former HEGRA CT3 telescope, and the MAGIC collaboration for their support.\\
The \textit{Fermi} LAT Collaboration acknowledges generous ongoing support
from a number of agencies and institutes that have supported both the
development and the operation of the LAT as well as scientific data analysis.
These include the National Aeronautics and Space Administration and the
Department of Energy in the United States, the Commissariat \`a l'Energie Atomique
and the Centre National de la Recherche Scientifique / Institut National de Physique
Nucl\'eaire et de Physique des Particules in France, the Agenzia Spaziale Italiana
and the Istituto Nazionale di Fisica Nucleare in Italy, the Ministry of Education,
Culture, Sports, Science and Technology (MEXT), High Energy Accelerator Research
Organization (KEK) and Japan Aerospace Exploration Agency (JAXA) in Japan, and
the K.~A.~Wallenberg Foundation, the Swedish Research Council and the
Swedish National Space Board in Sweden.
 
Additional support for science analysis during the operations phase is gratefully
acknowledged from the Istituto Nazionale di Astrofisica in Italy and the Centre
National d'\'Etudes Spatiales in France. This work performed in part under DOE
Contract DE-AC02-76SF00515.\\
This work made use of data from the \textit{NuSTAR} mission, a project led by the California Institute of Technology, managed by the Jet Propulsion Laboratory, and funded by the National Aeronautics and Space Administration. We thank the \textit{NuSTAR} Operations, Software, and Calibration teams for support with the execution and analysis of these observations. This research has made use of the \textit{NuSTAR} Data Analysis Software (NuSTARDAS) jointly developed by the ASI Science Data Center (ASDC; Italy) and the California Institute of Technology (USA).
This research has also made use of the XRT Data Analysis Software (XRTDAS) developed under the responsibility of the ASI Science Data Center (ASDC), Italy.\\
D.P acknowledges support from the Deutsche Forschungsgemeinschaft (DFG, German Research Foundation) under Germany's Excellence Strategy – EXC-2094 – 390783311.\\
M.\,B. acknowledges support from the YCAA Prize Postdoctoral Fellowship and from the Black Hole Initiative at Harvard University, which is funded in part by the Gordon and Betty Moore Foundation (grant GBMF8273) and in part by the John Templeton Foundation.\\
This publication makes use of data obtained at the Mets\"ahovi Radio Observatory, operated by Aalto University in Finland.\\
This research has made use of data from the OVRO 40-m monitoring program \citep{2011ApJS..194...29R} which is supported in part by NASA grants NNX08AW31G, NNX11A043G, and NNX14AQ89G and NSF grants AST-0808050 and AST-1109911.\\
I.A. acknowledges financial support from the Spanish "Ministerio de Ciencia e Innovaci\'on" (MCINN) through the "Center of Excellence Severo Ochoa" award for the Instituto de Astrof\'isica de Andaluc\'ia-CSIC (SEV-2017-0709). Acquisition and reduction of the MAPCAT data was supported in part by MICINN through grants AYA2016-80889-P and PID2019-107847RB-C44. The MAPCAT observations were carried out at the German-Spanish Calar Alto Observatory, which is jointly operated by Junta de Andaluc\'ia and Consejo Superior de Investigaciones Cient\'ificas. C.C. acknowledges support from the European Research Council (ERC) under the European Union Horizon 2020 research and innovation program under the grant agreement No 771282.\\
This research was partially supported by the Bulgarian National Science Fund of the Ministry of Education and Science under grants KP-06-H28/3 (2018), KP-06-H38/4 (2019) and KP-06-KITAJ/2 (2020).\\
We acknowledge support by Bulgarian National Science Fund under grant DN18-10/2017 and National RI Roadmap Projects DO1-277/16.12.2019 and DO1-268/16.12.2019 of the Ministry of Education and Science of the Republic of Bulgaria.\\
This research was supported by the Ministry of Education, Science and Technological Development of the Republic of Serbia (contract No 451-03-68/2020-14/200002). GD acknowledges observing grant support from the Institute of Astronomy and Rozhen NAO BAS through the bilateral joint research project "Gaia Celestial Reference Frame (CRF) and fast variable astronomical objects" (2020-2022, head - G.~Damljanovic).\\
The BU group was supported in part by NASA Fermi guest investigator program grants 80NSSC19K1505 and 80NSSC20K1566.\\
This study was based in part on observations conducted using the 1.8 m Perkins Telescope Observatory (PTO) in Arizona, which is owned and operated by Boston University.\\
This article is partly based on observations made with the LCOGT Telescopes, one of whose nodes is located at the Observatorios de Canarias del IAC on the island of Tenerife in the Observatorio del Teide.\\ 
This article is also based partly on data obtained with the STELLA robotic telescopes in Tenerife, an AIP facility jointly operated by AIP and IAC.\\
The Abastumani team acknowledges financial support by the Shota Rustaveli National Science Foundation under contract FR-19-6174.\\
Based on observations with the Medicina telescope operated by INAF - Istituto di Radioastronomia.
\end{acknowledgements}


\bibliographystyle{aa}
\bibliography{bibliography_paper.bib} 

\begin{thebibliography}{130}
\expandafter\ifx\csname natexlab\endcsname\relax\def\natexlab#1{#1}\fi

\bibitem[{{Abdo} {et~al.}(2010{\natexlab{a}}){Abdo}, {Ackermann}, {Agudo},
  {Ajello}, {Aller}, {Aller}, {Angelakis}, {Arkharov}, {Axelsson}, {Bach},
  {Baldini}, {Ballet}, {Barbiellini}, {Bastieri}, {Baughman}, {Bechtol},
  {Bellazzini}, {Benitez}, {Berdyugin}, {Berenji}, {Blandford}, {Bloom},
  {Boettcher}, {Bonamente}, {Borgland}, {Bregeon}, {Brez}, {Brigida}, {Bruel},
  {Burnett}, {Burrows}, {Buson}, {Caliandro}, {Calzoletti}, {Cameron},
  {Capalbi}, {Caraveo}, {Carosati}, {Casandjian}, {Cavazzuti}, {Cecchi},
  {{\c{C}}elik}, {Charles}, {Chaty}, {Chekhtman}, {Chen}, {Chiang},
  {Chincarini}, {Ciprini}, {Claus}, {Cohen-Tanugi}, {Colafrancesco},
  {Cominsky}, {Conrad}, {Costamante}, {Cutini}, {D'ammando}, {Deitrick},
  {D'Elia}, {Dermer}, {de Angelis}, {de Palma}, {Digel}, {Donnarumma}, {Silva},
  {Drell}, {Dubois}, {Dultzin}, {Dumora}, {Falcone}, {Farnier}, {Favuzzi},
  {Fegan}, {Focke}, {Forn{\'e}}, {Fortin}, {Frailis}, {Fuhrmann}, {Fukazawa},
  {Funk}, {Fusco}, {G{\'o}mez}, {Gargano}, {Gasparrini}, {Gehrels}, {Germani},
  {Giebels}, {Giglietto}, {Giommi}, {Giordano}, {Giuliani}, {Glanzman},
  {Godfrey}, {Grenier}, {Gronwall}, {Grove}, {Guillemot}, {Guiriec}, {Gurwell},
  {Hadasch}, {Hanabata}, {Harding}, {Hayashida}, {Hays}, {Healey}, {Heidt},
  {Hiriart}, {Horan}, {Hoversten}, {Hughes}, {Itoh}, {Jackson},
  {J{\'o}hannesson}, {Johnson}, {Johnson}, {Jorstad}, {Kadler}, {Kamae},
  {Katagiri}, {Kataoka}, {Kawai}, {Kennea}, {Kerr}, {Kimeridze},
  {Kn{\"o}dlseder}, {Kocian}, {Kopatskaya}, {Koptelova}, {Konstantinova},
  {Kovalev}, {Kovalev}, {Kurtanidze}, {Kuss}, {Lande}, {Larionov}, {Latronico},
  {Leto}, {Lindfors}, {Longo}, {Loparco}, {Lott}, {Lovellette}, {Lubrano},
  {Madejski}, {Makeev}, {Marchegiani}, {Marscher}, {Marshall}, {Max-Moerbeck},
  {Mazziotta}, {McConville}, {McEnery}, {Meurer}, {Michelson}, {Mitthumsiri},
  {Mizuno}, {Moiseev}, {Monte}, {Monzani}, {Morselli}, {Moskalenko}, {Murgia},
  {Nestoras}, {Nilsson}, {Nizhelsky}, {Nolan}, {Norris}, {Nuss}, {Ohsugi},
  {Ojha}, {Omodei}, {Orlando}, {Ormes}, {Osborne}, {Ozaki}, {Pacciani},
  {Padovani}, {Pagani}, {Page}, {Paneque}, {Panetta}, {Parent}, {Pasanen},
  {Pavlidou}, {Pelassa}, {Pepe}, {Perri}, {Pesce-Rollins}, {Piranomonte},
  {Piron}, {Pittori}, {Porter}, {Puccetti}, {Rahoui}, {Rain{\`o}}, {Raiteri},
  {Rando}, {Razzano}, {Reimer}, {Reimer}, {Reposeur}, {Richards}, {Ritz},
  {Rochester}, {Rodriguez}, {Romani}, {Ros}, {Roth}, {Roustazadeh}, {Ryde},
  {Sadrozinski}, {Sadun}, {Sanchez}, {Sander}, {Saz Parkinson}, {Scargle},
  {Sellerholm}, {Sgr{\`o}}, {Shaw}, {Sigua}, {Siskind}, {Smith}, {Smith},
  {Spandre}, {Spinelli}, {Starck}, {Stevenson}, {Stratta}, {Strickman},
  {Suson}, {Tajima}, {Takahashi}, {Takahashi}, {Takalo}, {Tanaka}, {Thayer},
  {Thayer}, {Thompson}, {Tibaldo}, {Torres}, {Tosti}, {Tramacere}, {Uchiyama},
  {Usher}, {Vasileiou}, {Verrecchia}, {Vilchez}, {Villata}, {Vitale}, {Waite},
  {Wang}, {Winer}, {Wood}, {Ylinen}, {Zensus}, {Zhekanis}, \&
  {Ziegler}}]{2010ApJ...716...30A}
{Abdo}, A.~A., {Ackermann}, M., {Agudo}, I., {et~al.} 2010{\natexlab{a}}, \apj,
  716, 30

\bibitem[{{Abdo} {et~al.}(2010{\natexlab{b}}){Abdo}, {Ackermann}, {Ajello},
  {Antolini}, {Baldini}, {Ballet}, {Barbiellini}, {Bastieri}, {Bechtol},
  {Bellazzini}, {Berenji}, {Blandford}, {Bloom}, {Bonamente}, {Borgland},
  {Bouvier}, {Bregeon}, {Brez}, {Brigida}, {Bruel}, {Buehler}, {Burnett},
  {Buson}, {Caliand ro}, {Cameron}, {Caraveo}, {Carrigan}, {Casandjian},
  {Cavazzuti}, {Cecchi}, {{\c{C}}elik}, {Chekhtman}, {Cheung}, {Chiang},
  {Ciprini}, {Claus}, {Cohen-Tanugi}, {Cominsky}, {Conrad}, {Costamante},
  {Cutini}, {Dermer}, {de Angelis}, {de Palma}, {Silva}, {Drell}, {Dubois},
  {Dumora}, {Farnier}, {Favuzzi}, {Fegan}, {Focke}, {Fortin}, {Frailis},
  {Fukazawa}, {Funk}, {Fusco}, {Gargano}, {Gasparrini}, {Gehrels}, {Germani},
  {Giebels}, {Giglietto}, {Giommi}, {Giordano}, {Glanzman}, {Godfrey},
  {Grenier}, {Grondin}, {Grove}, {Guiriec}, {Hadasch}, {Hayashida}, {Hays},
  {Healey}, {Horan}, {Hughes}, {Itoh}, {J{\'o}hannesson}, {Johnson}, {Johnson},
  {Kamae}, {Katagiri}, {Kataoka}, {Kawai}, {Kn{\"o}dlseder}, {Kuss}, {Lande},
  {Larsson}, {Latronico}, {Lemoine-Goumard}, {Longo}, {Loparco}, {Lott},
  {Lovellette}, {Lubrano}, {Madejski}, {Makeev}, {Massaro}, {Mazziotta},
  {McEnery}, {Michelson}, {Mitthumsiri}, {Mizuno}, {Moiseev}, {Monte},
  {Monzani}, {Morselli}, {Moskalenko}, {Mueller}, {Murgia}, {Nolan}, {Norris},
  {Nuss}, {Ohno}, {Ohsugi}, {Omodei}, {Orlando}, {Ormes}, {Ozaki}, {Panetta},
  {Parent}, {Pelassa}, {Pepe}, {Pesce-Rollins}, {Piron}, {Porter}, {Rain{\`o}},
  {Rando}, {Razzano}, {Reimer}, {Reimer}, {Ritz}, {Rodriguez}, {Romani},
  {Roth}, {Ryde}, {Sadrozinski}, {Sand er}, {Scargle}, {Sgr{\`o}}, {Shaw},
  {Smith}, {Spandre}, {Spinelli}, {Starck}, {Strickman}, {Suson}, {Takahashi},
  {Takahashi}, {Tanaka}, {Thayer}, {Thayer}, {Thompson}, {Tibaldo}, {Torres},
  {Tosti}, {Tramacere}, {Uchiyama}, {Usher}, {Vasileiou}, {Vilchez}, {Vitale},
  {Waite}, {Wallace}, {Wang}, {Winer}, {Wood}, {Yang}, {Ylinen}, \&
  {Ziegler}}]{2010ApJ...722..520A}
{Abdo}, A.~A., {Ackermann}, M., {Ajello}, M., {et~al.} 2010{\natexlab{b}},
  \apj, 722, 520

\bibitem[{{Abdo} {et~al.}(2011){Abdo}, {Ackermann}, {Ajello}, {Baldini},
  {Ballet}, {Barbiellini}, {Bastieri}, {Bechtol}, {Bellazzini}, {Berenji},
  {Bland ford}, {Bloom}, {Bonamente}, {Borgland }, {Bouvier}, {Bregeon},
  {Brez}, {Brigida}, {Bruel}, {Buehler}, {Buson}, {Caliandro}, {Cameron},
  {Cannon}, {Caraveo}, {Carrigan}, {Casandjian}, {Cavazzuti}, {Cecchi},
  {{\c{C}}elik}, {Charles}, {Chekhtman}, {Chiang}, {Ciprini}, {Claus},
  {Cohen-Tanugi}, {Conrad}, {Cutini}, {de Angelis}, {de Palma}, {Dermer},
  {Silva}, {Drell}, {Dubois}, {Dumora}, {Escande}, {Favuzzi}, {Fegan}, {Finke},
  {Focke}, {Fortin}, {Frailis}, {Fuhrmann}, {Fukazawa}, {Fukuyama}, {Funk},
  {Fusco}, {Gargano}, {Gasparrini}, {Gehrels}, {Georganopoulos}, {Germani},
  {Giebels}, {Giglietto}, {Giommi}, {Giordano}, {Giroletti}, {Glanzman},
  {Godfrey}, {Grenier}, {Guiriec}, {Hadasch}, {Hayashida}, {Hays}, {Horan},
  {Hughes}, {J{\'o}hannesson}, {Johnson}, {Johnson}, {Kadler}, {Kamae},
  {Katagiri}, {Kataoka}, {Kn{\"o}dlseder}, {Kuss}, {Lande}, {Latronico}, {Lee},
  {Longo}, {Loparco}, {Lott}, {Lovellette}, {Lubrano}, {Madejski}, {Makeev},
  {Max-Moerbeck}, {Mazziotta}, {McEnery}, {Mehault}, {Michelson},
  {Mitthumsiri}, {Mizuno}, {Monte}, {Monzani}, {Morselli}, {Moskalenko},
  {Murgia}, {Nakamori}, {Naumann-Godo}, {Nishino}, {Nolan}, {Norris}, {Nuss},
  {Ohsugi}, {Okumura}, {Omodei}, {Orlando}, {Ormes}, {Ozaki}, {Paneque},
  {Panetta}, {Parent}, {Pavlidou}, {Pearson}, {Pelassa}, {Pepe},
  {Pesce-Rollins}, {Pierbattista}, {Piron}, {Porter}, {Rain{\`o}}, {Rando},
  {Razzano}, {Readhead}, {Reimer}, {Reimer}, {Reyes}, {Richards}, {Ritz},
  {Roth}, {Sadrozinski}, {Sanchez}, {Sander}, {Sgr{\`o}}, {Siskind}, {Smith},
  {Spand re}, {Spinelli}, {Stawarz}, {Stevenson}, {Strickman}, {Suson},
  {Takahashi}, {Takahashi}, {Tanaka}, {Thayer}, {Thayer}, {Thompson},
  {Tibaldo}, {Torres}, {Tosti}, {Tramacere}, {Troja}, {Usher}, {Vandenbroucke},
  {Vasileiou}, {Vianello}, {Vilchez}, {Vitale}, {Waite}, {Wang}, {Wehrle},
  {Winer}, {Wood}, {Yang}, {Yatsu}, {Ylinen}, {Zensus}, {Ziegler}, {Fermi LAT
  Collaboration}, {Aleksi{\'c}}, {Antonelli}, {Antoranz}, {Backes}, {Barrio},
  {Becerra Gonz{\'a}lez}, {Bednarek}, {Berdyugin}, {Berger}, {Bernardini},
  {Biland}, {Blanch}, {Bock}, {Boller}, {Bonnoli}, {Bordas}, {Borla Tridon},
  {Bosch-Ramon}, {Bose}, {Braun}, {Bretz}, {Camara}, {Carmona}, {Carosi},
  {Colin}, {Colombo}, {Contreras}, {Cortina}, {Covino}, {Dazzi}, {de Angelis},
  {De Cea del Pozo}, {Delgado Mendez}, {De Lotto}, {De Maria}, {De Sabata},
  {Diago Ortega}, {Doert}, {Dom{\'\i}nguez}, {Dominis Prester}, {Dorner},
  {Doro}, {Elsaesser}, {Ferenc}, {Fonseca}, {Font}, {Garc{\'\i}a L{\'o}pez},
  {Garczarczyk}, {Gaug}, {Giavitto}, {Godinovi}, {Hadasch}, {Herrero},
  {Hildebrand}, {H{\"o}hne-M{\"o}nch}, {Hose}, {Hrupec}, {Jogler}, {Klepser},
  {Kr{\"a}henb{\"u}hl}, {Kranich}, {Krause}, {La Barbera}, {Leonardo},
  {Lindfors}, {Lombardi}, {L{\'o}pez}, {Lorenz}, {Majumdar}, {Makariev},
  {Maneva}, {Mankuzhiyil}, {Mannheim}, {Maraschi}, {Mariotti}, {Mart{\'\i}nez},
  {Mazin}, {Meucci}, {Mirand a}, {Mirzoyan}, {Miyamoto}, {Mold{\'o}n},
  {Moralejo}, {Nieto}, {Nilsson}, {Orito}, {Oya}, {Paoletti}, {Paredes},
  {Partini}, {Pasanen}, {Pauss}, {Pegna}, {Perez-Torres}, {Persic}, {Peruzzo},
  {Pochon}, {Prada}, {Prada Moroni}, {Prand ini}, {Puchades}, {Puljak},
  {Reichardt}, {Rhode}, {Rib{\'o}}, {Rico}, {Rissi}, {R{\"u}gamer}, {Saggion},
  {Saito}, {Saito}, {Salvati}, {S{\'a}nchez-Conde}, {Satalecka}, {Scalzotto},
  {Scapin}, {Schultz}, {Schweizer}, {Shayduk}, {Shore}, {Sierpowska-Bartosik},
  {Sillanp{\"a}{\"a}}, {Sitarek}, {Sobczynska}, {Spanier}, {Spiro}, {Stamerra},
  {Steinke}, {Storz}, {Strah}, {Struebig}, {Suric}, {Takalo}, {Tavecchio},
  {Temnikov}, {Terzi{\'c}}, {Tescaro}, {Teshima}, {Vankov}, {Wagner},
  {Weitzel}, {Zabalza}, {Zandanel}, {Zanin}, {MAGIC Collaboration}, {Villata},
  {Raiteri}, {Aller}, {Aller}, {Chen}, {Jordan}, {Koptelova}, {Kurtanidze},
  {L{\"a}hteenm{\"a}ki}, {McBreen}, {Larionov}, {Lin}, {Nikolashvili},
  {Reinthal}, {Angelakis}, {Capalbi}, {Carrami{\~n}ana}, {Carrasco}, {Cassaro},
  {Cesarini}, {Falcone}, {Gurwell}, {Hovatta}, {Kovalev}, {Kovalev},
  {Krichbaum}, {Krimm}, {Lister}, {Moody}, {Maccaferri}, {Mori}, {Nestoras},
  {Orlati}, {Pace}, {Pagani}, {Pearson}, {Perri}, {Piner}, {Ros}, {Sadun},
  {Sakamoto}, {Tammi}, \& {Zook}}]{2011ApJ...736..131A}
{Abdo}, A.~A., {Ackermann}, M., {Ajello}, M., {et~al.} 2011, \apj, 736, 131

\bibitem[{{Abdollahi} {et~al.}(2020){Abdollahi}, {Acero}, {Ackermann},
  {Ajello}, {Atwood}, {Axelsson}, {Baldini}, {Ballet}, {Barbiellini},
  {Bastieri}, {Becerra Gonzalez}, {Bellazzini}, {Berretta}, {Bissaldi}, {Bland
  ford}, {Bloom}, {Bonino}, {Bottacini}, {Brandt}, {Bregeon}, {Bruel},
  {Buehler}, {Burnett}, {Buson}, {Cameron}, {Caputo}, {Caraveo}, {Casandjian},
  {Castro}, {Cavazzuti}, {Charles}, {Chaty}, {Chen}, {Cheung}, {Chiaro},
  {Ciprini}, {Cohen-Tanugi}, {Cominsky}, {Coronado-Bl{\'a}zquez}, {Costantin},
  {Cuoco}, {Cutini}, {D'Ammando}, {DeKlotz}, {de la Torre Luque}, {de Palma},
  {Desai}, {Digel}, {Di Lalla}, {Di Mauro}, {Di Venere}, {Dom{\'\i}nguez},
  {Dumora}, {Fana Dirirsa}, {Fegan}, {Ferrara}, {Franckowiak}, {Fukazawa},
  {Funk}, {Fusco}, {Gargano}, {Gasparrini}, {Giglietto}, {Giommi}, {Giordano},
  {Giroletti}, {Glanzman}, {Green}, {Grenier}, {Griffin}, {Grondin}, {Grove},
  {Guiriec}, {Harding}, {Hayashi}, {Hays}, {Hewitt}, {Horan},
  {J{\'o}hannesson}, {Johnson}, {Kamae}, {Kerr}, {Kocevski}, {Kovac'evic'},
  {Kuss}, {Landriu}, {Larsson}, {Latronico}, {Lemoine-Goumard}, {Li},
  {Liodakis}, {Longo}, {Loparco}, {Lott}, {Lovellette}, {Lubrano}, {Madejski},
  {Maldera}, {Malyshev}, {Manfreda}, {Marchesini}, {Marcotulli},
  {Mart{\'\i}-Devesa}, {Martin}, {Massaro}, {Mazziotta}, {McEnery}, {Mereu},
  {Meyer}, {Michelson}, {Mirabal}, {Mizuno}, {Monzani}, {Morselli},
  {Moskalenko}, {Negro}, {Nuss}, {Ojha}, {Omodei}, {Orienti}, {Orlando},
  {Ormes}, {Palatiello}, {Paliya}, {Paneque}, {Pei}, {Pe{\~n}a-Herazo},
  {Perkins}, {Persic}, {Pesce-Rollins}, {Petrosian}, {Petrov}, {Piron}, {Poon},
  {Porter}, {Principe}, {Rain{\`o}}, {Rando}, {Razzano}, {Razzaque}, {Reimer},
  {Reimer}, {Remy}, {Reposeur}, {Romani}, {Saz Parkinson}, {Schinzel},
  {Serini}, {Sgr{\`o}}, {Siskind}, {Smith}, {Spandre}, {Spinelli}, {Strong},
  {Suson}, {Tajima}, {Takahashi}, {Tak}, {Thayer}, {Thompson}, {Tibaldo},
  {Torres}, {Torresi}, {Valverde}, {Van Klaveren}, {van Zyl}, {Wood},
  {Yassine}, \& {Zaharijas}}]{2020ApJS..247...33A}
{Abdollahi}, S., {Acero}, F., {Ackermann}, M., {et~al.} 2020, \apjs, 247, 33

\bibitem[{{Abeysekara} {et~al.}(2020){Abeysekara}, {Benbow}, {Bird}, {Brill},
  {Brose}, {Buchovecky}, {Buckley}, {Christiansen}, {Chromey}, {Daniel},
  {Dumm}, {Falcone}, {Feng}, {Finley}, {Fortson}, {Furniss}, {Galante}, {Gent},
  {Gillanders}, {Giuri}, {Gueta}, {Hassan}, {Hervet}, {Holder}, {Hughes},
  {Humensky}, {Johnson}, {Kaaret}, {Kar}, {Kelley-Hoskins}, {Kertzman},
  {Kieda}, {Krause}, {Krennrich}, {Kumar}, {Lang}, {Moriarty}, {Mukherjee},
  {Nelson}, {Nieto}, {Nievas-Rosillo}, {O'Brien}, {Ong}, {Otte}, {Park},
  {Petrashyk}, {Pichel}, {Pohl}, {Prado}, {Pueschel}, {Quinn}, {Ragan},
  {Reynolds}, {Richards}, {Roache}, {Rovero}, {Rulten}, {Sadeh}, {Santander},
  {Sembroski}, {Shahinyan}, {Stevenson}, {Sushch}, {Tyler}, {Vassiliev},
  {Wakely}, {Weinstein}, {Wells}, {Wilcox}, {Wilhelm}, {Williams}, {Zitzer},
  {Acciari}, {Ansoldi}, {Antonelli}, {Arbet Engels}, {Baack}, {Babi{\'c}},
  {Banerjee}, {Barres de Almeida}, {Barrio}, {Becerra Gonz{\'a}lez},
  {Bednarek}, {Bellizzi}, {Bernardini}, {Berti}, {Besenrieder},
  {Bhattacharyya}, {Bigongiari}, {Biland}, {Blanch}, {Bonnoli}, {Busetto},
  {Carosi}, {Ceribella}, {Chai}, {Cikota}, {Colak}, {Colin}, {Colombo},
  {Contreras}, {Cortina}, {Covino}, {D'Elia}, {Da Vela}, {Dazzi}, {De Angelis},
  {De Lotto}, {Delfino}, {Delgado}, {Di Pierro}, {Do Souto Espi{\~n}era},
  {Dominis Prester}, {Dorner}, {Doro}, {Einecke}, {Elsaesser}, {Fallah
  Ramazani}, {Fattorini}, {Fern{\'a}ndez-Barral}, {Ferrara}, {Fidalgo},
  {Foffano}, {Fonseca}, {Font}, {Fruck}, {Galindo}, {Gallozzi}, {Garc{\'\i}a
  L{\'o}pez}, {Garczarczyk}, {Gasparyan}, {Gaug}, {Godinovi{\'c}}, {Green},
  {Guberman}, {Hadasch}, {Hahn}, {Herrera}, {Hoang}, {Hrupec}, {Inoue},
  {Ishio}, {Iwamura}, {Kubo}, {Kushida}, {Lamastra}, {Lelas}, {Leone},
  {Lindfors}, {Lombardi}, {Longo}, {L{\'o}pez}, {L{\'o}pez-Coto},
  {L{\'o}pez-Oramas}, {Machado de Oliveira Fraga}, {Maggio}, {Majumdar},
  {Makariev}, {Mallamaci}, {Maneva}, {Manganaro}, {Mannheim}, {Maraschi},
  {Mariotti}, {Mart{\'\i}nez}, {Masuda}, {Mazin}, {Miceli}, {Minev}, {Miranda},
  {Mirzoyan}, {Molina}, {Moralejo}, {Morcuende}, {Moreno}, {Moretti},
  {Munar-Adrover}, {Neustroev}, {Niedzwiecki}, {Nievas Rosillo}, {Nigro},
  {Nilsson}, {Ninci}, {Nishijima}, {Noda}, {Nogu{\'e}s}, {N{\"o}the}, {Paiano},
  {Palacio}, {Palatiello}, {Paneque}, {Paoletti}, {Paredes}, {Pe{\~n}il},
  {Peresano}, {Persic}, {Prada Moroni}, {Prand ini}, {Puljak}, {Rhode},
  {Rib{\'o}}, {Rico}, {Righi}, {Rugliancich}, {Saha}, {Sahakyan}, {Saito},
  {Satalecka}, {Schweizer}, {Sitarek}, {{\v{S}}nidari{\'c}}, {Sobczynska},
  {Somero}, {Stamerra}, {Strom}, {Strzys}, {Sun}, {Suri{\'c}}, {Tavecchio},
  {Temnikov}, {Terzi{\'c}}, {Teshima}, {Torres-Alb{\`a}}, {Tsujimoto}, {van
  Scherpenberg}, {Vanzo}, {Vazquez Acosta}, {Vovk}, {Will}, {Zari{\'c}},
  {Aller}, {Aller}, {Carini}, {Horan}, {Jordan}, {Jorstad}, {Kurtanidze},
  {Kurtanidze}, {L{\"a}hteenm{\"a}ki}, {Larionov}, {Larionova}, {Madejski},
  {Marscher}, {Max-Moerbeck}, {Moody}, {Morozova}, {Nikolashvili}, {Raiteri},
  {Readhead}, {Richards}, {Sadun}, {Sakamoto}, {Sigua}, {Smith}, {Talvikki},
  {Tammi}, {Tornikoski}, {Troitsky}, \& {Villata}}]{2020ApJ...890...97A}
{Abeysekara}, A.~U., {Benbow}, W., {Bird}, R., {et~al.} 2020, \apj, 890, 97

\bibitem[{{Acciari} {et~al.}(2021){Acciari}, {Ansoldi}, \&
  {Antonelli}}]{Mrk421_2015-16}
{Acciari}, V.~A., {Ansoldi}, S., \& {Antonelli}, L.~A. 2021, MNRAS, 000, 0

\bibitem[{{Acciari} {et~al.}(2020){Acciari}, {Ansoldi}, {Antonelli}, {Arbet
  Engels}, {Baack}, {Babi{\'c}}, {Banerjee}, {Barres de Almeida}, {Barrio},
  {Becerra Gonz{\'a}lez}, {Bednarek}, {Bellizzi}, {Bernardini}, {Berti},
  {Besenrieder}, {Bhattacharyya}, {Bigongiari}, {Biland}, {Blanch}, {Bonnoli},
  {Bo{\v{s}}njak}, {Busetto}, {Carosi}, {Ceribella}, {Cerruti}, {Chai},
  {Chilingarian}, {Cikota}, {Colak}, {Colin}, {Colombo}, {Contreras},
  {Cortina}, {Covino}, {D'Elia}, {da Vela}, {Dazzi}, {de Angelis}, {de Lotto},
  {Del Puppo}, {Delfino}, {Delgado}, {Depaoli}, {di Pierro}, {di Venere}, {Do
  Souto Espi{\~n}eira}, {Prester}, {Donini}, {Dorner}, {Doro}, {Elsaesser},
  {Ramazani}, {Fattorini}, {Ferrara}, {Foffano}, {Fonseca}, {Font}, {Fruck},
  {Fukami}, {Garc{\'\i}a L{\'o}pez}, {Garczarczyk}, {Gasparyan}, {Gaug},
  {Giglietto}, {Giordano}, {Gliwny}, {Godinovi{\'c}}, {Green}, {Hadasch},
  {Hahn}, {Hassan}, {Herrera}, {Hoang}, {Hrupec}, {H{\"u}tten}, {Inada},
  {Inoue}, {Ishio}, {Iwamura}, {Jouvin}, {Kajiwara}, {Kerszberg}, {Kobayashi},
  {Kubo}, {Kushida}, {Lamastra}, {Lelas}, {Leone}, {Lindfors}, {Lombardi},
  {Longo}, {L{\'o}pez}, {L{\'o}pez-Coto}, {L{\'o}pez-Oramas}, {Loporchio},
  {Machado de Oliveira Fraga}, {Maggio}, {Majumdar}, {Makariev}, {Mallamaci},
  {Maneva}, {Manganaro}, {Mannheim}, {Maraschi}, {Mariotti}, {Mart{\'\i}nez},
  {Mazin}, {Mender}, {Mi{\'c}anovi{\'c}}, {Miceli}, {Miener}, {Minev},
  {Miranda}, {Mirzoyan}, {Molina}, {Moralejo}, {Morcuende}, {Moreno},
  {Moretti}, {Munar-Adrover}, {Neustroev}, {Nigro}, {Nilsson}, {Ninci},
  {Nishijima}, {Noda}, {Nogu{\'e}s}, {Nozaki}, {Ohtani}, {Oka}, {Otero-Santos},
  {Palatiello}, {Paneque}, {Paoletti}, {Paredes}, {Pavleti{\'c}}, {Pe{\~n}il},
  {Peresano}, {Persic}, {Moroni}, {Prandini}, {Puljak}, {Rhode}, {Rib{\'o}},
  {Rico}, {Righi}, {Rugliancich}, {Saha}, {Sahakyan}, {Saito}, {Sakurai},
  {Satalecka}, {Schleicher}, {Schmidt}, {Schweizer}, {Sitarek},
  {{\v{S}}nidari{\'c}}, {Sobczynska}, {Spolon}, {Stamerra}, {Strom}, {Strzys},
  {Suda}, {Suri{\'c}}, {Takahashi}, {Tavecchio}, {Temnikov}, {Terzi{\'c}},
  {Teshima}, {Torres-Alb{\`a}}, {Tosti}, {van Scherpenberg}, {Vanzo}, {Vazquez
  Acosta}, {Ventura}, {Verguilov}, {Vigorito}, {Vitale}, {Vovk}, {Will},
  {Zari{\'c}}, {MAGIC Collaboration}, {Finke}, {D'Ammando}, {Balokovi{\'c}},
  {Madejski}, {Mori}, {Puccetti}, {Leto}, {Perri}, {Verrecchia}, {Villata},
  {Raiteri}, {Agudo}, {Bachev}, {Berdyugin}, {Blinov}, {Chanishvili}, {Chen},
  {Chigladze}, {Damljanovic}, {Eswaraiah}, {Grishina}, {Ibryamov}, {Jordan},
  {Jorstad}, {Joshi}, {Kopatskaya}, {Kurtanidze}, {Kurtanidze}, {Larionova},
  {Larionova}, {Larionov}, {Latev}, {Lin}, {Marscher}, {Mokrushina},
  {Morozova}, {Nikolashvili}, {Semkov}, {Smith}, {Strigachev}, {Troitskaya},
  {Troitsky}, {Vince}, {Barnes}, {G{\"u}ver}, {Moody}, {Sadun}, {Hovatta},
  {Richards}, {Max-Moerbeck}, {Readhead}, {L{\"a}hteenm{\"a}ki}, {Tornikoski},
  {Tammi}, {Ramakrishnan}, \& {Reinthal}}]{2020ApJS..248...29A}
{Acciari}, V.~A., {Ansoldi}, S., {Antonelli}, L.~A., {et~al.} 2020, \apjs, 248,
  29

\bibitem[{{Acciari} {et~al.}(2014){Acciari}, {Arlen}, {Aune}, {Benbow}, {Bird},
  {Bouvier}, {Bradbury}, {Buckley}, {Bugaev}, {de la Calle Perez},
  {Carter-Lewis}, {Cesarini}, {Ciupik}, {Collins-Hughes}, {Connolly}, {Cui},
  {Duke}, {Dumm}, {Falcone}, {Federici}, {Fegan}, {Fegan}, {Finley},
  {Finnegan}, {Fortson}, {Gaidos}, {Galante}, {Gall}, {Gibbs}, {Gillanders},
  {Griffin}, {Grube}, {Gyuk}, {Hanna}, {Horan}, {Humensky}, {Kaaret},
  {Kertzman}, {Khassen}, {Kieda}, {Krawczynski}, {Krennrich}, {Lang},
  {McEnery}, {Madhavan}, {Moriarty}, {Nelson}, {O'Faol{\'a}in de Bhr{\'o}ithe},
  {Ong}, {Orr}, {Otte}, {Perkins}, {Petry}, {Pichel}, {Pohl}, {Quinn}, {Ragan},
  {Reynolds}, {Roache}, {Rovero}, {Schroedter}, {Sembroski}, {Smith},
  {Telezhinsky}, {Theiling}, {Toner}, {Tyler}, {Varlotta}, {Vivier}, {Wakely},
  {Ward}, {Weekes}, {Weinstein}, {Welsing}, {Williams}, \&
  {Wissel}}]{2014APh....54....1A}
{Acciari}, V.~A., {Arlen}, T., {Aune}, T., {et~al.} 2014, Astroparticle
  Physics, 54, 1

\bibitem[{{Ackermann} {et~al.}(2012){Ackermann}, {Ajello}, {Albert},
  {Allafort}, {Atwood}, {Axelsson}, {Baldini}, {Ballet}, {Barbiellini},
  {Bastieri}, {Bechtol}, {Bellazzini}, {Bissaldi}, {Blandford}, {Bloom},
  {Bogart}, {Bonamente}, {Borgland }, {Bottacini}, {Bouvier}, {Brandt},
  {Bregeon}, {Brigida}, {Bruel}, {Buehler}, {Burnett}, {Buson}, {Caliandro},
  {Cameron}, {Caraveo}, {Casandjian}, {Cavazzuti}, {Cecchi}, {{\c{C}}elik},
  {Charles}, {Chaves}, {Chekhtman}, {Cheung}, {Chiang}, {Ciprini}, {Claus},
  {Cohen-Tanugi}, {Conrad}, {Corbet}, {Cutini}, {D'Ammando}, {Davis}, {de
  Angelis}, {DeKlotz}, {de Palma}, {Dermer}, {Digel}, {Silva}, {Drell},
  {Drlica-Wagner}, {Dubois}, {Favuzzi}, {Fegan}, {Ferrara}, {Focke}, {Fortin},
  {Fukazawa}, {Funk}, {Fusco}, {Gargano}, {Gasparrini}, {Gehrels}, {Giebels},
  {Giglietto}, {Giordano}, {Giroletti}, {Glanzman}, {Godfrey}, {Grenier},
  {Grove}, {Guiriec}, {Hadasch}, {Hayashida}, {Hays}, {Horan}, {Hou}, {Hughes},
  {Jackson}, {Jogler}, {J{\'o}hannesson}, {Johnson}, {Johnson}, {Johnson},
  {Kamae}, {Katagiri}, {Kataoka}, {Kerr}, {Kn{\"o}dlseder}, {Kuss}, {Lande},
  {Larsson}, {Latronico}, {Lavalley}, {Lemoine-Goumard}, {Longo}, {Loparco},
  {Lott}, {Lovellette}, {Lubrano}, {Mazziotta}, {McConville}, {McEnery},
  {Mehault}, {Michelson}, {Mitthumsiri}, {Mizuno}, {Moiseev}, {Monte},
  {Monzani}, {Morselli}, {Moskalenko}, {Murgia}, {Naumann-Godo}, {Nemmen},
  {Nishino}, {Norris}, {Nuss}, {Ohno}, {Ohsugi}, {Okumura}, {Omodei},
  {Orienti}, {Orlando}, {Ormes}, {Paneque}, {Panetta}, {Perkins},
  {Pesce-Rollins}, {Pierbattista}, {Piron}, {Pivato}, {Porter}, {Racusin},
  {Rain{\`o}}, {Rand o}, {Razzano}, {Razzaque}, {Reimer}, {Reimer}, {Reposeur},
  {Reyes}, {Ritz}, {Rochester}, {Romoli}, {Roth}, {Sadrozinski}, {Sanchez},
  {Saz Parkinson}, {Sbarra}, {Scargle}, {Sgr{\`o}}, {Siegal-Gaskins},
  {Siskind}, {Spand re}, {Spinelli}, {Stephens}, {Suson}, {Tajima},
  {Takahashi}, {Tanaka}, {Thayer}, {Thayer}, {Thompson}, {Tibaldo},
  {Tinivella}, {Tosti}, {Troja}, {Usher}, {Vandenbroucke}, {Van Klaveren},
  {Vasileiou}, {Vianello}, {Vitale}, {Waite}, {Wallace}, {Winer}, {Wood},
  {Wood}, {Wood}, {Yang}, \& {Zimmer}}]{2012ApJS..203....4A}
{Ackermann}, M., {Ajello}, M., {Albert}, A., {et~al.} 2012, \apjs, 203, 4

\bibitem[{{Ahnen} {et~al.}(2017{\natexlab{a}}){Ahnen}, {Ansoldi}, {Antonelli},
  {Antoranz}, {Babic}, {Banerjee}, {Bangale}, {Barres de Almeida}, {Barrio},
  {Becerra Gonz{\'a}lez}, {Bednarek}, {Bernardini}, {Berti}, {Biasuzzi},
  {Biland}, {Blanch}, {Bonnefoy}, {Bonnoli}, {Borracci}, {Bretz}, {Buson},
  {Carosi}, {Chatterjee}, {Clavero}, {Colin}, {Colombo}, {Contreras},
  {Cortina}, {Covino}, {Da Vela}, {Dazzi}, {De Angelis}, {De Lotto}, {de
  O{\~n}a Wilhelmi}, {Di Pierro}, {Doert}, {Dom{\'\i}nguez}, {Dominis Prester},
  {Dorner}, {Doro}, {Einecke}, {Eisenacher Glawion}, {Elsaesser},
  {Engelkemeier}, {Fallah Ramazani}, {Fern{\'a}ndez-Barral}, {Fidalgo},
  {Fonseca}, {Font}, {Frantzen}, {Fruck}, {Galindo}, {Garc{\'\i}a L{\'o}pez},
  {Garczarczyk}, {Garrido Terrats}, {Gaug}, {Giammaria}, {Godinovi{\'c}},
  {Gonz{\'a}lez Mu{\~n}oz}, {Gora}, {Guberman}, {Hadasch}, {Hahn}, {Hanabata},
  {Hayashida}, {Herrera}, {Hose}, {Hrupec}, {Hughes}, {Idec}, {Kodani},
  {Konno}, {Kubo}, {Kushida}, {La Barbera}, {Lelas}, {Lindfors}, {Lombardi},
  {Longo}, {L{\'o}pez}, {L{\'o}pez-Coto}, {Majumdar}, {Makariev}, {Mallot},
  {Maneva}, {Manganaro}, {Mannheim}, {Maraschi}, {Marcote}, {Mariotti},
  {Mart{\'\i}nez}, {Mazin}, {Menzel}, {Mirand a}, {Mirzoyan}, {Moralejo},
  {Moretti}, {Nakajima}, {Neustroev}, {Niedzwiecki}, {Nievas Rosillo},
  {Nilsson}, {Nishijima}, {Noda}, {Nogu{\'e}s}, {Overkemping}, {Paiano},
  {Palacio}, {Palatiello}, {Paneque}, {Paoletti}, {Paredes}, {Paredes-Fortuny},
  {Pedaletti}, {Peresano}, {Perri}, {Persic}, {Poutanen}, {Prada Moroni},
  {Prandini}, {Puljak}, {Reichardt}, {Rhode}, {Rib{\'o}}, {Rico}, {Rodriguez
  Garcia}, {Saito}, {Satalecka}, {Schr{\"o}der}, {Schultz}, {Schweizer},
  {Shore}, {Sillanp{\"a}{\"a}}, {Sitarek}, {Snidaric}, {Sobczynska},
  {Stamerra}, {Steinbring}, {Strzys}, {Suri{\'c}}, {Takalo}, {Tavecchio},
  {Temnikov}, {Terzi{\'c}}, {Tescaro}, {Teshima}, {Thaele}, {Torres}, {Toyama},
  {Treves}, {Vanzo}, {Verguilov}, {Vovk}, {Ward}, {Will}, {Wu}, {Zanin},
  {Abeysekara}, {Archambault}, {Archer}, {Benbow}, {Bird}, {Buchovecky},
  {Buckley}, {Bugaev}, {Connolly}, {Cui}, {Dickinson}, {Falcone}, {Feng},
  {Finley}, {Fleischhack}, {Flinders}, {Fortson}, {Gillanders}, {Griffin},
  {Grube}, {H{\"u}tten}, {Hanna}, {Holder}, {Humensky}, {Kaaret}, {Kar},
  {Kelley-Hoskins}, {Kertzman}, {Kieda}, {Krause}, {Krennrich}, {Lang},
  {Maier}, {McCann}, {Moriarty}, {Mukherjee}, {Nieto}, {O'Brien}, {Ong},
  {Otte}, {Park}, {Perkins}, {Pichel}, {Pohl}, {Popkow}, {Pueschel}, {Quinn},
  {Ragan}, {Reynolds}, {Richards}, {Roache}, {Rovero}, {Rulten}, {Sadeh},
  {Santander}, {Sembroski}, {Shahinyan}, {Telezhinsky}, {Tucci}, {Tyler},
  {Wakely}, {Weinstein}, {Wilcox}, {Wilhelm}, {Williams}, {Zitzer}, {Razzaque},
  {Villata}, {Raiteri}, {Aller}, {Aller}, {Larionov}, {Arkharov}, {Blinov},
  {Efimova}, {Grishina}, {Hagen-Thorn}, {Kopatskaya}, {Larionova}, {Larionova},
  {Morozova}, {Troitsky}, {Ligustri}, {Calcidese}, {Berdyugin}, {Kurtanidze},
  {Nikolashvili}, {Kimeridze}, {Sigua}, {Kurtanidze}, {Chigladze}, {Chen},
  {Koptelova}, {Sakamoto}, {Sadun}, {Moody}, {Pace}, {Pearson}, {Yatsu},
  {Mori}, {Carraminyana}, {Carrasco}, {de la Fuente}, {Norris}, {Smith},
  {Wehrle}, {Gurwell}, {Zook}, {Pagani}, {Perri}, {Capalbi}, {Cesarini},
  {Krimm}, {Kovalev}, {Kovalev}, {Ros}, {Pushkarev}, {Lister}, {Sokolovsky},
  {Kadler}, {Piner}, {L{\"a}hteenm{\"a}ki}, {Tornikoski}, {Angelakis},
  {Krichbaum}, {Nestoras}, {Fuhrmann}, {Zensus}, {Cassaro}, {Orlati},
  {Maccaferri}, {Leto}, {Giroletti}, {Richards}, {Max-Moerbeck}, \&
  {Readhead}}]{2017A&A...603A..31A}
{Ahnen}, M.~L., {Ansoldi}, S., {Antonelli}, L.~A., {et~al.} 2017{\natexlab{a}},
  \aap, 603, A31

\bibitem[{{Ahnen} {et~al.}(2016){Ahnen}, {Ansoldi}, {Antonelli}, {Antoranz},
  {Babic}, {Banerjee}, {Bangale}, {Barres de Almeida}, {Barrio}, {Becerra
  Gonz{\'a}lez}, {Bednarek}, {Bernardini}, {Biasuzzi}, {Biland}, {Blanch},
  {Bonnefoy}, {Bonnoli}, {Borracci}, {Bretz}, {Buson}, {Carosi}, {Chatterjee},
  {Clavero}, {Colin}, {Colombo}, {Contreras}, {Cortina}, {Covino}, {Da Vela},
  {Dazzi}, {De Angelis}, {De Lotto}, {de O{\~n}a Wilhelmi}, {Di Pierro},
  {Dom{\'\i}nguez}, {Dominis Prester}, {Dorner}, {Doro}, {Einecke}, {Eisenacher
  Glawion}, {Elsaesser}, {Fern{\'a}ndez-Barral}, {Fidalgo}, {Fonseca}, {Font},
  {Frantzen}, {Fruck}, {Galindo}, {Garc{\'\i}a L{\'o}pez}, {Garczarczyk},
  {Garrido Terrats}, {Gaug}, {Giammaria}, {Godinovi{\'c}}, {Gonz{\'a}lez
  Mu{\~n}oz}, {Gora}, {Guberman}, {Hadasch}, {Hahn}, {Hanabata}, {Hayashida},
  {Herrera}, {Hose}, {Hrupec}, {Hughes}, {Idec}, {Kodani}, {Konno}, {Kubo},
  {Kushida}, {La Barbera}, {Lelas}, {Lindfors}, {Lombardi}, {Longo},
  {L{\'o}pez}, {L{\'o}pez-Coto}, {Majumdar}, {Makariev}, {Mallot}, {Maneva},
  {Manganaro}, {Mannheim}, {Maraschi}, {Marcote}, {Mariotti}, {Mart{\'\i}nez},
  {Mazin}, {Menzel}, {Mirand a}, {Mirzoyan}, {Moralejo}, {Moretti}, {Nakajima},
  {Neustroev}, {Niedzwiecki}, {Nievas Rosillo}, {Nilsson}, {Nishijima}, {Noda},
  {Nogu{\'e}s}, {Orito}, {Overkemping}, {Paiano}, {Palacio}, {Palatiello},
  {Paneque}, {Paoletti}, {Paredes}, {Paredes-Fortuny}, {Pedaletti}, {Perri},
  {Persic}, {Poutanen}, {Prada Moroni}, {Prandini}, {Puljak}, {Rhode},
  {Rib{\'o}}, {Rico}, {Rodriguez Garcia}, {Saito}, {Satalecka}, {Schultz},
  {Schweizer}, {Shore}, {Sillanp{\"a}{\"a}}, {Sitarek}, {Snidaric},
  {Sobczynska}, {Stamerra}, {Steinbring}, {Strzys}, {Takalo}, {Takami},
  {Tavecchio}, {Temnikov}, {Terzi{\'c}}, {Tescaro}, {Teshima}, {Thaele},
  {Torres}, {Toyama}, {Treves}, {Verguilov}, {Vovk}, {Ward}, {Will}, {Wu}, \&
  {Zanin}}]{2016A&A...593A..91A}
{Ahnen}, M.~L., {Ansoldi}, S., {Antonelli}, L.~A., {et~al.} 2016, \aap, 593,
  A91

\bibitem[{{Ahnen} {et~al.}(2017{\natexlab{b}}){Ahnen}, {Ansoldi}, {Antonelli},
  {Arcaro}, {Babi{\'c}}, {Banerjee}, {Bangale}, {Barres de Almeida}, {Barrio},
  {Becerra Gonz{\'a}lez}, {Bednarek}, {Bernardini}, {Berti}, {Bhattacharyya},
  {Biasuzzi}, {Biland }, {Blanch}, {Bonnefoy}, {Bonnoli}, {Carosi}, {Carosi},
  {Chatterjee}, {Colin}, {Colombo}, {Contreras}, {Cortina}, {Covino}, {Cumani},
  {Da Vela}, {Dazzi}, {De Angelis}, {De Lotto}, {de O{\~n}a Wilhelmi}, {Di
  Pierro}, {Doert}, {Dom{\'\i}nguez}, {Dominis Prester}, {Dorner}, {Doro},
  {Einecke}, {Eisenacher Glawion}, {Elsaesser}, {Engelkemeier}, {Fallah
  Ramazani}, {Fern{\'a}ndez-Barral}, {Fidalgo}, {Fonseca}, {Font}, {Fruck},
  {Galindo}, {Garc{\'\i}a L{\'o}pez}, {Garczarczyk}, {Gaug}, {Giammaria},
  {Godinovi{\'c}}, {Gora}, {Griffiths}, {Guberman}, {Hadasch}, {Hahn},
  {Hassan}, {Hayashida}, {Herrera}, {Hose}, {Hrupec}, {Hughes}, {Ishio},
  {Konno}, {Kubo}, {Kushida}, {Kuve{\v{z}}di{\'c}}, {Lelas}, {Lindfors},
  {Lombardi}, {Longo}, {L{\'o}pez}, {Maggio}, {Majumdar}, {Makariev}, {Maneva},
  {Manganaro}, {Mannheim}, {Maraschi}, {Mariotti}, {Mart{\'\i}nez}, {Mazin},
  {Menzel}, {Minev}, {Mirzoyan}, {Moralejo}, {Moreno}, {Moretti}, {Neustroev},
  {Niedzwiecki}, {Nievas Rosillo}, {Nilsson}, {Ninci}, {Nishijima}, {Noda},
  {Nogu{\'e}s}, {Paiano}, {Palacio}, {Paneque}, {Paoletti}, {Paredes},
  {Paredes-Fortuny}, {Pedaletti}, {Peresano}, {Perri}, {Persic}, {Prada
  Moroni}, {Prandini}, {Puljak}, {Garcia}, {Reichardt}, {Rhode}, {Rib{\'o}},
  {Rico}, {Rugliancich}, {Saito}, {Satalecka}, {Schroeder}, {Schweizer},
  {Sillanp{\"a}{\"a}}, {Sitarek}, {{\v{S}}nidari{\'c}}, {Sobczynska},
  {Stamerra}, {Strzys}, {Suri{\'c}}, {Takalo}, {Tavecchio}, {Temnikov},
  {Terzi{\'c}}, {Tescaro}, {Teshima}, {Torres}, {Torres-Alb{\`a}}, {Treves},
  {Vanzo}, {Vazquez Acosta}, {Vovk}, {Ward}, {Will}, \&
  {Zari{\'c}}}]{2017APh....94...29A}
{Ahnen}, M.~L., {Ansoldi}, S., {Antonelli}, L.~A., {et~al.} 2017{\natexlab{b}},
  Astroparticle Physics, 94, 29

\bibitem[{{Aleksi{\'c}} {et~al.}(2012){Aleksi{\'c}}, {Alvarez}, {Antonelli},
  {Antoranz}, {Asensio}, {Backes}, {Barrio}, {Bastieri}, {Becerra
  Gonz{\'a}lez}, {Bednarek}, {Berdyugin}, {Berger}, {Bernardini}, {Biland},
  {Blanch}, {Bock}, {Boller}, {Bonnoli}, {Borla Tridon}, {Braun}, {Bretz},
  {Ca{\~n}ellas}, {Carmona}, {Carosi}, {Colin}, {Colombo}, {Contreras},
  {Cortina}, {Cossio}, {Covino}, {Dazzi}, {De Angelis}, {De Caneva}, {De Cea
  del Pozo}, {De Lotto}, {Delgado Mendez}, {Diago Ortega}, {Doert},
  {Dom{\'\i}nguez}, {Dominis Prester}, {Dorner}, {Doro}, {Elsaesser}, {Ferenc},
  {Fonseca}, {Font}, {Fruck}, {Garc{\'\i}a L{\'o}pez}, {Garczarczyk},
  {Garrido}, {Giavitto}, {Godinovi{\'c}}, {Hadasch}, {H{\"a}fner}, {Herrero},
  {Hildebrand}, {H{\"o}hne-M{\"o}nch}, {Hose}, {Hrupec}, {Huber}, {Jogler},
  {Kellermann}, {Klepser}, {Kr{\"a}henb{\"u}hl}, {Krause}, {La Barbera},
  {Lelas}, {Leonardo}, {Lindfors}, {Lombardi}, {L{\'o}pez}, {L{\'o}pez},
  {Lorenz}, {Makariev}, {Maneva}, {Mankuzhiyil}, {Mannheim}, {Maraschi},
  {Mariotti}, {Mart{\'\i}nez}, {Mazin}, {Meucci}, {Miranda}, {Mirzoyan},
  {Miyamoto}, {Mold{\'o}n}, {Moralejo}, {Munar-Adrover}, {Nieto}, {Nilsson},
  {Orito}, {Oya}, {Paneque}, {Paoletti}, {Pardo}, {Paredes}, {Partini},
  {Pasanen}, {Pauss}, {Perez-Torres}, {Persic}, {Peruzzo}, {Pilia}, {Pochon},
  {Prada}, {Prada Moroni}, {Prandini}, {Puljak}, {Reichardt}, {Reinthal},
  {Rhode}, {Rib{\'o}}, {Rico}, {R{\"u}gamer}, {Saggion}, {Saito}, {Saito},
  {Salvati}, {Satalecka}, {Scalzotto}, {Scapin}, {Schultz}, {Schweizer},
  {Shayduk}, {Shore}, {Sillanp{\"a}{\"a}}, {Sitarek}, {Sobczynska}, {Spanier},
  {Spiro}, {Stamerra}, {Steinke}, {Storz}, {Strah}, {Suri{\'c}}, {Takalo},
  {Takami}, {Tavecchio}, {Temnikov}, {Terzi{\'c}}, {Tescaro}, {Teshima},
  {Tibolla}, {Torres}, {Treves}, {Uellenbeck}, {Vankov}, {Vogler}, {Wagner},
  {Weitzel}, {Zabalza}, {Zandanel}, \& {Zanin}}]{2012A&A...542A.100A}
{Aleksi{\'c}}, J., {Alvarez}, E.~A., {Antonelli}, L.~A., {et~al.} 2012, \aap,
  542, A100

\bibitem[{{Aleksi{\'c}} {et~al.}(2016){Aleksi{\'c}}, {Ansoldi}, {Antonelli},
  {Antoranz}, {Babic}, {Bangale}, {Barcel{\'o}}, {Barrio}, {Becerra
  Gonz{\'a}lez}, {Bednarek}, {Bernardini}, {Biasuzzi}, {Biland}, {Bitossi},
  {Blanch}, {Bonnefoy}, {Bonnoli}, {Borracci}, {Bretz}, {Carmona}, {Carosi},
  {Cecchi}, {Colin}, {Colombo}, {Contreras}, {Corti}, {Cortina}, {Covino}, {Da
  Vela}, {Dazzi}, {De Angelis}, {De Caneva}, {De Lotto}, {de O{\~n}a Wilhelmi},
  {Delgado Mendez}, {Dettlaff}, {Dominis Prester}, {Dorner}, {Doro}, {Einecke},
  {Eisenacher}, {Elsaesser}, {Fidalgo}, {Fink}, {Fonseca}, {Font}, {Frantzen},
  {Fruck}, {Galindo}, {Garc{\'\i}a L{\'o}pez}, {Garczarczyk}, {Garrido
  Terrats}, {Gaug}, {Giavitto}, {Godinovi{\'c}}, {Gonz{\'a}lez Mu{\~n}oz},
  {Gozzini}, {Haberer}, {Hadasch}, {Hanabata}, {Hayashida}, {Herrera},
  {Hildebrand }, {Hose}, {Hrupec}, {Idec}, {Illa}, {Kadenius}, {Kellermann},
  {Knoetig}, {Kodani}, {Konno}, {Krause}, {Kubo}, {Kushida}, {La Barbera},
  {Lelas}, {Lemus}, {Lewandowska}, {Lindfors}, {Lombardi}, {Longo},
  {L{\'o}pez}, {L{\'o}pez-Coto}, {L{\'o}pez-Oramas}, {Lorca}, {Lorenz},
  {Lozano}, {Makariev}, {Mallot}, {Maneva}, {Mankuzhiyil}, {Mannheim},
  {Maraschi}, {Marcote}, {Mariotti}, {Mart{\'\i}nez}, {Mazin}, {Menzel},
  {Miranda}, {Mirzoyan}, {Moralejo}, {Munar-Adrover}, {Nakajima}, {Negrello},
  {Neustroev}, {Niedzwiecki}, {Nilsson}, {Nishijima}, {Noda}, {Orito},
  {Overkemping}, {Paiano}, {Palatiello}, {Paneque}, {Paoletti}, {Paredes},
  {Paredes-Fortuny}, {Persic}, {Poutanen}, {Prada Moroni}, {Prandini},
  {Puljak}, {Reinthal}, {Rhode}, {Rib{\'o}}, {Rico}, {Rodriguez Garcia},
  {R{\"u}gamer}, {Saito}, {Saito}, {Satalecka}, {Scalzotto}, {Scapin},
  {Schultz}, {Schlammer}, {Schmidl}, {Schweizer}, {Shore}, {Sillanp{\"a}{\"a}},
  {Sitarek}, {Snidaric}, {Sobczynska}, {Spanier}, {Stamerra}, {Steinbring},
  {Storz}, {Strzys}, {Takalo}, {Takami}, {Tavecchio}, {Tejedor}, {Temnikov},
  {Terzi{\'c}}, {Tescaro}, {Teshima}, {Thaele}, {Tibolla}, {Torres}, {Toyama},
  {Treves}, {Vogler}, {Wetteskind}, {Will}, \& {Zanin}}]{2016APh....72...76A}
{Aleksi{\'c}}, J., {Ansoldi}, S., {Antonelli}, L.~A., {et~al.} 2016,
  Astroparticle Physics, 72, 76

\bibitem[{{Aleksi{\'c}} {et~al.}(2015{\natexlab{a}}){Aleksi{\'c}}, {Ansoldi},
  {Antonelli}, {Antoranz}, {Babic}, {Bangale}, {Barres de Almeida}, {Barrio},
  {Becerra Gonz{\'a}lez}, {Bednarek}, {Berger}, {Bernardini}, {Biland},
  {Blanch}, {Bock}, {Bonnefoy}, {Bonnoli}, {Borracci}, {Bretz}, {Carmona},
  {Carosi}, {Carreto Fidalgo}, {Colin}, {Colombo}, {Contreras}, {Cortina},
  {Covino}, {Da Vela}, {Dazzi}, {De Angelis}, {De Caneva}, {De Lotto}, {Delgado
  Mendez}, {Doert}, {Dom{\'\i}nguez}, {Dominis Prester}, {Dorner}, {Doro},
  {Einecke}, {Eisenacher}, {Elsaesser}, {Farina}, {Ferenc}, {Fonseca}, {Font},
  {Frantzen}, {Fruck}, {Garc{\'\i}a L{\'o}pez}, {Garczarczyk}, {Garrido
  Terrats}, {Gaug}, {Giavitto}, {Godinovi{\'c}}, {Gonz{\'a}lez Mu{\~n}oz},
  {Gozzini}, {Hadamek}, {Hadasch}, {Herrero}, {Hildebrand}, {Hose}, {Hrupec},
  {Idec}, {Kadenius}, {Kellermann}, {Knoetig}, {Krause}, {Kushida}, {La
  Barbera}, {Lelas}, {Lewandowska}, {Lindfors}, {Longo}, {Lombardi},
  {L{\'o}pez}, {L{\'o}pez-Coto}, {L{\'o}pez-Oramas}, {Lorenz}, {Lozano},
  {Makariev}, {Mallot}, {Maneva}, {Mankuzhiyil}, {Mannheim}, {Maraschi},
  {Marcote}, {Mariotti}, {Mart{\'\i}nez}, {Mazin}, {Menzel}, {Meucci},
  {Miranda}, {Mirzoyan}, {Moralejo}, {Munar-Adrover}, {Nakajima},
  {Niedzwiecki}, {Nilsson}, {Nowak}, {Orito}, {Overkemping}, {Paiano},
  {Palatiello}, {Paneque}, {Paoletti}, {Paredes}, {Paredes-Fortuny}, {Partini},
  {Persic}, {Prada}, {Prada Moroni}, {Prand ini}, {Preziuso}, {Puljak},
  {Reinthal}, {Rhode}, {Rib{\'o}}, {Rico}, {RodriguezGarcia}, {R{\"u}gamer},
  {Saggion}, {Saito}, {Salvati}, {Satalecka}, {Scalzotto}, {Scapin}, {Schultz},
  {Schweizer}, {Shore}, {Sillanp{\"a}{\"a}}, {Sitarek}, {Snidaric},
  {Sobczynska}, {Spanier}, {Stamatescu}, {Stamerra}, {Steinbring}, {Storz},
  {Sun}, {Suri{\'c}}, {Takalo}, {Tavecchio}, {Temnikov}, {Terzi{\'c}},
  {Tescaro}, {Teshima}, {Thaele}, {Tibolla}, {Torres}, {Toyama}, {Treves},
  {Uellenbeck}, {Vogler}, {Wagner}, {Zandanel}, {Zanin}, {MAGIC Collaboration},
  {Archambault}, {Behera}, {Beilicke}, {Benbow}, {Bird}, {Buckley}, {Bugaev},
  {Cerruti}, {Chen}, {Ciupik}, {Collins-Hughes}, {Cui}, {Dumm}, {Eisch},
  {Falcone}, {Federici}, {Feng}, {Finley}, {Fleischhack}, {Fortin}, {Fortson},
  {Furniss}, {Griffin}, {Griffiths}, {Grube}, {Gyuk}, {Hanna}, {Holder},
  {Hughes}, {Humensky}, {Johnson}, {Kaaret}, {Kertzman}, {Khassen}, {Kieda},
  {Krawczynski}, {Krennrich}, {Kumar}, {Lang}, {Maier}, {McArthur}, {Meagher},
  {Moriarty}, {Mukherjee}, {Ong}, {Otte}, {Park}, {Pichel}, {Pohl}, {Popkow},
  {Prokoph}, {Quinn}, {Ragan}, {Rajotte}, {Reynolds}, {Richards}, {Roache},
  {Rovero}, {Sembroski}, {Shahinyan}, {Staszak}, {Telezhinsky}, {Theiling},
  {Tucci}, {Tyler}, {Varlotta}, {Wakely}, {Weekes}, {Weinstein}, {Welsing},
  {Wilhelm}, {Williams}, {Zitzer}, {VERITAS Collaboration}, {Villata},
  {Raiteri}, {Aller}, {Aller}, {Chen}, {Jordan}, {Koptelova}, {Kurtanidze},
  {L{\"a}hteenm{\"a}ki}, {McBreen}, {Larionov}, {Lin}, {Nikolashvili},
  {Angelakis}, {Capalbi}, {Carrami{\~n}ana}, {Carrasco}, {Cassaro}, {Cesarini},
  {Fuhrmann}, {Giroletti}, {Hovatta}, {Krichbaum}, {Krimm}, {Max-Moerbeck},
  {Moody}, {Maccaferri}, {Mori}, {Nestoras}, {Orlati}, {Pace}, {Pearson},
  {Perri}, {Readhead}, {Richards}, {Sadun}, {Sakamoto}, {Tammi}, {Tornikoski},
  {Yatsu}, \& {Zook}}]{2015A&A...576A.126A}
{Aleksi{\'c}}, J., {Ansoldi}, S., {Antonelli}, L.~A., {et~al.}
  2015{\natexlab{a}}, \aap, 576, A126

\bibitem[{{Aleksi{\'c}} {et~al.}(2015{\natexlab{b}}){Aleksi{\'c}}, {Ansoldi},
  {Antonelli}, {Antoranz}, {Babic}, {Bangale}, {Barres de Almeida}, {Barrio},
  {Becerra Gonz{\'a}lez}, {Bednarek}, {Berger}, {Bernardini}, {Biland},
  {Blanch}, {Bock}, {Bonnefoy}, {Bonnoli}, {Borracci}, {Bretz}, {Carmona},
  {Carosi}, {Carreto Fidalgo}, {Colin}, {Colombo}, {Contreras}, {Cortina},
  {Covino}, {da Vela}, {Dazzi}, {de Angelis}, {de Caneva}, {de Lotto}, {Delgado
  Mendez}, {Doert}, {Dom{\'\i}nguez}, {Dominis Prester}, {Dorner}, {Doro},
  {Einecke}, {Eisenacher}, {Elsaesser}, {Farina}, {Ferenc}, {Fonseca}, {Font},
  {Frantzen}, {Fruck}, {Garc{\'\i}a L{\'o}pez}, {Garczarczyk}, {Garrido
  Terrats}, {Gaug}, {Giavitto}, {Godinovi{\'c}}, {Gonz{\'a}lez Mu{\~n}oz},
  {Gozzini}, {Hadamek}, {Hadasch}, {Herrero}, {Hildebrand}, {Hose}, {Hrupec},
  {Idec}, {Kadenius}, {Kellermann}, {Knoetig}, {Krause}, {Kushida}, {La
  Barbera}, {Lelas}, {Lewandowska}, {Lindfors}, {Lombardi}, {L{\'o}pez},
  {L{\'o}pez-Coto}, {L{\'o}pez-Oramas}, {Lorenz}, {Lozano}, {Makariev},
  {Mallot}, {Maneva}, {Mankuzhiyil}, {Mannheim}, {Maraschi}, {Marcote},
  {Mariotti}, {Mart{\'\i}nez}, {Mazin}, {Menzel}, {Meucci}, {Miranda},
  {Mirzoyan}, {Moralejo}, {Munar-Adrover}, {Nakajima}, {Niedzwiecki},
  {Nilsson}, {Nowak}, {Orito}, {Overkemping}, {Paiano}, {Palatiello},
  {Paneque}, {Paoletti}, {Paredes}, {Paredes-Fortuny}, {Partini}, {Persic},
  {Prada}, {Prada Moroni}, {Prandini}, {Preziuso}, {Puljak}, {Reinthal},
  {Rhode}, {Rib{\'o}}, {Rico}, {Rodriguez Garcia}, {R{\"u}gamer}, {Saggion},
  {Saito}, {Saito}, {Salvati}, {Satalecka}, {Scalzotto}, {Scapin}, {Schultz},
  {Schweizer}, {Shore}, {Sillanp{\"a}{\"a}}, {Sitarek}, {Snidaric},
  {Sobczynska}, {Spanier}, {Stamatescu}, {Stamerra}, {Steinbring}, {Storz},
  {Sun}, {Suri{\'c}}, {Takalo}, {Tavecchio}, {Temnikov}, {Terzi{\'c}},
  {Tescaro}, {Teshima}, {Thaele}, {Tibolla}, {Torres}, {Toyama}, {Treves},
  {Uellenbeck}, {Vogler}, {Wagner}, {Zandanel}, {Zanin}, {MAGIC Collaboration},
  {Behera}, {Beilicke}, {Benbow}, {Berger}, {Bird}, {Bouvier}, {Bugaev},
  {Cerruti}, {Chen}, {Ciupik}, {Collins-Hughes}, {Cui}, {Duke}, {Dumm},
  {Falcone}, {Federici}, {Feng}, {Finley}, {Fortson}, {Furniss}, {Galante},
  {Gillanders}, {Griffin}, {Griffiths}, {Grube}, {Gyuk}, {Hanna}, {Holder},
  {Johnson}, {Kaaret}, {Kertzman}, {Kieda}, {Krawczynski}, {Lang}, {Madhavan},
  {Maier}, {Majumdar}, {Meagher}, {Moriarty}, {Mukherjee}, {Nieto},
  {O'Faol{\'a}in de Bhr{\'o}ithe}, {Ong}, {Otte}, {Pichel}, {Pohl}, {Popkow},
  {Prokoph}, {Quinn}, {Rajotte}, {Ratliff}, {Reyes}, {Reynolds}, {Richards},
  {Roache}, {Sembroski}, {Shahinyan}, {Sheidaei}, {Smith}, {Staszak},
  {Telezhinsky}, {Theiling}, {Tyler}, {Varlotta}, {Vincent}, {Wakely},
  {Weekes}, {Welsing}, {Williams}, {Zajczyk}, {Zitzer}, {VERITAS
  Collaboration}, {Villata}, {Raiteri}, {Ajello}, {Perri}, {Aller}, {Aller},
  {Larionov}, {Efimova}, {Konstantinova}, {Kopatskaya}, {Chen}, {Koptelova},
  {Hsiao}, {Kurtanidze}, {Nikolashvili}, {Kimeridze}, {Jordan}, {Leto},
  {Buemi}, {Trigilio}, {Umana}, {L{\"a}hteenm{\"a}ki}, {Nieppola},
  {Tornikoski}, {Sainio}, {Kadenius}, {Giroletti}, {Cesarini}, {Fuhrmann},
  {Kovalev}, \& {Kovalev}}]{2015A&A...573A..50A}
{Aleksi{\'c}}, J., {Ansoldi}, S., {Antonelli}, L.~A., {et~al.}
  2015{\natexlab{b}}, \aap, 573, A50

\bibitem[{{Aleksi{\'c}} {et~al.}(2015{\natexlab{c}}){Aleksi{\'c}}, {Ansoldi},
  {Antonelli}, {Antoranz}, {Babic}, {Bangale}, {Barres de Almeida}, {Barrio},
  {Becerra Gonz{\'a}lez}, {Bednarek}, {Bernardini}, {Biasuzzi}, {Biland},
  {Blanch}, {Boller}, {Bonnefoy}, {Bonnoli}, {Borracci}, {Bretz}, {Carmona},
  {Carosi}, {Colin}, {Colombo}, {Contreras}, {Cortina}, {Covino}, {Da Vela},
  {Dazzi}, {De Angelis}, {De Caneva}, {De Lotto}, {de O{\~n}a Wilhelmi},
  {Delgado Mendez}, {Dominis Prester}, {Dorner}, {Doro}, {Einecke},
  {Eisenacher}, {Elsaesser}, {Fonseca}, {Font}, {Frantzen}, {Fruck}, {Galindo},
  {Garc{\'\i}a L{\'o}pez}, {Garczarczyk}, {Garrido Terrats}, {Gaug},
  {Godinovi{\'c}}, {Gonz{\'a}lez Mu{\~n}oz}, {Gozzini}, {Hadasch}, {Hanabata},
  {Hayashida}, {Herrera}, {Hildebrand }, {Hose}, {Hrupec}, {Hughes}, {Idec},
  {Kadenius}, {Kellermann}, {Knoetig}, {Kodani}, {Konno}, {Krause}, {Kubo},
  {Kushida}, {La Barbera}, {Lelas}, {Lewand owska}, {Lindfors}, {Lombardi},
  {L{\'o}pez}, {L{\'o}pez-Coto}, {L{\'o}pez-Oramas}, {Lorenz}, {Lozano},
  {Makariev}, {Mallot}, {Maneva}, {Mankuzhiyil}, {Mannheim}, {Maraschi},
  {Marcote}, {Mariotti}, {Mart{\'\i}nez}, {Mazin}, {Menzel}, {Miranda},
  {Mirzoyan}, {Moralejo}, {Munar-Adrover}, {Nakajima}, {Niedzwiecki},
  {Nilsson}, {Nishijima}, {Noda}, {Orito}, {Overkemping}, {Paiano},
  {Palatiello}, {Paneque}, {Paoletti}, {Paredes}, {Paredes-Fortuny}, {Persic},
  {Prada Moroni}, {Prandini}, {Puljak}, {Reinthal}, {Rhode}, {Rib{\'o}},
  {Rico}, {Rodriguez Garcia}, {R{\"u}gamer}, {Saito}, {Saito}, {Satalecka},
  {Scalzotto}, {Scapin}, {Schultz}, {Schweizer}, {Sun}, {Shore},
  {Sillanp{\"a}{\"a}}, {Sitarek}, {Snidaric}, {Sobczynska}, {Spanier},
  {Stamatescu}, {Stamerra}, {Steinbring}, {Steinke}, {Storz}, {Strzys},
  {Takalo}, {Takami}, {Tavecchio}, {Temnikov}, {Terzi{\'c}}, {Tescaro},
  {Teshima}, {Thaele}, {Tibolla}, {Torres}, {Toyama}, {Treves}, {Uellenbeck},
  {Vogler}, {Zanin}, {MAGIC Collaboration}, {Archambault}, {Archer},
  {Beilicke}, {Benbow}, {Berger}, {Bird}, {Biteau}, {Buckley}, {Bugaev},
  {Cerruti}, {Chen}, {Ciupik}, {Collins-Hughes}, {Cui}, {Eisch}, {Falcone},
  {Feng}, {Finley}, {Fortin}, {Fortson}, {Furniss}, {Galante}, {Gillanders},
  {Griffin}, {Gyuk}, {H{\r{a}}kansson}, {Holder}, {Johnson}, {Kaaret}, {Kar},
  {Kertzman}, {Kieda}, {Lang}, {McArthur}, {McCann}, {Meagher}, {Millis},
  {Moriarty}, {Ong}, {Otte}, {Perkins}, {Pichel}, {Pohl}, {Popkow}, {Prokoph},
  {Pueschel}, {Ragan}, {Reyes}, {Reynolds}, {Richards}, {Roache}, {Rovero},
  {Sembroski}, {Shahinyan}, {Staszak}, {Telezhinsky}, {Tucci}, {Tyler},
  {Varlotta}, {Wakely}, {Welsing}, {Wilhelm}, {Williams}, {VERITAS
  Collaboration}, {Buson}, {Finke}, {Villata}, {Raiteri}, {Aller}, {Aller},
  {Cesarini}, {Chen}, {Gurwell}, {Jorstad}, {Kimeridze}, {Koptelova},
  {Kurtanidze}, {Kurtanidze}, {L{\"a}hteenm{\"a}ki}, {Larionov}, {Larionova},
  {Lin}, {McBreen}, {Moody}, {Morozova}, {Marscher}, {Max-Moerbeck},
  {Nikolashvili}, {Perri}, {Readhead}, {Richards}, {Ros}, {Sadun}, {Sakamoto},
  {Sigua}, {Smith}, {Tornikoski}, {Troitsky}, {Wehrle}, \&
  {Jordan}}]{2015A&A...578A..22A}
{Aleksi{\'c}}, J., {Ansoldi}, S., {Antonelli}, L.~A., {et~al.}
  2015{\natexlab{c}}, \aap, 578, A22

\bibitem[{{Anderhub} {et~al.}(2013){Anderhub}, {Backes}, {Biland}, {Boccone},
  {Braun}, {Bretz}, {Bu{\ss}}, {Cadoux}, {Commichau}, {Djambazov}, {Dorner},
  {Einecke}, {Eisenacher}, {Gendotti}, {Grimm}, {von Gunten}, {Haller},
  {Hildebrand}, {Horisberger}, {Huber}, {Kim}, {Knoetig}, {K{\"o}hne},
  {Kr{\"a}henb{\"u}hl}, {Krumm}, {Lee}, {Lorenz}, {Lustermann}, {Lyard},
  {Mannheim}, {Meharga}, {Meier}, {Montaruli}, {Neise}, {Nessi-Tedaldi},
  {Overkemping}, {Paravac}, {Pauss}, {Renker}, {Rhode}, {Ribordy}, {R{\"o}ser},
  {Stucki}, {Schneider}, {Steinbring}, {Temme}, {Thaele}, {Tobler}, {Viertel},
  {Vogler}, {Walter}, {Warda}, {Weitzel}, \&
  {Z{\"a}nglein}}]{2013JInst...8P6008A}
{Anderhub}, H., {Backes}, M., {Biland}, A., {et~al.} 2013, Journal of
  Instrumentation, 8, P06008

\bibitem[{{Arnaud}(1996)}]{1996ASPC..101...17A}
{Arnaud}, K.~A. 1996, in Astronomical Society of the Pacific Conference Series,
  Vol. 101, Astronomical Data Analysis Software and Systems V, ed. G.~H.
  {Jacoby} \& J.~{Barnes}, 17

\bibitem[{{Asano} {et~al.}(2014){Asano}, {Takahara}, {Kusunose}, {Toma}, \&
  {Kakuwa}}]{2014ApJ...780...64A}
{Asano}, K., {Takahara}, F., {Kusunose}, M., {Toma}, K., \& {Kakuwa}, J. 2014,
  \apj, 780, 64

\bibitem[{{Atwood} {et~al.}(2009){Atwood}, {Abdo}, {Ackermann}, {Althouse},
  {Anderson}, {Axelsson}, {Baldini}, {Ballet}, {Band}, {Barbiellini},
  {Bartelt}, {Bastieri}, {Baughman}, {Bechtol}, {B{\'e}d{\'e}r{\`e}de},
  {Bellardi}, {Bellazzini}, {Berenji}, {Bignami}, {Bisello}, {Bissaldi},
  {Blandford}, {Bloom}, {Bogart}, {Bonamente}, {Bonnell}, {Borgland },
  {Bouvier}, {Bregeon}, {Brez}, {Brigida}, {Bruel}, {Burnett}, {Busetto},
  {Caliandro}, {Cameron}, {Caraveo}, {Carius}, {Carlson}, {Casandjian},
  {Cavazzuti}, {Ceccanti}, {Cecchi}, {Charles}, {Chekhtman}, {Cheung},
  {Chiang}, {Chipaux}, {Cillis}, {Ciprini}, {Claus}, {Cohen-Tanugi},
  {Condamoor}, {Conrad}, {Corbet}, {Corucci}, {Costamante}, {Cutini}, {Davis},
  {Decotigny}, {DeKlotz}, {Dermer}, {de Angelis}, {Digel}, {do Couto e Silva},
  {Drell}, {Dubois}, {Dumora}, {Edmonds}, {Fabiani}, {Farnier}, {Favuzzi},
  {Flath}, {Fleury}, {Focke}, {Funk}, {Fusco}, {Gargano}, {Gasparrini},
  {Gehrels}, {Gentit}, {Germani}, {Giebels}, {Giglietto}, {Giommi}, {Giordano},
  {Glanzman}, {Godfrey}, {Grenier}, {Grondin}, {Grove}, {Guillemot}, {Guiriec},
  {Haller}, {Harding}, {Hart}, {Hays}, {Healey}, {Hirayama}, {Hjalmarsdotter},
  {Horn}, {Hughes}, {J{\'o}hannesson}, {Johansson}, {Johnson}, {Johnson},
  {Johnson}, {Johnson}, {Kamae}, {Katagiri}, {Kataoka}, {Kavelaars}, {Kawai},
  {Kelly}, {Kerr}, {Klamra}, {Kn{\"o}dlseder}, {Kocian}, {Komin}, {Kuehn},
  {Kuss}, {Landriu}, {Latronico}, {Lee}, {Lee}, {Lemoine-Goumard}, {Lionetto},
  {Longo}, {Loparco}, {Lott}, {Lovellette}, {Lubrano}, {Madejski}, {Makeev},
  {Marangelli}, {Massai}, {Mazziotta}, {McEnery}, {Menon}, {Meurer},
  {Michelson}, {Minuti}, {Mirizzi}, {Mitthumsiri}, {Mizuno}, {Moiseev},
  {Monte}, {Monzani}, {Moretti}, {Morselli}, {Moskalenko}, {Murgia},
  {Nakamori}, {Nishino}, {Nolan}, {Norris}, {Nuss}, {Ohno}, {Ohsugi}, {Omodei},
  {Orlando}, {Ormes}, {Paccagnella}, {Paneque}, {Panetta}, {Parent}, {Pearce},
  {Pepe}, {Perazzo}, {Pesce-Rollins}, {Picozza}, {Pieri}, {Pinchera}, {Piron},
  {Porter}, {Poupard}, {Rain{\`o}}, {Rando}, {Rapposelli}, {Razzano}, {Reimer},
  {Reimer}, {Reposeur}, {Reyes}, {Ritz}, {Rochester}, {Rodriguez}, {Romani},
  {Roth}, {Russell}, {Ryde}, {Sabatini}, {Sadrozinski}, {Sanchez}, {Sand er},
  {Sapozhnikov}, {Parkinson}, {Scargle}, {Schalk}, {Scolieri}, {Sgr{\`o}},
  {Share}, {Shaw}, {Shimokawabe}, {Shrader}, {Sierpowska-Bartosik}, {Siskind},
  {Smith}, {Smith}, {Spandre}, {Spinelli}, {Starck}, {Stephens}, {Strickman},
  {Strong}, {Suson}, {Tajima}, {Takahashi}, {Takahashi}, {Tanaka}, {Tenze},
  {Tether}, {Thayer}, {Thayer}, {Thompson}, {Tibaldo}, {Tibolla}, {Torres},
  {Tosti}, {Tramacere}, {Turri}, {Usher}, {Vilchez}, {Vitale}, {Wang},
  {Watters}, {Winer}, {Wood}, {Ylinen}, \& {Ziegler}}]{2009ApJ...697.1071A}
{Atwood}, W.~B., {Abdo}, A.~A., {Ackermann}, M., {et~al.} 2009, \apj, 697, 1071

\bibitem[{{Balokovi{\'c}} {et~al.}(2016){Balokovi{\'c}}, {Paneque}, {Madejski},
  {Furniss}, {Chiang}, {Ajello}, {Alexander}, {Barret}, {Blandford}, {Boggs},
  {Christensen}, {Craig}, {Forster}, {Giommi}, {Grefenstette}, {Hailey},
  {Harrison}, {Hornstrup}, {Kitaguchi}, {Koglin}, {Madsen}, {Mao}, {Miyasaka},
  {Mori}, {Perri}, {Pivovaroff}, {Puccetti}, {Rana}, {Stern}, {Tagliaferri},
  {Urry}, {Westergaard}, {Zhang}, {Zoglauer}, {NuSTAR Team}, {Archambault},
  {Archer}, {Barnacka}, {Benbow}, {Bird}, {Buckley}, {Bugaev}, {Cerruti},
  {Chen}, {Ciupik}, {Connolly}, {Cui}, {Dickinson}, {Dumm}, {Eisch}, {Falcone},
  {Feng}, {Finley}, {Fleischhack}, {Fortson}, {Griffin}, {Griffiths}, {Grube},
  {Gyuk}, {Huetten}, {H{\r{a}}kansson}, {Holder}, {Humensky}, {Johnson},
  {Kaaret}, {Kertzman}, {Khassen}, {Kieda}, {Krause}, {Krennrich}, {Lang},
  {Maier}, {McArthur}, {Meagher}, {Moriarty}, {Nelson}, {Nieto}, {Ong}, {Park},
  {Pohl}, {Popkow}, {Pueschel}, {Reynolds}, {Richards}, {Roache}, {Santander},
  {Sembroski}, {Shahinyan}, {Smith}, {Staszak}, {Telezhinsky}, {Todd}, {Tucci},
  {Tyler}, {Vincent}, {Weinstein}, {Wilhelm}, {Williams}, {Zitzer}, {VERITAS
  Collaboration}, {Ahnen}, {Ansoldi}, {Antonelli}, {Antoranz}, {Babic},
  {Banerjee}, {Bangale}, {Barres de Almeida}, {Barrio}, {Becerra Gonz{\'a}lez},
  {Bednarek}, {Bernardini}, {Biasuzzi}, {Biland}, {Blanch}, {Bonnefoy},
  {Bonnoli}, {Borracci}, {Bretz}, {Carmona}, {Carosi}, {Chatterjee}, {Clavero},
  {Colin}, {Colombo}, {Contreras}, {Cortina}, {Covino}, {Da Vela}, {Dazzi}, {De
  Angelis}, {De Lotto}, {de O{\~n}a Wilhelmi}, {Delgado Mendez}, {Di Pierro},
  {Dominis Prester}, {Dorner}, {Doro}, {Einecke}, {Elsaesser},
  {Fern{\'a}ndez-Barral}, {Fidalgo}, {Fonseca}, {Font}, {Frantzen}, {Fruck},
  {Galindo}, {Garc{\'\i}a L{\'o}pez}, {Garczarczyk}, {Garrido Terrats}, {Gaug},
  {Giammaria}, {Glawion (Eisenacher}, {Godinovi{\'c}}, {Gonz{\'a}lez
  Mu{\~n}oz}, {Guberman}, {Hahn}, {Hanabata}, {Hayashida}, {Herrera}, {Hose},
  {Hrupec}, {Hughes}, {Idec}, {Kodani}, {Konno}, {Kubo}, {Kushida}, {La
  Barbera}, {Lelas}, {Lindfors}, {Lombardi}, {Longo}, {L{\'o}pez},
  {L{\'o}pez-Coto}, {L{\'o}pez-Oramas}, {Lorenz}, {Majumdar}, {Makariev},
  {Mallot}, {Maneva}, {Manganaro}, {Mannheim}, {Maraschi}, {Marcote},
  {Mariotti}, {Mart{\'\i}nez}, {Mazin}, {Menzel}, {Miranda}, {Mirzoyan},
  {Moralejo}, {Moretti}, {Nakajima}, {Neustroev}, {Niedzwiecki}, {Nievas
  Rosillo}, {Nilsson}, {Nishijima}, {Noda}, {Orito}, {Overkemping}, {Paiano},
  {Palacio}, {Palatiello}, {Paoletti}, {Paredes}, {Paredes-Fortuny}, {Persic},
  {Poutanen}, {Prada Moroni}, {Prandini}, {Puljak}, {Rhode}, {Rib{\'o}},
  {Rico}, {Rodriguez Garcia}, {Saito}, {Satalecka}, {Scapin}, {Schultz},
  {Schweizer}, {Shore}, {Sillanp{\"a}{\"a}}, {Sitarek}, {Snidaric},
  {Sobczynska}, {Stamerra}, {Steinbring}, {Strzys}, {Takalo}, {Takami},
  {Tavecchio}, {Temnikov}, {Terzi{\'c}}, {Tescaro}, {Teshima}, {Thaele},
  {Torres}, {Toyama}, {Treves}, {Verguilov}, {Vovk}, {Ward}, {Will}, {Wu},
  {Zanin}, {MAGIC Collaboration}, {Perkins}, {Verrecchia}, {Leto},
  {B{\"o}ttcher}, {Villata}, {Raiteri}, {Acosta-Pulido}, {Bachev}, {Berdyugin},
  {Blinov}, {Carnerero}, {Chen}, {Chinchilla}, {Damljanovic}, {Eswaraiah},
  {Grishina}, {Ibryamov}, {Jordan}, {Jorstad}, {Joshi}, {Kopatskaya},
  {Kurtanidze}, {Kurtanidze}, {Larionova}, {Larionova}, {Larionov}, {Latev},
  {Lin}, {Marscher}, {Mokrushina}, {Morozova}, {Nikolashvili}, {Semkov},
  {Smith}, {Strigachev}, {Troitskaya}, {Troitsky}, {Vince}, {Barnes},
  {G{\"u}ver}, {Moody}, {Sadun}, {Sun}, {Hovatta}, {Richards}, {Max-Moerbeck},
  {Readhead}, {L{\"a}hteenm{\"a}ki}, {Tornikoski}, {Tammi}, {Ramakrishnan},
  {Reinthal}, {Angelakis}, {Fuhrmann}, {Myserlis}, {Karamanavis}, {Sievers},
  {Ungerechts}, \& {Zensus}}]{2016ApJ...819..156B}
{Balokovi{\'c}}, M., {Paneque}, D., {Madejski}, G., {et~al.} 2016, \apj, 819,
  156

\bibitem[{{Biland} {et~al.}(2014){Biland}, {Bretz}, {Bu{\ss}}, {Commichau},
  {Djambazov}, {Dorner}, {Einecke}, {Eisenacher}, {Freiwald}, {Grimm}, {von
  Gunten}, {Haller}, {Hempfling}, {Hildebrand}, {Hughes}, {Horisberger},
  {Knoetig}, {Kr{\"a}henb{\"u}hl}, {Lustermann}, {Lyard}, {Mannheim}, {Meier},
  {Mueller}, {Neise}, {Overkemping}, {Paravac}, {Pauss}, {Rhode}, {R{\"o}ser},
  {Stucki}, {Steinbring}, {Temme}, {Thaele}, {Vogler}, {Walter}, \&
  {Weitzel}}]{2014JInst...9P0012B}
{Biland}, A., {Bretz}, T., {Bu{\ss}}, J., {et~al.} 2014, Journal of
  Instrumentation, 9, P10012

\bibitem[{{Biteau} {et~al.}(2020){Biteau}, {Prandini}, {Costamante}, {Lemoine},
  {Padovani}, {Pueschel}, {Resconi}, {Tavecchio}, {Taylor}, \&
  {Zech}}]{2020NatAs...4..124B}
{Biteau}, J., {Prandini}, E., {Costamante}, L., {et~al.} 2020, Nature
  Astronomy, 4, 124

\bibitem[{B{\l}a{\.{z}}ejowski {et~al.}(2005)B{\l}a{\.{z}}ejowski, Blaylock,
  Bond, Bradbury, Buckley, Carter-Lewis, Celik, Cogan, Cui, Daniel, Duke,
  Falcone, Fegan, Fegan, Finley, Fortson, Gammell, Gibbs, Gillanders, Grube,
  Gutierrez, Hall, Hanna, Holder, Horan, Humensky, Kenny, Kertzman, Kieda,
  Kildea, Knapp, Kosack, Krawczynski, Krennrich, Lang, LeBohec, Linton,
  Lloyd-Evans, Maier, Mendoza, Milovanovic, Moriarty, Nagai, Ong, Power-Mooney,
  Quinn, Quinn, Ragan, Reynolds, Rebillot, Rose, Schroedter, Sembroski, Swordy,
  Syson, Valcarel, Vassiliev, Wakely, Walker, Weekes, White, Zweerink,
  Mochejska, Smith, Aller, Aller, Terasranta, Boltwood, Sadun, Stanek, Adams,
  Foster, Hartman, Lai, Bottcher, Reimer, \& and}]{B_a_ejowski_2005}
B{\l}a{\.{z}}ejowski, M., Blaylock, G., Bond, I.~H., {et~al.} 2005, The
  Astrophysical Journal, 630, 130

\bibitem[{{B{\"o}ttcher} {et~al.}(2013){B{\"o}ttcher}, {Reimer}, {Sweeney}, \&
  {Prakash}}]{2013ApJ...768...54B}
{B{\"o}ttcher}, M., {Reimer}, A., {Sweeney}, K., \& {Prakash}, A. 2013, \apj,
  768, 54

\bibitem[{{Breeveld} {et~al.}(2011){Breeveld}, {Landsman}, {Holland}, {Roming},
  {Kuin}, \& {Page}}]{2011AIPC.1358..373B}
{Breeveld}, A.~A., {Landsman}, W., {Holland}, S.~T., {et~al.} 2011, in American
  Institute of Physics Conference Series, Vol. 1358, Gamma Ray Bursts 2010, ed.
  J.~E. {McEnery}, J.~L. {Racusin}, \& N.~{Gehrels}, 373--376

\bibitem[{{Bretz}(2019)}]{2019APh...111...72B}
{Bretz}, T. 2019, Astroparticle Physics, 111, 72

\bibitem[{{Bretz} \& {Dorner}(2010)}]{2010apsp.conf..681B}
{Bretz}, T. \& {Dorner}, D. 2010, in Astroparticle, Particle and Space Physics,
  Detectors and Medical Physics Applications, ed. C.~{Leroy}, P.-G. {Rancoita},
  M.~{Barone}, A.~{Gaddi}, L.~{Price}, \& R.~{Ruchti}, 681--687

\bibitem[{{Burrows} {et~al.}(2005){Burrows}, {Hill}, {Nousek}, {Kennea},
  {Wells}, {Osborne}, {Abbey}, {Beardmore}, {Mukerjee}, {Short}, {Chincarini},
  {Campana}, {Citterio}, {Moretti}, {Pagani}, {Tagliaferri}, {Giommi},
  {Capalbi}, {Tamburelli}, {Angelini}, {Cusumano}, {Br{\"a}uninger}, {Burkert},
  \& {Hartner}}]{2005SSRv..120..165B}
{Burrows}, D.~N., {Hill}, J.~E., {Nousek}, J.~A., {et~al.} 2005, \ssr, 120, 165

\bibitem[{{Carnerero} {et~al.}(2017){Carnerero}, {Raiteri}, {Villata},
  {Acosta-Pulido}, {Larionov}, {Smith}, {D'Ammando}, {Agudo}, {Ar{\'e}valo},
  {Bachev}, {Barnes}, {Boeva}, {Bozhilov}, {Carosati}, {Casadio}, {Chen},
  {Damljanovic}, {Eswaraiah}, {Forn{\'e}}, {Gantchev}, {G{\'o}mez},
  {Gonz{\'a}lez-Morales}, {Gri{\~n}{\'o}n-Mar{\'\i}n}, {Grishina}, {Holden},
  {Ibryamov}, {Joner}, {Jordan}, {Jorstad}, {Joshi}, {Kopatskaya}, {Koptelova},
  {Kurtanidze}, {Kurtanidze}, {Larionova}, {Larionova}, {Latev}, {L{\'a}zaro},
  {Ligustri}, {Lin}, {Marscher}, {Mart{\'\i}nez-Lombilla}, {McBreen}, {Mihov},
  {Molina}, {Moody}, {Morozova}, {Nikolashvili}, {Nilsson}, {Ovcharov}, {Pace},
  {Panwar}, {Pastor Yabar}, {Pearson}, {Pinna}, {Protasio}, {Rizzi},
  {Redondo-Lorenzo}, {Rodr{\'\i}guez-Coira}, {Ros}, {Sadun}, {Savchenko},
  {Semkov}, {Slavcheva-Mihova}, {Smith}, {Strigachev}, {Troitskaya},
  {Troitsky}, {Vasilyev}, \& {Vince}}]{2017MNRAS.472.3789C}
{Carnerero}, M.~I., {Raiteri}, C.~M., {Villata}, M., {et~al.} 2017, \mnras,
  472, 3789

\bibitem[{{Cerruti} {et~al.}(2015){Cerruti}, {Zech}, {Boisson}, \&
  {Inoue}}]{2015MNRAS.448..910C}
{Cerruti}, M., {Zech}, A., {Boisson}, C., \& {Inoue}, S. 2015, \mnras, 448, 910

\bibitem[{{Chatterjee} {et~al.}(2008){Chatterjee}, {Jorstad}, {Marscher}, {Oh},
  {McHardy}, {Aller}, {Aller}, {Balonek}, {Miller}, {Ryle}, {Tosti},
  {Kurtanidze}, {Nikolashvili}, {Larionov}, \&
  {Hagen-Thorn}}]{2008ApJ...689...79C}
{Chatterjee}, R., {Jorstad}, S.~G., {Marscher}, A.~P., {et~al.} 2008, \apj,
  689, 79

\bibitem[{{Christie} {et~al.}(2019){Christie}, {Petropoulou}, {Sironi}, \&
  {Giannios}}]{2019MNRAS.482...65C}
{Christie}, I.~M., {Petropoulou}, M., {Sironi}, L., \& {Giannios}, D. 2019,
  \mnras, 482, 65

\bibitem[{{Christie} {et~al.}(2020){Christie}, {Petropoulou}, {Sironi}, \&
  {Giannios}}]{2020MNRAS.492..549C}
{Christie}, I.~M., {Petropoulou}, M., {Sironi}, L., \& {Giannios}, D. 2020,
  \mnras, 492, 549

\bibitem[{{Crumley} {et~al.}(2019){Crumley}, {Caprioli}, {Markoff}, \&
  {Spitkovsky}}]{2019MNRAS.485.5105C}
{Crumley}, P., {Caprioli}, D., {Markoff}, S., \& {Spitkovsky}, A. 2019, \mnras,
  485, 5105

\bibitem[{{de Vaucouleurs} {et~al.}(1991){de Vaucouleurs}, {de Vaucouleurs},
  {Corwin}, {Buta}, {Paturel}, \& {Fouque}}]{1991rc3..book.....D}
{de Vaucouleurs}, G., {de Vaucouleurs}, A., {Corwin}, Herold~G., J., {et~al.}
  1991, {Third Reference Catalogue of Bright Galaxies}

\bibitem[{{Dom{\'\i}nguez} {et~al.}(2011){Dom{\'\i}nguez}, {Primack},
  {Rosario}, {Prada}, {Gilmore}, {Faber}, {Koo}, {Somerville},
  {P{\'e}rez-Torres}, {P{\'e}rez-Gonz{\'a}lez}, {Huang}, {Davis},
  {Guhathakurta}, {Barmby}, {Conselice}, {Lozano}, {Newman}, \&
  {Cooper}}]{2011MNRAS.410.2556D}
{Dom{\'\i}nguez}, A., {Primack}, J.~R., {Rosario}, D.~J., {et~al.} 2011,
  \mnras, 410, 2556

\bibitem[{{Dorner} {et~al.}(2015){Dorner}, {Ahnen}, {Bergmann}, {Biland},
  {Balbo}, {Bretz}, {Buss}, {Einecke}, {Freiwald}, {Hempfling}, {Hildebrand},
  {Hughes}, {Lustermann}, {Mannheim}, {Meier}, {Mueller}, {Neise}, {Neronov},
  {Overkemping}, {Paravac}, {Pauss}, {Rhode}, {Steinbring}, {Temme}, {Thaele},
  {Toscano}, {Vogler}, {Walter}, \& {Wilbert}}]{2015arXiv150202582D}
{Dorner}, D., {Ahnen}, M.~L., {Bergmann}, M., {et~al.} 2015, arXiv e-prints,
  arXiv:1502.02582

\bibitem[{{Edelson} \& {Krolik}(1988)}]{1988ApJ...333..646E}
{Edelson}, R.~A. \& {Krolik}, J.~H. 1988, \apj, 333, 646

\bibitem[{{Finke}(2013)}]{2013ApJ...763..134F}
{Finke}, J.~D. 2013, \apj, 763, 134

\bibitem[{{Finke} {et~al.}(2008){Finke}, {Dermer}, \&
  {B{\"o}ttcher}}]{2008ApJ...686..181F}
{Finke}, J.~D., {Dermer}, C.~D., \& {B{\"o}ttcher}, M. 2008, \apj, 686, 181

\bibitem[{{Fitzpatrick}(1999)}]{1999PASP..111...63F}
{Fitzpatrick}, E.~L. 1999, \pasp, 111, 63

\bibitem[{{Fossati} {et~al.}(2008){Fossati}, {Buckley}, {Bond}, {Bradbury},
  {Carter-Lewis}, {Chow}, {Cui}, {Falcone}, {Finley}, {Gaidos}, {Grube},
  {Holder}, {Horan}, {Horns}, {Jordan}, {Kieda}, {Kildea}, {Krawczynski},
  {Krennrich}, {Lang}, {LeBohec}, {Lee}, {Moriarty}, {Ong}, {Petry}, {Quinn},
  {Sembroski}, {Wakely}, \& {Weekes}}]{2008ApJ...677..906F}
{Fossati}, G., {Buckley}, J.~H., {Bond}, I.~H., {et~al.} 2008, \apj, 677, 906

\bibitem[{{Gaidos} {et~al.}(1996){Gaidos}, {Akerlof}, {Biller}, {Boyle},
  {Breslin}, {Buckley}, {Carter-Lewis}, {Catanese}, {Cawley}, {Fegan},
  {Finley}, {Gordo}, {Hillas}, {Krennrich}, {Lamb}, {Lessard}, {McEnery},
  {Masterson}, {Mohanty}, {Moriarty}, {Quinn}, {Rodgers}, {Rose}, {Samuelson},
  {Schubnell}, {Sembroski}, {Srinivasan}, {Weekes}, {Wilson}, \&
  {Zweerink}}]{1996Natur.383..319G}
{Gaidos}, J.~A., {Akerlof}, C.~W., {Biller}, S., {et~al.} 1996, \nat, 383, 319

\bibitem[{{Ghisellini} \& {Maraschi}(1996)}]{1996ASPC..110..436G}
{Ghisellini}, G. \& {Maraschi}, L. 1996, Astronomical Society of the Pacific
  Conference Series, Vol. 110, {High energy variability and blazar emission
  models.}, ed. H.~R. {Miller}, J.~R. {Webb}, \& J.~C. {Noble}, 436--449

\bibitem[{{Ghisellini} {et~al.}(2017){Ghisellini}, {Righi}, {Costamante}, \&
  {Tavecchio}}]{2017MNRAS.469..255G}
{Ghisellini}, G., {Righi}, C., {Costamante}, L., \& {Tavecchio}, F. 2017,
  \mnras, 469, 255

\bibitem[{{Ghisellini} {et~al.}(2005){Ghisellini}, {Tavecchio}, \&
  {Chiaberge}}]{2005A&A...432..401G}
{Ghisellini}, G., {Tavecchio}, F., \& {Chiaberge}, M. 2005, \aap, 432, 401

\bibitem[{{Giannios}(2013)}]{2013MNRAS.431..355G}
{Giannios}, D. 2013, \mnras, 431, 355

\bibitem[{{Giannios} {et~al.}(2010){Giannios}, {Uzdensky}, \&
  {Begelman}}]{2010MNRAS.402.1649G}
{Giannios}, D., {Uzdensky}, D.~A., \& {Begelman}, M.~C. 2010, \mnras, 402, 1649

\bibitem[{{Giebels} {et~al.}(2007){Giebels}, {Dubus}, \&
  {Kh{\'e}lifi}}]{2007A&A...462...29G}
{Giebels}, B., {Dubus}, G., \& {Kh{\'e}lifi}, B. 2007, \aap, 462, 29

\bibitem[{{Giroletti} {et~al.}(2006){Giroletti}, {Giovannini}, {Taylor}, \&
  {Falomo}}]{2006ApJ...646..801G}
{Giroletti}, M., {Giovannini}, G., {Taylor}, G.~B., \& {Falomo}, R. 2006, \apj,
  646, 801

\bibitem[{{Giroletti} \& {Righini}(2020)}]{2020MNRAS.492.2807G}
{Giroletti}, M. \& {Righini}, S. 2020, \mnras, 492, 2807

\bibitem[{{Hervet} {et~al.}(2019){Hervet}, {Williams}, {Falcone}, \&
  {Kaur}}]{2019ApJ...877...26H}
{Hervet}, O., {Williams}, D.~A., {Falcone}, A.~D., \& {Kaur}, A. 2019, \apj,
  877, 26

\bibitem[{{Hildebrand} {et~al.}(2017){Hildebrand}, {Ahnen}, {Balbo}, {Biland},
  {Bretz}, {Buss}, {Dorner}, {Einecke}, {Elsaesser}, {Herbst}, {Mahlke},
  {Mannheim}, {Neise}, {Neronov}, {Noethe}, {Paravac}, {Pauss}, {Rhode},
  {Sliusar}, {Temme}, {Walter}, {Adam}, {Baack}, {Blank}, {Bruegge},
  {Dmytriiev}, {Hempfling}, {Kortmann}, {Linhoff}, {Oberkirch}, {Schleicher},
  {Schulz}, {Shukla}, {Thaele}, \& {Mueller}}]{2017ICRC...35..779H}
{Hildebrand}, D., {Ahnen}, M.~L., {Balbo}, M., {et~al.} 2017, in International
  Cosmic Ray Conference, Vol. 301, 35th International Cosmic Ray Conference
  (ICRC2017), 779

\bibitem[{{Hovatta} {et~al.}(2015){Hovatta}, {Petropoulou}, {Richards},
  {Giannios}, {Wiik}, {Balokovi{\'c}}, {L{\"a}hteenm{\"a}ki}, {Lott},
  {Max-Moerbeck}, {Ramakrishnan}, \& {Readhead}}]{2015MNRAS.448.3121H}
{Hovatta}, T., {Petropoulou}, M., {Richards}, J.~L., {et~al.} 2015, \mnras,
  448, 3121

\bibitem[{{Jorstad} {et~al.}(2005){Jorstad}, {Marscher}, {Lister}, {Stirling},
  {Cawthorne}, {Gear}, {G{\'o}mez}, {Stevens}, {Smith}, {Forster}, \&
  {Robson}}]{2005AJ....130.1418J}
{Jorstad}, S.~G., {Marscher}, A.~P., {Lister}, M.~L., {et~al.} 2005, \aj, 130,
  1418

\bibitem[{{Kalberla} {et~al.}(2005){Kalberla}, {Burton}, {Hartmann}, {Arnal},
  {Bajaja}, {Morras}, \& {P{\"o}ppel}}]{2005A&A...440..775K}
{Kalberla}, P.~M.~W., {Burton}, W.~B., {Hartmann}, D., {et~al.} 2005, \aap,
  440, 775

\bibitem[{{Kardashev}(1962)}]{1962SvA.....6..317K}
{Kardashev}, N.~S. 1962, \sovast, 6, 317

\bibitem[{{Kastendieck} {et~al.}(2011){Kastendieck}, {Ashley}, \&
  {Horns}}]{2011A&A...531A.123K}
{Kastendieck}, M.~A., {Ashley}, M.~C.~B., \& {Horns}, D. 2011, \aap, 531, A123

\bibitem[{{Katarzy{\'n}ski} {et~al.}(2005){Katarzy{\'n}ski}, {Ghisellini},
  {Tavecchio}, {Maraschi}, {Fossati}, \& {Mastichiadis}}]{2005A&A...433..479K}
{Katarzy{\'n}ski}, K., {Ghisellini}, G., {Tavecchio}, F., {et~al.} 2005, \aap,
  433, 479

\bibitem[{{Katarzy{\'n}ski} \& {Walczewska}(2010)}]{2010A&A...510A..63K}
{Katarzy{\'n}ski}, K. \& {Walczewska}, K. 2010, \aap, 510, A63

\bibitem[{{Krawczynski} {et~al.}(2004){Krawczynski}, {Hughes}, {Horan},
  {Aharonian}, {Aller}, {Aller}, {Boltwood}, {Buckley}, {Coppi}, {Fossati},
  {G{\"o}tting}, {Holder}, {Horns}, {Kurtanidze}, {Marscher}, {Nikolashvili},
  {Remillard}, {Sadun}, \& {Schr{\"o}der}}]{2004ApJ...601..151K}
{Krawczynski}, H., {Hughes}, S.~B., {Horan}, D., {et~al.} 2004, \apj, 601, 151

\bibitem[{Krimm {et~al.}(2013)Krimm, Holland, Corbet, Pearlman, Romano, Kennea,
  Bloom, Barthelmy, Baumgartner, Cummings, Gehrels, Lien, Markwardt, Palmer,
  Sakamoto, Stamatikos, \& Ukwatta}]{Krimm_2013}
Krimm, H.~A., Holland, S.~T., Corbet, R. H.~D., {et~al.} 2013, The
  Astrophysical Journal Supplement Series, 209, 14

\bibitem[{{Levinson} \& {Rieger}(2011)}]{2011ApJ...730..123L}
{Levinson}, A. \& {Rieger}, F. 2011, \apj, 730, 123

\bibitem[{{Li} {et~al.}(2006){Li}, {Jha}, {Filippenko}, {Bloom}, {Pooley},
  {Foley}, \& {Perley}}]{2006PASP..118...37L}
{Li}, W., {Jha}, S., {Filippenko}, A.~V., {et~al.} 2006, \pasp, 118, 37

\bibitem[{{Lichti} {et~al.}(2008){Lichti}, {Bottacini}, {Ajello}, {Charlot},
  {Collmar}, {Falcone}, {Horan}, {Huber}, {von Kienlin}, {L{\"a}hteenm{\"a}ki},
  {Lindfors}, {Morris}, {Nilsson}, {Petry}, {R{\"u}ger}, {Sillanp{\"a}{\"a}},
  {Spanier}, \& {Tornikoski}}]{2008A&A...486..721L}
{Lichti}, G.~G., {Bottacini}, E., {Ajello}, M., {et~al.} 2008, \aap, 486, 721

\bibitem[{{Lico} {et~al.}(2014){Lico}, {Giroletti}, {Orienti}, {G{\'o}mez},
  {Casadio}, {D'Ammando}, {Blasi}, {Cotton}, {Edwards}, {Fuhrmann}, {Jorstad},
  {Kino}, {Kovalev}, {Krichbaum}, {Marscher}, {Paneque}, {Piner}, \&
  {Sokolovsky}}]{2014A&A...571A..54L}
{Lico}, R., {Giroletti}, M., {Orienti}, M., {et~al.} 2014, \aap, 571, A54

\bibitem[{{Longair}(2011)}]{2011hea..book.....L}
{Longair}, M.~S. 2011, {High Energy Astrophysics}

\bibitem[{Madejski {et~al.}(1999)Madejski, Sikora, Jaffe, B{\l}a{\.{z}}ejowski,
  Jahoda, \& Moderski}]{Madejski_1999}
Madejski, G.~M., Sikora, M., Jaffe, T., {et~al.} 1999, The Astrophysical
  Journal, 521, 145

\bibitem[{{Madsen} {et~al.}(2017){Madsen}, {Beardmore}, {Forster}, {Guainazzi},
  {Marshall}, {Miller}, {Page}, \& {Stuhlinger}}]{2017AJ....153....2M}
{Madsen}, K.~K., {Beardmore}, A.~P., {Forster}, K., {et~al.} 2017, \aj, 153, 2

\bibitem[{{Madsen} {et~al.}(2015){Madsen}, {Harrison}, {Markwardt}, {An},
  {Grefenstette}, {Bachetti}, {Miyasaka}, {Kitaguchi}, {Bhalerao}, {Boggs},
  {Christensen}, {Craig}, {Forster}, {Fuerst}, {Hailey}, {Perri}, {Puccetti},
  {Rana}, {Stern}, {Walton}, {J{\o}rgen Westergaard}, \&
  {Zhang}}]{2015ApJS..220....8M}
{Madsen}, K.~K., {Harrison}, F.~A., {Markwardt}, C.~B., {et~al.} 2015, \apjs,
  220, 8

\bibitem[{{MAGIC Collaboration} {et~al.}(2020{\natexlab{a}}){MAGIC
  Collaboration}, {Acciari}, {Ansoldi}, {Antonelli}, {Arbet Engels},
  {Babi{\'c}}, {Banerjee}, {Barres de Almeida}, {Barrio}, {Becerra
  Gonz{\'a}lez}, {Bednarek}, {Bellizzi}, {Bernardini}, {Berti}, {Besenrieder},
  {Bhattacharyya}, {Bigongiari}, {Blanch}, {Bonnoli}, {Bo{\v{s}}njak},
  {Busetto}, {Carosi}, {Ceribella}, {Cerruti}, {Chai}, {Chilingaryan},
  {Cikota}, {Colak}, {Colin}, {Colombo}, {Contreras}, {Cortina}, {Covino},
  {D'Elia}, {da Vela}, {Dazzi}, {de Angelis}, {de Lotto}, {Delfino}, {Delgado},
  {Depaoli}, {di Pierro}, {di Venere}, {Do Souto Espi{\~n}eira}, {Dominis
  Prester}, {Donini}, {Doro}, {Elsaesser}, {Fallah Ramazani}, {Fattorini},
  {Ferrara}, {Foffano}, {Fonseca}, {Font}, {Fruck}, {Fukami}, {Garc{\'\i}a
  L{\'o}pez}, {Garczarczyk}, {Gasparyan}, {Gaug}, {Giglietto}, {Giordano},
  {Godinovi{\'c}}, {Gliwny}, {Green}, {Hadasch}, {Hahn}, {Herrera}, {Hoang},
  {Hrupec}, {H{\"u}tten}, {Inada}, {Inoue}, {Ishio}, {Iwamura}, {Jouvin},
  {Kajiwara}, {Kerszberg}, {Kobayashi}, {Kubo}, {Kushida}, {Lamastra}, {Lelas},
  {Leone}, {Lindfors}, {Lombardi}, {Longo}, {L{\'o}pez}, {L{\'o}pez-Coto},
  {L{\'o}pez-Oramas}, {Loporchio}, {Machado de Oliveira Fraga}, {Maggio},
  {Majumdar}, {Makariev}, {Mallamaci}, {Maneva}, {Manganaro}, {Maraschi},
  {Mariotti}, {Mart{\'\i}nez}, {Mazin}, {Mender}, {Mi{\'c}anovi{\'c}},
  {Miceli}, {Miener}, {Minev}, {Miranda}, {Mirzoyan}, {Molina}, {Moralejo},
  {Morcuende}, {Moreno}, {Moretti}, {Munar-Adrover}, {Neustroev}, {Nigro},
  {Nilsson}, {Ninci}, {Nishijima}, {Noda}, {Nogu{\'e}s}, {Nozaki}, {Ohtani},
  {Oka}, {Otero-Santos}, {Paiano}, {Palatiello}, {Paneque}, {Paoletti},
  {Paredes}, {Pavleti{\'c}}, {Pe{\~n}il}, {Peresano}, {Persic}, {Prada Moroni},
  {Prandini}, {Puljak}, {Rib{\'o}}, {Rico}, {Righi}, {Rugliancich}, {Saha},
  {Sahakyan}, {Saito}, {Sakurai}, {Satalecka}, {Schleicher}, {Schmidt},
  {Schweizer}, {Sitarek}, {{\v{S}}nidari{\'c}}, {Sobczynska}, {Spolon},
  {Stamerra}, {Strom}, {Strzys}, {Suda}, {Suri{\'c}}, {Takahashi}, {Tavecchio},
  {Temnikov}, {Terzi{\'c}}, {Teshima}, {Torres-Alb{\`a}}, {Tosti}, {van
  Scherpenberg}, {Vanzo}, {Vazquez Acosta}, {Ventura}, {Verguilov}, {Vigorito},
  {Vitale}, {Vovk}, {Will}, {Zari{\'c}}, {Baack}, {Fact Collaboration},
  {Balbo}, {Beck}, {Biederbeck}, {Biland}, {Blank}, {Bretz}, {Bruegge},
  {Bulinski}, {Buss}, {Doerr}, {Dorner}, {Hildebrand}, {Iotov}, {Klinger},
  {Mannheim}, {Achim Mueller}, {Neise}, {Neronov}, {N{\"o}the}, {Paravac},
  {Rhode}, {Schleicher}, {Sedlaczek}, {Shukla}, {Sliusar}, {Tani}, {Theissen},
  {Walter}, {Acosta Pulido}, {Mwl Collaborators}, {Filippenko}, {Hovatta},
  {Kiehlmann}, {Larionov}, {Max-Moerbeck}, {Raiteri}, {Readhead},
  {{\v{S}}egon}, {Villata}, \& {Zheng}}]{2020MNRAS.496.3912M}
{MAGIC Collaboration}, {Acciari}, V.~A., {Ansoldi}, S., {et~al.}
  2020{\natexlab{a}}, \mnras, 496, 3912

\bibitem[{{MAGIC Collaboration} {et~al.}(2020{\natexlab{b}}){MAGIC
  Collaboration}, {Acciari}, {Ansoldi}, {Antonelli}, {Asano}, {Babi{\'c}},
  {Banerjee}, {Baquero}, {Barres de Almeida}, {Barrio}, {Becerra Gonz{\'a}lez},
  {Bednarek}, {Bellizzi}, {Bernardini}, {Bernardos}, {Berti}, {Besenrieder},
  {Bhattacharyya}, {Bigongiari}, {Blanch}, {Bonnoli}, {Bo{\v{s}}njak},
  {Busetto}, {Carosi}, {Ceribella}, {Cerruti}, {Chai}, {Chilingarian},
  {Cikota}, {Colak}, {Colombo}, {Contreras}, {Cortina}, {Covino}, {D'Amico},
  {D'Elia}, {Da Vela}, {Dazzi}, {De Angelis}, {De Lotto}, {Delfino}, {Delgado},
  {Delgado Mendez}, {Depaoli}, {Di Girolamo}, {Di Pierro}, {Di Venere}, {Do
  Souto Espi{\~n}eira}, {Dominis Prester}, {Donini}, {Doro}, {Fallah Ramazani},
  {Fattorini}, {Ferrara}, {Foffano}, {Fonseca}, {Font}, {Fruck}, {Fukami},
  {Garc{\'\i}a L{\'o}pez}, {Garczarczyk}, {Gasparyan}, {Gaug}, {Giglietto},
  {Giordano}, {Gliwny}, {Godinovi{\'c}}, {Green}, {Green}, {Hadasch}, {Hahn},
  {Heckmann}, {Herrera}, {Hoang}, {Hrupec}, {H{\"u}tten}, {Inada}, {Inoue},
  {Ishio}, {Iwamura}, {Jormanainen}, {Jouvin}, {Kajiwara}, {Karjalainen},
  {Kerszberg}, {Kobayashi}, {Kubo}, {Kushida}, {Lamastra}, {Lelas}, {Leone},
  {Lindfors}, {Lombardi}, {Longo}, {L{\'o}pez}, {L{\'o}pez-Coto},
  {L{\'o}pez-Oramas}, {Loporchio}, {Machado de Oliveira Fraga}, {Maggio},
  {Majumdar}, {Makariev}, {Mallamaci}, {Maneva}, {Manganaro}, {Maraschi},
  {Mariotti}, {Mart{\'\i}nez}, {Mazin}, {Mender}, {Mi{\'c}anovi{\'c}},
  {Miceli}, {Miener}, {Minev}, {Miranda}, {Mirzoyan}, {Molina}, {Moralejo},
  {Morcuende}, {Moreno}, {Moretti}, {Munar-Adrover}, {Neustroev}, {Nigro},
  {Nilsson}, {Ninci}, {Nishijima}, {Noda}, {Nozaki}, {Ohtani}, {Oka},
  {Otero-Santos}, {Palatiello}, {Paneque}, {Paoletti}, {Paredes},
  {Pavleti{\'c}}, {Pe{\~n}il}, {Perennes}, {Persic}, {Prada Moroni},
  {Prandini}, {Priyadarshi}, {Puljak}, {Rhode}, {Rib{\'o}}, {Rico}, {Righi},
  {Rugliancich}, {Saha}, {Sahakyan}, {Saito}, {Sakurai}, {Satalecka},
  {Schleicher}, {Schmidt}, {Schweizer}, {Sitarek}, {{\v{S}}nidari{\'c}},
  {Sobczynska}, {Spolon}, {Stamerra}, {Strom}, {Strzys}, {Suda}, {Suri{\'c}},
  {Takahashi}, {Tavecchio}, {Temnikov}, {Terzi{\'c}}, {Teshima},
  {Torres-Alb{\`a}}, {Tosti}, {Truzzi}, {van Scherpenberg}, {Vanzo}, {Vazquez
  Acosta}, {Ventura}, {Verguilov}, {Vigorito}, {Vitale}, {Vovk}, {Will},
  {Zari{\'c}}, {FACT collaboration}, {:}, {Arbet-Engels}, {Baack}, {Balbo},
  {Beck}, {Biederbeck}, {Biland}, {Bretz}, {Bruegge}, {Buss}, {Dorner},
  {Elsaesser}, {Hildebrand}, {Iotov}, {Klinger}, {Mannheim}, {Neise},
  {Neronov}, {Noethe}, {Paravac}, {Rhode}, {Schleicher}, {Sliusar}, {Theissen},
  {Walter}, {Collaborators}, {:}, {Valverde}, {Horan}, {Giroletti}, {Perri},
  {Verrecchia}, {Leto}, {Sadun}, {Moody}, {Joner}, {Marscher}, {Jorstad},
  {L{\"a}hteenm{\"a}ki}, {Tornikoski}, {Ramakrishnan}, {J{\"a}rvel{\"a}},
  {Vera}, {Righini}, \& {Lien}}]{2020arXiv201201348M}
{MAGIC Collaboration}, {Acciari}, V.~A., {Ansoldi}, S., {et~al.}
  2020{\natexlab{b}}, arXiv e-prints, arXiv:2012.01348

\bibitem[{{MAGIC Collaboration} {et~al.}(2020{\natexlab{c}}){MAGIC
  Collaboration}, {Acciari}, {Ansoldi}, {Antonelli}, {Babi{\'c}}, {Banerjee},
  {Barres de Almeida}, {Barrio}, {Becerra Gonz{\'a}lez}, {Bednarek},
  {Bernardini}, {Berti}, {Besenrieder}, {Bhattacharyya}, {Bigongiari},
  {Blanch}, {Bonnoli}, {Busetto}, {Carosi}, {Ceribella}, {Cikota}, {Colak},
  {Colin}, {Colombo}, {Contreras}, {Cortina}, {Covino}, {D'Elia}, {da Vela},
  {Dazzi}, {de Angelis}, {de Lotto}, {Delfino}, {Delgado}, {di Pierro}, {Do
  Souto Espi{\~n}era}, {Dom{\'\i}nguez}, {Dominis Prester}, {Doro}, {Fallah
  Ramazani}, {Fattorini}, {Fern{\'a}ndez-Barral}, {Ferrara}, {Fidalgo},
  {Foffano}, {Fonseca}, {Font}, {Fruck}, {Galindo}, {Gallozzi}, {Garc{\'\i}a
  L{\'o}pez}, {Garczarczyk}, {Gasparyan}, {Gaug}, {Giammaria}, {Godinovi{\'c}},
  {Guberman}, {Hadasch}, {Hahn}, {Hassan}, {Herrera}, {Hoang}, {Hrupec},
  {Inoue}, {Ishio}, {Iwamura}, {Kubo}, {Kushida}, {Kuve{\v{z}}di{\'c}},
  {Lamastra}, {Lelas}, {Leone}, {Lindfors}, {Lombardi}, {Longo}, {L{\'o}pez},
  {L{\'o}pez-Oramas}, {Machado de Oliveira Fraga}, {Maggio}, {Majumdar},
  {Makariev}, {Mallamaci}, {Maneva}, {Manganaro}, {Maraschi}, {Mariotti},
  {Mart{\'\i}nez}, {Masuda}, {Mazin}, {Minev}, {Miranda}, {Mirzoyan}, {Molina},
  {Moralejo}, {Moreno}, {Moretti}, {Munar-Adrover}, {Neustroev}, {Niedzwiecki},
  {Nievas Rosillo}, {Nigro}, {Nilsson}, {Ninci}, {Nishijima}, {Noda},
  {Nogu{\'e}s}, {Paiano}, {Palacio}, {Paneque}, {Paoletti}, {Paredes},
  {Pedaletti}, {Pe{\~n}il}, {Peresano}, {Persic}, {Prada Moroni}, {Prand ini},
  {Puljak}, {Garcia}, {Rib{\'o}}, {Rico}, {Righi}, {Rugliancich}, {Saha},
  {Sahakyan}, {Saito}, {Satalecka}, {Schweizer}, {Sitarek},
  {{\v{S}}nidari{\'c}}, {Sobczynska}, {Somero}, {Stamerra}, {Strzys},
  {Suri{\'c}}, {Tavecchio}, {Temnikov}, {Terzi{\'c}}, {Teshima},
  {Torres-Alb{\`a}}, {Tsujimoto}, {van Scherpenberg}, {Vanzo}, {Vazquez
  Acosta}, {Vovk}, {Will}, {Zari{\'c}}, {Fact Collaboration}, {Arbet-Engels},
  {Baack}, {Balbo}, {Biland}, {Blank}, {Bretz}, {Bruegge}, {Bulinski}, {Buss},
  {Doerr}, {Dorner}, {Einecke}, {Elsaesser}, {Hildebrand}, {Linhoff},
  {Mannheim}, {Mueller}, {Neise}, {Neronov}, {Noethe}, {Paravac}, {Rhode},
  {Schleicher}, {Schulz}, {Sedlaczek}, {Shukla}, {Sliusar}, {von Willert},
  {Walter}, {Wendel}, {Tramacere}, {Lien}, {Perri}, {Verrecchia}, {Armas
  Padilla}, {Leto}, {L{\"a}hteenm{\"a}ki}, {Tornikoski}, \&
  {Tammi}}]{2020A&A...637A..86M}
{MAGIC Collaboration}, {Acciari}, V.~A., {Ansoldi}, S., {et~al.}
  2020{\natexlab{c}}, \aap, 637, A86

\bibitem[{{Mahlke} {et~al.}(2017){Mahlke}, {Bretz}, {Adam}, {Ahnen}, {Baack},
  {Balbo}, {Biland}, {Blank}, {Bruegge}, {Buss}, {Dmytriiev}, {Dorner},
  {Einecke}, {Elsaesser}, {Hempfling}, {Herbst}, {Hildebrand}, {Kortmann},
  {Linhoff}, {Mahlke}, {Mannheim}, {Mueller}, {Neise}, {Neronov}, {Noethe},
  {Oberkirch}, {Paravac}, {Pauss}, {Rhode}, {Schleicher}, {Schulz}, {Shukla},
  {Sliusar}, {Temme}, {Thaele}, \& {Walter}}]{2017ICRC...35..612M}
{Mahlke}, M., {Bretz}, T., {Adam}, J., {et~al.} 2017, in International Cosmic
  Ray Conference, Vol. 301, 35th International Cosmic Ray Conference
  (ICRC2017), 612

\bibitem[{{Malizia} {et~al.}(2000){Malizia}, {Capalbi}, {Fiore}, {Giommi},
  {Gand olfi}, {Tesseri}, {Antonelli}, {Butler}, {Celidonio}, {Coletta}, {Di
  Ciolo}, {Muller}, {Piro}, {Rebecchi}, {Ricci}, {Ricci}, {Smith}, \&
  {Torroni}}]{2000MNRAS.312..123M}
{Malizia}, A., {Capalbi}, M., {Fiore}, F., {et~al.} 2000, \mnras, 312, 123

\bibitem[{{Mannheim}(1993)}]{1993A&A...269...67M}
{Mannheim}, K. 1993, \aap, 269, 67

\bibitem[{{Maraschi} {et~al.}(1992){Maraschi}, {Ghisellini}, \&
  {Celotti}}]{1992ApJ...397L...5M}
{Maraschi}, L., {Ghisellini}, G., \& {Celotti}, A. 1992, \apjl, 397, L5

\bibitem[{{Markwardt} {et~al.}(2007){Markwardt}, {Barthelmy}, {Cummings},
  {Hullinger}, {Krimm}, \& {Parsons}}]{Markwardt07}
{Markwardt}, C.~B., {Barthelmy}, S.~D., {Cummings}, J.~C., {et~al.} 2007,
  http://swift.gsfc.nasa.gov/analysis/bat{\_}swguide{\_}v6{\_}3.pdf

\bibitem[{{Massaro} {et~al.}(2004{\natexlab{a}}){Massaro}, {Perri}, {Giommi},
  \& {Nesci}}]{2004A&A...413..489M}
{Massaro}, E., {Perri}, M., {Giommi}, P., \& {Nesci}, R. 2004{\natexlab{a}},
  \aap, 413, 489

\bibitem[{{Massaro} {et~al.}(2004{\natexlab{b}}){Massaro}, {Perri}, {Giommi},
  {Nesci}, \& {Verrecchia}}]{2004A&A...422..103M}
{Massaro}, E., {Perri}, M., {Giommi}, P., {Nesci}, R., \& {Verrecchia}, F.
  2004{\natexlab{b}}, \aap, 422, 103

\bibitem[{{Massaro} {et~al.}(2006){Massaro}, {Tramacere}, {Perri}, {Giommi}, \&
  {Tosti}}]{2006A&A...448..861M}
{Massaro}, E., {Tramacere}, A., {Perri}, M., {Giommi}, P., \& {Tosti}, G. 2006,
  \aap, 448, 861

\bibitem[{{Mattox} {et~al.}(1996){Mattox}, {Bertsch}, {Chiang}, {Dingus},
  {Digel}, {Esposito}, {Fierro}, {Hartman}, {Hunter}, {Kanbach}, {Kniffen},
  {Lin}, {Macomb}, {Mayer-Hasselwander}, {Michelson}, {von Montigny},
  {Mukherjee}, {Nolan}, {Ramanamurthy}, {Schneid}, {Sreekumar}, {Thompson}, \&
  {Willis}}]{1996ApJ...461..396M}
{Mattox}, J.~R., {Bertsch}, D.~L., {Chiang}, J., {et~al.} 1996, \apj, 461, 396

\bibitem[{{Max-Moerbeck} {et~al.}(2014){Max-Moerbeck}, {Richards}, {Hovatta},
  {Pavlidou}, {Pearson}, \& {Readhead}}]{2014MNRAS.445..437M}
{Max-Moerbeck}, W., {Richards}, J.~L., {Hovatta}, T., {et~al.} 2014, \mnras,
  445, 437

\bibitem[{{Moderski} {et~al.}(2003){Moderski}, {Sikora}, \&
  {B{\l}a{\.z}ejowski}}]{2003A&A...406..855M}
{Moderski}, R., {Sikora}, M., \& {B{\l}a{\.z}ejowski}, M. 2003, \aap, 406, 855

\bibitem[{{M{\"u}cke} \& {Protheroe}(2001)}]{2001APh....15..121M}
{M{\"u}cke}, A. \& {Protheroe}, R.~J. 2001, Astroparticle Physics, 15, 121

\bibitem[{{Nilsson} {et~al.}(2018){Nilsson}, {Lindfors}, {Takalo}, {Reinthal},
  {Berdyugin}, {Sillanp{\"a}{\"a}}, {Ciprini}, {Halkola}, {Hein{\"a}m{\"a}ki},
  {Hovatta}, {Kadenius}, {Nurmi}, {Ostorero}, {Pasanen}, {Rekola}, {Saarinen},
  {Sainio}, {Tuominen}, {Villforth}, {Vornanen}, \&
  {Zaprudin}}]{2018A&A...620A.185N}
{Nilsson}, K., {Lindfors}, E., {Takalo}, L.~O., {et~al.} 2018, \aap, 620, A185

\bibitem[{{Nilsson} {et~al.}(2007){Nilsson}, {Pasanen}, {Takalo}, {Lindfors},
  {Berdyugin}, {Ciprini}, \& {Pforr}}]{nilsson2007}
{Nilsson}, K., {Pasanen}, M., {Takalo}, L.~O., {et~al.} 2007, \aap, 475, 199

\bibitem[{{Padovani} \& {Giommi}(1995)}]{1995ApJ...444..567P}
{Padovani}, P. \& {Giommi}, P. 1995, \apj, 444, 567

\bibitem[{{Petropoulou} {et~al.}(2016){Petropoulou}, {Giannios}, \&
  {Sironi}}]{2016MNRAS.462.3325P}
{Petropoulou}, M., {Giannios}, D., \& {Sironi}, L. 2016, \mnras, 462, 3325

\bibitem[{{Pian} {et~al.}(1998){Pian}, {Vacanti}, {Tagliaferri}, {Ghisellini},
  {Maraschi}, {Treves}, {Urry}, {Fiore}, {Giommi}, {Palazzi}, {Chiappetti}, \&
  {Sambruna}}]{1998ApJ...492L..17P}
{Pian}, E., {Vacanti}, G., {Tagliaferri}, G., {et~al.} 1998, \apjl, 492, L17

\bibitem[{{Poutanen} {et~al.}(2008){Poutanen}, {Zdziarski}, \&
  {Ibragimov}}]{2008MNRAS.389.1427P}
{Poutanen}, J., {Zdziarski}, A.~A., \& {Ibragimov}, A. 2008, \mnras, 389, 1427

\bibitem[{Press {et~al.}(2007)Press, Teukolsky, Vetterling, \&
  Flannery}]{press2007numerical}
Press, W., Teukolsky, S., Vetterling, W., \& Flannery, B. 2007, Numerical
  Recipes: The Art of Scientific Computing, 3rd edn. (Cambridge University
  Press)

\bibitem[{{Ptitsyna} \& {Neronov}(2016)}]{2016A&A...593A...8P}
{Ptitsyna}, K. \& {Neronov}, A. 2016, \aap, 593, A8

\bibitem[{{Punch} {et~al.}(1992){Punch}, {Akerlof}, {Cawley}, {Chantell},
  {Fegan}, {Fennell}, {Gaidos}, {Hagan}, {Hillas}, {Jiang}, {Kerrick}, {Lamb},
  {Lawrence}, {Lewis}, {Meyer}, {Mohanty}, {O'Flaherty}, {Reynolds}, {Rovero},
  {Schubnell}, {Sembroski}, {Weekes}, {Whitaker}, \&
  {Wilson}}]{1992Natur.358..477P}
{Punch}, M., {Akerlof}, C.~W., {Cawley}, M.~F., {et~al.} 1992, \nat, 358, 477

\bibitem[{{Raiteri} {et~al.}(2017){Raiteri}, {Villata}, {Acosta-Pulido},
  {Agudo}, {Arkharov}, {Bachev}, {Baida}, {Ben{\'{\i}}tez}, {Borman},
  {Boschin}, {Bozhilov}, {Butuzova}, {Calcidese}, {Carnerero}, {Carosati},
  {Casadio}, {Castro-Segura}, {Chen}, {Damljanovic}, {D'Ammando}, {di Paola},
  {Echevarr{\'{\i}}a}, {Efimova}, {Ehgamberdiev}, {Espinosa}, {Fuentes},
  {Giunta}, {G{\'o}mez}, {Grishina}, {Gurwell}, {Hiriart}, {Jermak}, {Jordan},
  {Jorstad}, {Joshi}, {Kopatskaya}, {Kuratov}, {Kurtanidze}, {Kurtanidze},
  {L{\"a}hteenm{\"a}ki}, {Larionov}, {Larionova}, {Larionova}, {L{\'a}zaro},
  {Lin}, {Malmrose}, {Marscher}, {Matsumoto}, {McBreen}, {Michel}, {Mihov},
  {Minev}, {Mirzaqulov}, {Mokrushina}, {Molina}, {Moody}, {Morozova},
  {Nazarov}, {Nikolashvili}, {Ohlert}, {Okhmat}, {Ovcharov}, {Pinna},
  {Polakis}, {Protasio}, {Pursimo}, {Redondo-Lorenzo}, {Rizzi},
  {Rodriguez-Coira}, {Sadakane}, {Sadun}, {Samal}, {Savchenko}, {Semkov},
  {Skiff}, {Slavcheva-Mihova}, {Smith}, {Steele}, {Strigachev}, {Tammi},
  {Thum}, {Tornikoski}, {Troitskaya}, {Troitsky}, {Vasilyev}, \&
  {Vince}}]{raiteri2017}
{Raiteri}, C.~M., {Villata}, M., {Acosta-Pulido}, J.~A., {et~al.} 2017, \nat,
  552, 374

\bibitem[{{Raiteri} {et~al.}(2007){Raiteri}, {Villata}, {Larionov}, {Pursimo},
  {Ibrahimov}, {Nilsson}, {Aller}, {Kurtanidze}, {Foschini}, {Ohlert},
  {Papadakis}, {Sumitomo}, {Volvach}, {Aller}, {Arkharov}, {Bach}, {Berdyugin},
  {B{\"o}ttcher}, {Buemi}, {Calcidese}, {Charlot}, {Delgado S{\'a}nchez}, {di
  Paola}, {Djupvik}, {Dolci}, {Efimova}, {Fan}, {Forn{\'e}}, {Gomez}, {Gupta},
  {Hagen-Thorn}, {Hooks}, {Hovatta}, {Ishii}, {Kamada}, {Konstantinova},
  {Kopatskaya}, {Kovalev}, {Kovalev}, {L{\"a}hteenm{\"a}ki}, {Lanteri}, {Le
  Campion}, {Lee}, {Leto}, {Lin}, {Lindfors}, {Mingaliev}, {Mizoguchi},
  {Nicastro}, {Nikolashvili}, {Nishiyama}, {{\"O}stman}, {Ovcharov},
  {P{\"a}{\"a}kk{\"o}nen}, {Pasanen}, {Pian}, {Rector}, {Ros}, {Sadakane},
  {Selj}, {Semkov}, {Sharapov}, {Somero}, {Stanev}, {Strigachev}, {Takalo},
  {Tanaka}, {Tavani}, {Torniainen}, {Tornikoski}, {Trigilio}, {Umana},
  {Vercellone}, {Valcheva}, {Volvach}, \& {Yamanaka}}]{rai2007}
{Raiteri}, C.~M., {Villata}, M., {Larionov}, V.~M., {et~al.} 2007, \aap, 473,
  819

\bibitem[{{Richards} {et~al.}(2011){Richards}, {Max-Moerbeck}, {Pavlidou},
  {King}, {Pearson}, {Readhead}, {Reeves}, {Shepherd}, {Stevenson},
  {Weintraub}, {Fuhrmann}, {Angelakis}, {Zensus}, {Healey}, {Romani}, {Shaw},
  {Grainge}, {Birkinshaw}, {Lancaster}, {Worrall}, {Taylor}, {Cotter}, \&
  {Bustos}}]{2011ApJS..194...29R}
{Richards}, J.~L., {Max-Moerbeck}, W., {Pavlidou}, V., {et~al.} 2011, \apjs,
  194, 29

\bibitem[{{Romanova} \& {Lovelace}(1992)}]{1992A&A...262...26R}
{Romanova}, M.~M. \& {Lovelace}, R.~V.~E. 1992, \aap, 262, 26

\bibitem[{{Roming} {et~al.}(2005){Roming}, {Kennedy}, {Mason}, {Nousek}, {Ahr},
  {Bingham}, {Broos}, {Carter}, {Hancock}, {Huckle}, {Hunsberger}, {Kawakami},
  {Killough}, {Koch}, {McLelland}, {Smith}, {Smith}, {Soto}, {Boyd},
  {Breeveld}, {Holland}, {Ivanushkina}, {Pryzby}, {Still}, \&
  {Stock}}]{2005SSRv..120...95R}
{Roming}, P. W.~A., {Kennedy}, T.~E., {Mason}, K.~O., {et~al.} 2005, \ssr, 120,
  95

\bibitem[{{Schlafly} \& {Finkbeiner}(2011)}]{2011ApJ...737..103S}
{Schlafly}, E.~F. \& {Finkbeiner}, D.~P. 2011, \apj, 737, 103

\bibitem[{Schleicher {et~al.}(2019)Schleicher, Arbet-Engels, Baack, Balbo,
  Biland, Blank, Bretz, Bruegge, Bulinski, Buss, Doerr, Dorner, Elsaesser,
  Grischagin, Hildebrand, Linhoff, Mannheim, Mueller, Neise, Neronov, Noethe,
  Paravac, Rhode, Schulz, Sedlaczek, Shukla, Sliusar, Willert, \&
  Walter}]{galaxies7020062}
Schleicher, B., Arbet-Engels, A., Baack, D., {et~al.} 2019, Galaxies, 7

\bibitem[{{Schlickeiser}(1985)}]{1985A&A...143..431S}
{Schlickeiser}, R. 1985, \aap, 143, 431

\bibitem[{{Schlickeiser} {et~al.}(2010){Schlickeiser}, {B{\"o}ttcher}, \&
  {Menzler}}]{2010A&A...519A...9S}
{Schlickeiser}, R., {B{\"o}ttcher}, M., \& {Menzler}, U. 2010, \aap, 519, A9

\bibitem[{{Sironi} {et~al.}(2015){Sironi}, {Petropoulou}, \&
  {Giannios}}]{2015MNRAS.450..183S}
{Sironi}, L., {Petropoulou}, M., \& {Giannios}, D. 2015, \mnras, 450, 183

\bibitem[{{Sironi} \& {Spitkovsky}(2014)}]{2014ApJ...783L..21S}
{Sironi}, L. \& {Spitkovsky}, A. 2014, \apjl, 783, L21

\bibitem[{{Stawarz} \& {Petrosian}(2008)}]{2008ApJ...681.1725S}
{Stawarz}, {\L}. \& {Petrosian}, V. 2008, \apj, 681, 1725

\bibitem[{{Tanihata} {et~al.}(2004){Tanihata}, {Kataoka}, {Takahashi}, \&
  {Madejski}}]{2004ApJ...601..759T}
{Tanihata}, C., {Kataoka}, J., {Takahashi}, T., \& {Madejski}, G.~M. 2004,
  \apj, 601, 759

\bibitem[{{Tavecchio} \& {Ghisellini}(2016)}]{2016MNRAS.456.2374T}
{Tavecchio}, F. \& {Ghisellini}, G. 2016, \mnras, 456, 2374

\bibitem[{{Tavecchio} {et~al.}(2010){Tavecchio}, {Ghisellini}, {Ghirlanda},
  {Foschini}, \& {Maraschi}}]{2010MNRAS.401.1570T}
{Tavecchio}, F., {Ghisellini}, G., {Ghirlanda}, G., {Foschini}, L., \&
  {Maraschi}, L. 2010, \mnras, 401, 1570

\bibitem[{{Tavecchio} {et~al.}(1998){Tavecchio}, {Maraschi}, \&
  {Ghisellini}}]{1998ApJ...509..608T}
{Tavecchio}, F., {Maraschi}, L., \& {Ghisellini}, G. 1998, \apj, 509, 608

\bibitem[{{Teraesranta} {et~al.}(1998){Teraesranta}, {Tornikoski}, {Mujunen},
  {Karlamaa}, {Valtonen}, {Henelius}, {Urpo}, {Lainela}, {Pursimo}, {Nilsson},
  {Wiren}, {Laehteenmaeki}, {Korpi}, {Rekola}, {Heinaemaeki}, {Hanski},
  {Nurmi}, {Kokkonen}, {Keinaenen}, {Joutsamo}, {Oksanen}, {Pietilae},
  {Valtaoja}, {Valtonen}, \& {Koenoenen}}]{1998A&AS..132..305T}
{Teraesranta}, H., {Tornikoski}, M., {Mujunen}, A., {et~al.} 1998, \aaps, 132,
  305

\bibitem[{{Timmer} \& {Koenig}(1995)}]{1995A&A...300..707T}
{Timmer}, J. \& {Koenig}, M. 1995, \aap, 300, 707

\bibitem[{{Tramacere} {et~al.}(2007{\natexlab{a}}){Tramacere}, {Giommi},
  {Massaro}, {Perri}, {Nesci}, {Colafrancesco}, {Tagliaferri}, {Chincarini},
  {Falcone}, {Burrows}, {Roming}, {McMath Chester}, \&
  {Gehrels}}]{2007A&A...467..501T}
{Tramacere}, A., {Giommi}, P., {Massaro}, E., {et~al.} 2007{\natexlab{a}},
  \aap, 467, 501

\bibitem[{{Tramacere} {et~al.}(2009){Tramacere}, {Giommi}, {Perri},
  {Verrecchia}, \& {Tosti}}]{2009A&A...501..879T}
{Tramacere}, A., {Giommi}, P., {Perri}, M., {Verrecchia}, F., \& {Tosti}, G.
  2009, \aap, 501, 879

\bibitem[{{Tramacere} {et~al.}(2011){Tramacere}, {Massaro}, \&
  {Taylor}}]{2011ApJ...739...66T}
{Tramacere}, A., {Massaro}, E., \& {Taylor}, A.~M. 2011, \apj, 739, 66

\bibitem[{{Tramacere} {et~al.}(2007{\natexlab{b}}){Tramacere}, {Massaro}, \&
  {Cavaliere}}]{2007A&A...466..521T}
{Tramacere}, A., {Massaro}, F., \& {Cavaliere}, A. 2007{\natexlab{b}}, \aap,
  466, 521

\bibitem[{{Urry} \& {Padovani}(1995)}]{1995PASP..107..803U}
{Urry}, C.~M. \& {Padovani}, P. 1995, \pasp, 107, 803

\bibitem[{{Uttley} {et~al.}(2003){Uttley}, {Edelson}, {McHardy}, {Peterson}, \&
  {Markowitz}}]{2003ApJ...584L..53U}
{Uttley}, P., {Edelson}, R., {McHardy}, I.~M., {Peterson}, B.~M., \&
  {Markowitz}, A. 2003, \apjl, 584, L53

\bibitem[{{Uttley} {et~al.}(2002){Uttley}, {McHardy}, \&
  {Papadakis}}]{2002MNRAS.332..231U}
{Uttley}, P., {McHardy}, I.~M., \& {Papadakis}, I.~E. 2002, \mnras, 332, 231

\bibitem[{{Vaughan} {et~al.}(2003){Vaughan}, {Edelson}, {Warwick}, \&
  {Uttley}}]{2003MNRAS.345.1271V}
{Vaughan}, S., {Edelson}, R., {Warwick}, R.~S., \& {Uttley}, P. 2003, \mnras,
  345, 1271

\bibitem[{{Villata} {et~al.}(2006){Villata}, {Raiteri}, {Balonek}, {Aller},
  {Jorstad}, {Kurtanidze}, {Nicastro}, {Nilsson}, {Aller}, {Arai}, {Arkharov},
  \& {Bach}}]{villata2006}
{Villata}, M., {Raiteri}, C.~M., {Balonek}, T.~J., {et~al.} 2006, \aap, 453,
  817

\bibitem[{{Villata} {et~al.}(2009){Villata}, {Raiteri}, {Gurwell}, {Larionov},
  {Kurtanidze}, {Aller}, {L{\"a}hteenm{\"a}ki}, {Chen}, {Nilsson}, {Agudo},
  {Aller}, {Arkharov}, {Bach}, {Bachev}, {Beltrame}, {Ben{\'{\i}}tez}, {Buemi},
  {B{\"o}ttcher}, {Calcidese}, {Capezzali}, {Carosati}, {da Rio}, {di Paola},
  {Dolci}, {Dultzin}, {Forn{\'e}}, {G{\'o}mez}, {Hagen-Thorn}, {Halkola},
  {Heidt}, {Hiriart}, {Hovatta}, {Hsiao}, {Jorstad}, {Kimeridze},
  {Konstantinova}, {Kopatskaya}, {Koptelova}, {Leto}, {Ligustri}, {Lindfors},
  {Lopez}, {Marscher}, {Mommert}, {Mujica}, {Nikolashvili}, {Palma}, {Pasanen},
  {Roca-Sogorb}, {Ros}, {Roustazadeh}, {Sadun}, {Saino}, {Sigua}, {Sorcia},
  {Takalo}, {Tornikoski}, {Trigilio}, {Turchetti}, \& {Umana}}]{villata2009}
{Villata}, M., {Raiteri}, C.~M., {Gurwell}, M.~A., {et~al.} 2009, \aap, 504, L9

\bibitem[{{Villata} {et~al.}(2002){Villata}, {Raiteri}, {Kurtanidze},
  {Nikolashvili}, {Ibrahimov}, {Papadakis}, {Tsinganos}, {Sadakane}, {Okada},
  {Takalo}, {Sillanp{\"a}{\"a}}, {Tosti}, {Ciprini}, {Frasca}, {Marilli},
  {Robb}, {Noble}, {Jorstad}, {Hagen-Thorn}, {Larionov}, {Nesci}, {Maesano},
  {Schwartz}, {Basler}, {Gorham}, {Iwamatsu}, {Kato}, {Pullen},
  {Ben{\'{\i}}tez}, {de Diego}, {Moilanen}, {Oksanen}, {Rodriguez}, {Sadun},
  {Kelly}, {Carini}, {Miller}, {Catalano}, {Dultzin-Hacyan}, {Fan}, {Ishioka},
  {Karttunen}, {Kein{\"a}nen}, {Kudryavtseva}, {Lainela}, {Lanteri},
  {Larionova}, {Matsumoto}, {Mattox}, {Montagni}, {Nucciarelli}, {Ostorero},
  {Papamastorakis}, {Pasanen}, {Sobrito}, \& {Uemura}}]{villata2002}
{Villata}, M., {Raiteri}, C.~M., {Kurtanidze}, O.~M., {et~al.} 2002, \aap, 390,
  407

\bibitem[{{Villata} {et~al.}(2008){Villata}, {Raiteri}, {Larionov},
  {Kurtanidze}, {Nilsson}, {Aller}, {Tornikoski}, {Volvach}, {Aller}, \&
  {Arkharov}}]{villata2008}
{Villata}, M., {Raiteri}, C.~M., {Larionov}, V.~M., {et~al.} 2008, \aap, 481,
  L79

\bibitem[{{Wendel} {et~al.}(2021){Wendel}, {Becerra}, {Paneque}, \&
  {Mannheim}}]{Wendel2021}
{Wendel}, C., {Becerra}, J., {Paneque}, D., \& {Mannheim}, K. 2021, \aap
  [\eprint[arXiv]{2011.XXXX}]

\bibitem[{{Zabalza}(2015)}]{naima}
{Zabalza}, V. 2015, Proc.~of International Cosmic Ray Conference 2015, 922

\bibitem[{{Zanin} {et~al.}(2013){Zanin}, {Carmona}, {Sitarek}, {Colin},
  {Frantzen}, {Gaug}, {Lombardi}, {Lopez}, {Moralejo}, {Satalecka}, {Scapin},
  \& {Stamatescu}}]{zanin2013}
{Zanin}, R., {Carmona}, E., {Sitarek}, J., {et~al.} 2013, in Proceedings of the
  33rd International Cosmic Ray Conference (ICRC2013): Rio de Janeiro, Brazil,
  July 2-9, 2013, 0773

\bibitem[{Zhang {et~al.}(1999)Zhang, Celotti, Treves, Chiappetti, Ghisellini,
  Maraschi, Pian, Tagliaferri, Tavecchio, \& Urry}]{Zhang_1999}
Zhang, Y.~H., Celotti, A., Treves, A., {et~al.} 1999, The Astrophysical
  Journal, 527, 719

\end{thebibliography}
\nocite{B_a_ejowski_2005}
\nocite{2000MNRAS.312..123M}
\nocite{2004A&A...413..489M}

\begin{appendix}
\section{MAGIC analysis results}
\label{appendix_magic}
In this Section we report in detail the results of the MAGIC analysis. Table~\ref{tab:MAGIC_spectral_param} lists the nightly fluxes in the 0.2-1\,TeV and >1\,TeV bands for each observation. The parameters of the log-parabolic (normalisation energy is 300\,GeV) or power-law fits above 100\,GeV are given for each day. A log-parabolic fit is applied in cases where it provides a better description of the data at a significance above $3\sigma$ in comparison to a power-law function. All fits are performed after correcting for the EBL absorption.\par 

The two nights MJD~57757 and MJD~57789 show a relatively poor $\chi^2/dof$ in comparison with the other nights, even with a log-parabolic model. Namely, the fits yield $\chi^2/dof = 34.5/12$ and $\chi^2/dof = 28.1/14$, which corresponds to a p-value of $6\times10^{-4}$ and $10^{-2}$, respectively. We found that a better fit could be obtained using a power-law with a exponential cutoff:
\begin{equation}
    \frac{dN}{dE} = f_0 \left(\frac{E}{E_0}\right)^{-\alpha} \, \exp{\left(-E/E_c \right)}
\end{equation}
For MJD~57757, the best-fit parameters are $f_0 = (1.23 \pm 0.06) \times 10^{-9}$\,cm$^{-2}$\,s$^{-1}$\,TeV$^{-1}$, $\alpha = 1.86 \pm 0.06$, and $E_C = 1.7 \pm 0.3$\,TeV. The associated $\chi^2/dof$ is $\chi^2/dof = 25.6/12$.\par

For MJD~57789, the best-fit parameters are $f_0 = (1.04 \pm 0.04) \times 10^{-9}$\,cm$^{-2}$\,s$^{-1}$\,TeV$^{-1}$, $\alpha = 1.87 \pm 0.06$, and $E_C = 2.87 \pm 0.6$\,TeV. The associated $\chi^2/dof$ is $\chi^2/dof = 23.5/14$.

\begin{table*}
\caption{\label{tab:MAGIC_spectral_param}MAGIC analysis results.} 
\centering
\begin{tabular}{ l c c c c c c c c }     
\hline\hline 
 MJD start & MJD end  & $F_{0.2-1\text{\,TeV}}$ & $F_{>1\text{\,TeV}}$ & $f_0$ & $\alpha$ & $\beta$ & $\chi^2$/dof\\  
 &   & $[10^{-10} \mathrm{cm}^{-2} \mathrm{s}^{-1}]$ & $[10^{-11} \mathrm{cm}^{-2} \mathrm{s}^{-1}]$ & $[10^{-10} \mathrm{cm}^{-2} \mathrm{s}^{-1} \mathrm{TeV}^{-1}]$ &  &  &  \\
\hline\hline   
57727.248 & 57727.269 & 1.11 $\pm$ 0.15 & 0.17 $\pm$ 0.12 & 3.08 $\pm$ 0.31 & 2.77 $\pm$ 0.11 &  --  & 12.6/8 \\
57729.244 & 57729.269 & 0.83 $\pm$ 0.13 & 0.00 $\pm$ 0.10 & 2.65 $\pm$ 0.28 & 2.42 $\pm$ 0.15 &  --  & 13.7/6 \\
57749.217 & 57749.273 & 1.50 $\pm$ 0.14 & 0.46 $\pm$ 0.14 & 5.54 $\pm$ 0.41 & 2.53 $\pm$ 0.09 & 0.97 $\pm$ 0.24 & 7.2/9 \\
57751.236 & 57751.278 & 1.62 $\pm$ 0.18 & 0.44 $\pm$ 0.17 & 5.63 $\pm$ 0.45 & 2.42 $\pm$ 0.08 & 0.64 $\pm$ 0.22 & 12.3/8 \\
57753.184 & 57753.226 & 1.55 $\pm$ 0.14 & 0.89 $\pm$ 0.19 & 4.79 $\pm$ 0.23 & 2.39 $\pm$ 0.06 &  --  & 8.8/9 \\
57756.183 & 57756.258 & 2.38 $\pm$ 0.14 & 1.37 $\pm$ 0.20 & 7.99 $\pm$ 0.34 & 2.20 $\pm$ 0.05 & 0.53 $\pm$ 0.10 & 7.5/11 \\
57757.019 & 57757.281 & 3.82 $\pm$ 0.10 & 3.17 $\pm$ 0.14 & 12.10 $\pm$ 0.20 & 2.15 $\pm$ 0.02 & 0.28 $\pm$ 0.03 & 10.9/13 \\
57762.157 & 57762.185 & 2.75 $\pm$ 0.21 & 2.26 $\pm$ 0.36 & 7.79 $\pm$ 0.35 & 2.20 $\pm$ 0.05 &  --  & 20.1/11 \\
57763.154 & 57763.196 & 1.81 $\pm$ 0.14 & 1.02 $\pm$ 0.19 & 5.29 $\pm$ 0.25 & 2.29 $\pm$ 0.06 &  --  & 27.2/10 \\
57771.127 & 57771.279 & 1.49 $\pm$ 0.07 & 0.72 $\pm$ 0.08 & 5.19 $\pm$ 0.19 & 2.38 $\pm$ 0.05 & 0.47 $\pm$ 0.11 & 15.6/12 \\
57772.038 & 57772.059 & 2.41 $\pm$ 0.21 & 1.95 $\pm$ 0.29 & 9.78 $\pm$ 0.20 & 2.09 $\pm$ 0.03 & 0.25 $\pm$ 0.04 & 14.7/14 \\
57776.190 & 57776.228 & 1.86 $\pm$ 0.15 & 0.76 $\pm$ 0.18 & 6.55 $\pm$ 0.40 & 2.30 $\pm$ 0.06 & 0.47 $\pm$ 0.15 & 19.6/9 \\
57778.189 & 57778.276 & 2.16 $\pm$ 0.11 & 1.81 $\pm$ 0.18 & 6.70 $\pm$ 0.37 & 2.07 $\pm$ 0.08 & 0.47 $\pm$ 0.14 & 7.9/10 \\
57780.185 & 57780.234 & 2.09 $\pm$ 0.14 & 1.44 $\pm$ 0.21 & 7.22 $\pm$ 0.35 & 2.21 $\pm$ 0.06 & 0.44 $\pm$ 0.12 & 9.3/11 \\
57782.177 & 57782.203 & 0.87 $\pm$ 0.13 & 0.36 $\pm$ 0.16 & 2.95 $\pm$ 0.26 & 2.60 $\pm$ 0.11 &  --  & 5.8/9 \\
57785.002 & 57785.140 & 3.38 $\pm$ 0.14 & 3.01 $\pm$ 0.22 & 10.60 $\pm$ 0.30 & 2.09 $\pm$ 0.03 & 0.36 $\pm$ 0.06 & 34.5/12 \\
57788.039 & 57788.067 & 5.85 $\pm$ 0.29 & 7.39 $\pm$ 0.59 & 17.60 $\pm$ 0.55 & 1.95 $\pm$ 0.04 & 0.19 $\pm$ 0.05 & 10.7/14 \\
57789.025 & 57789.099 & 3.01 $\pm$ 0.13 & 3.44 $\pm$ 0.26 & 9.68 $\pm$ 0.28 & 2.00 $\pm$ 0.04 & 0.29 $\pm$ 0.06 & 28.1/14 \\
57789.969 & 57790.146 & 1.63 $\pm$ 0.08 & 1.07 $\pm$ 0.09 & 5.35 $\pm$ 0.20 & 2.18 $\pm$ 0.05 & 0.39 $\pm$ 0.09 & 8.8/11 \\
57791.026 & 57791.145 & 1.08 $\pm$ 0.07 & 0.45 $\pm$ 0.08 & 3.78 $\pm$ 0.20 & 2.45 $\pm$ 0.07 & 0.50 $\pm$ 0.16 & 4.3/11 \\
57792.160 & 57792.230 & 0.99 $\pm$ 0.08 & 0.48 $\pm$ 0.10 & 3.09 $\pm$ 0.18 & 2.53 $\pm$ 0.07 &  --  & 14.7/13 \\
57800.065 & 57800.257 & 1.97 $\pm$ 0.07 & 1.01 $\pm$ 0.08 & 6.69 $\pm$ 0.20 & 2.29 $\pm$ 0.04 & 0.50 $\pm$ 0.09 & 16.7/11 \\
57802.178 & 57802.199 & 1.56 $\pm$ 0.16 & 0.46 $\pm$ 0.17 & 4.54 $\pm$ 0.33 & 2.57 $\pm$ 0.10 &  --  & 12.7/10 \\
57803.086 & 57803.166 & 1.93 $\pm$ 0.11 & 1.21 $\pm$ 0.16 & 6.36 $\pm$ 0.26 & 2.24 $\pm$ 0.04 & 0.37 $\pm$ 0.09 & 9.3/12 \\
57806.091 & 57806.155 & 1.13 $\pm$ 0.08 & 0.63 $\pm$ 0.12 & 3.75 $\pm$ 0.17 & 2.44 $\pm$ 0.06 &  --  & 8.6/10 \\
57807.108 & 57807.136 & 1.30 $\pm$ 0.14 & 0.66 $\pm$ 0.19 & 3.86 $\pm$ 0.24 & 2.38 $\pm$ 0.08 &  --  & 18.0/9 \\
57809.082 & 57809.092 & 1.46 $\pm$ 0.26 & 1.37 $\pm$ 0.50 & 4.47 $\pm$ 0.45 & 2.23 $\pm$ 0.12 &  --  & 10.6/9 \\
57811.103 & 57811.124 & 1.30 $\pm$ 0.15 & 0.93 $\pm$ 0.27 & 3.81 $\pm$ 0.28 & 2.20 $\pm$ 0.08 &  --  & 9.6/10 \\
57812.931 & 57813.264 & 1.39 $\pm$ 0.06 & 0.98 $\pm$ 0.08 & 4.75 $\pm$ 0.16 & 2.09 $\pm$ 0.04 & 0.47 $\pm$ 0.09 & 14.0/10 \\
57815.088 & 57815.103 & 1.50 $\pm$ 0.21 & 2.51 $\pm$ 0.55 & 4.80 $\pm$ 0.40 & 2.01 $\pm$ 0.09 &  --  & 7.7/11 \\
57816.863 & 57816.956 & 1.12 $\pm$ 0.19 & 0.59 $\pm$ 0.08 & 3.46 $\pm$ 0.48 & 2.14 $\pm$ 0.18 &  --  & 2.8/7 \\
57820.029 & 57820.196 & 1.07 $\pm$ 0.07 & 0.41 $\pm$ 0.08 & 3.66 $\pm$ 0.22 & 2.16 $\pm$ 0.10 & 0.95 $\pm$ 0.24 & 10.8/9 \\
57833.101 & 57833.122 & 2.43 $\pm$ 0.21 & 1.42 $\pm$ 0.27 & 8.42 $\pm$ 0.57 & 2.11 $\pm$ 0.09 & 0.63 $\pm$ 0.19 & 3.8/9 \\
57835.142 & 57835.162 & 1.18 $\pm$ 0.16 & 0.82 $\pm$ 0.20 & 3.63 $\pm$ 0.33 & 2.39 $\pm$ 0.11 &  --  & 7.1/10 \\
57838.027 & 57838.048 & 0.83 $\pm$ 0.13 & 0.39 $\pm$ 0.16 & 2.44 $\pm$ 0.24 & 2.51 $\pm$ 0.11 &  --  & 24.5/10 \\
57839.907 & 57840.233 & 0.83 $\pm$ 0.04 & 0.29 $\pm$ 0.05 & 2.87 $\pm$ 0.13 & 2.49 $\pm$ 0.05 & 0.50 $\pm$ 0.13 & 10.3/8 \\
57841.020 & 57841.039 & 1.25 $\pm$ 0.17 & 0.79 $\pm$ 0.26 & 3.48 $\pm$ 0.30 & 2.19 $\pm$ 0.10 &  --  & 6.0/10 \\
57841.866 & 57841.887 & 1.09 $\pm$ 0.16 & 0.35 $\pm$ 0.15 & 2.99 $\pm$ 0.36 & 2.80 $\pm$ 0.16 &  --  & 8.9/9 \\
57843.951 & 57843.972 & 1.15 $\pm$ 0.16 & 0.76 $\pm$ 0.25 & 3.10 $\pm$ 0.27 & 2.20 $\pm$ 0.10 &  --  & 10.4/10 \\
57846.038 & 57846.080 & 0.59 $\pm$ 0.08 & 0.21 $\pm$ 0.09 & 1.88 $\pm$ 0.16 & 2.59 $\pm$ 0.10 &  --  & 20.6/11 \\
57846.945 & 57847.029 & 0.52 $\pm$ 0.07 & 0.45 $\pm$ 0.12 & 1.83 $\pm$ 0.15 & 2.37 $\pm$ 0.10 &  --  & 7.3/9 \\
57847.958 & 57848.037 & 0.82 $\pm$ 0.07 & 0.26 $\pm$ 0.08 & 2.93 $\pm$ 0.22 & 2.45 $\pm$ 0.11 & 0.69 $\pm$ 0.26 & 8.1/9 \\
57861.964 & 57862.009 & 1.88 $\pm$ 0.13 & 1.52 $\pm$ 0.22 & 5.51 $\pm$ 0.22 & 2.22 $\pm$ 0.04 &  --  & 15.8/12 \\
57864.954 & 57865.003 & 1.75 $\pm$ 0.13 & 1.01 $\pm$ 0.18 & 5.73 $\pm$ 0.34 & 2.24 $\pm$ 0.06 & 0.43 $\pm$ 0.14 & 6.5/9 \\
57866.952 & 57867.005 & 1.08 $\pm$ 0.10 & 0.83 $\pm$ 0.15 & 3.10 $\pm$ 0.17 & 2.21 $\pm$ 0.06 &  --  & 14.9/10 \\
57889.925 & 57889.956 & 1.15 $\pm$ 0.12 & 0.89 $\pm$ 0.20 & 3.25 $\pm$ 0.23 & 2.18 $\pm$ 0.08 &  --  & 10.7/10 \\
57890.930 & 57890.968 & 1.32 $\pm$ 0.11 & 1.10 $\pm$ 0.18 & 4.05 $\pm$ 0.22 & 2.21 $\pm$ 0.06 &  --  & 19.8/13 \\
57891.939 & 57891.967 & 1.08 $\pm$ 0.12 & 0.75 $\pm$ 0.18 & 3.64 $\pm$ 0.26 & 2.37 $\pm$ 0.08 &  --  & 11.7/11 \\
\hline 
\end{tabular}
\tablefoot{The start and end of each observation are listed in the first two columns. The 0.2-1\,keV and >1\,TeV fluxes are given in the third and fourth column. The parameters of the log-parabolic (normalisation energy is 300\,GeV) or power-law fits above 100\,GeV with their corresponding $\chi^2/dof$ are also listed. All fits are performed after correcting for the EBL absorption.} 
\end{table*}

\section{\textit{NuSTAR} analysis results}
\label{NuSTAR_tables}
\begin{figure}[h!]
    \centering
    \begin{subfigure}[b]{0.497\textwidth}
       \centering
       \includegraphics[width=\textwidth]{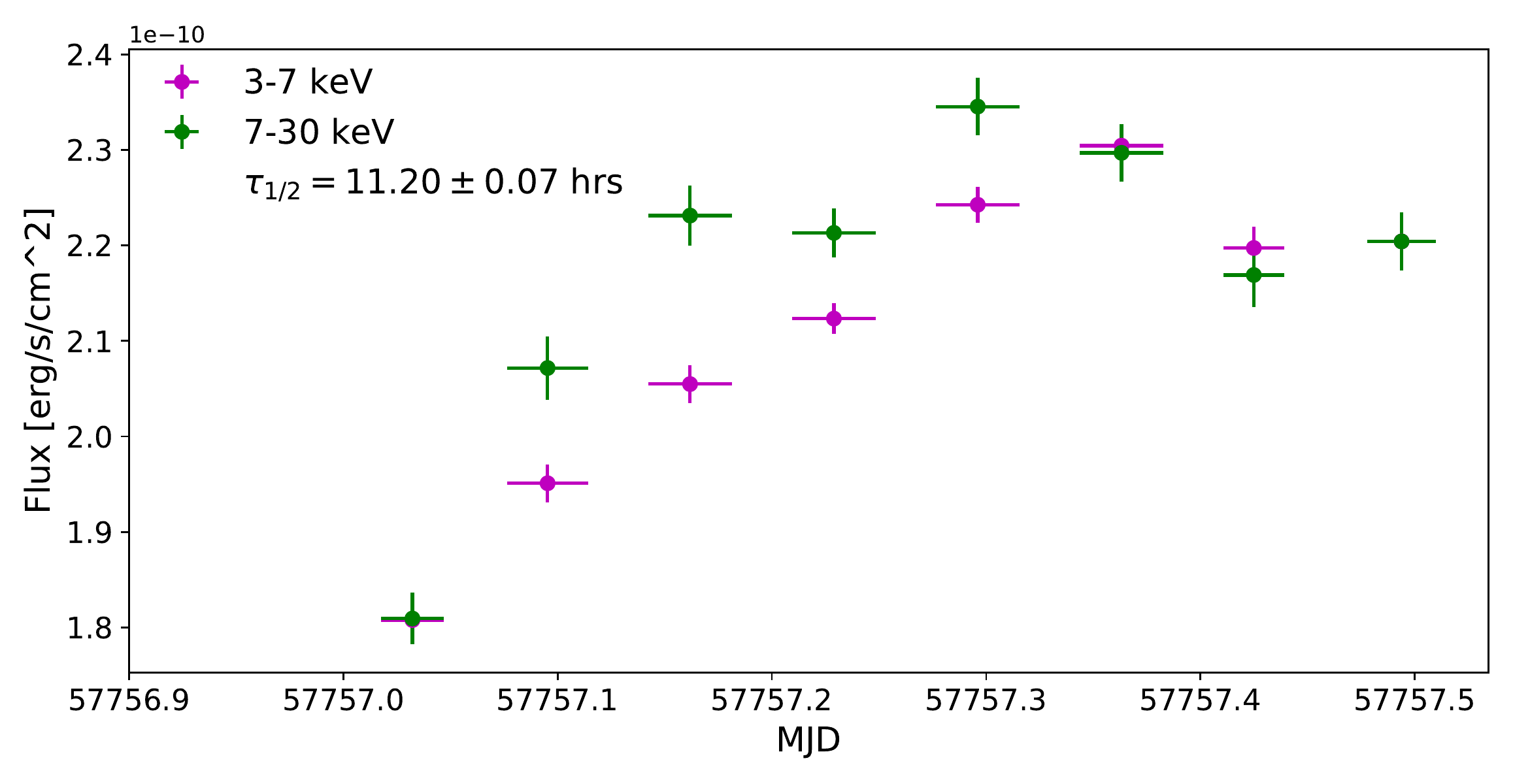}
    \end{subfigure}
    \begin{subfigure}[b]{0.497\textwidth}
       \centering
       \includegraphics[width=\textwidth]{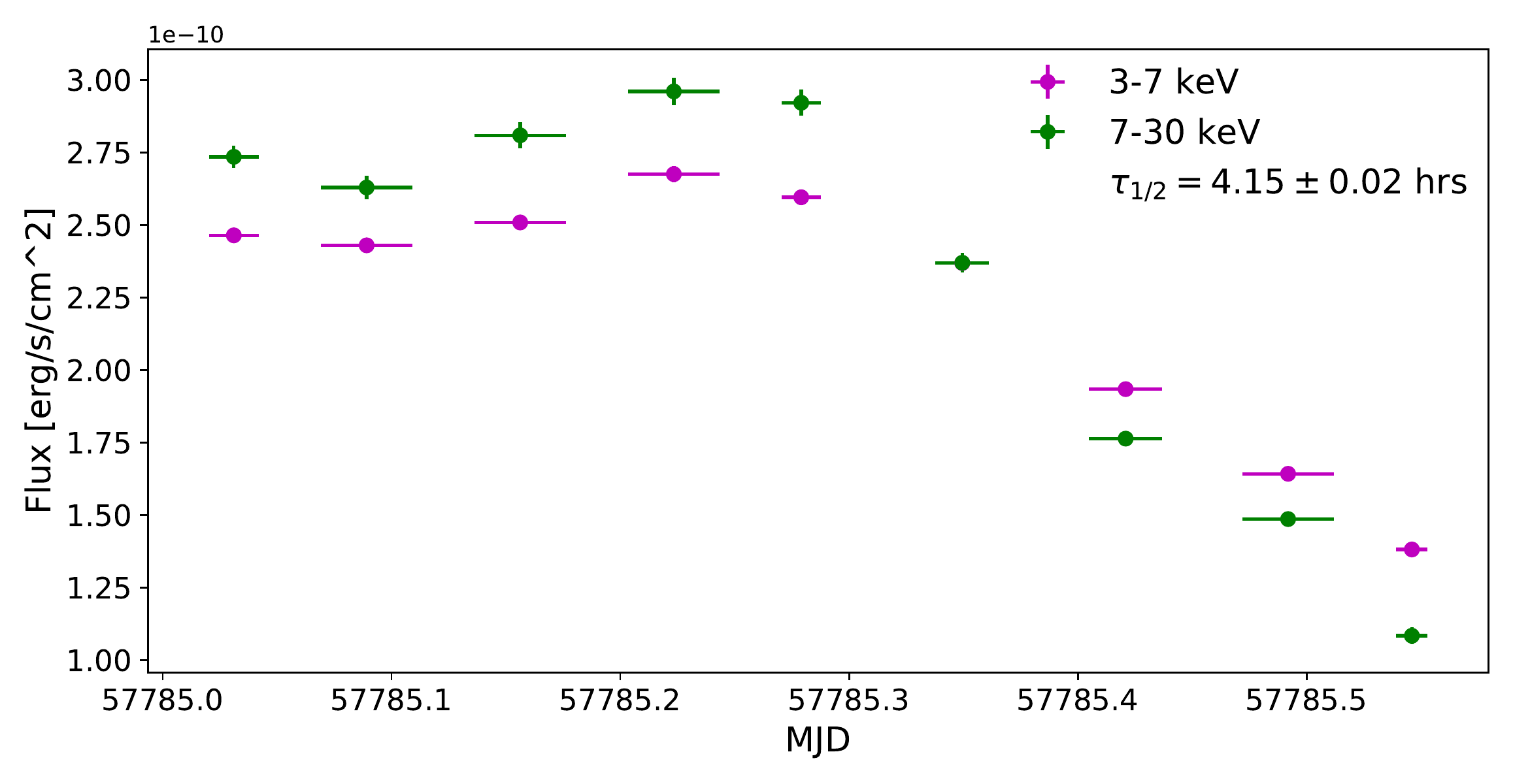}
    \end{subfigure}
    
    \begin{subfigure}[b]{0.497\textwidth}
       \centering
       \includegraphics[width=\textwidth]{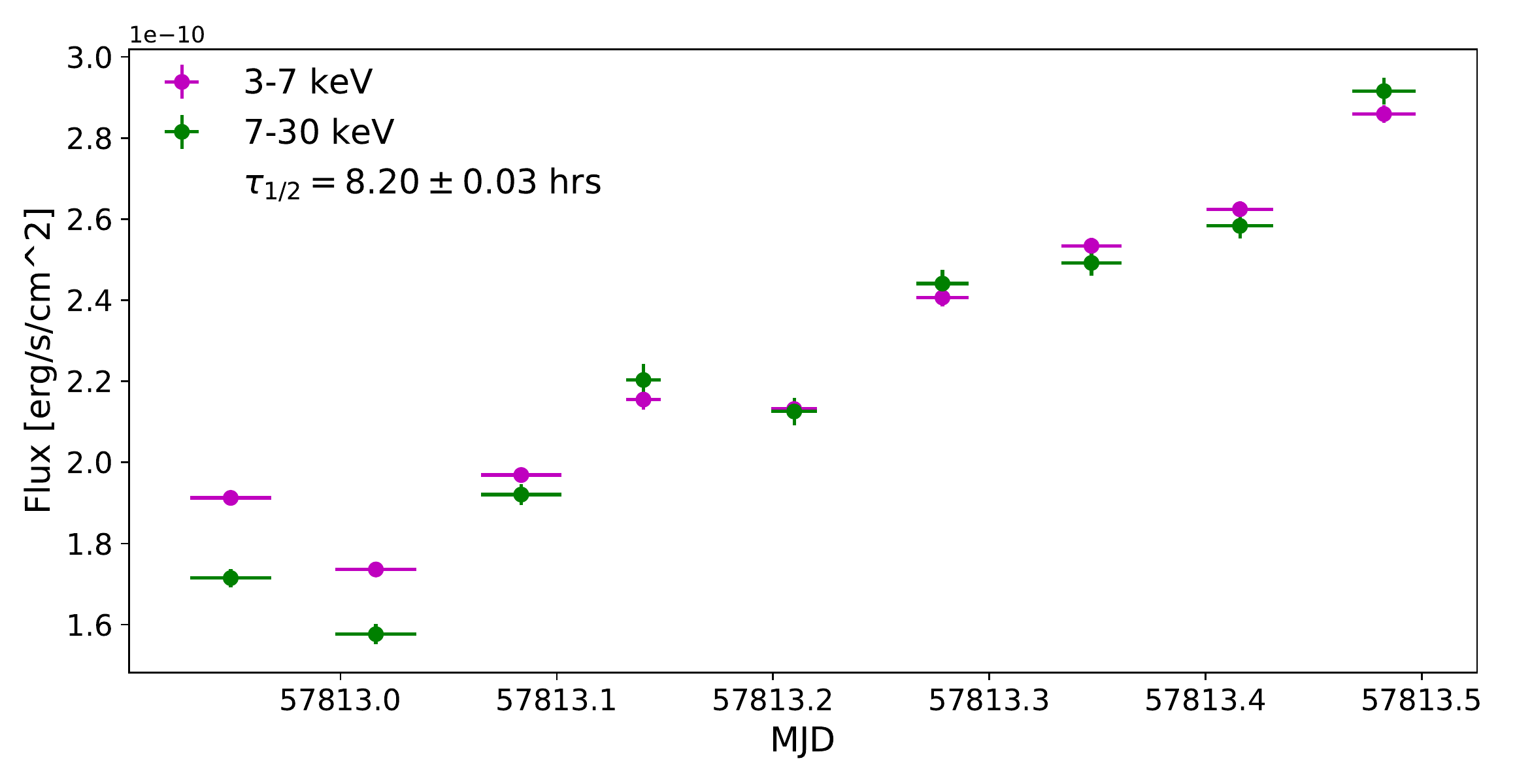}
    \end{subfigure}
    \begin{subfigure}[b]{0.497\textwidth}
       \centering
       \includegraphics[width=\textwidth]{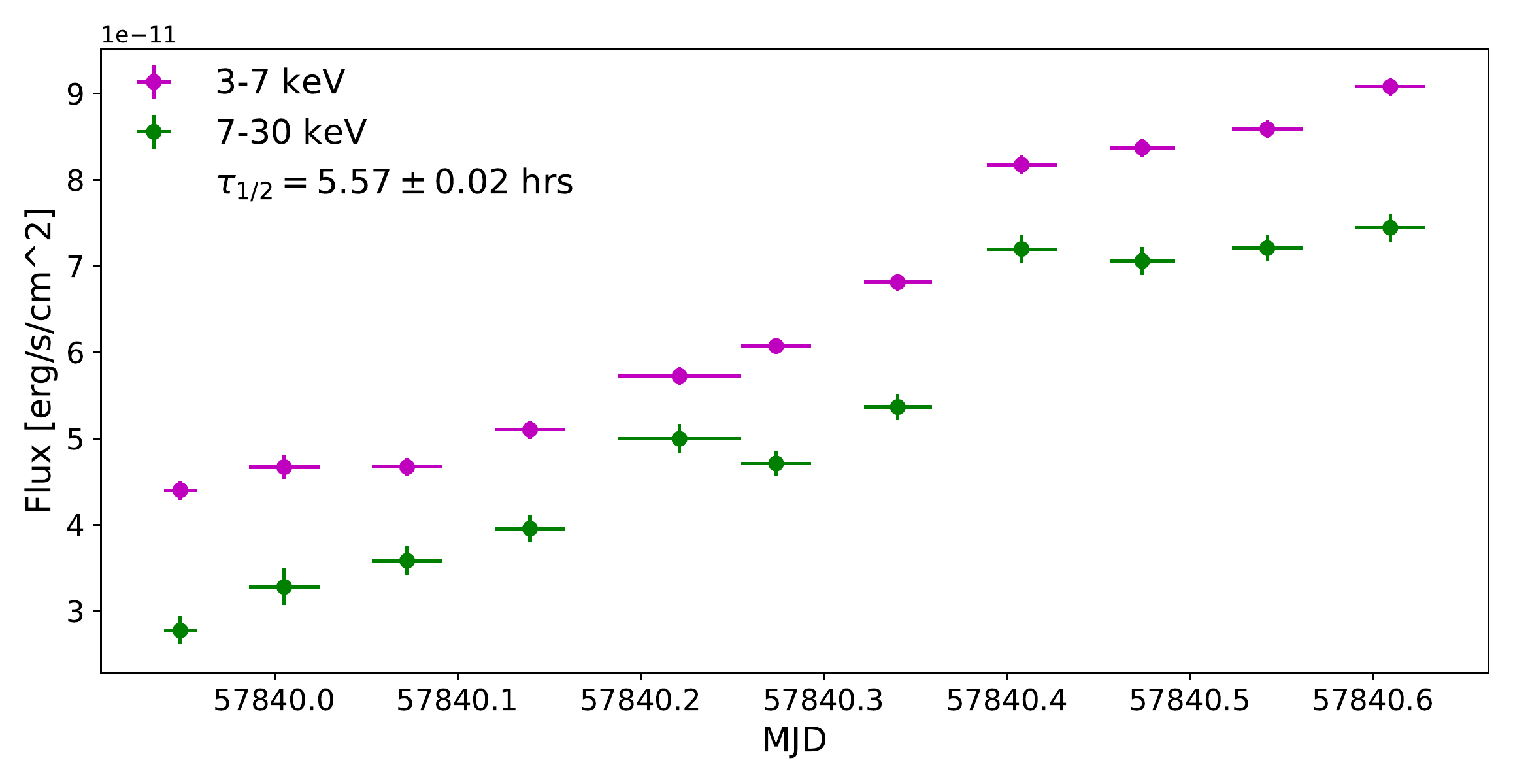}
    \end{subfigure}
\caption{\textit{NuSTAR} orbit-wise 3-7\,keV and 7-30\,keV light curves of the four observations on MJD~57757, MJD~57785, MJD~57813 and MJD~57840. The flux doubling/halving time is indicated in the legend. It is computed based on the 7-30\,keV flux using the prescription of \citet{Zhang_1999}.}
\label{NUSTAR_orbit_wise_LC}
\end{figure}

In this Section, we report in more detail on the orbit-wise light curves from the \textit{NuSTAR} observations. They are plotted in Fig.~\ref{NUSTAR_orbit_wise_LC}. Each light curve is incompatible with a constant flux. In order to quantify the variability time scale, we use the prescription of \citet{Zhang_1999}, which provides an estimate on the flux doubling/halving time $t_{1/2}$. For any consecutive flux measurement $i$ and $j$, $t_{1/2}$ is defined as the minimum of:
\begin{equation}
    t^{i,j}_{1/2} = \left| \frac{F_{j}+F_{i}}{2} \frac{T_{j}-T_{i}}{F_{j}-F_{i}} \right|
\end{equation}
where $F_{i,j}$ is the flux and $T_{i,j}$ is the time. The uncertainty is obtained by propagating the error of $F_{i,j}$. We compute $t_{1/2}$ in the 7-30\,keV band as it shows the strongest variability. The resulting value for each day is shown in the respective plot from Fig.~\ref{NUSTAR_orbit_wise_LC}. It ranges from ${\sim}4$\,hrs to  ${\sim}11$\,hrs.\par 

Table~\ref{tab:nustar_orbitwise_spectral_param} list the flux values along with the spectral parameters of the log-parabola fits. 

\clearpage

\begin{table*}[h!]
\caption{\label{tab:nustar_orbitwise_spectral_param}\textit{NuSTAR} orbit-wise analysis results of the four pointing on MJD~57757, MJD~57785, MJD~57813 and MJD~57840.}
\centering
\begin{tabular}{ l c c c c c c c c c }     
\hline\hline 
 MJD start & MJD end & $F_{3-7\text{\,keV}}$ & $F_{7-30\text{\,keV}}$ & $\alpha$ & $\beta$ & $\chi^2$/dof\\
  &  & $[10^{-10} \mathrm{erg} \, \mathrm{cm}^{-2} \mathrm{s}^{-1}]$ & $[10^{-10} \mathrm{erg} \, \mathrm{cm}^{-2} \mathrm{s}^{-1}]$ &   &  & \\
\hline\hline  
57757.0175 & 57757.0322 & 1.81 $\pm$ 0.02 & 1.81 $\pm$ 0.03 & 2.26 $\pm$ 0.09 & 0.11 $\pm$ 0.05 & 1.11\\
57757.0765 & 57757.0953 & 1.95 $\pm$ 0.02 & 2.07 $\pm$ 0.03 & 2.24 $\pm$ 0.09 & 0.10 $\pm$ 0.05 & 0.97\\
57757.1423 & 57757.1618 & 2.05 $\pm$ 0.02 & 2.23 $\pm$ 0.03 & 2.14 $\pm$ 0.08 & 0.15 $\pm$ 0.04 & 1.00\\
57757.2095 & 57757.2289 & 2.12 $\pm$ 0.02 & 2.21 $\pm$ 0.03 & 2.07 $\pm$ 0.07 & 0.21 $\pm$ 0.04 & 0.95\\
57757.2766 & 57757.2961 & 2.24 $\pm$ 0.02 & 2.35 $\pm$ 0.03 & 2.12 $\pm$ 0.07 & 0.19 $\pm$ 0.04 & 1.01\\
57757.3437 & 57757.3632 & 2.30 $\pm$ 0.02 & 2.30 $\pm$ 0.03 & 2.02 $\pm$ 0.08 & 0.26 $\pm$ 0.04 & 1.01\\
57757.4108 & 57757.4249 & 2.20 $\pm$ 0.02 & 2.17 $\pm$ 0.03 & 1.92 $\pm$ 0.09 & 0.31 $\pm$ 0.05 & 0.98\\
57757.4780 & 57757.4939 & 2.20 $\pm$ 0.02 & 2.20 $\pm$ 0.03 & 2.01 $\pm$ 0.08 & 0.26 $\pm$ 0.05 & 0.99\\

\hline  
57785.0203 & 57785.0311 & 2.46 $\pm$ 0.02 & 2.74 $\pm$ 0.04 & 1.94 $\pm$ 0.08 & 0.24 $\pm$ 0.05 & 1.02\\
57785.0692 & 57785.0891 & 2.43 $\pm$ 0.02 & 2.63 $\pm$ 0.04 & 1.99 $\pm$ 0.09 & 0.23 $\pm$ 0.05 & 0.93\\
57785.1363 & 57785.1563 & 2.51 $\pm$ 0.03 & 2.81 $\pm$ 0.04 & 1.96 $\pm$ 0.09 & 0.23 $\pm$ 0.05 & 0.97\\
57785.2035 & 57785.2234 & 2.68 $\pm$ 0.03 & 2.96 $\pm$ 0.05 & 2.03 $\pm$ 0.10 & 0.20 $\pm$ 0.06 & 0.89\\
57785.2706 & 57785.2791 & 2.60 $\pm$ 0.03 & 2.92 $\pm$ 0.04 & 1.98 $\pm$ 0.09 & 0.21 $\pm$ 0.05 & 0.90\\
57785.3377 & 57785.3495 & 2.37 $\pm$ 0.02 & 2.37 $\pm$ 0.03 & 1.96 $\pm$ 0.09 & 0.29 $\pm$ 0.05 & 1.01\\
57785.4049 & 57785.4209 & 1.93 $\pm$ 0.02 & 1.76 $\pm$ 0.02 & 2.06 $\pm$ 0.09 & 0.28 $\pm$ 0.05 & 0.96\\
57785.4720 & 57785.4919 & 1.64 $\pm$ 0.01 & 1.49 $\pm$ 0.02 & 1.96 $\pm$ 0.09 & 0.35 $\pm$ 0.05 & 1.01\\
57785.5391 & 57785.5460 & 1.38 $\pm$ 0.02 & 1.08 $\pm$ 0.03 & 1.88 $\pm$ 0.17 & 0.47 $\pm$ 0.10 & 1.01\\

\hline  
57812.9305 & 57812.9492 & 1.91 $\pm$ 0.02 & 1.71 $\pm$ 0.02 & 2.15 $\pm$ 0.08 & 0.24 $\pm$ 0.05 & 0.93\\
57812.9977 & 57813.0163 & 1.74 $\pm$ 0.02 & 1.58 $\pm$ 0.02 & 2.08 $\pm$ 0.09 & 0.27 $\pm$ 0.05 & 0.92\\
57813.0650 & 57813.0835 & 1.97 $\pm$ 0.02 & 1.92 $\pm$ 0.03 & 2.17 $\pm$ 0.08 & 0.18 $\pm$ 0.05 & 0.98\\
57813.1321 & 57813.1401 & 2.15 $\pm$ 0.03 & 2.20 $\pm$ 0.04 & 1.98 $\pm$ 0.11 & 0.26 $\pm$ 0.06 & 0.96\\
57813.1993 & 57813.2098 & 2.13 $\pm$ 0.02 & 2.13 $\pm$ 0.03 & 1.94 $\pm$ 0.10 & 0.30 $\pm$ 0.06 & 1.05\\
57813.2663 & 57813.2783 & 2.41 $\pm$ 0.02 & 2.44 $\pm$ 0.03 & 1.84 $\pm$ 0.09 & 0.35 $\pm$ 0.05 & 1.01\\
57813.3333 & 57813.3473 & 2.53 $\pm$ 0.02 & 2.49 $\pm$ 0.03 & 1.94 $\pm$ 0.08 & 0.30 $\pm$ 0.04 & 0.97\\
57813.4005 & 57813.4159 & 2.62 $\pm$ 0.02 & 2.58 $\pm$ 0.03 & 1.90 $\pm$ 0.07 & 0.34 $\pm$ 0.04 & 1.03\\
57813.4678 & 57813.4826 & 2.86 $\pm$ 0.02 & 2.92 $\pm$ 0.03 & 1.86 $\pm$ 0.07 & 0.33 $\pm$ 0.04 & 1.02\\

\hline  
57839.9397 & 57839.9485 & 0.44 $\pm$ 0.01 & 0.28 $\pm$ 0.02 & 2.96 $\pm$ 0.34 & -0.03 $\pm$ 0.21 & 0.91\\
57839.9861 & 57840.0053 & 0.47 $\pm$ 0.01 & 0.33 $\pm$ 0.02 & 2.84 $\pm$ 0.37 & -0.05 $\pm$ 0.23 & 0.93\\
57840.0533 & 57840.0724 & 0.47 $\pm$ 0.01 & 0.36 $\pm$ 0.02 & 2.81 $\pm$ 0.29 & -0.04 $\pm$ 0.18 & 0.93\\
57840.1204 & 57840.1395 & 0.51 $\pm$ 0.01 & 0.40 $\pm$ 0.02 & 2.42 $\pm$ 0.24 & 0.18 $\pm$ 0.15 & 1.02\\
57840.1875 & 57840.2212 & 0.57 $\pm$ 0.01 & 0.50 $\pm$ 0.02 & 2.41 $\pm$ 0.20 & 0.11 $\pm$ 0.12 & 0.89\\
57840.2550 & 57840.2740 & 0.61 $\pm$ 0.01 & 0.47 $\pm$ 0.01 & 2.30 $\pm$ 0.18 & 0.23 $\pm$ 0.11 & 0.96\\
57840.3218 & 57840.3405 & 0.68 $\pm$ 0.01 & 0.54 $\pm$ 0.02 & 2.29 $\pm$ 0.17 & 0.22 $\pm$ 0.10 & 1.10\\
57840.3890 & 57840.4081 & 0.82 $\pm$ 0.01 & 0.72 $\pm$ 0.02 & 2.28 $\pm$ 0.15 & 0.17 $\pm$ 0.09 & 0.94\\
57840.4561 & 57840.4739 & 0.84 $\pm$ 0.01 & 0.71 $\pm$ 0.02 & 2.29 $\pm$ 0.13 & 0.20 $\pm$ 0.08 & 1.04\\
57840.5232 & 57840.5423 & 0.86 $\pm$ 0.01 & 0.72 $\pm$ 0.02 & 2.25 $\pm$ 0.13 & 0.23 $\pm$ 0.08 & 0.89\\
57840.5903 & 57840.6095 & 0.91 $\pm$ 0.01 & 0.74 $\pm$ 0.02 & 2.29 $\pm$ 0.13 & 0.21 $\pm$ 0.08 & 1.00\\
\hline 
\end{tabular}
\tablefoot{The start and end of the observations are listed in the first two columns. The second and third column are the 3-7\,keV and 7-30\,keV fluxes. The fifth and sixth column give the best-fit spectral parameters of the log-parabolic fits. The corresponding $\chi^2/dof$ are listed in the last column.}
\end{table*}

\clearpage

\section{Multiwavelength light curves during the simultaneous MAGIC/\textit{NuSTAR}/\textit{Swift} observations}
\label{sect:MAGIC_nustar_lc}
\begin{figure}[h!]
   \centering
   \includegraphics[width=1\columnwidth]{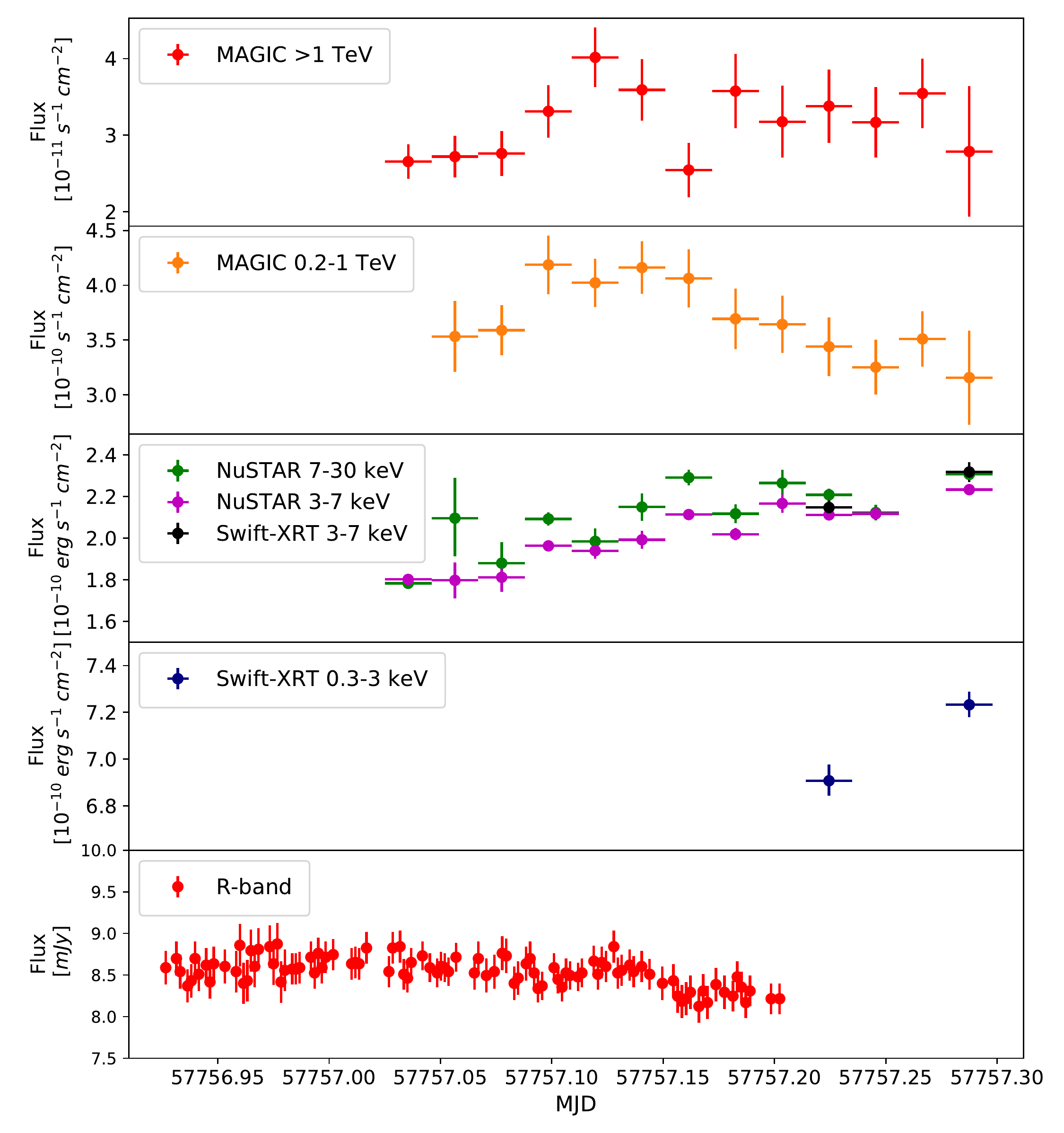}
   \caption{MWL light curves during the simultaneous MAGIC/\textit{NuSTAR}/\textit{Swift} observations on MJD~57757. The first two panels from the top are the MAGIC light curves in the 0.2-1\,TeV and >1\,TeV bands with 30\,min binning. The third and fourth panels from the top show the \textit{NuSTAR} and \textit{Swift}-XRT light curves in the 0.3-3\,keV, 3-7\,keV and 7-30\,keV bands with 30-minute binning. The bottom panel shows the R-band observations provided by the WEBT-GASP community.}
    \label{NUSTAR_MAGIC_sim_LC_first}
\end{figure}

\begin{figure}[h!]
   \centering
   \includegraphics[width=1\columnwidth]{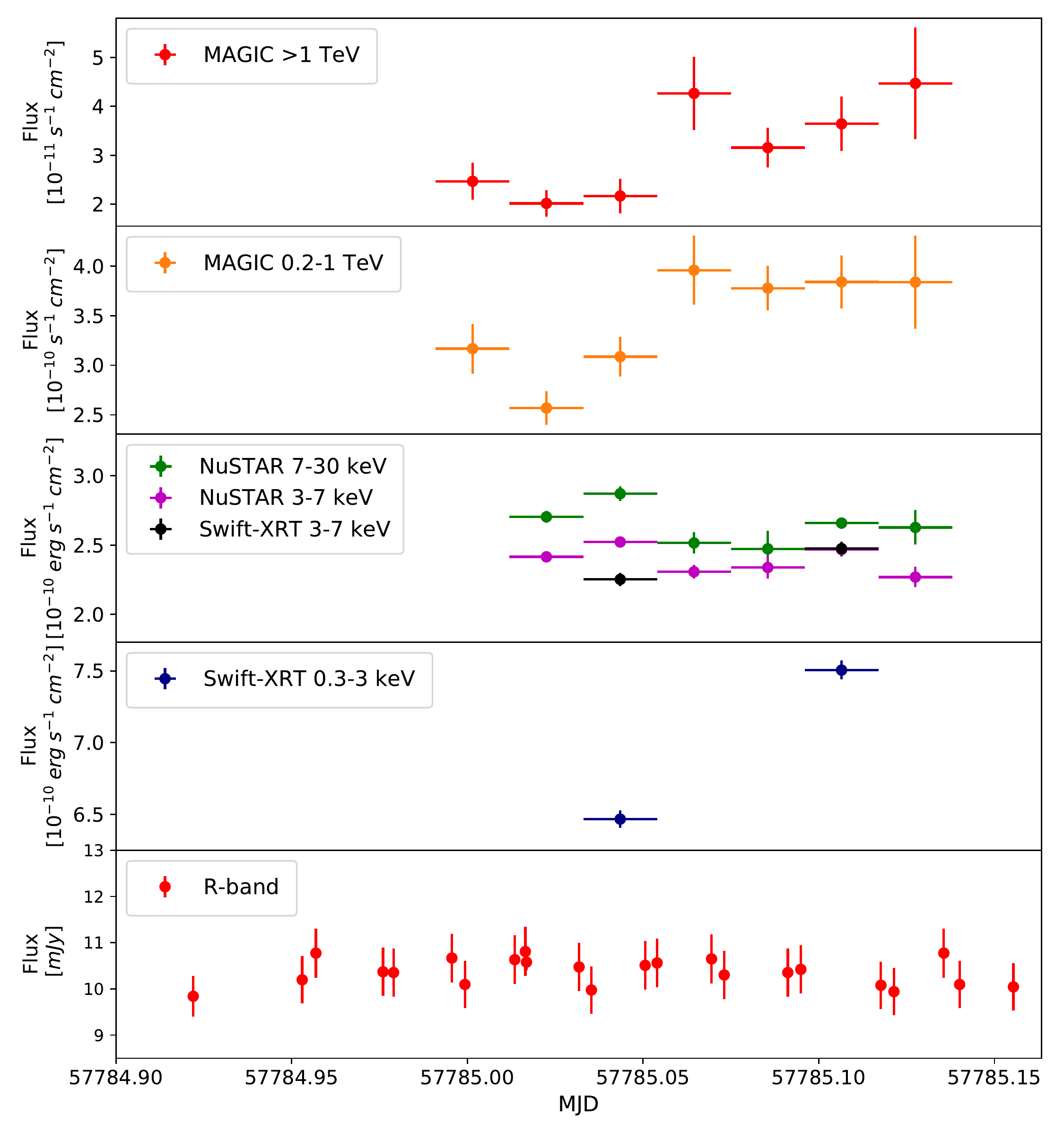}
   \caption{Same description as in Fig.~\ref{NUSTAR_MAGIC_sim_LC_first} for MJD~57785.}
    \label{NUSTAR_MAGIC_sim_LC_second}
\end{figure}

\begin{figure}[h!]
   \centering
   \includegraphics[width=1\columnwidth]{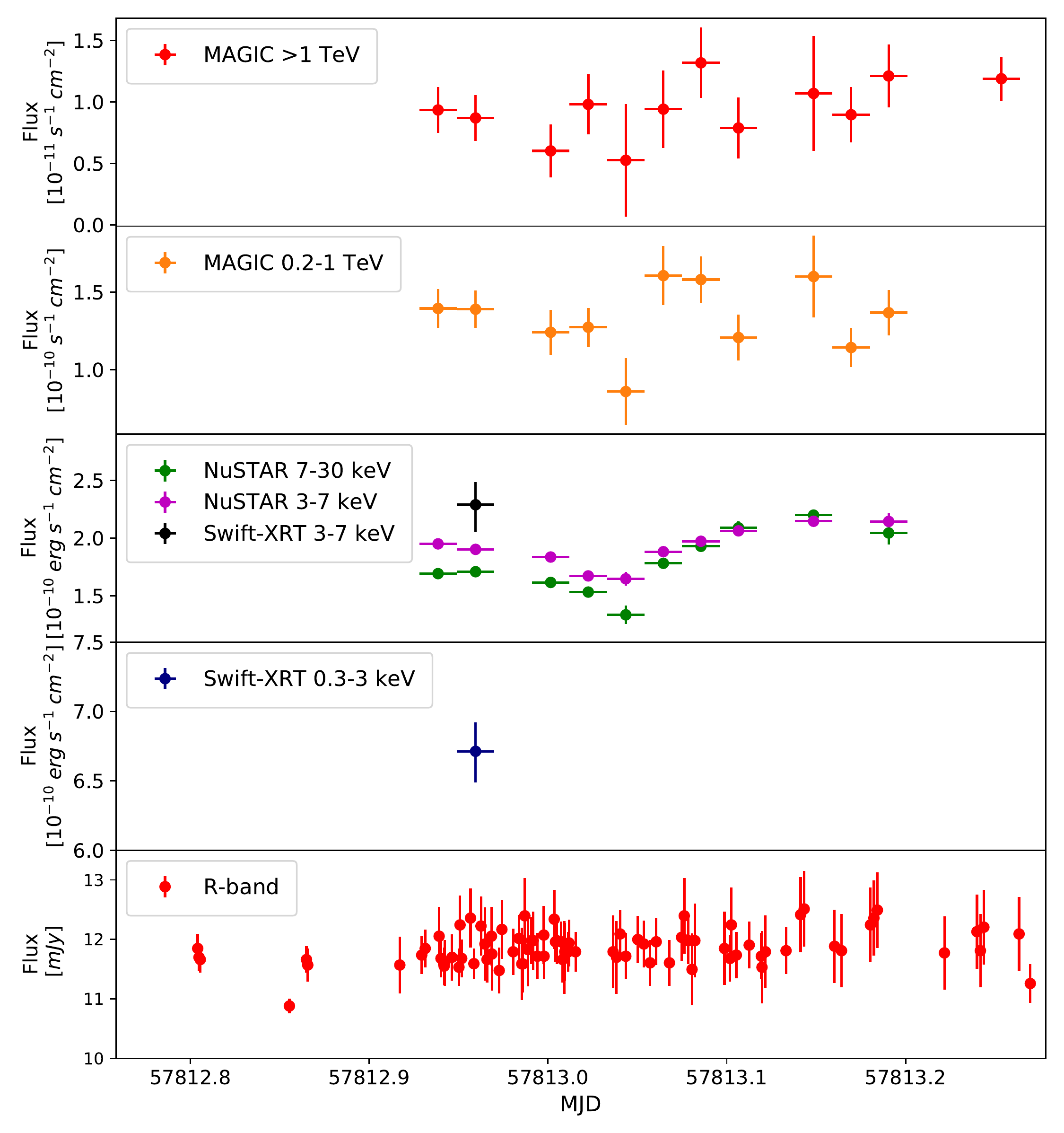}
   \caption{Same description as in Fig.~\ref{NUSTAR_MAGIC_sim_LC_first} for MJD~57813.}
    \label{NUSTAR_MAGIC_sim_LC_third}
\end{figure}

\begin{figure}[h!]
   \centering
   \includegraphics[width=1\columnwidth]{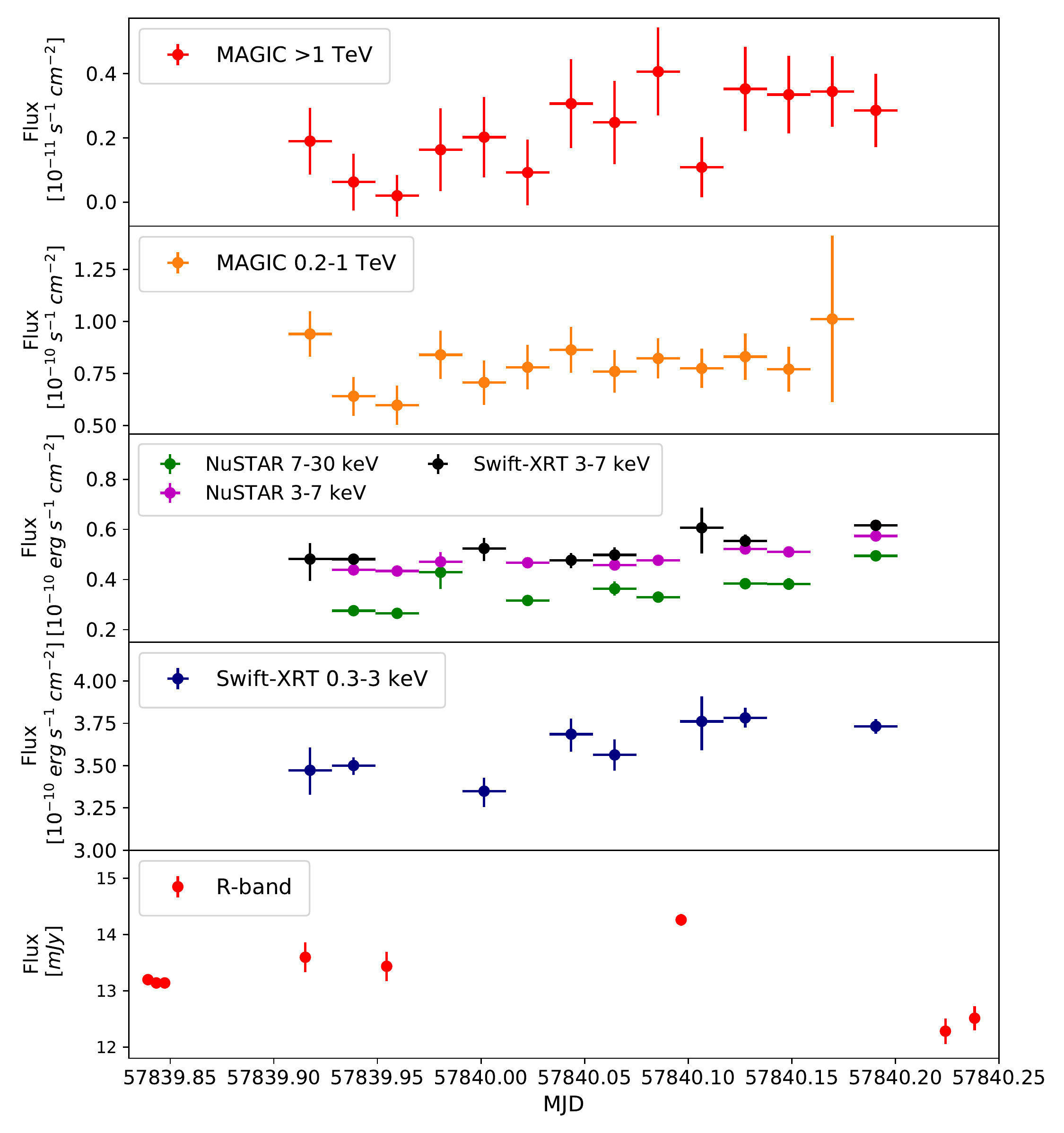}
   \caption{Same description as in Fig.~\ref{NUSTAR_MAGIC_sim_LC_first} for MJD~57840.}
    \label{NUSTAR_MAGIC_sim_LC_fourth}
\end{figure}

In this section, the MWL light curves during the four simultaneous MAGIC/\textit{NuSTAR}/\textit{Swift} observations are shown in Fig.~\ref{NUSTAR_MAGIC_sim_LC_first} to Fig.~\ref{NUSTAR_MAGIC_sim_LC_fourth}. The MAGIC, \textit{NuSTAR}, \textit{Swift} fluxes are computed in identical 30-minute time bins.

\clearpage

\section{\textit{Swift}-XRT analysis results}
\label{SwiftResults}
In this Section we report in detail the results of the \textit{Swift}-XRT analysis. Table~\ref{tab:swift_spectral_param} lists the fluxes in the 0.3-2\,keV, 2-10\,keV and 3-7\,keV bands for each observation. The best-fit power-law index ($\Gamma$) as well as the best-fit parameters $\alpha$ and $\beta$ from the log-parabolic fits (pivot energy of 1\,keV) are also given.

\begin{table*}
\caption{\label{tab:swift_spectral_param}\textit{Swift}-XRT analysis results.} 
\tiny
\centering
\begin{tabular}{ l c c c c c c c c c c c}     
\hline\hline 
 MJD & $F_{0.3-2\text{\,keV}}$ & $F_{2-10\text{\,keV}}$ & $F_{3-7\text{\,keV}}$ & $\Gamma$ & $\chi^2$/dof & $\alpha$ & $\beta$ & $\chi^2$/dof \\
  & $[10^{-10} \mathrm{erg} \, \mathrm{cm}^{-2} \mathrm{s}^{-1}]$ & $[10^{-10} \mathrm{erg} \, \mathrm{cm}^{-2} \mathrm{s}^{-1}]$ & $[10^{-10} \mathrm{erg}\, \mathrm{cm}^{-2} \mathrm{s}^{-1}]$ &  &  &  &  &\\
\hline\hline  
57719.2630 & 1.78 $\pm$ 0.03 & 0.34 $\pm$ 0.02 & 0.16 $\pm$ 0.01 & 2.79 $\pm$ 0.02 & 151/158 & 2.76 $\pm$ 0.02 & 0.21 $\pm$ 0.08 & 144/157\\
57721.2552 & 1.75 $\pm$ 0.04 & 0.52 $\pm$ 0.03 & 0.26 $\pm$ 0.02 & 2.58 $\pm$ 0.03 & 169/157 & 2.58 $\pm$ 0.03 & 0.04 $\pm$ 0.08 & 168/156\\
57723.2478 & 1.62 $\pm$ 0.03 & 0.33 $\pm$ 0.02 & 0.16 $\pm$ 0.01 & 2.78 $\pm$ 0.02 & 132/148 & 2.76 $\pm$ 0.03 & 0.11 $\pm$ 0.08 & 130/147\\
57724.4422 & 1.88 $\pm$ 0.04 & 0.41 $\pm$ 0.02 & 0.20 $\pm$ 0.01 & 2.72 $\pm$ 0.02 & 179/160 & 2.70 $\pm$ 0.03 & 0.17 $\pm$ 0.08 & 174/159\\
57725.2497 & 2.32 $\pm$ 0.03 & 0.60 $\pm$ 0.02 & 0.30 $\pm$ 0.01 & 2.63 $\pm$ 0.02 & 220/191 & 2.60 $\pm$ 0.02 & 0.17 $\pm$ 0.06 & 212/190\\
57726.4369 & 3.03 $\pm$ 0.08 & 0.78 $\pm$ 0.03 & 0.38 $\pm$ 0.02 & 2.64 $\pm$ 0.02 & 228/188 & 2.57 $\pm$ 0.02 & 0.26 $\pm$ 0.06 & 209/187\\
57727.2465 & 2.86 $\pm$ 0.04 & 0.71 $\pm$ 0.03 & 0.35 $\pm$ 0.01 & 2.66 $\pm$ 0.02 & 197/201 & 2.63 $\pm$ 0.02 & 0.18 $\pm$ 0.05 & 185/200\\
57728.4954 & 2.41 $\pm$ 0.04 & 0.64 $\pm$ 0.03 & 0.31 $\pm$ 0.02 & 2.61 $\pm$ 0.02 & 227/183 & 2.57 $\pm$ 0.02 & 0.22 $\pm$ 0.07 & 215/182\\
57729.2381 & 2.02 $\pm$ 0.03 & 0.51 $\pm$ 0.02 & 0.25 $\pm$ 0.01 & 2.64 $\pm$ 0.02 & 200/182 & 2.61 $\pm$ 0.02 & 0.19 $\pm$ 0.06 & 190/181\\
57730.4214 & 1.61 $\pm$ 0.03 & 0.35 $\pm$ 0.02 & 0.17 $\pm$ 0.01 & 2.70 $\pm$ 0.02 & 159/159 & 2.65 $\pm$ 0.03 & 0.32 $\pm$ 0.08 & 140/158\\
57731.2858 & 2.06 $\pm$ 0.04 & 0.58 $\pm$ 0.03 & 0.29 $\pm$ 0.02 & 2.59 $\pm$ 0.02 & 184/165 & 2.53 $\pm$ 0.03 & 0.24 $\pm$ 0.08 & 174/164\\
57732.4804 & 1.55 $\pm$ 0.04 & 0.26 $\pm$ 0.02 & 0.12 $\pm$ 0.01 & 2.89 $\pm$ 0.03 & 130/133 & 2.83 $\pm$ 0.03 & 0.25 $\pm$ 0.10 & 124/132\\
57733.2108 & 2.14 $\pm$ 0.04 & 0.53 $\pm$ 0.02 & 0.26 $\pm$ 0.01 & 2.68 $\pm$ 0.02 & 182/179 & 2.66 $\pm$ 0.02 & 0.09 $\pm$ 0.06 & 180/178\\
57735.2044 & 1.88 $\pm$ 0.06 & 0.34 $\pm$ 0.02 & 0.16 $\pm$ 0.01 & 2.82 $\pm$ 0.02 & 192/154 & 2.73 $\pm$ 0.03 & 0.40 $\pm$ 0.09 & 169/153\\
57737.0035 & 1.94 $\pm$ 0.05 & 0.46 $\pm$ 0.03 & 0.22 $\pm$ 0.02 & 2.66 $\pm$ 0.02 & 122/149 & 2.62 $\pm$ 0.03 & 0.27 $\pm$ 0.08 & 111/148\\
57739.9222 & 3.14 $\pm$ 0.08 & 1.38 $\pm$ 0.08 & 0.71 $\pm$ 0.04 & 2.38 $\pm$ 0.03 & 144/147 & 2.39 $\pm$ 0.03 & -0.03 $\pm$ 0.09 & 143/146\\
57744.1243 & 3.43 $\pm$ 0.05 & 1.05 $\pm$ 0.04 & 0.52 $\pm$ 0.02 & 2.53 $\pm$ 0.01 & 239/220 & 2.47 $\pm$ 0.02 & 0.27 $\pm$ 0.05 & 208/219\\
57744.5739 & 3.45 $\pm$ 0.05 & 1.23 $\pm$ 0.04 & 0.62 $\pm$ 0.02 & 2.45 $\pm$ 0.01 & 229/235 & 2.39 $\pm$ 0.02 & 0.25 $\pm$ 0.05 & 199/234\\
57747.1845 & 4.01 $\pm$ 0.05 & 1.71 $\pm$ 0.05 & 0.88 $\pm$ 0.03 & 2.36 $\pm$ 0.01 & 299/257 & 2.31 $\pm$ 0.02 & 0.17 $\pm$ 0.04 & 280/256\\
57749.1769 & 4.60 $\pm$ 0.05 & 1.91 $\pm$ 0.06 & 0.98 $\pm$ 0.03 & 2.38 $\pm$ 0.01 & 249/236 & 2.34 $\pm$ 0.02 & 0.16 $\pm$ 0.05 & 237/235\\
57749.4980 & 4.50 $\pm$ 0.05 & 1.79 $\pm$ 0.05 & 0.91 $\pm$ 0.03 & 2.40 $\pm$ 0.01 & 279/256 & 2.35 $\pm$ 0.02 & 0.18 $\pm$ 0.04 & 261/255\\
57751.2394 & 4.29 $\pm$ 0.04 & 1.88 $\pm$ 0.05 & 0.99 $\pm$ 0.03 & 2.33 $\pm$ 0.01 & 316/242 & 2.28 $\pm$ 0.02 & 0.20 $\pm$ 0.04 & 290/241\\
57753.1643 & 5.41 $\pm$ 0.04 & 3.53 $\pm$ 0.06 & 1.85 $\pm$ 0.03 & 2.13 $\pm$ 0.01 & 412/351 & 2.08 $\pm$ 0.01 & 0.15 $\pm$ 0.03 & 386/350\\
57754.4787 & 6.44 $\pm$ 0.05 & 5.04 $\pm$ 0.08 & 2.67 $\pm$ 0.05 & 2.03 $\pm$ 0.01 & 496/382 & 1.97 $\pm$ 0.01 & 0.16 $\pm$ 0.03 & 464/381\\
57755.2234 & 6.70 $\pm$ 0.06 & 4.34 $\pm$ 0.08 & 2.29 $\pm$ 0.05 & 2.12 $\pm$ 0.01 & 392/348 & 2.04 $\pm$ 0.01 & 0.22 $\pm$ 0.03 & 337/347\\
57756.1531 & 4.88 $\pm$ 0.05 & 2.70 $\pm$ 0.06 & 1.39 $\pm$ 0.03 & 2.22 $\pm$ 0.01 & 301/297 & 2.17 $\pm$ 0.02 & 0.16 $\pm$ 0.04 & 280/296\\
57757.0076 & 5.82 $\pm$ 0.02 & 4.21 $\pm$ 0.03 & 2.20 $\pm$ 0.02 & 2.067 $\pm$ 0.004 & 1032/589 & 1.99 $\pm$ 0.01 & 0.20 $\pm$ 0.01 & 757/588\\
57758.1411 & 6.82 $\pm$ 0.17 & 4.69 $\pm$ 0.30 & 2.44 $\pm$ 0.15 & 2.09 $\pm$ 0.03 & 77/97 & 2.04 $\pm$ 0.05 & 0.15 $\pm$ 0.10 & 74/96\\
57759.1376 & 8.14 $\pm$ 0.06 & 6.61 $\pm$ 0.10 & 3.38 $\pm$ 0.05 & 1.99 $\pm$ 0.01 & 518/406 & 1.89 $\pm$ 0.01 & 0.26 $\pm$ 0.03 & 422/405\\
57760.0711 & 8.58 $\pm$ 0.04 & 8.83 $\pm$ 0.08 & 4.70 $\pm$ 0.04 & 1.87 $\pm$ 0.01 & 919/564 & 1.77 $\pm$ 0.01 & 0.21 $\pm$ 0.02 & 751/563\\
57761.4544 & 9.42 $\pm$ 0.06 & 8.48 $\pm$ 0.14 & 4.50 $\pm$ 0.07 & 1.93 $\pm$ 0.01 & 512/412 & 1.83 $\pm$ 0.01 & 0.26 $\pm$ 0.03 & 415/411\\
57763.2594 & 7.37 $\pm$ 0.06 & 4.94 $\pm$ 0.09 & 2.59 $\pm$ 0.05 & 2.10 $\pm$ 0.01 & 518/379 & 2.02 $\pm$ 0.01 & 0.24 $\pm$ 0.03 & 442/378\\
57765.2574 & 5.00 $\pm$ 0.03 & 3.41 $\pm$ 0.06 & 1.79 $\pm$ 0.03 & 2.11 $\pm$ 0.01 & 400/393 & 2.08 $\pm$ 0.01 & 0.10 $\pm$ 0.03 & 386/392\\
57772.0299 & 5.57 $\pm$ 0.04 & 3.75 $\pm$ 0.07 & 2.04 $\pm$ 0.04 & 2.10 $\pm$ 0.01 & 453/357 & 2.02 $\pm$ 0.01 & 0.22 $\pm$ 0.03 & 395/356\\
57774.3547 & 5.29 $\pm$ 0.04 & 4.00 $\pm$ 0.08 & 2.10 $\pm$ 0.04 & 2.07 $\pm$ 0.01 & 368/354 & 2.08 $\pm$ 0.01 & -0.03 $\pm$ 0.03 & 367/353\\
57776.2013 & 7.73 $\pm$ 0.09 & 4.64 $\pm$ 0.13 & 2.42 $\pm$ 0.07 & 2.16 $\pm$ 0.01 & 266/260 & 2.08 $\pm$ 0.02 & 0.24 $\pm$ 0.04 & 234/259\\
57778.4635 & 5.06 $\pm$ 0.06 & 2.38 $\pm$ 0.06 & 1.23 $\pm$ 0.03 & 2.31 $\pm$ 0.01 & 328/257 & 2.24 $\pm$ 0.02 & 0.21 $\pm$ 0.04 & 300/256\\
57780.1863 & 7.34 $\pm$ 0.09 & 3.69 $\pm$ 0.11 & 1.91 $\pm$ 0.06 & 2.27 $\pm$ 0.01 & 300/245 & 2.21 $\pm$ 0.02 & 0.18 $\pm$ 0.05 & 284/244\\
57782.1802 & 2.13 $\pm$ 0.05 & 0.92 $\pm$ 0.04 & 0.47 $\pm$ 0.02 & 2.36 $\pm$ 0.02 & 204/195 & 2.28 $\pm$ 0.03 & 0.23 $\pm$ 0.06 & 190/194\\
57784.1748 & 4.02 $\pm$ 0.08 & 3.61 $\pm$ 0.13 & 1.77 $\pm$ 0.06 & 1.98 $\pm$ 0.02 & 207/221 & 2.02 $\pm$ 0.03 & -0.09 $\pm$ 0.06 & 205/220\\
57784.9791 & 5.77 $\pm$ 0.05 & 4.50 $\pm$ 0.09 & 2.38 $\pm$ 0.05 & 2.03 $\pm$ 0.01 & 422/367 & 1.99 $\pm$ 0.02 & 0.12 $\pm$ 0.03 & 404/366\\
57785.0364 & 5.66 $\pm$ 0.06 & 4.51 $\pm$ 0.10 & 2.25 $\pm$ 0.05 & 2.03 $\pm$ 0.01 & 305/330 & 2.01 $\pm$ 0.02 & 0.05 $\pm$ 0.03 & 302/329\\
57785.1031 & 6.21 $\pm$ 0.06 & 4.64 $\pm$ 0.09 & 2.47 $\pm$ 0.05 & 2.06 $\pm$ 0.01 & 372/332 & 2.01 $\pm$ 0.02 & 0.12 $\pm$ 0.03 & 357/331\\
57785.1685 & 6.27 $\pm$ 0.07 & 4.86 $\pm$ 0.09 & 2.55 $\pm$ 0.05 & 2.04 $\pm$ 0.01 & 357/342 & 2.01 $\pm$ 0.02 & 0.08 $\pm$ 0.03 & 350/341\\
57785.2351 & 7.05 $\pm$ 0.09 & 5.50 $\pm$ 0.14 & 2.44 $\pm$ 0.06 & 2.04 $\pm$ 0.01 & 300/289 & 2.01 $\pm$ 0.02 & 0.07 $\pm$ 0.04 & 296/288\\
57785.3018 & 6.40 $\pm$ 0.06 & 4.64 $\pm$ 0.09 & 2.44 $\pm$ 0.05 & 2.07 $\pm$ 0.01 & 427/360 & 2.03 $\pm$ 0.01 & 0.12 $\pm$ 0.03 & 410/359\\
57785.3677 & 5.88 $\pm$ 0.06 & 4.24 $\pm$ 0.08 & 2.15 $\pm$ 0.04 & 2.08 $\pm$ 0.01 & 313/333 & 2.06 $\pm$ 0.02 & 0.07 $\pm$ 0.03 & 308/332\\
57785.4335 & 5.60 $\pm$ 0.05 & 3.42 $\pm$ 0.07 & 1.87 $\pm$ 0.04 & 2.16 $\pm$ 0.01 & 352/334 & 2.09 $\pm$ 0.02 & 0.20 $\pm$ 0.03 & 312/333\\
57785.4996 & 5.70 $\pm$ 0.06 & 3.66 $\pm$ 0.08 & 2.20 $\pm$ 0.05 & 2.14 $\pm$ 0.01 & 308/309 & 2.10 $\pm$ 0.02 & 0.13 $\pm$ 0.03 & 294/308\\
57786.2307 & 3.25 $\pm$ 0.07 & 1.33 $\pm$ 0.04 & 0.68 $\pm$ 0.02 & 2.40 $\pm$ 0.02 & 227/218 & 2.37 $\pm$ 0.02 & 0.11 $\pm$ 0.05 & 223/217\\
57788.0371 & 6.40 $\pm$ 0.06 & 6.18 $\pm$ 0.13 & 3.27 $\pm$ 0.07 & 1.92 $\pm$ 0.01 & 362/338 & 1.90 $\pm$ 0.02 & 0.05 $\pm$ 0.03 & 360/337\\
57790.1500 & 5.02 $\pm$ 0.05 & 2.84 $\pm$ 0.06 & 1.48 $\pm$ 0.04 & 2.22 $\pm$ 0.01 & 268/305 & 2.20 $\pm$ 0.02 & 0.08 $\pm$ 0.03 & 263/304\\
57792.2097 & 3.62 $\pm$ 0.05 & 1.33 $\pm$ 0.04 & 0.68 $\pm$ 0.02 & 2.47 $\pm$ 0.02 & 254/227 & 2.46 $\pm$ 0.02 & 0.04 $\pm$ 0.05 & 253/226\\
57801.1365 & 3.64 $\pm$ 0.05 & 1.65 $\pm$ 0.05 & 0.85 $\pm$ 0.03 & 2.33 $\pm$ 0.01 & 257/240 & 2.27 $\pm$ 0.02 & 0.19 $\pm$ 0.05 & 238/239\\
57803.0594 & 5.08 $\pm$ 0.06 & 2.82 $\pm$ 0.07 & 1.47 $\pm$ 0.04 & 2.23 $\pm$ 0.01 & 300/294 & 2.22 $\pm$ 0.02 & 0.04 $\pm$ 0.04 & 299/293\\
57805.1207 & 6.31 $\pm$ 0.06 & 3.58 $\pm$ 0.07 & 1.87 $\pm$ 0.04 & 2.21 $\pm$ 0.01 & 380/323 & 2.17 $\pm$ 0.02 & 0.13 $\pm$ 0.03 & 363/322\\
57807.1131 & 5.91 $\pm$ 0.06 & 2.86 $\pm$ 0.06 & 1.48 $\pm$ 0.03 & 2.30 $\pm$ 0.01 & 313/314 & 2.26 $\pm$ 0.01 & 0.13 $\pm$ 0.03 & 295/313\\
57810.3549 & 5.32 $\pm$ 0.05 & 3.07 $\pm$ 0.07 & 1.60 $\pm$ 0.04 & 2.20 $\pm$ 0.01 & 337/304 & 2.16 $\pm$ 0.02 & 0.12 $\pm$ 0.04 & 325/303\\
57812.9520 & 5.67 $\pm$ 0.27 & 4.55 $\pm$ 0.61 & 2.28 $\pm$ 0.21 & 2.00 $\pm$ 0.06 & 22/30 & 1.96 $\pm$ 0.09 & 0.14 $\pm$ 0.21 & 22/29\\
57819.0605 & 4.98 $\pm$ 0.05 & 3.07 $\pm$ 0.07 & 1.62 $\pm$ 0.04 & 2.18 $\pm$ 0.01 & 307/302 & 2.18 $\pm$ 0.02 & 0.01 $\pm$ 0.04 & 307/301\\
57826.9093 & 6.02 $\pm$ 0.05 & 3.42 $\pm$ 0.07 & 1.78 $\pm$ 0.04 & 2.20 $\pm$ 0.01 & 356/325 & 2.14 $\pm$ 0.02 & 0.18 $\pm$ 0.03 & 325/324\\
57828.9601 & 6.26 $\pm$ 0.06 & 4.06 $\pm$ 0.08 & 2.13 $\pm$ 0.04 & 2.14 $\pm$ 0.01 & 352/344 & 2.11 $\pm$ 0.01 & 0.08 $\pm$ 0.03 & 346/343\\
57831.0791 & 7.10 $\pm$ 0.06 & 4.74 $\pm$ 0.09 & 2.49 $\pm$ 0.05 & 2.12 $\pm$ 0.01 & 390/361 & 2.07 $\pm$ 0.01 & 0.13 $\pm$ 0.03 & 370/360\\
57833.0863 & 7.64 $\pm$ 0.11 & 3.97 $\pm$ 0.11 & 2.06 $\pm$ 0.07 & 2.25 $\pm$ 0.01 & 239/239 & 2.17 $\pm$ 0.02 & 0.24 $\pm$ 0.05 & 212/238\\
57835.1363 & 4.57 $\pm$ 0.05 & 1.68 $\pm$ 0.05 & 0.85 $\pm$ 0.02 & 2.44 $\pm$ 0.01 & 285/258 & 2.41 $\pm$ 0.02 & 0.16 $\pm$ 0.04 & 270/257\\
57838.0569 & 3.62 $\pm$ 0.05 & 1.03 $\pm$ 0.04 & 0.51 $\pm$ 0.02 & 2.58 $\pm$ 0.02 & 257/208 & 2.55 $\pm$ 0.02 & 0.18 $\pm$ 0.05 & 245/207\\

\hline 
\end{tabular}
\tablefoot{For each observation, the start time in MJD is given as well as the 0.3-2\,keV, 2-10\,keV and 3-7\,keV fluxes. The observations have a typical exposure of about 1\,ks. The best-fit power-law indices $\Gamma$ are listed in the fifth column with the corresponding $\chi^2$/dof in the sixth column. The best-fit parameters $\alpha$ and $\beta$ from the log-parabolic fits with a pivot energy fixed at 1\,keV are also given with their corresponding $\chi^2$/dof. The analysis includes a photoelectric absorption by a fixed column density of $N_{\rm H}=1.92\times10^{20}$\,cm$^{-2}$ \citep[][]{2005A&A...440..775K}.} 
\end{table*}

\begin{table*}
\ContinuedFloat
\caption{continued.} 
\tiny
\centering
\begin{tabular}{ l c c c c c c c c c c c}     
\hline\hline 
 MJD & $F_{0.3-2\text{\,keV}}$ & $F_{2-10\text{\,keV}}$ & $F_{3-7\text{\,keV}}$ & $\Gamma$ & $\chi^2$/dof & $\alpha$ & $\beta$ & $\chi^2$/dof \\
  & $[10^{-10} \mathrm{erg} \, \mathrm{cm}^{-2} \mathrm{s}^{-1}]$ & $[10^{-10} \mathrm{erg} \, \mathrm{cm}^{-2} \mathrm{s}^{-1}]$ & $[10^{-10} \mathrm{erg}\, \mathrm{cm}^{-2} \mathrm{s}^{-1}]$ &  &  &  &  &\\
\hline\hline  
57839.9271 & 3.16 $\pm$ 0.05 & 0.96 $\pm$ 0.03 & 0.48 $\pm$ 0.02 & 2.54 $\pm$ 0.02 & 259/222 & 2.50 $\pm$ 0.02 & 0.19 $\pm$ 0.05 & 243/221\\
57840.0035 & 3.33 $\pm$ 0.04 & 1.21 $\pm$ 0.04 & 0.59 $\pm$ 0.02 & 2.46 $\pm$ 0.01 & 262/239 & 2.44 $\pm$ 0.02 & 0.11 $\pm$ 0.04 & 256/238\\
57840.0483 & 3.33 $\pm$ 0.07 & 0.94 $\pm$ 0.04 & 0.47 $\pm$ 0.02 & 2.60 $\pm$ 0.02 & 238/200 & 2.57 $\pm$ 0.02 & 0.11 $\pm$ 0.06 & 234/199\\
57840.1157 & 3.56 $\pm$ 0.06 & 1.12 $\pm$ 0.04 & 0.55 $\pm$ 0.02 & 2.54 $\pm$ 0.02 & 172/198 & 2.51 $\pm$ 0.02 & 0.12 $\pm$ 0.06 & 167/197\\
57840.1833 & 3.32 $\pm$ 0.04 & 1.20 $\pm$ 0.04 & 0.61 $\pm$ 0.02 & 2.46 $\pm$ 0.01 & 250/229 & 2.43 $\pm$ 0.02 & 0.12 $\pm$ 0.05 & 244/228\\
57840.3234 & 3.37 $\pm$ 0.04 & 1.41 $\pm$ 0.04 & 0.72 $\pm$ 0.02 & 2.38 $\pm$ 0.01 & 268/241 & 2.35 $\pm$ 0.02 & 0.11 $\pm$ 0.05 & 262/240\\
57841.0430 & 5.01 $\pm$ 0.05 & 2.51 $\pm$ 0.06 & 1.29 $\pm$ 0.03 & 2.28 $\pm$ 0.01 & 291/282 & 2.25 $\pm$ 0.02 & 0.11 $\pm$ 0.04 & 282/281\\
57841.8530 & 3.81 $\pm$ 0.07 & 1.86 $\pm$ 0.08 & 1.01 $\pm$ 0.05 & 2.31 $\pm$ 0.02 & 171/189 & 2.30 $\pm$ 0.03 & 0.02 $\pm$ 0.06 & 171/188\\
57843.9154 & 3.61 $\pm$ 0.05 & 1.69 $\pm$ 0.05 & 0.87 $\pm$ 0.03 & 2.32 $\pm$ 0.02 & 227/233 & 2.31 $\pm$ 0.02 & 0.07 $\pm$ 0.05 & 225/232\\
57844.3651 & 3.24 $\pm$ 0.04 & 1.43 $\pm$ 0.04 & 0.74 $\pm$ 0.02 & 2.33 $\pm$ 0.01 & 265/250 & 2.26 $\pm$ 0.02 & 0.24 $\pm$ 0.05 & 233/249\\
57846.0920 & 2.90 $\pm$ 0.04 & 1.18 $\pm$ 0.04 & 0.60 $\pm$ 0.02 & 2.38 $\pm$ 0.01 & 279/236 & 2.33 $\pm$ 0.02 & 0.21 $\pm$ 0.05 & 258/235\\
57848.0853 & 3.23 $\pm$ 0.06 & 1.50 $\pm$ 0.06 & 0.77 $\pm$ 0.03 & 2.30 $\pm$ 0.02 & 237/192 & 2.24 $\pm$ 0.03 & 0.23 $\pm$ 0.06 & 223/191\\
57858.1898 & 8.75 $\pm$ 0.06 & 6.95 $\pm$ 0.12 & 3.68 $\pm$ 0.06 & 2.00 $\pm$ 0.01 & 529/387 & 1.89 $\pm$ 0.01 & 0.28 $\pm$ 0.03 & 419/386\\
57860.2471 & 10.47 $\pm$ 0.06 & 10.22 $\pm$ 0.13 & 5.42 $\pm$ 0.07 & 1.90 $\pm$ 0.01 & 618/475 & 1.84 $\pm$ 0.01 & 0.15 $\pm$ 0.02 & 572/474\\
57861.7240 & 5.75 $\pm$ 0.05 & 3.40 $\pm$ 0.08 & 1.77 $\pm$ 0.04 & 2.18 $\pm$ 0.01 & 328/311 & 2.11 $\pm$ 0.02 & 0.20 $\pm$ 0.03 & 291/310\\
57863.2481 & 6.17 $\pm$ 0.06 & 3.58 $\pm$ 0.07 & 1.87 $\pm$ 0.04 & 2.19 $\pm$ 0.01 & 361/327 & 2.13 $\pm$ 0.02 & 0.18 $\pm$ 0.03 & 327/326\\
57864.8394 & 6.44 $\pm$ 0.05 & 3.80 $\pm$ 0.07 & 1.98 $\pm$ 0.04 & 2.18 $\pm$ 0.01 & 359/346 & 2.13 $\pm$ 0.01 & 0.17 $\pm$ 0.03 & 325/345\\
57866.6973 & 6.38 $\pm$ 0.05 & 4.43 $\pm$ 0.08 & 2.33 $\pm$ 0.05 & 2.09 $\pm$ 0.01 & 368/347 & 2.01 $\pm$ 0.02 & 0.20 $\pm$ 0.03 & 323/346\\
57867.1568 & 5.96 $\pm$ 0.05 & 3.98 $\pm$ 0.08 & 2.09 $\pm$ 0.04 & 2.11 $\pm$ 0.01 & 363/336 & 2.05 $\pm$ 0.02 & 0.17 $\pm$ 0.03 & 334/335\\
57868.3548 & 6.19 $\pm$ 0.05 & 3.94 $\pm$ 0.07 & 2.06 $\pm$ 0.04 & 2.13 $\pm$ 0.01 & 475/366 & 2.04 $\pm$ 0.01 & 0.25 $\pm$ 0.03 & 396/365\\
57869.2978 & 6.18 $\pm$ 0.06 & 3.94 $\pm$ 0.09 & 2.07 $\pm$ 0.05 & 2.14 $\pm$ 0.01 & 342/303 & 2.09 $\pm$ 0.02 & 0.15 $\pm$ 0.04 & 325/302\\
57870.2135 & 5.81 $\pm$ 0.08 & 2.92 $\pm$ 0.10 & 1.51 $\pm$ 0.06 & 2.26 $\pm$ 0.02 & 241/218 & 2.17 $\pm$ 0.03 & 0.29 $\pm$ 0.06 & 211/217\\
57871.3427 & 5.45 $\pm$ 0.05 & 3.12 $\pm$ 0.06 & 1.63 $\pm$ 0.03 & 2.19 $\pm$ 0.01 & 395/322 & 2.12 $\pm$ 0.02 & 0.23 $\pm$ 0.03 & 343/321\\
57872.2050 & 5.80 $\pm$ 0.05 & 3.14 $\pm$ 0.07 & 1.63 $\pm$ 0.04 & 2.23 $\pm$ 0.01 & 345/304 & 2.18 $\pm$ 0.02 & 0.17 $\pm$ 0.04 & 320/303\\
57873.8733 & 6.35 $\pm$ 0.06 & 4.31 $\pm$ 0.08 & 2.26 $\pm$ 0.04 & 2.11 $\pm$ 0.01 & 367/355 & 2.06 $\pm$ 0.01 & 0.15 $\pm$ 0.03 & 340/354\\
57874.0754 & 7.24 $\pm$ 0.06 & 4.83 $\pm$ 0.08 & 2.53 $\pm$ 0.05 & 2.12 $\pm$ 0.01 & 426/365 & 2.06 $\pm$ 0.01 & 0.15 $\pm$ 0.03 & 397/364\\
57880.1794 & 5.74 $\pm$ 0.05 & 2.74 $\pm$ 0.05 & 1.41 $\pm$ 0.03 & 2.30 $\pm$ 0.01 & 374/325 & 2.25 $\pm$ 0.01 & 0.16 $\pm$ 0.03 & 345/324\\
57887.8889 & 6.13 $\pm$ 0.05 & 4.93 $\pm$ 0.08 & 2.59 $\pm$ 0.04 & 2.02 $\pm$ 0.01 & 444/392 & 2.00 $\pm$ 0.01 & 0.06 $\pm$ 0.03 & 439/391\\
57892.2072 & 6.04 $\pm$ 0.05 & 5.24 $\pm$ 0.10 & 2.77 $\pm$ 0.05 & 1.97 $\pm$ 0.01 & 354/341 & 1.91 $\pm$ 0.02 & 0.15 $\pm$ 0.03 & 333/340\\
57894.3911 & 5.85 $\pm$ 0.05 & 4.66 $\pm$ 0.08 & 2.45 $\pm$ 0.04 & 2.02 $\pm$ 0.01 & 395/378 & 1.99 $\pm$ 0.01 & 0.09 $\pm$ 0.03 & 383/377\\
57895.1868 & 5.44 $\pm$ 0.05 & 3.93 $\pm$ 0.08 & 2.07 $\pm$ 0.04 & 2.06 $\pm$ 0.01 & 422/345 & 1.99 $\pm$ 0.02 & 0.20 $\pm$ 0.03 & 377/344\\
57898.1778 & 7.09 $\pm$ 0.05 & 4.94 $\pm$ 0.08 & 2.60 $\pm$ 0.04 & 2.09 $\pm$ 0.01 & 441/376 & 2.03 $\pm$ 0.01 & 0.17 $\pm$ 0.03 & 402/375\\
57901.4399 & 6.82 $\pm$ 0.05 & 6.65 $\pm$ 0.11 & 3.52 $\pm$ 0.05 & 1.90 $\pm$ 0.01 & 489/413 & 1.85 $\pm$ 0.01 & 0.14 $\pm$ 0.03 & 463/412\\
57908.1586 & 3.70 $\pm$ 0.04 & 2.31 $\pm$ 0.05 & 1.20 $\pm$ 0.03 & 2.17 $\pm$ 0.01 & 263/282 & 2.16 $\pm$ 0.02 & 0.04 $\pm$ 0.04 & 261/281\\
57915.6590 & 4.39 $\pm$ 0.05 & 2.13 $\pm$ 0.05 & 1.10 $\pm$ 0.03 & 2.29 $\pm$ 0.01 & 302/274 & 2.25 $\pm$ 0.02 & 0.14 $\pm$ 0.04 & 287/273\\
57922.4419 & 4.27 $\pm$ 0.05 & 2.11 $\pm$ 0.05 & 1.09 $\pm$ 0.03 & 2.28 $\pm$ 0.01 & 297/275 & 2.24 $\pm$ 0.02 & 0.16 $\pm$ 0.04 & 280/274\\
57925.0849 & 4.50 $\pm$ 0.04 & 2.10 $\pm$ 0.05 & 1.08 $\pm$ 0.03 & 2.31 $\pm$ 0.01 & 284/285 & 2.27 $\pm$ 0.02 & 0.15 $\pm$ 0.04 & 267/284\\
57927.3541 & 3.88 $\pm$ 0.04 & 1.24 $\pm$ 0.04 & 0.62 $\pm$ 0.02 & 2.49 $\pm$ 0.01 & 334/241 & 2.43 $\pm$ 0.02 & 0.31 $\pm$ 0.05 & 283/240\\
57928.8789 & 3.73 $\pm$ 0.04 & 1.20 $\pm$ 0.04 & 0.60 $\pm$ 0.02 & 2.49 $\pm$ 0.01 & 302/235 & 2.42 $\pm$ 0.02 & 0.31 $\pm$ 0.05 & 254/234\\
57929.8125 & 3.91 $\pm$ 0.05 & 1.47 $\pm$ 0.04 & 0.75 $\pm$ 0.02 & 2.43 $\pm$ 0.01 & 298/253 & 2.38 $\pm$ 0.02 & 0.20 $\pm$ 0.04 & 274/252\\
57931.6004 & 5.05 $\pm$ 0.04 & 2.96 $\pm$ 0.06 & 1.55 $\pm$ 0.03 & 2.18 $\pm$ 0.01 & 359/323 & 2.13 $\pm$ 0.02 & 0.17 $\pm$ 0.03 & 328/322\\

\hline 
\end{tabular}
\end{table*}
\clearpage

\section{Evaluation of the statistical significance of the UV/optical versus X-ray anti-correlation}
\label{anti_corr_significance}
In this section, we describe the details of the procedure used to assess the statistical significance of the UV/optical versus X-ray anti-correlation.\par 

First of all, the power spectral density (PSD) of the light curves are estimated \citep[e.g.,][]{2014MNRAS.445..437M}. The PSD gives a measure of the strength of the variability as function of the temporal frequency. It is a powerful and widely used tool to characterise the temporal flux behaviour on a large range of time scales. For this study, we adopt the \textit{multiple fragments variance function} (MFVF) from \citet{2011A&A...531A.123K} to estimate the PSD. The MFVF method has the advantage of not relying on any interpolation and re-binning as required in other widely used PSD estimation method \citep[e.g., the PSRESP method described in][]{2002MNRAS.332..231U}. Because our light curves can show strong variability (especially in the X-rays) and have a very irregular sampling as well as large gaps, applying some interpolation or re-binning might introduce additional systematic effects. We assume here as a PSD model a simple power-law shape, i.e. $P(\nu) \propto \nu^{-\beta}$. This simple parametrisation was found to describe well Mrk~421, as well as other blazar light curves on a broad energy range \citep[e.g.,][] {2002MNRAS.332..231U, 2008ApJ...689...79C,  2010ApJ...722..520A, 2015A&A...576A.126A}. $\beta$ usually ranges from 1 (referred to as pink noise) to 2 (referred to as red noise). We carry an estimation of $\beta$ similarly to \citet{2018A&A...620A.185N}. We summarise below the procedure.\par 

Based on the assumed PSD model, light curves are simulated following the method described in \citet{1995A&A...300..707T}. The light curves are simulated on timescales 100 times longer than the observations in order to take into account red-noise leakage. We then apply the exact same sampling patterns as the real observations to include the distortions of the PSD caused by the sampling. Furthermore, distortions due to aliasing effects are included by generating light curves with a time resolution matching the typical exposure time of the data \citep{2002MNRAS.332..231U}. Finally, the simulated light curves are rescaled to match the flux variance of the observations, and Gaussian noise corresponding to the measurement uncertainties is added. The simulated light curves can now be considered as \textit{realistic} light curves.\par 

For a fixed $\beta$, we simulate 5000 \textit{realistic} light curves and the MFVF is computed for each of them. The MFVF is characterised down to a minimal temporal frequency $f_{min}=1/T$ up to a maximal frequency $f_{max}=1/{\Delta t_{0}}$, where $T$ is the total length of the light curve and $\Delta t_{0}$ is the shortest time scale variability probed. In our case, $T$ lies between ${\approx}150$\,days and ${\approx}180$\,days depending on the light curves, while we fix for all light curves $\Delta t_{0}=1$\,day, which is the typical shortest cadence of the observations. We then bin the MFVF in 7 frequency bins. For each frequency bin $f_i$, the MFVF probability density function $p(\beta, f_i)$ is estimated using a Gaussian kernel density estimation. Finally, from the measured light curve, a log likelihood function is computed as follows:

\begin{equation}
    \mathcal{L(\beta)} = \sum_{i=0}^N{ \mathrm{ln}(p(\beta, f_i))}
\end{equation}
where $N$ is the number of frequency bins. $p(\beta, f_i)$ relates to the probability of measuring a MFVF in a frequency bin $f_i$ assuming a power-law index $\beta$ for the PSD. The best-fit index $\beta_{fit}$ maximises $\mathcal{L}$. For this, we scan a range of $\beta$ from 0.7 to 2.1, with 0.05 steps. The resulting best-fit indices for the optical (R-band), \textit{Swift}-UVOT/W2, \textit{Swift}-XRT(0.3-2\,keV) and \textit{Swift}-XRT(2-10\,keV) light curves are summarised in Table~\ref{tab:best_fit_beta}. The uncertainty on $\beta_{fit}$ is estimated by performing a large number of fits on simulated light curves that have a true $\beta$ equal to $\beta_{fit}$. The uncertainties are computed from the 68\% containment around the mean of the distribution of the resulting best-fit indices.\par

\begin{table}[h]
\caption{\label{tab:best_fit_beta}PSD index best-fit $\beta_{fit}$ for each energy band.} 

\centering
\begin{tabular}{ l c c}     
\hline\hline
 Instrument & $\beta_{fit}$ \\  
\hline
\hline
R-band & $1.50 \pm 0.25$ \\
\textit{Swift}-UVOT W2 & $1.45 \pm 0.25$ \\
\textit{Swift}-XRT (0.3-2\,keV) & $1.45 \pm 0.20$ \\
\textit{Swift}-XRT (2-10\,keV) & $1.30 \pm 0.20$ \\
\hline
\end{tabular}
\end{table}

The R-band and UV light curves yield $\beta_{fit}$ values compatible within uncertainties. This is expected given the proximity in energy. \citet{2018A&A...620A.185N} found similar results based on an optical light curve spanning ${\approx}10$ years. In the two X-ray band, $\beta_{fit}$ agree within uncertainties and the values are consistent with what \citet{2015A&A...576A.126A} reported.\par 

The significance of the DCF coefficients between two observed light curves can now be determined. With the same procedure described above, we simulate $2 \times 10^4$ \textit{realistic} and uncorrelated light curves for each energy band assuming the best-fit PSD index $\beta_{fit}$ in the PSD model. The DCF coefficients between the two sets of simulated light curves are computed, and at each time lag the distribution of the DCF coefficients is drawn to extract the $2\sigma$ and $3\sigma$ confidence intervals.\par

In Fig.~\ref{DCF_xray_optical}, the significance bands show a spike at a time lag of ${\sim}30$\,days in the case of the R-band versus X-ray. We found that this feature is caused by the sampling of the two light curves and is not a statistical artefact. Around this time lag, due to the sampling of the two light curves, the DCF is highly dominated by measurements performed on MJD~57757 and MJD~57785 (corresponding to two of the MAGIC/\textit{NuSTAR}/\textit{Swift} simultaneous observations). On those two dates the optical measurements (from GASP-WEBT) have a dense and fine binning on a sub-hour time scale, allowing many pairs contributing to the overall DCF (see Fig.~\ref{NUSTAR_MAGIC_sim_LC_first} and Fig.~\ref{NUSTAR_MAGIC_sim_LC_third}). On such a time scale, the temporal properties of the optical flux lead to an almost constant flux behaviour, but also potentially to an overall decrease or increase of the flux. When correlated to the \textit{Swift}-XRT data, there is therefore a sizable probability of detecting a spurious (anti-)correlation around this date, which would then dominate the total DCF.

\section{Investigation of the temporal behaviour of the UV/optical versus X-ray anti-correlation}
\label{sect:DCF_before_after_57760}
From a visual inspection of Fig.~\ref{MWL}, UV/optical versus X-ray anti-correlation is mainly visible for a period of roughly 40\,days between MJD~57720 to MJD~57760. In this section, we investigate the impact of this time period on the correlated behaviour derived over the entire MWL campaign.\par 

\begin{figure}[h!]
        \centering
        \begin{subfigure}[b]{0.79\columnwidth}
            \centering
            \includegraphics[width=\linewidth]{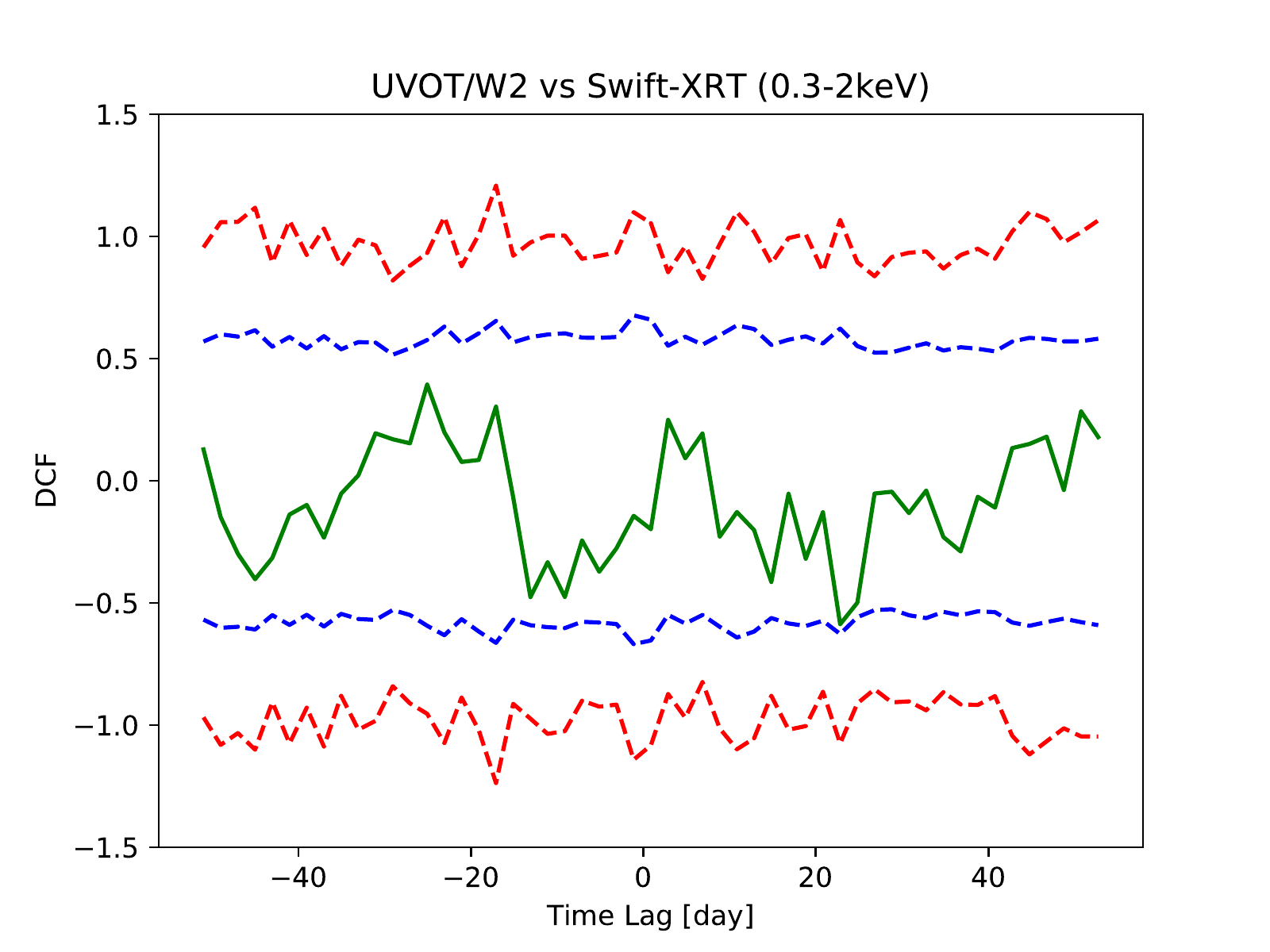}
        \vspace{-0.77cm}
        \end{subfigure}
        \begin{subfigure}[b]{0.79\columnwidth}  
            \centering 
            \includegraphics[width=\linewidth]{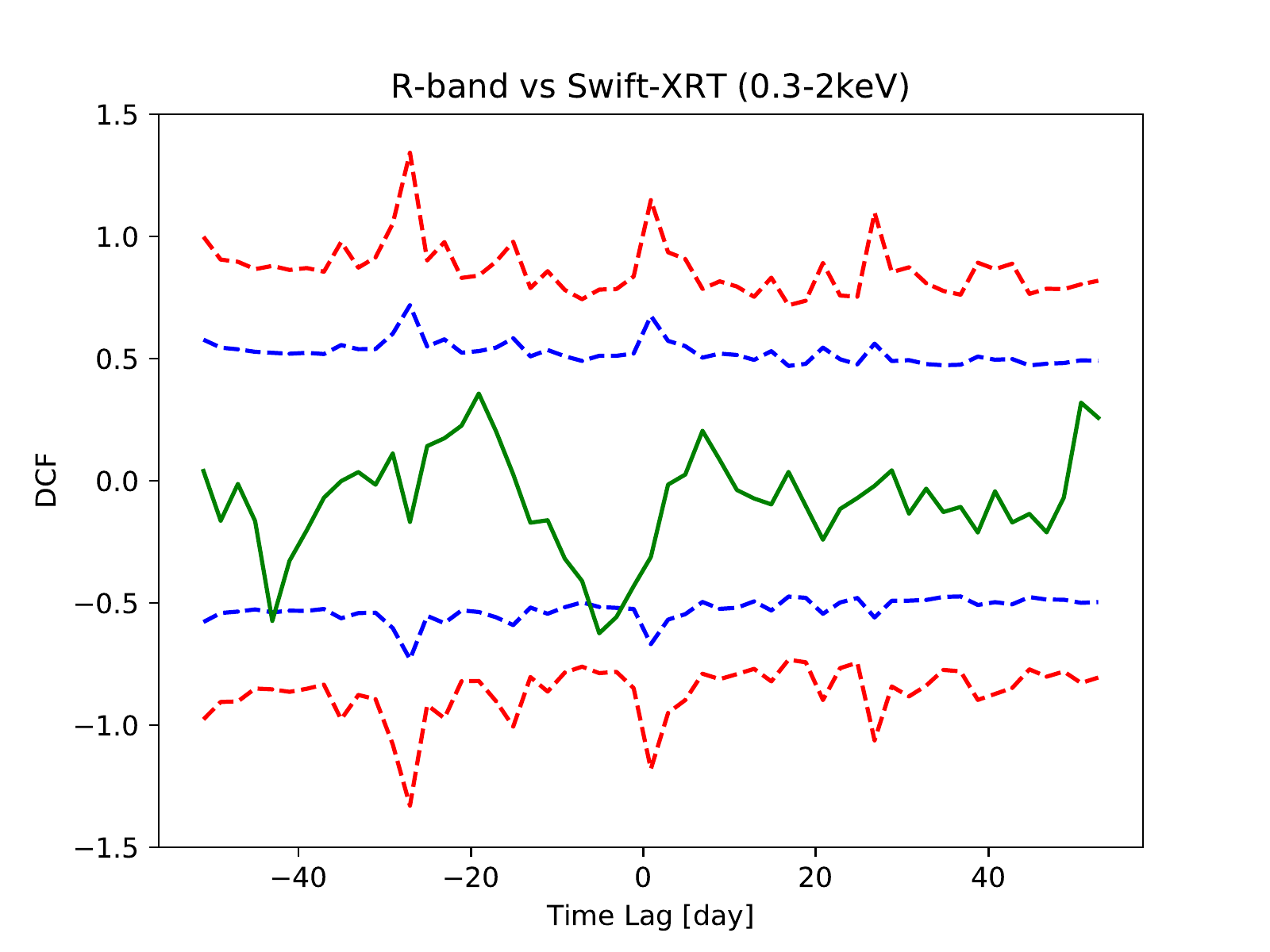}
        \vspace{-0.77cm}
        \end{subfigure}
        \begin{subfigure}[b]{0.79\columnwidth}
            \centering
            \includegraphics[width=\linewidth]{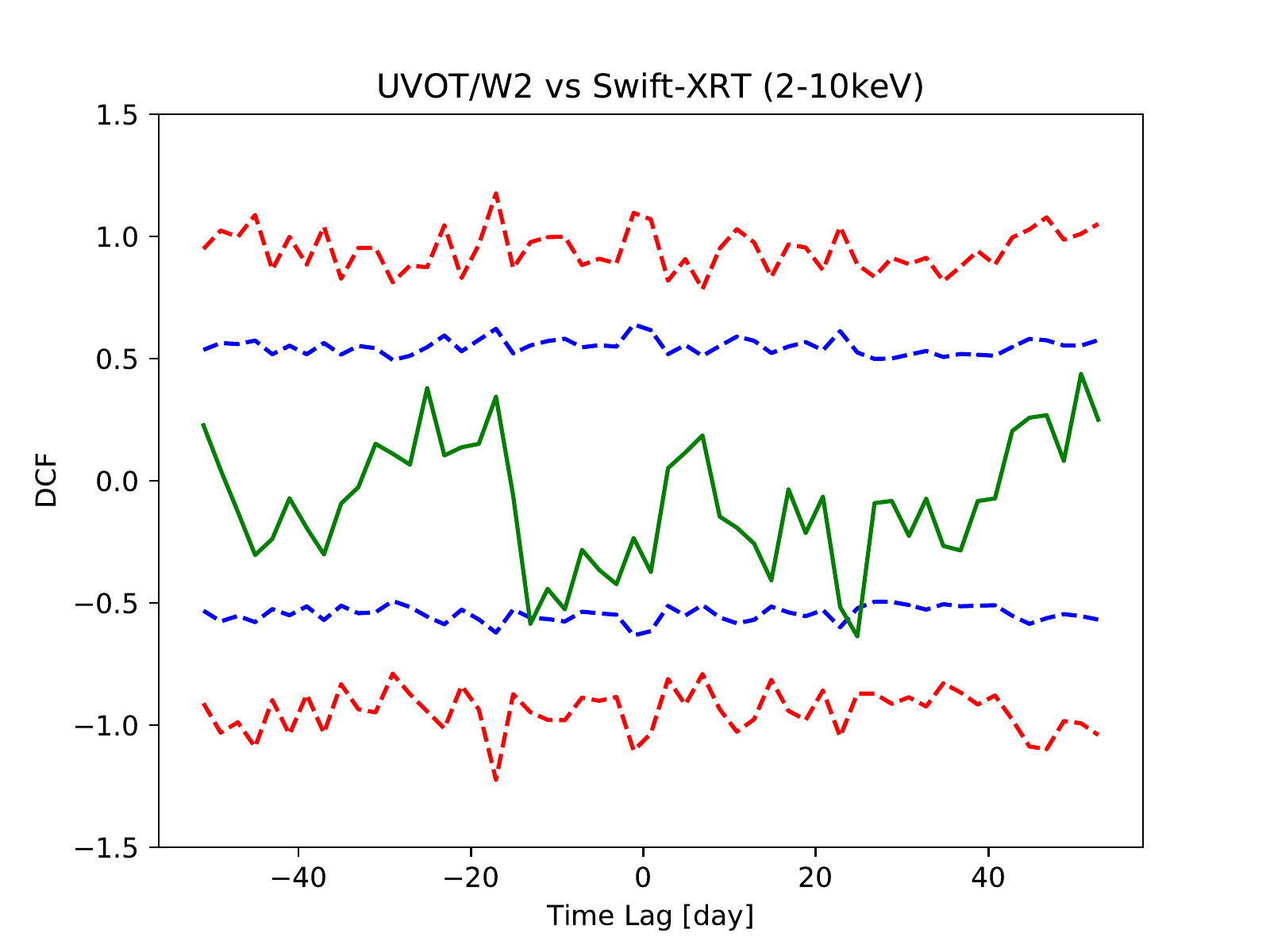}
        \vspace{-0.77cm}
        \end{subfigure}
        \begin{subfigure}[b]{0.79\columnwidth}  
            \centering 
            \includegraphics[width=\linewidth]{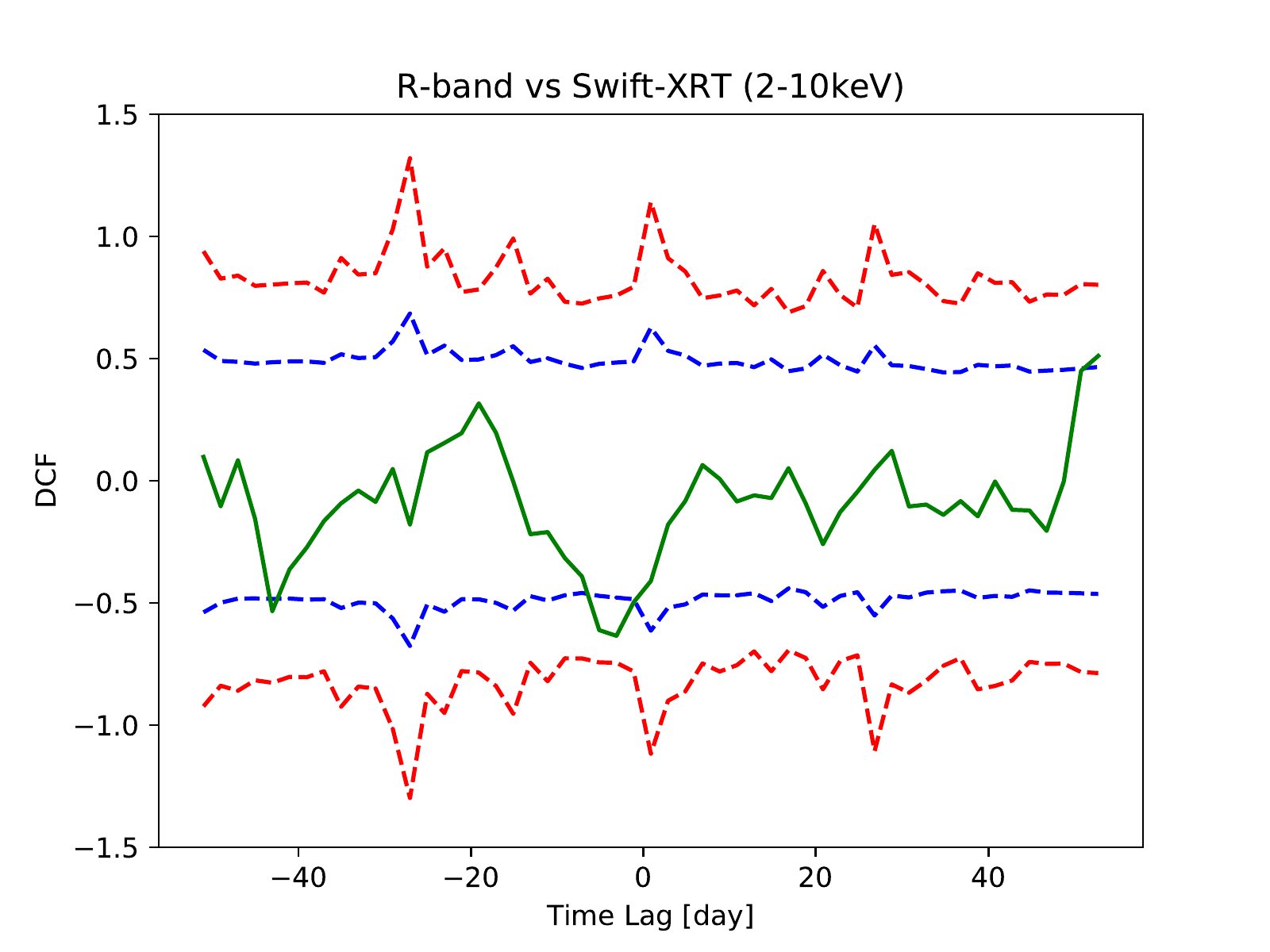}
        \vspace{-0.5cm}
        \end{subfigure}
        \caption{DCF between X-ray (0.3-2\,keV and 2-10\,keV) and UV/optical (\textit{Swift}-UVW2 and R-band) when removing data before MJD~57760. The blue- and red-dashed-lines indicate the $2\sigma$ and $3\sigma$ confidence intervals estimated from the Monte Carlo simulations, as described in the text.}  
\label{DCF_xray_optical_after_57760}
\end{figure}

\begin{figure}[h!]
        \centering
        \begin{subfigure}[b]{0.79\columnwidth}  
            \centering
            \includegraphics[width=\linewidth]{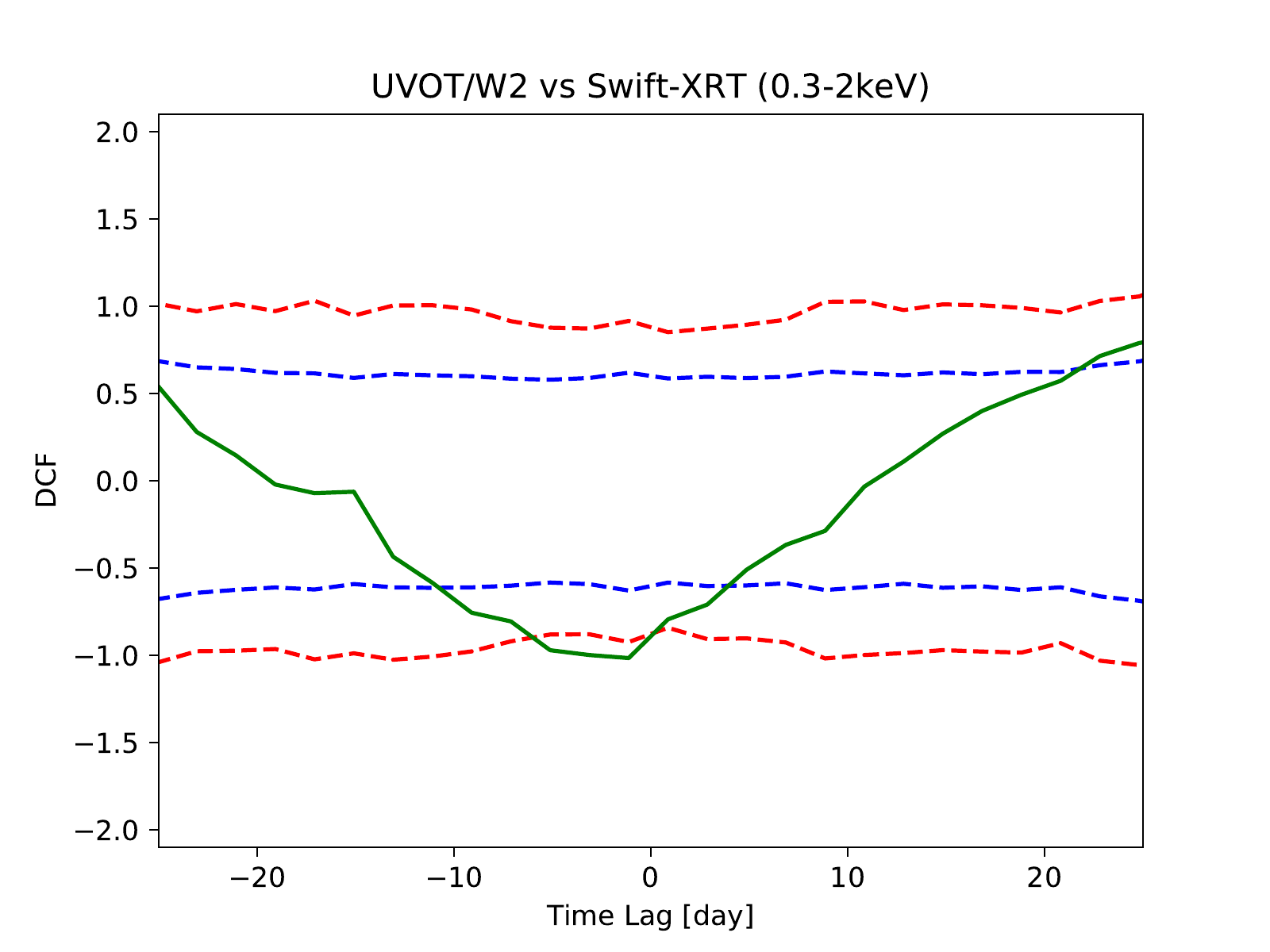}
        \vspace{-0.77cm}
        \end{subfigure}
        \begin{subfigure}[b]{0.79\columnwidth}  
            \centering 
            \includegraphics[width=\linewidth]{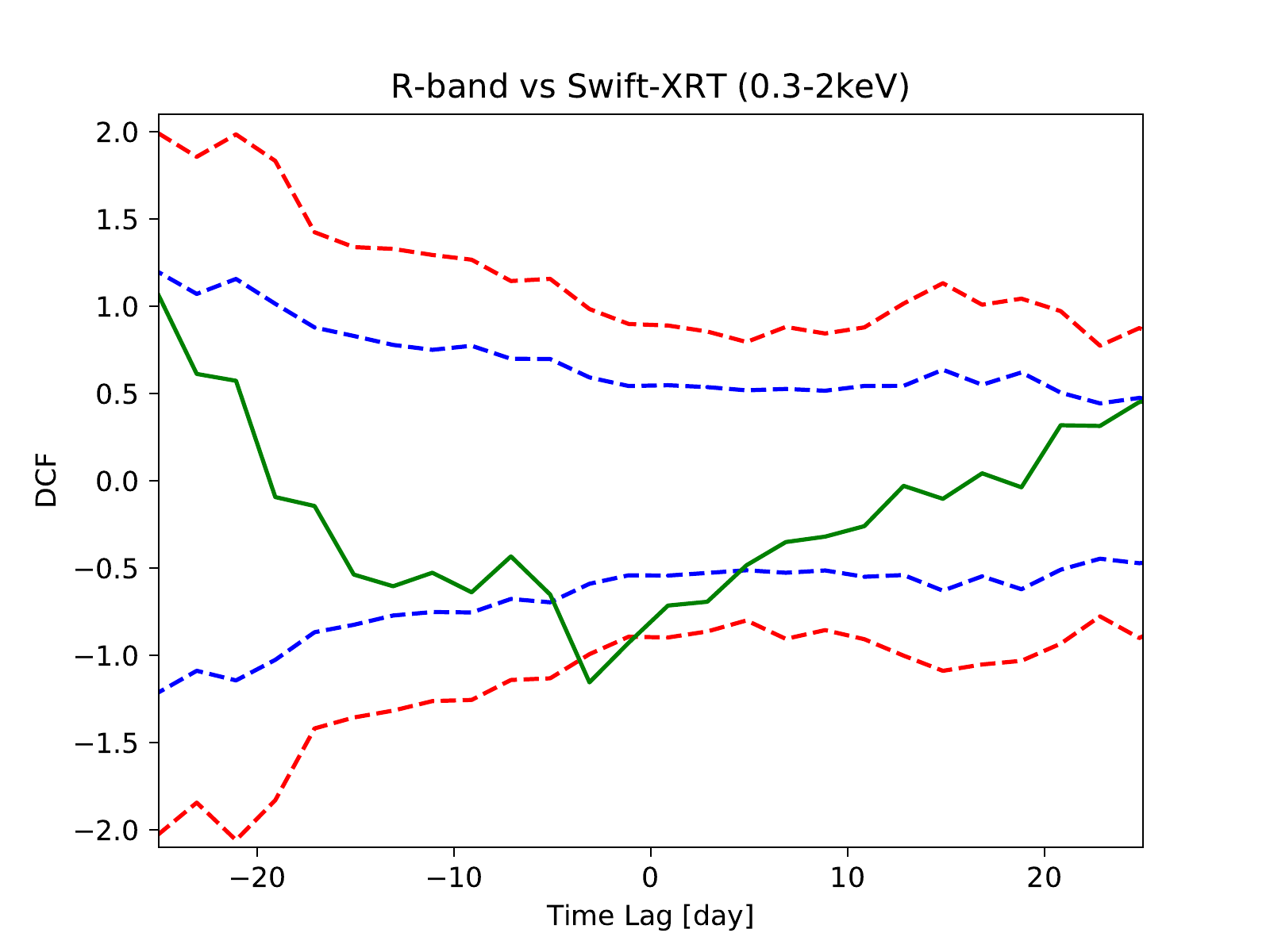}
            \vspace{-0.77cm}
        \end{subfigure}
        \begin{subfigure}[b]{0.79\columnwidth}  
            \centering 
            \includegraphics[width=\linewidth]{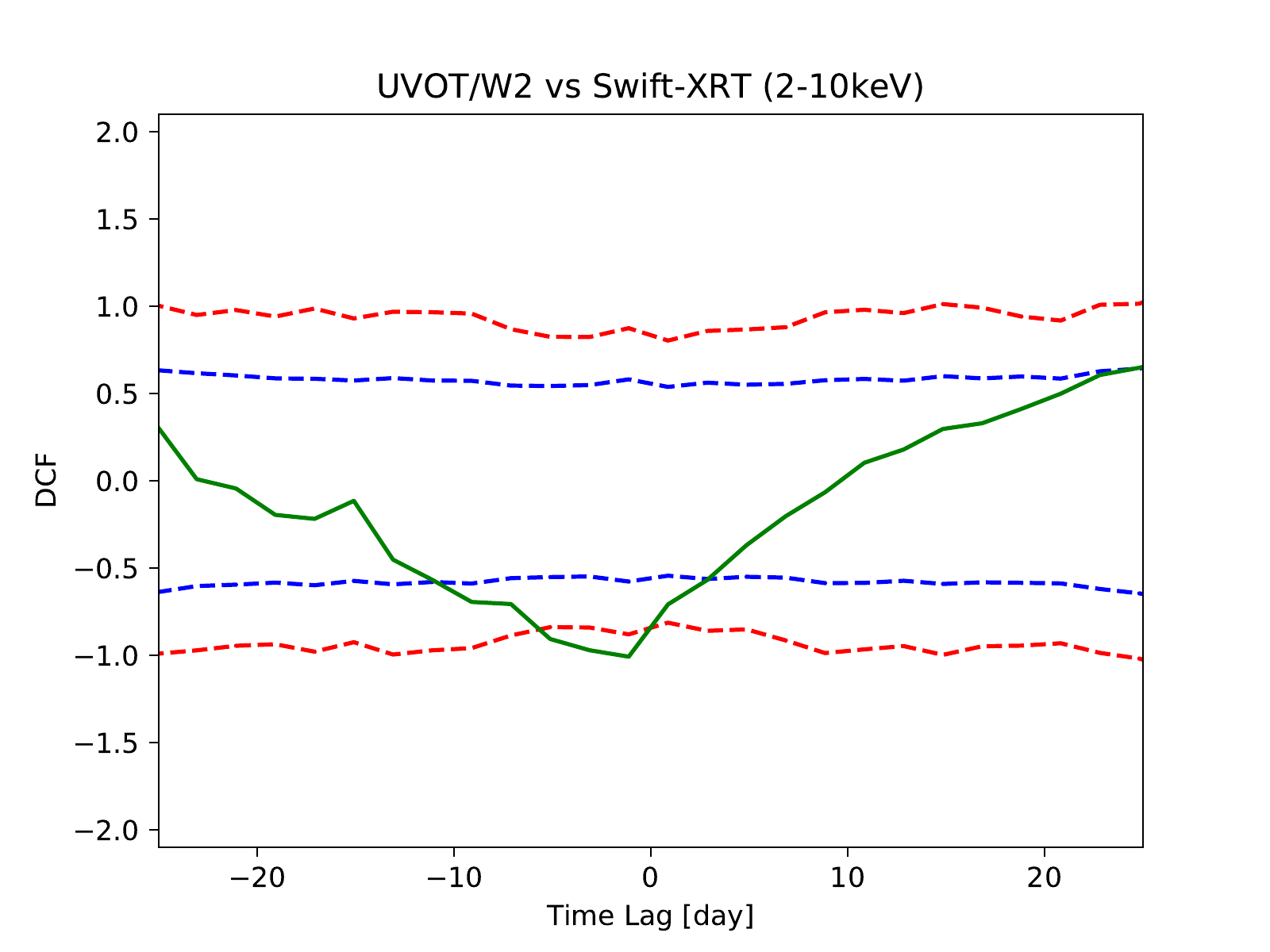}
        \vspace{-0.77cm}
        \end{subfigure}
        \begin{subfigure}[b]{0.79\columnwidth}  
            \centering 
            \includegraphics[width=\linewidth]{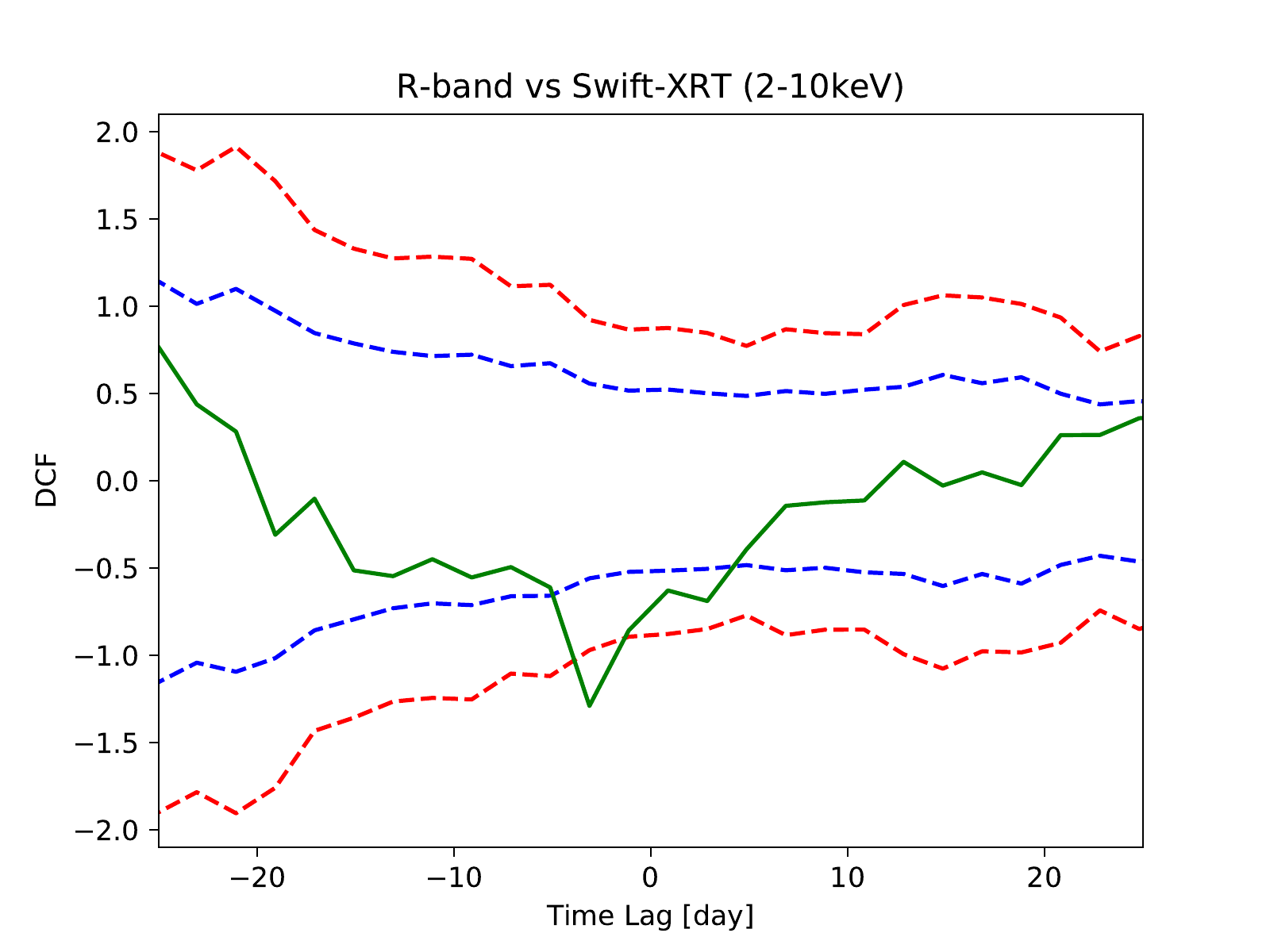}
        \end{subfigure}
        \caption{Same description as for Fig.~\ref{DCF_xray_optical_after_57760} but this time data after MJD~57760 are removed.}  
\label{DCF_xray_optical_before_57760}
\end{figure}

We repeat the study described in Sect.~\ref{sect:synch_peak} and Appendix~\ref{sect:DCF_before_after_57760} and compute the DCF significance by ignoring data before MJD~57760. The results are shown in Fig.~\ref{DCF_xray_optical_after_57760}. While a hint at the level of $2-3\sigma$ remains in R-band versus X-rays, the significance is below $2\sigma$ in the UV versus X-rays. We repeat again the exercise, but this time ignoring data after MJD~57760. The results are shown in Fig~\ref{DCF_xray_optical_before_57760}. For this time period, the significance is always above $3\sigma$, and the DCF value at the peak is also higher.\par 
\vspace{-0.4cm}

\end{appendix}

\end{document}